\documentclass[11pt]{article} 

\usepackage[a4paper, top=0.95in, bottom=1in, right=0.85in, left=0.85in]{geometry} 

\usepackage{setspace}
\usepackage{titletoc}

\usepackage[T1]{fontenc}       
\usepackage[utf8]{inputenc}    
\usepackage{lmodern}           
\usepackage{microtype}         

\usepackage{amsmath, amssymb, amsthm, amsfonts} 
\usepackage{mathtools}                         
\usepackage{bbm}                               
\usepackage{bm}                                
\usepackage{commath}                           
\usepackage{nicefrac}                          
\usepackage{xfrac}                             
\usepackage{dsfont}                            
\usepackage{mathrsfs}                          
\usepackage{verbatim}                          

\usepackage{mathrsfs}        
\DeclareMathAlphabet{\mathscr}{OMS}{rsfs}{m}{n}  

\theoremstyle{plain}
\newtheorem{theorem}{Theorem}[section]

\newtheorem{corollary}[theorem]{Corollary}

\theoremstyle{definition}

\newtheorem{remark}[theorem]{Remark}

\usepackage{graphicx}         
\usepackage{caption}          
\usepackage{subcaption}       
\usepackage{booktabs}         
\usepackage{multirow}         
\usepackage{array}            
\usepackage{float}            
\usepackage{wrapfig}          
\usepackage{siunitx}
\usepackage[normalem]{ulem}
\usepackage{arydshln}         

\usepackage[authoryear]{natbib}  
\usepackage{hyperref}         
\hypersetup{
    colorlinks=true,          
    linkcolor=blue,           
    citecolor=blue,           
    urlcolor=blue             
}

\usepackage{etoolbox}

\usepackage{tikz}             
\usetikzlibrary{
    decorations.pathmorphing, 
    positioning,              
    arrows,                   
    decorations.pathreplacing,
    calc                     
}

\usepackage{pgfplots}
\usepgfplotslibrary{fillbetween} 
\pgfplotsset{compat=1.17} 

\usepackage{algorithm}        
\usepackage{algpseudocode}    

\usepackage{colortbl}      
\usepackage{hhline}        
\usepackage{calc}          
\usepackage{tabularx}      
\usepackage{comment}       

\usepackage{xcolor}           
\usepackage{enumitem}         
\usepackage{xspace}           
\usepackage{fancyvrb}         
\usepackage{tcolorbox}        

\usepackage{changepage}       

\usepackage{todonotes}        

\newcommand{\diff}{{\mathrm d}}       
 
\newcommand{\expit}{\mathrm{expit}}  
\newcommand{\binomial}{\mathrm{Binomial}}  
\newcommand{\uniform}{\mathrm{Uniform}}  

\newcommand{\E}{\mathbb{E}}           
\newcommand{\I}{\mathbb{I}}
\newcommand{\R}{\mathbb{R}}
\newcommand{\N}{\mathcal{N}}

\newcommand{\M}{\mathcal{M}}
\newcommand{\calQ}{\mathcal{Q}}

\def\p{\mathrm{p}}  
\def\P{\mathrm{P}}  
\def\Pn{\mathrm{P}_{\! \mathrm{n}} } 
\def\Pnk{\mathrm{P}_{\! \mathrm{n}, \mathrm{k}} } 
\def\Rem{\mathrm{R}_{2} }
\def\Q{\mathrm{Q}}  
\def\q{\mathrm{q}} 
\def\sp{\ \! } 

\DeclareMathOperator{\argmin}{argmin}
\DeclareMathOperator{\argmax}{argmax}

\newcommand{\smallO}{o_\P}

\DeclareMathOperator{\pa}{pa}

\newcommand{\red}{\textcolor{red}}
\newcommand{\blue}{\textcolor{blue}}

\allowdisplaybreaks        

\makeatletter
\newcommand{\pushright}[1]{\ifmeasuring@#1\else\omit\hfill$\displaystyle#1$\fi\ignorespaces}
\newcommand{\pushleft}[1]{\ifmeasuring@#1\else\omit$\displaystyle#1$\hfill\fi\ignorespaces}
\makeatother

\newcommand{\thickline}{\noalign{\hrule height 1pt} } 

\title{Flexible Nonparametric Inference for Causal Effects under the \\[-0.8em] Front-Door Model}

\author{
  Anna Guo, David Benkeser, Razieh Nabi \\[0.25em]
  {Department of Biostatistics and Bioinformatics, Emory University, Atlanta, GA, USA} \\[0.0em]
    {$^{*}$Corresponding author: \texttt{razieh.nabi@emory.edu}}
}

\date{}

\usepackage{eso-pic}

\newcommand{\publishedversion}{%
\AddToShipoutPictureFG*{%
\AtPageUpperLeft{%
\raisebox{-0.55in}[0pt][0pt]{%
\hspace{0.85in}%
\begin{minipage}[t]{0.75\textwidth}
\small
\setstretch{1.0}
\textit{Journal of the Royal Statistical Society Series B: Statistical Methodology} \\ {Published: Sept 2026}\\
\end{minipage}%
}%
}%
}%
}

\begin{document}

\publishedversion

\maketitle

\abstract{Evaluating causal treatment effects in observational studies requires addressing confounding. While the \textit{back-door} criterion enables identification through adjustment for observed covariates, it fails in the presence of unmeasured confounding. The \textit{front-door} criterion offers an alternative by leveraging variables that fully mediate the treatment effect and are unaffected by unmeasured confounders of the treatment-outcome pair. We develop novel one-step and targeted minimum loss-based estimators for both the \textit{average treatment effect} and the \textit{average treatment effect on the treated} under front-door assumptions. Our estimators are built on multiple parameterizations of the observed data distribution, including approaches that avoid modeling the mediator density entirely, and are compatible with flexible, machine learning-based nuisance estimation. We establish conditions for root-$n$ consistency and asymptotic linearity by deriving second-order remainder bounds. We also develop flexible tests for assessing identification assumptions, including a doubly robust testing procedure, within a semiparametric extension of the front-door model that encodes \textit{generalized (Verma) independence} constraints. We further show how these constraints can be leveraged to improve the efficiency of causal effect estimators. Simulation studies confirm favorable finite-sample performance, and real-data applications in education and emergency medicine illustrate the practical utility of our methods. 
}


\section{Introduction} 
\label{sec:intro}

Two key causal parameters are the average treatment effect (ATE), capturing the population-level effect, and the average treatment effect on the treated (ATT), capturing the effect among treated units. When all confounders are observed, these effects are typically identified using the \textit{back-door} criterion, which adjusts for covariates that block all non-causal paths between the treatment and outcome \citep{pearl09causality}. Under this criterion, the ATE is identified via the g-formula \citep{robins86new, hahn1998role} and/or inverse probability of treatment weighting (IPTW)  \citep{hirano2003efficient}. A rich literature exists for estimating these functionals using plug-in, IPTW, augmented IPTW, and targeted minimum loss-based estimators (TMLEs)  \citep{bickel1993efficient, van2000asymptotic, tsiatis2007semiparametric, robins1994estimation, van2011targeted, double17chernozhukov}. 

Identifying a sufficient back-door adjustment set is not always feasible due to unmeasured confounding. Alternative strategies include instrumental variable methods \citep{balke94counter}, sensitivity analyses \citep{robins2000sensitivity, nabi2024semiparametric}, and bounds analysis \citep{manski1990nonparametric}. Other approaches include those based on causal graphical models that enable reasoning about identification using independence constraints between counterfactual and observed variables \citep{tian02general, thomas13swig}. These models underlie \textit{sound} and \textit{complete} algorithms for identifying causal parameters from observed data \citep{shpitser06id, huang06do, bhattacharya2022semiparametric, richardson2023nested}.

The \textit{front-door} criterion enables causal identification in the presence of unmeasured confounding \citep{pearl1995causal}. It requires one or more mediators satisfying two conditions: (i) no unmeasured confounding between the treatment and mediators nor between mediators and outcome, and (ii) the effect of treatment on the outcome is fully mediated through the mediators. When these hold, average causal effects are identifiable from observed data. When the full mediation assumption (ii) is violated, \citet{fulcher19robust} proposed the \textit{population intervention indirect effect}, which relaxes this assumption by introducing an additional \textit{cross-world counterfactual} independence. Although the estimand differs, the underlying identification strategy remains closely related. Empirical studies suggest the front-door criterion can yield reliable estimates in real-world settings where unmeasured confounding is expected \citep{glynn2018front, bellemare2019paper, fulcher19robust, bhattacharya2022testability, piccininni2023effect, wen2024causal}. 

A nonparametric efficient estimator of the front-door functional was proposed by \citet{fulcher19robust}, who developed a \textit{one-step estimator} based on parametric working models for key nuisance components: the outcome regression, mediator conditional density, and the treatment conditional probability. This estimator is \textit{doubly robust} and marked an important contribution to front-door estimation. However, several gaps in its applicability remain. 
First, the estimator is implemented using parametric working models. While \citet{fulcher19robust} note that the efficient influence function can, in principle, be combined with flexible, data-adaptive methods and briefly mention rate conditions (e.g., $o_P(n^{-1/4})$ convergence of nuisance estimators), they do not develop estimation strategies tailored to such settings or provide general conditions for valid inference with flexible machine learning approaches. 
Second, the approach is functionally restricted to settings with a single mediator, as it requires estimation of the mediator density. Yet in practice, multiple mediators often arise—whether to satisfy identification assumptions under full mediation or to capture complex indirect pathways under partial mediation—making density estimation impractical.  
Third, the estimator can produce estimates outside the natural parameter space, which is problematic for binary or bounded continuous outcomes. 
Recent work by \citet{wen2024causal} addresses some of these issues by introducing TMLE-based estimators for a related target parameter, using a reparameterization that avoids direct modeling of the mediator density. Their approach improves practical feasibility, particularly with continuous mediators. However, their estimand differs from the standard ATE front-door functional considered here, and it remains unclear how to incorporate flexible nuisance estimation into that framework.

The front-door criterion enables identification under unmeasured treatment-outcome confounding, assuming no unmeasured confounding between treatment and mediator(s), between mediator(s) and outcome, and no direct effect of treatment on the outcome. These assumptions are not testable in a nonparametrically saturated model. \citet{bhattacharya2022testability} described the use of an auxiliary \textit{anchor} variable, a baseline covariate associated with treatment (and possibly mediator) but not a direct cause of the outcome, to assess the front-door assumptions. The presence of such an anchor induces a testable \textit{Verma constraint}—a generalized independence relation in the observed data distribution \citep{verma1990equivalence}—that encodes the absence of a direct effect of the anchor variable on the outcome. Parametric tests for this constraint have been proposed, but they rely on strong modeling assumptions and are limited in flexibility. 

This work extends the foundational contributions of \citet{fulcher19robust, wen2024causal}, and \citet{bhattacharya2022testability} in several respects. 
First, we propose a suite of robust and efficient estimators for the ATE front-door functional based on three parameterizations of the observed data distribution that enable scalable inference with multivariate mediators of mixed types (Section~\ref{sec:est_ATE}). 
Second, we develop efficient estimators of the ATT by deriving and leveraging its efficient influence function, complementing the proposal of \citet{fulcher19robust} who estimate the ATT by rescaling their PIIE estimator (Section~4). 
Third, we derive second-order remainder terms for all ATE and ATT estimators and establish conditions for root-$n$ consistency and asymptotic linearity under flexible, data-adaptive nuisance estimation (Section~\ref{sec:asymptotic}). Characterizing these remainder terms lays the foundation for additional work in increasing the robustness of confidence interval and hypothesis test construction \citep{van2014targeted,benkeser2017doubly}. 
Fourth, we evaluate the validity of the front-door model with an anchor variable by developing flexible tests based on weighted risk minimization, along with a novel doubly robust testing procedure (Sections~\ref{subsec:weighted_risk_min} and \ref{subsec:dr_CATE_test}). 
We further show how the Verma constraint can be exploited to improve efficiency of causal effect estimators (Section~\ref{subsec:verma}).
Finally, we demonstrate the practical utility of our methods through simulation studies (Section~\ref{sec:sims}) and two real-world applications: one analyzing the effect of early academic performance on later income and another evaluating the impact of mobile stroke unit deployment on clinical outcomes in emergency medicine (Section~\ref{sec:real_data}). 


\section{Causal front-door model} 
\label{sec:prelim}

Let $A$ denote a binary treatment, with $A=1$ indicating treatment and $A=0$ control, and $Y$ denote the observed outcome. Let $Y^a$ denote the potential outcome under treatment level $a \in \{0, 1\}$ \citep{neyman23app, rubin74potential}. We write $\P$ for distributions and $\p$ for densities, assuming continuous variables admit Lebesgue densities (though this is not required). The ATE and ATT are defined as $\text{ATE} \coloneqq \E(Y^1 - Y^0)$ and $\text{ATT} \coloneqq \E(Y^1 - Y^0 \! \mid \! A=1)$, where $\E(Y^a) = \int \! y \sp \p(Y^a =y) \sp \diff y$ and $\E(Y^a \! \mid \! A = 1) = \int \! y \sp \p(Y^a =y \! \mid \! A = 1) \sp \diff y$.

Common identification approaches assume: (i) \textit{consistency} which states that $Y = A Y^1 + (1 - A) Y^0$; (ii) \textit{conditional ignorability} which assumes the existence of a set of observed pre-treatment covariates $X$ such that $Y^a \perp A \! \mid \! X$, for $a \in \{0, 1\}$; and (iii) \textit{positivity} which ensures that $\p(A=a \! \mid \! x) >0$ for $a \in \{0, 1\}$ and all $x$ in the support of $X$. Under assumptions (i)-(iii), the ATE and ATT are both identified via the \textit{back-door adjustment formulae} $\E\big(\E(Y \! \mid \! A = 1, X) - \E(Y \! \mid \! A = 0, X)\big)$ and $\E\big(\E(Y \! \mid \! A = 1, X) - \E(Y \! \mid \! A = 0, X) \! \mid \! A  = 1 \big)$, respectively. The ATT identification requires a weaker form of positivity: $\p(A = 0 \mid x) > 0$ for all $x$ with $\p(A = 1 \mid x) > 0$. This causal model corresponds to the DAG in Fig.~\ref{fig:graphs}(a) (without $A \leftarrow U \rightarrow Y$).

Various methods have been developed to infer the back-door adjustment formulae from observed data, including propensity score matching \citep{rosenbaum83propensity}, g-computation \citep{robins86new}, (stabilized) IPTW \citep{hernan2006estimating}, augmented IPTW \citep{robins1994estimation}, and TMLE \citep{van2006targeted}. However, in the presence of unmeasured confounders ($U$ in Fig.~\ref{fig:graphs}(a)), the ATE and ATT are no longer identifiable, and any inference based on the back-door adjustment formulae are likely to be biased.

As an alternative to the back-door, Pearl proposed the front-door model \citep{pearl1995causal}, which enables causal identification even in the presence of unmeasured confounders between treatment $A$ and outcome $Y$. It relies on mediators $M$ that intercept all directed paths from $A$ to $Y$ and share no unmeasured confounders with either. These conditions correspond to the absence of dashed gray edges in Fig.~\ref{fig:graphs}(b), where $U_{AM}$ and $U_{MY}$ denote unmeasured confounding between treatment-mediator and mediator-outcome, respectively. We consider a generalized front-door model that additionally allows observed common causes $X$ of $A$, $M$, and $Y$ (Fig.~\ref{fig:graphs}(c)).

\subsection{Identification of the ATE and ATT}

The identification assumptions for ATE in the front-door model based on observations of $O = (X, A, M, Y) \sim \P$ are: (i) \textit{consistency},  $M^a = M$ when $A = a$ and $Y^m = Y$ when $M = m$; (ii) \textit{conditional ignorability} which assumes the absence of unmeasured confounders between the treatment-mediator and mediator-outcome pairs, i.e., $M^a \perp A \! \mid \! X$ and $Y^m \perp M \! \mid \! A, X$; (iii) \textit{no direct effect} which assumes that $M$ intercepts all directed paths from $A$ to $Y$, i.e., $Y^{a, m} = Y^m$ for $a \in \{0, 1\}$ and all $m$ in the support of $M$; and (iv) \textit{positivity}, $\p(A = a \! \mid \! X=x)$ and $\p(M=m \! \mid \! A=a, X=x)$ are positive for all $(x, a, m)$ in the support of $(X, A, M)$. We denote by $\mathcal{M}$ the nonparametric model for $\P$, subject to the positivity conditions in (iv).

\begin{figure}[t] 
	\begin{center}
    \scalebox{0.7}{
    \begin{tikzpicture}[>=stealth, node distance=1.7cm]
        \tikzstyle{format} = [thick, circle, minimum size=1.0mm, inner sep=2pt]
        \tikzstyle{square} = [draw, thick, minimum size=4.5mm, inner sep=2pt]
    
    \begin{scope}[xshift=0cm, yshift=0cm]
		\path[->, thick]
		
		node[] (a) {$A$}
   		node[above right of=a, xshift=0cm, yshift=-0.25cm] (u) {\red{$U$}}
		node[right of=u, xshift=0.25cm, yshift=0cm] (x) {$X$}
		node[below right of=x, xshift=0cm, yshift=0.25cm] (y) {$Y$}
		
		(u) edge[blue] (a) 
		(u) edge[blue] (y) 
		(x) edge[blue] (a) 
		(x) edge[blue] (y) 
		(a) edge[blue] (y)

        node[below of=a, xshift=2.cm, yshift=0.85cm] (t1) {(a)} ;
		
	\end{scope}
   
	\begin{scope}[xshift=7.25cm, yshift=0cm]
		\path[->, thick]
		
		node[] (a) {$A$}
		node[right of=a, xshift=0.75cm] (m) {$M$}
		node[above left of=a, xshift=0.cm, yshift=-0.25cm] (u0) {\red{$U_{AM}$}}
		node[above of=m, yshift=-0.75cm] (u) {\red{$U$}}
		node[right of=m, xshift=0.75cm] (y) {$Y$}
		node[above right of=y, xshift=0.cm, yshift=-0.25cm] (u1) {\red{$U_{MY}$}}
		
		(u0) edge[gray, dashed] (a) 
		(u0) edge[gray, dashed] (m) 
		(u1) edge[gray, dashed] (m) 
		(u1) edge[gray, dashed] (y) 
		(a) edge[gray, dashed, bend right=15] (y) 
		(a) edge[blue] (m) 
		(m) edge[blue] (y) 
		(u) edge[blue] (y) 
		(u) edge[blue] (a)

        node[below of=a, xshift=2.5cm, yshift=0.85cm] (t2) {(b)} ;
		
	\end{scope}
				
	\begin{scope}[xshift=15.cm, yshift=0cm]
		\path[->, thick]
		
		node[] (a) {$A$}
		node[right of=a, xshift=0.75cm] (m) {$M$}
		node[above right of=m, yshift=-0.25cm] (x) {$X$}
		node[above left of=m, yshift=-0.25cm] (u) {\red{$U$}}
		node[right of=m, xshift=0.75cm] (y) {$Y$}
		
		(x) edge[blue, bend left=0] (a) 
		(x) edge[blue] (m) 
		(x) edge[blue, bend right=0] (y) 
		(a) edge[blue] (m) 
		(m) edge[blue] (y) 
		(u) edge[blue] (y) 
		(u) edge[blue] (a)

        node[below of=a, xshift=2.5cm, yshift=0.85cm] (t3) {(c)} ; 
		
	\end{scope}
	\end{tikzpicture}
	}
	\caption{(a) Example of a DAG with measured confounders $X$ and unmeasured confounders $U$; (b) The front-door DAG with unmeasured confounders $U$ between $A$ and $Y$ (dashed edges indicate assumptions); (c) The front-door DAG with the inclusion of measured confounders $X$.} 
	\label{fig:graphs}
	\end{center}
\end{figure}

Given that identification arguments and estimation techniques for $\E(Y^1)$ and $\E(Y^0)$ are similar, we explicitly consider $\E(Y^{a_0})$,  $a_0 \in \{0, 1\}$ to be the parameter of interest when studying the ATE. Under assumptions (i)-(iv), $\E(Y^{a_0})$ is identified by $\psi_{a_0}(\P)$ \citep{pearl1995causal}, where
\vspace{-0.25cm}
\begin{align}
    \psi_{a_0}(\P) \! = \! \iiint \! \sum_{a=0}^1 \sp y \sp \p(y \! \mid \! m, a, x) \sp \p(a \! \mid \! x) \sp \p(m \! \mid \! A=a_0, x) \sp  \p(x) \sp \diff y \sp \diff m \sp \diff x \sp .  
     \label{eq:id_ATE} 
\end{align}

\vspace{-0.35cm}
Under the same assumptions, the ATT can also be expressed as a functional of $\P$. To enable a formulation that naturally extends to the \textit{average treatment effect among controls} (ATC), defined as $\E(Y^1 - Y^0 \! \mid \! A = 0)$, we consider the general counterfactual quantity $\E(Y^{a_0} \! \mid \! A = 1-a_0)$, for $a_0 \in \{0,1\}$. Since $\E(Y^{a_0} \! \mid \! A = a_0)$ is identified by consistency as $\E(Y \! \mid \! A = a_0)$ and can be directly estimated via the subpopulation sample mean, we focus on the nontrivial term $\E(Y^{a_0} \! \mid \! A = 1-a_0)$, which is identified under the front-door model by the functional: 
\vspace{-0.25cm}
\begin{align}
    \beta_{a_0}(\P) \! = \! \iiint \! y \sp \p(y \! \mid \! m, A \! = \! 1-a_0, x) \sp \p(m \! \mid \! A \! = \! a_0,x) \sp \p(x \! \mid \! A \! = \! 1-a_0) \sp \diff y \sp \diff m \sp \diff x \sp . 
    \label{eq:id_ATT}
\end{align}%

\vspace{-0.35cm}
\noindent We note that above identification requires a weaker form of positivity: $\p(M = m \! \mid \! A=a, X=x) > 0$ for $a \in \{0, 1\}$, all $m$ in the support of $M$, and all $x$ such that $\p(X=x \! \mid \! A=1-a_0) > 0$.  

We adopt the terminology of ATE and ATT front-door functionals to refer to $\psi_{a_0}$ and $\beta_{a_0}$, respectively, with the understanding that these represent counterfactual means rather than effect contrasts. Under these formulations, the ATE, ATT, and ATC are identified as $\psi_{1}(\P) - \psi_{0}(\P)$, $\E(Y \! \mid \! A = 1) - \beta_{0}(\P)$, and $\beta_{1}(\P) - \E(Y \! \mid \! A = 0)$, respectively (see Appendix~\ref{app:model_id_proof} for proof).

\textbf{Alternative interpretations of the front-door functionals:} The ATE front-door functional in \eqref{eq:id_ATE} admits multiple, closely related causal interpretations beyond the full-mediation setting. In particular, it coincides with the \textit{population intervention indirect effect} (PIIE) of \cite{fulcher19robust}, defined as $\E(Y - Y^{A,M^{a_0}})$ and identifiable under a cross-world independence assumption rather than the no-direct-effect assumption. The ATT front-door functional in \eqref{eq:id_ATT} similarly recovers subgroup-specific PIIEs (among treated or controls); see Appendix~\ref{app:alternative_interpretations} for details.  \cite{wen2024causal} further regards the same functional as the \textit{average causal effect of an intervenable treatment component} $A_M$, namely $\E(Y^{a_M=1}-Y^{a_M=0})$, which is identified by the front-door formula even when $A$ itself is not manipulable. Thus, our estimators for both ATE and ATT continue to estimate meaningful indirect effects when the full mediation assumption is relaxed or when focusing on modifiable treatment components; see Appendix~\ref{app:alternative_interpretations} for details.

Our primary objective is to develop estimators for the front-door functionals in \eqref{eq:id_ATE} and \eqref{eq:id_ATT}, using $n$ i.i.d. observations of $O = (X, A, M, Y)$. We begin by reviewing existing estimation strategies for the ATE front-door functional and highlighting their limitations. In contrast, estimation results for the ATT front-door functional have received little to no prior attention.  

\subsection{Prior estimation for the ATE front‐door functional}
\label{subsec:one‐step}

Let $\Q=(\mu,\pi,f_M,\p_X)$ denote the collection of \textit{nuisance parameters}, where $\mu(m,a,x)=\E(Y \! \mid \! M=m,A=a,X=x)$, $\pi(a \! \mid \! x)=\P(A=a \! \mid  \! X=x)$, $f_M(m\mid a_0,x)=\p(M=m \! \mid \! A=a_0,X=x)$, and $\p_X(x)=\p(X=x)$. Then $\psi_{a_0}(\P)$ and $\beta_{a_0}(\P)$ can be written as $\psi_{a_0}(\Q)$ and $\beta_{a_0}(\Q)$ for fixed $a_0 \in \{0,1\}$; we suppress the subscript $a_0$ for notational simplicity. We also define: $\xi(M,X) \coloneqq \sum_{a = 0}^1 \mu(M,a,X) \sp \pi(a \! \mid \! X)$, $\eta(A, X) \coloneqq \int \mu(m, A, X) \sp f_M(m \! \mid \! a_0, X) \sp \diff m$, and $ \theta(X) \coloneqq \int \xi(m,X) \sp f_M(m \! \mid \! a_0, X) \sp \diff m$. The parameters $\xi$, $\eta$, and $\theta$ are indexed by elements of $\Q$. Thus, a particular choice of $\Q$ implies values for each of these parameters as well.  

An estimator of $\psi(\Q)$ could be constructed by generating estimates $\hat{\Q}$ of $\Q$ and plugging in: 
\vspace{-0.35cm}
\begin{align}
    \psi(\hat{\Q}) = \frac{1}{n} \sum_{i=1}^n \hat{\theta}(X_i) \sp , \qquad \quad \text{\small (plug-in estimator of \eqref{eq:id_ATE})}
    \label{eq:plugin_1} 
\end{align}

\vspace{-0.35cm}
\noindent where $\hat{\theta}(x) = \sum_{m} \hat{\xi}(m, x) \hat{f}_M(m \! \mid \! a_0, x)$ (if $M$ is discrete), $\hat{\xi}(m, x) = \sum_{a = 0}^1 \hat{\mu}(m, a, x) \hat{\pi}(a \! \mid \! x)$, and $\hat{\mu}, \hat{\pi}$, $\hat{f}_M$ are estimates of $\mu$, $\pi$, and $f_M$, respectively. If $M$ is continuous, $\hat{\theta}(x)$ is computed via numerical integration (or Monte Carlo approximation): $\hat{\theta}(x) = \int \hat{\xi}(m, x) \sp \hat{f}_M(m \! \mid \! a_0, x) \sp \diff m$. 

Given a $\P$-integrable function $f$ of the observed data $O$, let $\P f \coloneqq \int \! f(o) \sp \p(o) \sp \diff o$ and $\Pn f \coloneqq \frac{1}{n} \sum_{i = 1}^n f(O_i)$. A linear expansion of $\psi(\hat{\Q})$ yields $\psi(\hat{\Q}) = \psi(\Q) - \P \Phi(\hat{\Q}) + \Rem(\hat{\Q}, \Q)$, where $\Phi$ is a gradient of $\psi$ satisfying $\P \Phi(\Q) = 0$, and $\Rem(\hat{\Q}, \Q)$ denotes a second-order remainder term. Although multiple gradients may satisfy this expansion, the tangent space of our model is saturated, yielding a unique gradient, the efficient influence function (EIF) \citep{bickel1993efficient}.

The EIF for $\psi(\Q)$ in \eqref{eq:id_ATE} was provided by \cite{fulcher19robust} and can be written as a sum of four components (see Appendix~\ref{app:model_eif} for a proof)
\vspace{-0.25cm}
\begin{equation} \label{eq:eif_tangent_space}
\begin{aligned} 
\Phi(\Q)(O_i) 
    &= \sp \underbrace{\frac{f_M(M_i \mid a_0, X_i)}{f_M(M_i \mid A_i, X_i)} \left\{ Y_i - \mu(M_i, A_i, X_i) \right\}}_{\Phi_Y(\Q)(O_i)}
    \sp + \sp \underbrace{\frac{\mathbb{I}(A_i = a_0)}{\pi(a_0 \mid X_i)} \left\{\xi(M_i,X_i) - \theta(X_i) \right\}}_{\Phi_M(\Q)(O_i)}  \\ 
    &\hspace{0.5cm} + \underbrace{\left\{\eta(1, X_i) - \eta(0, X_i)\right\} \left\{A_i - \pi(1 \mid X_i) \right\}}_{\Phi_A(\Q)(O_i)}  
   \sp + \sp \underbrace{\theta(X_i) - \psi(\Q)}_{\Phi_X(\Q)(O_i)} \sp . 
\end{aligned} 
\end{equation}

\vspace{-0.35cm}
\noindent For our later use, we note that if $M$ is binary, $\Phi_M(\Q)$ can be rewritten (see Appendix~\ref{app:model_eif}),   
\vspace{-0.25cm}
\begin{align}
    \Phi_M(\Q)(O_i) &= \frac{\I(A_i = a_0)}{\pi(a_0 \mid X_i)} \left\{\xi(1, X_i) - \xi(0, X_i) \right\} \left\{ M_i - f_M(1 \mid a_0, X_i) \right\} \sp .
    \label{eq:eif_m_score_binary}
\end{align}

\vspace{-0.35cm}
The first-order bias of the plug-in estimator is $- \Pn \Phi(\hat{\Q})$ (see Appendix~\ref{app:one-step_overview}). When flexible nuisance estimation strategies are used (e.g., based on machine learning), this term may not have standard root-$n$ asymptotic behavior. This  motivates the one-step corrected plug-in estimator, denoted by $ \psi^+_1(\hat{\Q})$, to be $ \psi(\hat{\Q}) + \Pn \Phi(\hat{\Q})$. The one-step estimator takes the form: 
\vspace{-0.35cm}
\begin{equation}  \label{eq:one-step} 
\begin{aligned}
    \psi^+_1(\hat{\Q}) 
    \! &= \! \frac{1}{n} \! \sum_{i=1}^n \! \frac{\hat{f}_M(M_i \! \mid \! a_0, X_i)}{\hat{f}_M(M_i  \! \mid \! A_i, X_i)} \! \left\{ Y_i \! - \! \hat{\mu}(M_i, A_i, X_i) \right\}
   \! + \! \frac{\mathbb{I}(A_i \! = \! a_0)}{\hat{\pi}(a_0 \! \mid \! X_i)}\! \big\{\hat{\xi}(M_i,X_i) \! - \! \hat{\theta}(X_i) \big\} \! + \! \hat{\eta}(A_i, X_i) \sp , 
\end{aligned}
\end{equation}%

\vspace{-0.35cm}
\noindent  where $\hat{\eta}(a, x) \! = \! \int \! \hat{\mu}(m, a, x) \sp \hat{f}_M(m \! \mid \! a_0, x) \sp \diff m$. 

\cite{fulcher19robust} showed, under parametric working models, this estimator is both asymptotically normal and \textit{doubly robust}, meaning it is consistent for $\psi(\Q)$ if either $(\hat{\mu}, \hat{\pi})$ or $\hat{f}_M$ are consistent for their respective target parameters. However, this estimator requires estimating $f_M$, which may be high‐dimensional, and can produce estimates outside the parameter space, particularly for binary or bounded outcomes. These drawbacks motivate the development of alternative estimators, such as TMLEs which combine statistical efficiency with respect for parameter constraints. \citet{wen2024causal} addresses some of these concerns. But, notably, their target estimand differs slightly from the standard front-door functional in \eqref{eq:id_ATE}, as it marginalizes over the treatment variable early in the derivation, resulting in a decomposition that includes a direct plug-in component and a modified front-door term. While this alternative formulation is well-justified, its statistical structure and interpretation differ from the estimands considered here. They also do not establish detailed conditions under which flexible learning yields valid inference. 

We extend existing front-door estimation methods by proposing several novel doubly/multiply robust one-step estimators and TMLEs for both ATE and ATT, designed to address the limitations discussed above through flexible nuisance estimation and targeted learning. 

\section{Proposed estimators for the ATE front-door functional} 
\label{sec:est_ATE}

In this section, we present three representations of the EIF for the ATE functional in \eqref{eq:id_ATE}, each tied to a different parameterization of the observed data distribution and motivating distinct estimators. The \textit{first} uses the standard factorization and requires estimation of all components, including conditional densities (Section~\ref{subsec:est_with_density}). The \textit{second} and \textit{third} avoid direct conditional density estimations via density-ratio or regression-based reparameterizations (Section~\ref{subsec:est_without_density}). For each, we describe the nuisance components and develop corresponding one-step estimators and TMLEs.    

The TMLE construction starts from an initial plug‐in estimate $\psi(\hat{\Q})$, and updates $\hat{\Q}$ to yield $\hat{\Q}^\star$ by simultaneously (i) reducing empirical risk relative to $\hat{\Q}$ and (ii) solving the approximate-equation-solving property where $\Pn \Phi(\hat{\Q}^\star) = o_\p(n^{-1/2})$. Concretely, for each nuisance $\Q_j \in \Q$ we posit a one‐dimensional submodel through $\hat{\Q}_j$ with an associated loss  whose score recovers the corresponding EIF component.  Iterative minimization along these submodels yields $\hat{\Q}^\star$, and the final TMLE is $\psi(\hat{\Q}^\star)$.  For details see Appendix~\ref{app:tmle_overview} and \citet{van2011targeted}.

Throughout, we assume $Y$ is continuous and defer binary-outcome extensions to Appendix~\ref{app:tmle_binaryY}.  

\subsection{Estimation based on standard factorization}
\label{subsec:est_with_density} 

Consider the plug-in estimator in \eqref{eq:plugin_1}, where $\Q = (\mu, f_M, \pi, \p_X)$ denotes the nuisance functions under the standard factorization of $\P$. While the corresponding one-step estimator was reviewed in Section~\ref{subsec:one‐step}, we describe a TMLE here. We begin with initial estimates $\hat{\Q} = (\hat{\mu}, \hat{f}_M, \hat{\pi}, \hat{\p}_X)$, where $\mu$ and $\pi$ can be estimated via regression (including machine learning methods) and $\hat{p}_X$ is taken as the empirical distribution of $X$. Estimation of $f_M$ depends on the nature of the mediator. Here, we focus on direct estimation of the mediator density, which is most practical when $M$ is low-dimensional or discrete. For discrete mediators, standard categorical regression suffices; for continuous, low-dimensional mediators, one may use parametric or flexible conditional density estimators such as kernel methods or the highly adaptive LASSO \citep{hayfield2008nonparametric,benkeser2016highly}. 

Given an initial estimate $\hat{\Q}$, we outline the targeting step of the TMLE. We begin with binary $M$ and later extend the procedure to accommodate continuous mediators.  We assume $\Q$ belongs in a functional space $\calQ$, defined as the Cartesian product of each nuisance-functional space  $\M_{\Q_j}$. 

\vspace{0.15cm}
\underline{\bf Binary $M$.} Let $\hat{\Q}^{(t)} = (\hat{\mu}^{(t)}, \hat{f}^{(t)}_M, \hat{\pi}^{(t)}, \hat{\p}_X)$ denote the nuisance estimates at iteration $t$, with initialization $\hat{\Q}^{(0)} = \hat{\Q}$. Since the empirical distribution of $X$ satisfies $\Pn\Phi_X(\hat{\Q}^\star) = o_\p(n^{-1/2})$, there is no targeting of $\hat{\p}_X$. We therefore focus on updating $\hat{\Q}^{(t)} = (\hat{\mu}^{(t)}, \hat{f}^{(t)}_M, \hat{\pi}^{(t)})$ to ensure that $\Pn\Phi_Y(\hat{\Q}^\star)$, $\Pn\Phi_M(\hat{\Q}^\star)$, and $\Pn\Phi_A(\hat{\Q}^\star)$ are all $o_\p(n^{-1/2})$, where $\Phi_A$ and $\Phi_Y$ are defined in \eqref{eq:eif_tangent_space} and $\Phi_M$ is given in \eqref{eq:eif_m_score_binary} for binary $M$. We adopt an iterative procedure with a convergence threshold $C_n = o(n^{-1/2})$, repeating steps (1–4) while $|\Pn \Phi(\hat{\Q}^{(t)})| > C_n$.

\vspace{0.2cm}

\noindent \emph{Step 1: Define loss functions and submodels for $\hat{\pi}^{(t)}$, $\hat{f}^{(t)}_M$, and $\hat{\mu}^{(t)}$}, satisfying conditions (C1)–(C3).

\vspace{0.15cm}
For a given $\hat{\Q}^{(t)} \in \calQ$ and $\varepsilon_A, \varepsilon_M, \varepsilon_Y \in \mathbb{R}$, the parametric submodels are defined as: 

\vspace{-1.2cm}
\begin{align}
   \hat{\pi}\big(\varepsilon_A; \hat{\mu}^{(0)}, \hat{f}_M^{(t)}\big)(1 \mid X) 
    &= \operatorname{expit}\Big\{\operatorname{logit}\{ \hat{\pi}^{(t)}(1 \mid X) \}+ \varepsilon_A\big\{\hat{\eta}^{(t)}(1, X)-\hat{\eta}^{(t)}(0, X)\big\} \Big\} \sp ,
    \notag \\
    \hat{f}_M\big(\varepsilon_M; \hat{\mu}^{(0)}, \hat{\pi}^{(t)}\big)(1 \mid A, X) 
    &= \operatorname{expit}\Big\{\operatorname{logit}\big\{\hat{f}_M^{(t)}(1 \mid A, X) \big\} + \varepsilon_M \sp \frac{\hat{\xi}^{(t)}(1, X)-\hat{\xi}^{(t)}(0, X)}{\hat{\pi}^{(t)}(A \mid X)} \Big\} \sp , \notag \\
    \hat{\mu}(\varepsilon_Y)(M, A, X) 
    &= \hat{\mu}^{(t)}(M, A, X) + \varepsilon_Y \sp , \label{eq:submodels_binaryM}
\end{align}%

\vspace{-0.35cm}
\noindent where  
$\hat{\eta}^{(t)}(a^{\ast},X) \!=\! \sum_{m=0}^{1}\hat{\mu}^{(0)}(m,a^{\ast},X) \hat{f}_{m}^{(t)}(a_0,X)$ and $\hat{\xi}^{(t)}(m^{*},X) \!=\! \sum_{a=0}^{1}\hat{\mu}^{(0)}(m^{*},a,X) \hat{\pi}^{(t)}(a \! \mid \! X)$, for $a^*, m^*\in\{0,1\}$. 
Given $\tilde{\pi}\in\M_{\pi}$, $\tilde{f}_M \in \mathcal{M}_{f_M}$, $\tilde{\mu} \in \mathcal{M}_{\mu}$,  the loss functions are defined as: 

\vspace{-1.1cm}
\begin{equation} \label{eq:loss_functions} 
\begin{aligned} 
    &L_{A}(\tilde{\pi})(O) 
    = - \log \tilde{\pi}(A \mid X) \sp , \quad 
    L_M(\tilde{f}_M)(O) 
    = - \I(A = a_0) \log \tilde{f}_M(M \mid A, X) \sp , \\
    &\hspace{1.5cm} L_Y\big(\tilde{\mu}; \hat{f}_M^{(t)}\big)(O) 
    = \big\{\hat{f}_M^{(t)}(M \mid a_0, X)\big/\hat{f}_M^{(t)}(M \mid A, X)\big\} \{ Y - \tilde{\mu}(M, A, X) \}^2 \sp .  
\end{aligned}
\end{equation}

\vspace{-0.25cm}
\noindent See Appendix~\ref{app:tmle_loss+submodel} for a proof of validity of these submodel–loss function pairs under (C1)–(C3). 

We also considered targeting $\hat{\mu}$ using the expit submodel proposed by \citet{gruber2010targeted}, in which $Y$ is first rescaled to the unit interval. This nonlinear submodel has been shown to yield more stable estimates in sparse data settings with low Fisher information \citep{gruber2010targeted}. Details are provided in Appendix~\ref{app:tmle_nonlinear}. 

Because the submodel for $\hat{\mu}^{(t)}$ is linear in $\varepsilon_Y$, the quantities $\hat{\eta}^{(t)}(1,X) - \hat{\eta}^{(t)}(0,X)$ and $ \hat{\xi}^{(t)}(1,X) - \hat{\xi}^{(t)}(0,X)$ depend only on the initial estimate $\hat{\mu}^{(0)}$. Consequently, the submodels  $\hat{\pi}\big(\varepsilon_A; \hat{\mu}^{(t)}, \hat{f}_m^{(t)}\big)$ and $\hat{f}_M\big(\varepsilon_M; \hat{\mu}^{(t)}, \hat{\pi}^{(t)}\big)$ depend on $\hat{\mu}^{(t)}$ only through $\hat{\mu}^{(0)}$. 
We emphasize this by rewriting them as $\hat{\pi}(\varepsilon_A; \hat{\mu}^{(0)}, \hat{f}_M^{(t)})$ and $\hat{f}_M(\varepsilon_M; \hat{\mu}^{(0)}, \hat{\pi}^{(t)})$. Moreover, the loss functions for $\tilde{\pi}$ and $\tilde{f}_M$ are independent of $\hat{\mu}^{(t)}$. Therefore, updates to $\hat{\pi}$ and $\hat{f}_M$ can be performed iteratively without involving updated values of $\hat{\mu}$, which can instead be updated in a single step after finalizing $\hat{f}_M$ (due to its appearance in the loss function for $\tilde{\mu}$).

\noindent \emph{Step 2: Perform iterative risk minimization to obtain $\hat{\pi}^\star$ and $\hat{f}_M^\star$.} 

\noindent \emph{Step 2a: Update the estimate of $\pi$} by solving the empirical risk minimization
\vspace{-0.5cm}
\begin{align}
    \hat{\varepsilon}_A \sp = \sp \argmin_{\varepsilon_A \in \mathbb{R}} \sp \Pn L_A\Big(\hat{\pi}\big(\varepsilon_A; \hat{\mu}^{(0)}, \hat{f}_M^{(t)}\big)\Big) \sp . \label{eq:eps_pi}
\end{align}%

\vspace{-0.5cm}
\noindent This corresponds to fitting a logistic regression without an intercept term:

\vspace{-1.3cm}
\begin{align*}
   A\sim\mathrm{offset}\big(\operatorname{logit}\ \hat{\pi}^{(t)}(1\mid X)\big)+ \hat{H}_{A}^{(t)}\big(X\big) \sp , \sp \text{where } \sp \hat{H}_{A}^{(t)}(X) \coloneqq \hat{\eta}^{(t)}(1,X)-\hat{\eta}^{(t)}(0,X) \sp .
\end{align*}%

\vspace{-0.5cm}
\noindent The auxiliary variable $\hat{H}_{A}^{(t)}(X)$ is often referred to as the ``clever covariate.'' The coefficient on this covariate corresponds to $\hat{\varepsilon}_A$, the solution to \eqref{eq:eps_pi}. We update $\hat{\pi}^{(t+1)} = \pi(\hat{\varepsilon}_A; \hat{\mu}^{(0)}, \hat{f}_M^{(t)})$ and define $\hat{\Q}^\text{(temp)} = (\hat{\mu}, \hat{\pi}^{(t+1)}, \hat{f}_M^{(t)}, \hat{\p}_X)$. Condition (C3) then implies $\Pn \Phi_A(\hat{\Q}^\text{(temp)}) = o_\p(n^{-1/2})$. 

\noindent \emph{Step 2b: Update the estimate of $f_M$} by solving the empirical risk minimization
\vspace{-0.5cm}
\begin{align}
    \hat{\varepsilon}_M \sp = \sp \argmin_{\varepsilon_M \in \R} \sp \Pn L_M\Big(\hat{f}_M\big(\varepsilon_M; \hat{\mu}^{(0)}, \hat{\pi}^{(t+1)}\big)\Big) \sp .  
    \label{eq:eps_fM}
\end{align}%

\vspace{-0.5cm}
\noindent This corresponds to fitting a logistic regression without an intercept term:

\vspace{-1.3cm}
\begin{align*}
    M\sim\mathrm{offset}\big(\mathrm{logit}\sp \hat{f}_M^{(t)}(1\mid a_{0},X)\big)+\hat{H}_{M}^{(t)}\big(X\big) \sp , \sp \text{where } \sp \hat{H}_{M}^{(t)}\big(X\big) \coloneqq \frac{\hat{\xi}^{(t)}(1,X)-\hat{\xi}^{(t)}(0,X)}{\hat{\pi}^{(t+1)}(a_0\mid X)} \sp .
\end{align*}%

\vspace{-0.35cm}
\noindent The coefficient on the clever covariate  $\hat{H}_{M}^{(t)}(X)$ yields $\hat{\varepsilon}_M$, the solution to \eqref{eq:eps_fM}. Finally, we update $\hat{f}^{(t+1)}_M = \hat{f}_M(\hat{\varepsilon}_M; \hat{\mu}^{(0)}, \hat{\pi}^{(t+1)})$ and let $\hat{\Q}^{(t+1)} = (\hat{\mu}^{(0)}, \hat{\pi}^{(t+1)}, \hat{f}_M^{(t+1)}, \hat{\p}_X)$. 
Under condition (C3), this ensures $\Pn \Phi_M(\hat{\Q}^{(t+1)}) = \smallO(n^{-1/2})$. We increment $t$ and repeat \textit{Step 2} until convergence. 

Multiple iterations are required because updates to one nuisance parameter affect the auxiliary variable used in updating another. For example, while $\Pn \Phi_M(\hat{\Q}^{(t+1)}) = o_\p(n^{-1/2})$, the term $\Pn \Phi_A(\hat{\Q}^{(t+1)})$ may no longer satisfy this rate, as updating $\hat{f}_M$ changes the auxiliary variable $\hat{H}_A$, necessitating a new solution to \eqref{eq:eps_pi}. Likewise, updating $\hat{\pi}$ alters $\hat{H}_M$, requiring re-optimization of \eqref{eq:eps_fM}. This interdependence of updates and auxiliary variables underlies the need for iteration.

Assume convergence at iteration $t^\star$. Let $\hat{\pi}^\star = \hat{\pi}^{(t^\star)}$, $\hat{f}_M^\star = \hat{f}_M^{(t^\star)}$, and $\hat{\Q}^{(t^\star)} = (\hat{\mu}^{(0)}, \hat{\pi}^\star, \hat{f}_M^\star)$.

\noindent \emph{Step 3: Perform one-step risk minimization to obtain $\hat{\mu}^\star$}. 

\noindent \emph{Update the estimate of $\mu$} by solving the empirical risk minimization 
\vspace{-0.5cm}
\begin{align}
    \hat{\varepsilon}_Y \sp = \sp \argmin_{\varepsilon_Y \in \mathbb{R}} \sp \Pn L_Y\Big(\hat{\mu}(\varepsilon_Y); \hat{f}_M^\star \Big) \sp .  
    \label{eq:eps_mu}
\end{align}%

\vspace{-0.5cm}
\noindent This corresponds to fitting a weighted regression: 
\vspace{-0.5cm}
\begin{align*}
    Y\sim\mathrm{offset}(\hat{\mu}^{(0)})+1 \sp , \sp \text{with weight}={\hat{f}_M^\star(M\mid a_0, X)}/{\hat{f}_M^\star(M\mid A, X)} \sp .
\end{align*}

\vspace{-0.5cm}
\noindent The estimated intercept of this model corresponds to $\hat{\varepsilon}_Y$, as a solution to \eqref{eq:eps_mu}. We update $\hat{\mu}^\star = \hat{\mu}(\hat{\varepsilon}_Y; \hat{f}_M^\star)$ and define $\hat{\Q}^\star =(\hat{\mu}^\star,\hat{\pi}^\star,\hat{f}_M^\star)$. Condition (C3) then implies $\Pn\Phi_{Y}(\hat{\Q}^\star)=0$. 

\noindent \emph{Step 4: Evaluate the plug-in estimator in \eqref{eq:plugin_1} using the updated nuisance estimates $\hat{\Q}^\star$:} 
\vspace{-0.35cm}
\begin{align}
    \psi_1(\hat{\Q}^\star) = \frac{1}{n} \sum_{i=1}^n \hat{\theta}^\star(X_i) \sp , 
    \label{eq:tmle_1} 
\end{align}%

\vspace{-0.35cm}
\noindent where $\hat{\theta}^\star(x) \! = \! \sum_{m=0}^1 \hat{\xi}^\star(m, x) \hat{f}_M^\star(m \! \mid \! a_0, x)$ and $\hat{\xi}^\star(m, x) \! = \! \sum_{a=0}^1 \hat{\mu}^\star(m, a, x) \hat{\pi}^\star(a \!  \mid \! x).$

\begin{remark}
    The iterative updates of $\hat{\pi}$ and $\hat{f}_M$ can be avoided by using the empirical distribution of $(A, X)$. This ensures that $\Pn[\Phi_A(\hat{\Q}^\star) + \Phi_X(\hat{\Q}^\star)] = o_\p(n^{-1/2})$, leading to the modified TMLE: 
    \vspace{-0.35cm}
    \begin{align}
        \psi_{1, \text{mod}}(\hat{\Q}^\star) &= \frac{1}{n} \sum_{i = 1}^n \sp \sum_{m=0}^{1} \hat{\mu}^\star(m, A_i, X_i) \sp \hat{f}_M^\star(m\mid a_0, X_i) \sp . 
        \label{eq:tmle_1_mod} 
    \end{align}%

    \vspace{-0.35cm}
    \noindent Here, $\hat{f}_M^\star$ and $\hat{\mu}^\star$ are obtained by solving the respective optimization problems in \eqref{eq:eps_fM} and \eqref{eq:eps_mu} sequentially, using a flexible estimate of $\pi$ to compute the auxiliary variable $\hat{H}_M$. This approach, however, introduces a potential inconsistency: it combines two estimates of $\p(A \! \mid \! X)$—one implicit in the empirical distribution and another derived from a regression model for $\pi(A \! \mid \! X)$ used in constructing $\hat{H}_M$. Despite this incompatibility, the discrepancy is typically negligible.
\end{remark} 

\underline{\bf Continuous $M$.} 
If $M$ is continuous, the TMLE largely mirrors the binary case, but with additional complexities due to $f_M$ being a conditional probability density function. In this case, we propose to use the following submodel, 

\vspace{-1.25cm}
\begin{align}
    \hat{f}_M(\varepsilon_M; \hat{\mu}^{(0)},  \hat{\pi}^{(t)})(M \mid a_0, X) = \hat{f}_M^{(t)}(M\mid a_0,X) \Big\{ 1 + \varepsilon_M \sp\frac{\hat{\xi}^{(t)}(M,X)-\hat{\theta}^{(t)}(X)}{\hat{\pi}^{(t)}(a_0\mid X)} \Big\} \sp , \label{eq:submodel_cont} 
\end{align}%

\vspace{-0.35cm}
\noindent where 
$\hat{\xi}^{(t)}(M,X) = \sum_{a=0}^{1}\hat{\mu}^{(0)}(M,a,X) \sp \hat{\pi}^{(t)}(a \! \mid \! X)$ and $\hat{\theta}^{(t)}(X) = \int \hat{\xi}^{(t)}(m,X) \hat{f}^{(t)}_M(m \! \mid \! a_0, X) \sp \diff m$. To ensure validity as a submodel of $\mathcal{M}_{f_M}$, the range of $\varepsilon_M$ must be restricted. Appendix~\ref{app:tmle_valid_submodel} presents a more general alternative, but at higher computational cost.

The empirical risk minimization problem in \eqref{eq:eps_fM} requires a grid search or other numerical optimization methods. Upon convergence, condition (C3) ensures that $\Pn\Phi_M(\hat{\Q}^\star) = o_\p(n^{-1/2})$. The full TMLE procedure for computing $\psi_1(\hat{\Q}^\star)$ is summarized in Appendix~\ref{app:tmle_alg}.

The submodel in \eqref{eq:submodel_cont} also extends to multivariate mediators. However, flexibly estimating $f_M$ in high dimensions presents significant theoretical and computational challenges. To mitigate this, we consider alternative strategies that avoid direct estimation of the conditional mediator density.

\subsection{Estimation without density modeling}
\label{subsec:est_without_density}  

To bypass mediator density estimation we may reinterpret $\theta(X)$ as a quantity estimable via \textit{sequential regression}. Note that $\theta(X) = \E( \xi(M, X) \! \mid \! A=a_0, X)$. This representation suggests an alternative plug-in estimator of the ATE front-door functional in \eqref{eq:id_ATE}. We first generate estimates $\hat{\mu}$ and $\hat{\pi}$, then define the \emph{pseudo-outcome} variable $\hat{\xi}(M_i,X_i)=\sum_{a=0}^{1} \hat{\mu}(M_i, a, X_i) \sp \hat{\pi}(a \! \mid \! X_i).$ To estimate $\theta$, we regress the pseudo-outcome on $X$ using only data points where $A_i = a_0$. This replaces the conditional density estimation with a sequential regression task. We denote this estimate of $\theta$ via $\hat{\gamma}$ to distinguish it from the one used previously. Finally, the plug-in estimator can be computed by marginalizing $\hat{\gamma}$ over the empirical distribution of $X$, 

\vspace{-1.25cm}
\begin{align}
    \psi_2(\hat{\Q}) = \frac{1}{n} \sum_{i=1}^n \hat{\gamma}(X_i) \sp . 
    \label{eq:plugin_2} 
\end{align}

\vspace{-0.35cm}
To implement a one-step estimator or TMLE using this plug-in formulation, we must still consider $f_M$, as it enters $\Phi_Y(\Q)$ via the \emph{density ratio} ${f_M(M \! \mid \! A=a_0,X)}/{f_M(M \! \mid \! A,X)}$, denoted $f_M^r(M, A, X)$. In multivariate settings, estimating this ratio directly is often more tractable than estimating $f_M$ itself. Several flexible methods exist for direct ratio estimation \citep{sugiyama2007direct, kanamori2009least, yamada2013relative, SUGIYAMA201044}. Alternatively, Bayes' theorem yields a reformulation of $f_M^r$ as: 

\vspace{-1.35cm}
\begin{align}
f_M^{r}(M, A, X) = \frac{\lambda(a_0 \mid X, M)}{\lambda(A \mid X, M)} \times \frac{\pi(A \mid X)}{\pi(a_0 \mid X)} \sp ,  \label{eq:bayes} 
\end{align}%

\vspace{-0.35cm}
\noindent where $\lambda(a \! \mid \! x, m) \coloneqq \p(A = a \! \mid \! X = x, M = m)$. 
This representation enables density ratio estimation through binary regressions for $\lambda$ and $\pi$, offering a practical and flexible alternative to direct ratio estimation. It naturally accommodates multivariate mediators and supports a wide range of tools for binary regression, from logistic regression to machine learning. This reparameterization strategy parallels approaches proposed in prior literature on mediation analysis \citep{zheng2012targeted, diaz2021nonparametric}.

Similarly, we can adopt a sequential regression approach to estimate $\eta$. Since $\eta(A, X) = A \kappa_1(X) + (1-A) \kappa_0(X)$, where $\kappa_a(X) \coloneqq \E\big( \mu(M, a, X) \! \mid \! A = a_0, X \big)$, we compute $\hat{\mu}(M_i, a, X_i)$ for all $i$ and regress this outcome on $X$ using only observations with $A_i = a_0$, yielding $\hat{\kappa}_a$. Repeating this for $a = \{0, 1\}$ gives $\hat{\eta}(A, X) = A \hat{\kappa}_1(X) + (1-A) \hat{\kappa}_0(X)$.

Let $\hat{\Q} = (\hat{\mu}, \hat{\kappa}_a, \hat{f}_M^r, \hat{\pi}, \hat{\gamma},\hat{\p}_X)$ denote the revised set of nuisance estimates, where $\hat{f}_M$ is replaced by components that avoid conditional density estimation. The one-step estimator is 

\vspace{-1.25cm}
\begin{align}
\psi_2^+(\hat{\Q}) &= \frac{1}{n} \sum_{i=1}^n \Big\{ \hat{\gamma}(X_i) +  \hat{f}_M^r(M_i, A_i, X_i) \{Y_i - \hat{\mu}(M_i, A_i, X_i)\}  \label{eq:one-step_2} \\
&\hspace{1.4cm} + \frac{\mathbb{I}(A_i = a_0)}{\hat{\pi}(a_0 \mid X_i)} \{ \hat{\xi}(M_i, X_i) - \hat{\gamma}(X_i) \} + \{ \hat{\kappa}_1(X_i) - \hat{\kappa}_0(X_i) \} \{A_i - \hat{\pi}(1 \mid X_i)\} \Big\} \sp .  \notag 
\end{align}

\vspace{-0.35cm}
\noindent To differentiate the two approaches for estimating $f_M^r$ in $\psi_2^+(\hat{\Q})$, we define $\psi_{2a}^+(\hat{\Q})$ for direct density ratio estimation, and  $\psi_{2b}^+(\hat{\Q})$ for the regression-based method via $\hat{\lambda}$ and $\hat{\pi}$. 

Given $\hat{\Q}$, we next construct a TMLE based on the sequential regression and density ratio parameterization, following the procedure in Section~\ref{subsec:est_with_density} with key modifications outlined below.

\emph{Submodels and loss functions}. The submodel for $\hat{\mu}$ remains linear with corresponding loss $L_Y(\tilde{\mu};\hat{f}_M^r) = \hat{f}_M^r(M,A,X) \{Y - \tilde{\mu}(M, A, X)\}^2$. The submodel for $\hat{\pi}$ is defined as in \eqref{eq:submodels_multivariate}, indexed by $\hat{\kappa}_1(X) - \hat{\kappa}_0(X)$, with standard negative log likelihood loss. In addition, we introduce a linear submodel for $\hat{\gamma}$: $\hat{\gamma}(\varepsilon_\gamma)(X) = \hat{\gamma}(X) + \varepsilon_\gamma$, with loss $L_{\gamma}(\tilde{\gamma};\hat{\pi},\hat{\xi})(O) = \frac{\I(A=a_0)}{\hat{\pi}(a_0\mid X)}\big\{\hat{\xi}(M,X)-\tilde{\gamma}(X)\big\}^2.$ See Appendix~\ref{app:tmle_loss+submodel} for a proof of submodel–loss validity under (C1)–(C3). 

\emph{Targeting steps.} We first update $\hat{\mu}$ via weighted least squares regression with weight $\hat{f}_M^r(M,A,X)$ to obtain $\hat{\mu}^\star$. Next, using the updated $\hat{\mu}^\star$ to recompute $\hat{\kappa}$, we update $\hat{\pi}$ via logistic regression with no intercept and a single covariate $\hat{\kappa}_1(X) - \hat{\kappa}_0(X)$, yielding $\hat{\pi}^\star$. Then, using $\hat{\mu}^\star$ and $\hat{\pi}^\star$, we compute $\hat{\xi}^\star(M, X) = \sum_{a} \hat{\mu}^\star(M, a, X) \hat{\pi}^\star(a \! \mid \! X)$ and regress it on $X$ (restricted to $A = a_0$) to estimate $\hat{\gamma}$. An update via weighted regression yields $\hat{\gamma}^\star$. See more details in Appendix~\ref{app:tmle_multi}. 

\emph{Plug-in estimator.} Define $\hat{\Q}^\star = (\hat{\mu}^\star, \hat{\kappa}_a, \hat{f}_M^r, \hat{\pi}^\star, \hat{\gamma}^\star,\hat{\p}_X)$, and evaluate
\vspace{-0.25cm}
\begin{align}
    \psi_2(\hat{\Q}^\star) = \frac{1}{n} \sum_{i=1}^n \hat{\gamma}^\star(X_i) \sp . 
    \label{eq:tmle_2} 
\end{align}%

\vspace{-0.35cm}
\noindent The TMLE that avoids mediator density estimation is detailed in Algorithm~\ref{appalg:multi}, Appendix~\ref{app:tmle_alg}.

As in the one‐step case, we define TMLEs $\psi_{2a}(\hat{\Q}^\star)$ via direct ratio estimation of $f_M^r$ and $\psi_{2b}(\hat{\Q}^\star)$ via regression using $\hat{\lambda}$ and $\hat{\pi}$. 

\section{Proposed estimators for the ATT front-door functional}
\label{sec:est_ATT}

As with the ATE, the ATT functional \eqref{eq:id_ATT} admits two estimation strategies. First, under the standard factorization of $\P$, one can write $\beta(\Q) = \iint \sum_{a=0}^1 \frac{\I(a=a_1)}{\p(a)} \mu(m, a, x) \sp f_M(m \! \mid \! a_0, x) \sp \p(a,x) \sp \diff m \sp \diff x$, where $a_1=1-a_0$, and construct density-based estimators (plug-in, one-step and TMLE) by estimating $\mu$ and $f_M$ (with $\p(a)$ and $\p(a,x)$ replaced by empirical counterparts). Second, estimation of $f_M$ can be avoided via density-ratio or regression reparameterizations. We focus here on these regression-based approaches and defer density-based constructions to Appendices~\ref{app:att_density_est} and \ref{app:att_tmle_alg}.

Specifically, we rewrite \eqref{eq:id_ATT} as $\beta(\Q) \! = \! \int \sum_{a=0}^1 \frac{\I(a=a_1)}{\p(a)} \kappa_a(x) \sp \p(a,x) \sp \diff x$, where $\kappa_a(x) = \E(\mu(M, a, x) \! \mid \! A = a_0, x)$. 
Let $\hat{\Q} = (\hat{\mu}, \hat{\kappa}_{a_1}, \hat{\p}_A, \hat{\p}_{AX})$ denote the collection of nuisance estimates. Estimation procedures for $\hat{\mu}$ and $\hat{\kappa}_a$ are described in Section~\ref{sec:est_ATE}, while $\hat{\p}_A$ and $\hat{\p}_{AX}$ refer to empirical estimates of $\p(A)$ and $\p(A,X)$, respectively. This yields the following plug-in estimator:  
\vspace{-0.35cm}
\begin{align} 
    \beta(\hat{\Q})=\frac{1}{n}\sum_{i=1}^{n} \frac{\mathbb{I}(A_i=a_1)}{\hat{\p}_A(a_1)} \sp \hat{\kappa}_{a_1}(X_i) \sp . 
    \label{eq:att_plugin} 
\end{align}

\vspace{-0.35cm}
\noindent We build on this version of the plug-in to derive a one-step corrected estimator and a TMLE.  As a first step, we derive the EIF for $\beta(\Q)$, denoted $\Phi_{\beta}(\Q)$ (see Appendix~\ref{app:model_eif} for a proof): 

\vspace{-1.25cm}
\begin{align}
    \Phi_{\beta}(\Q)(O_i)&=\underbrace{\frac{\I(A_i=a_1)}{\p_A(a_1)} f^r_M(M_i, A_i, X_i) \sp \big\{Y_i-\mu(M_i,A_i,X_i)\big\}}_{\Phi_{\beta;Y}(\Q)(O_i)} 
    \label{eq:eif_att}
    \\
    &\hspace{0.1cm}+\underbrace{\frac{\I(A_i=a_0)}{\p_A(a_1)}\frac{\pi(a_1\mid X_i)}{\pi(a_0\mid X_i)}\big\{\mu(M_i,a_1,X_i)-\kappa_{a_1}(X_i)\big\}}_{\Phi_{\beta; M}(\Q)(O_i)}
    +\underbrace{\frac{\I(A_i=a_1)}{\p_A(a_1)}\big\{\kappa_{a_1}(X_i)-\beta(\Q)\big\}}_{\Phi_{\beta; AX}(\Q)(O_i)} \sp . \notag 
\end{align}

\vspace{-0.35cm}
\noindent Given $\hat{\Q} = (\hat{\mu}, \hat{\pi}, \hat{f}^r_M, \hat{\kappa}_{a_1}, \hat{\p}_A, \hat{\p}_{AX})$, the one-step correction of $\beta(\hat{\Q})$, denoted by $\beta^+(\hat{\Q})$, is  
\vspace{-0.25cm}
\begin{align}
    \beta^+(\hat{\Q}) &= \beta(\hat{\Q}) + \frac{1}{n} \sum_{i=1}^n \Big\{\frac{\I(A_i = a_1)}{\hat{\p}_A(a_1)} \hat{f}_M^r(M_i, A_i, X_i) \sp \big\{Y_i - \hat{\mu}(M_i, A_i, X_i)\big\} 
    \label{eq:oneste_p_Att_2} \\
    &\hspace{0.35cm} + \frac{\I(A_i  = a_0)}{\hat{\p}_A(a_1)} \frac{\hat{\pi}(a_1 \! \mid \! X_i)}{\hat{\pi}(a_0 \! \mid \! X_i)} \big\{\hat{\mu}(M_i, a_1, X_i) \! - \! \hat{\kappa}_{a_1}(X_i)\big\} + \frac{\I(A_i = a_1)}{\hat{\p}_A(a_1)} \big\{\hat{\kappa}_{a_1}(X_i) \! - \! \beta(\hat{\Q})\big\}  \Big\} \sp . \notag 
\end{align}

\vspace{-0.35cm}
\noindent As in the ATE case, $\hat{f}_M^r$ can be estimated either directly or based on \eqref{eq:bayes} using estimates $\hat{\lambda}$ and $\hat{\pi}$. The corresponding one-step estimators are denoted $\beta^+_a(\hat{\Q})$ and $\beta^+_b(\hat{\Q})$, respectively.  

We next describe a TMLE for the plug-in $\beta(\hat{\Q})$ in \eqref{eq:att_plugin}, assuming $Y$ is continuous; modifications for binary outcomes mirror those used for the TMLEs of the ATE and are omitted. It suffices for the updated $\hat{\Q}^\star$ to satisfy $\Pn \Phi_{\beta;Y}(\hat{\Q}^\star) = o_\p(n^{-1/2})$ and $\Pn \Phi_{\beta;M}(\hat{\Q}^\star) = o_\p(n^{-1/2})$, as the final term $\Pn \Phi_{\beta;AX}(\hat{\Q}^\star)$ vanishes when $\p_{AX}$ is estimated empirically. The TMLE updates $\hat{\mu}$ and $\hat{\kappa}_{a_1}$ using a single-step targeting procedure.

To target $\hat{\mu}$, we define a linear submodel (as in Section~\ref{sec:est_ATE}) and minimize the empirical risk:

\vspace{-1.25cm}
\begin{align}
\hat{\varepsilon}_Y = \argmin_{\varepsilon_Y \in \mathbb{R}} \frac{1}{n} \sum_{i = 1}^n \frac{\I(A_i = a_1)}{\hat{\p}_A(a_1)} \hat{f}_M^r(M_i, a_1, X_i) \big\{ Y_i - \hat{\mu}(\varepsilon_Y)(M_i, a_1, X_i) \big\}^2 \sp .
\label{eq:eps_mu_att}
\end{align}

\vspace{-0.35cm}
\noindent This corresponds to fitting a weighted regression of the outcome on an intercept-only submodel with offset $\hat{\mu}(M, a_1, X)$ and weights proportional to $\frac{\I(A = a_1)}{\hat{\p}_A(a_1)} \hat{f}_M^r(M, a_1, X)$. The updated estimate is $\hat{\mu}^\star(m, a, x) = \hat{\mu}(m, a, x) + \hat{\varepsilon}_Y$. 

Next, we update $\hat{\kappa}_{a_1}$ via a linear submodel $\hat{\kappa}_{a_1}(\varepsilon_\kappa)(x) = \hat{\kappa}_{a_1}(x) + \varepsilon_\kappa$, minimizing:
\vspace{-0.25cm}
\begin{align}
\hat{\varepsilon}_\kappa = \argmin_{\varepsilon_\kappa \in \mathbb{R}} \frac{1}{n} \sum_{i = 1}^n \frac{\I(A_i = a_0)}{\hat{\p}_A(a_1)} \frac{\hat{\pi}(a_1 \mid X_i)}{\hat{\pi}(a_0 \mid X_i)} \left\{ \hat{\mu}^\star(M_i, a_1, X_i) - \hat{\kappa}_{a_1}(\varepsilon_\kappa)(X_i) \right\}^2 \sp .
\label{eq:eps_kappa_att}
\end{align}

\vspace{-0.35cm}
\noindent This corresponds to fitting a weighted regression of $\hat{\mu}^\star(M, a_1, X)$ on an intercept with offset $\hat{\kappa}_{a_1}(X)$ and weights $\frac{\I(A = a_0)}{\hat{\p}_A(a_1)} \frac{\hat{\pi}(a_1 \mid X)}{\hat{\pi}(a_0 \mid X)}$. The updated function is given by $\hat{\kappa}_{a_1}^\star(x) = \hat{\kappa}_{a_1}(x) + \hat{\varepsilon}_\kappa$.

The TMLE is then defined as:  
\vspace{-0.75cm}
\begin{align}
    \beta(\hat{\Q}^\star) = \frac{1}{n} \sum_{i=1}^n \frac{\I(A_i=a_1)}{\hat{\p}_A(a_1)}\ \hat{\kappa}^\star_{a_1}(X_i) \sp . 
    \label{eq:tmle_1_att} 
\end{align}%

\vspace{-0.35cm} 
\noindent As above, $\hat{f}_M^r$ may be estimated either directly or via Bayes' rule, yielding TMLEs denoted by $\beta_a(\hat{\Q}^\star)$ and $\beta_b(\hat{\Q}^\star)$, respectively.

\section{Inference and asymptotic properties}
\label{sec:asymptotic}

We now establish the asymptotic properties of our estimators, presenting the expansion using TMLE notation with targeted estimates $\hat{\Q}^\star$. The same form and remainder bounds apply to one-step estimators, which we omit for brevity. Given a TMLE $\omega(\hat{\Q}^\star)$ and EIF $\Phi_\omega(\Q)$ for a parameter $\omega(\Q)$—either $\psi(\Q)$ or $\beta(\Q)$—its linear expansion takes the form: 
\vspace{-0.25cm}
\begin{align}
    \omega(\hat{\Q}^\star) - \omega(\Q) = \Pn \Phi_\omega(\Q) - \Pn \Phi_\omega(\hat{\Q}^\star) + (\Pn - \P) \big\{ \Phi_\omega(\hat{\Q}^\star) - \Phi_\omega(\Q) \big\} + \Rem(\hat{\Q}^\star, \Q) \sp . 
    \label{eq:expansion_plus_tmle}
\end{align}

\vspace{-0.35cm}
\noindent To establish asymptotic linearity, we require the following conditions:
\begin{itemize}
    \setlength{\itemindent}{-0.65cm}
    \item[] (A1) \emph{Donsker estimates}: $\Phi_\omega(\hat{\Q}^{\star}) \! -\! \Phi_\omega(\Q)$ falls in a $\P$-Donsker class with probability tending to 1;
    \item[] (A2) \emph{$L^2(\P)$-consistent influence function estimates}: $\P\{\Phi_\omega(\hat{\Q}^{\star}) - \Phi_\omega(\Q)\}^2 = o_\p(1)$;
    \item[] (A3) \emph{Successful targeting of nuisance parameters}: $\Pn \Phi_\omega(\hat{\Q}^{\star}) = o_\p(n^{-1/2})$.
\end{itemize}
Conditions (A1)–(A2) imply $(\Pn - \P)\{\Phi_\omega(\hat{\Q}^{\star}) - \Phi_\omega(\Q)\} = o_\p(n^{-1/2})$, so together with (A3), the expansion in \eqref{eq:expansion_plus_tmle} yields
$\omega(\hat{\Q}^{\star}) - \omega(\Q) = \Pn \Phi_\omega(\Q) + \Rem(\hat{\Q}^{\star}, \Q) + o_\p(n^{-1/2})$.
It remains to characterize $\Rem$ for each estimator, which we do in separate subsections below, followed by the corresponding asymptotic linearity theorems. Note that finite‐dimensional parametric models satisfy the Donsker condition (A1) \citep{vaart2023empirical}. In Section~\ref{subsec:sample_split}, we introduce sample splitting to relax (A1) for flexible nuisance estimators.

Throughout, we let $||f||\! = \! \! \sqrt{\P f^2}$ denote the $L^2(\P)$-norm of a $\P$-measurable function $f$. 

\subsection{ATE front-door functional estimators}
\label{subsec:asymp_ATE}

\subsubsection{$\psi_1(\hat{\Q}^\star)$: TMLE with standard factorization} 
\label{subsec:asymp_psi_1}

Consider the TMLE $\psi_1(\hat{\Q}^\star)$ from Section~\ref{subsec:est_with_density}, with $\hat{\Q}^\star = (\hat{\mu}^\star, \hat{f}_M^\star, \hat{\pi}^\star,\hat{\p}_X)$. Under regularity conditions detailed in Appendix~\ref{app:asym_ate:r2_psi1}, the $R_2$ remainder for $\psi_1(\hat{\Q}^\star)$ is bounded by: 
\vspace{-0.35cm}
\begin{align} 
    \Rem(\hat{\Q}^\star, \Q) & \leq C\Big\{ ||\hat{f}^\star_{M}-f_{M} || \times ||\hat{\mu}^\star-\mu|| +||\hat{f}^\star_{M}-f_{M} || \times || \hat{\pi}^{\star} - \pi || \Big\} \sp ,
    \label{eq:r2_1_propto}
\end{align}

\vspace{-0.35cm}
\noindent for some constant $C > 0$. The full expression of $\Rem(\hat{\Q}^\star, \Q)$ is provided in Appendix~\ref{app:asym_ate:r2_psi1}. This result paves the way for establishing asymptotic linearity of $\psi_1(\hat{\Q}^\star)$.
\begin{theorem}[Asymptotic linearity of $\psi_1(\hat{\Q}^\star)$] 
\label{thm:asymp_psi1} 
Suppose the nuisance estimates in $\hat{\Q}^{\star}$ have the following \emph{$L^2(\P)$ convergence rates}: 
$|| \hat{\pi}^{\star} - \pi || =\smallO(n^{-\frac{1}{k}})$, 
$|| \hat{f}_{M}^\star - f_{M} || =\smallO(n^{-\frac{1}{b}})$, 
$|| \hat{\mu}^{\star} - \mu|| = \smallO(n^{-\frac{1}{q}})$, 
and that the convergence exponents satisfy: 
\begin{enumerate}
    \item[] \textnormal{(A4.1)} \ 
    $\frac{1}{b} + \frac{1}{q} \geq \frac{1}{2}$ and 
    $\frac{1}{k} + \frac{1}{b} \geq \frac{1}{2}$.
\end{enumerate}
Under (A1)–(A3), (A4.1), and regularity conditions (outlined in Appendix~\ref{app:asym_ate:r2_psi1}), $\psi_1(\hat{\Q}^\star)$ is asymptotically linear: $\psi_1(\hat{\Q}^\star) - \psi(\Q) = \Pn \Phi(\Q) + o_\p(n^{-1/2})$, with influence function $\Phi(\Q)$. 
\end{theorem}
Condition (A4.1) ensures $\Rem(\hat{\Q}^\star, \Q) = o_\p(n^{-1/2})$ via the bound in \eqref{eq:r2_1_propto}. The cross-product structure allows nuisance estimates to converge at slower than root-$n$ rates, thereby allowing for a potentially wider application of flexible machine learning and statistical models than what is possible under the conditions imposed by \cite{fulcher19robust}. 

An immediate corollary of Theorem~\ref{thm:asymp_psi1} is that our TMLE inherits the double robustness properties of the one-step estimator proposed by \citet{fulcher19robust}. While their formulation is framed in terms of parametric working models, we restate the result using $L^2(\P)$-consistency for parsimony and alignment with the TMLEs below.
\begin{corollary}[Robustness of $\psi_1(\hat{\Q}^\star)$] \label{cor:robust_psi1}
$\psi_1(\hat{\Q}^\star)$ is consistent for $\psi(\Q)$ if either (i) $||\hat{\pi}^\star - \pi|| = o_\p(1) \sp \text{and} \sp || \hat{\mu}^\star - \mu || = o_\p(1)$, or (ii) $|| \hat{f}^\star_M - f_M || = o_\p(1)$, or both (i) and (ii) hold.
\end{corollary}

\subsubsection{$\psi_{2a}(\hat{\Q}^\star)$: TMLE with direct density ratio and sequential regression} \label{subsec:asymp_psi_2} 

Consider the TMLE $\psi_{2a}(\hat{\Q}^\star)$ from Section~\ref{subsec:est_without_density}, where $\hat{f}_M^r$ is obtained via direct density ratio estimation; thus $\hat{\Q}^\star = (\hat{\mu}^\star, \hat{\kappa}_a, \hat{f}_M^r, \hat{\pi}^\star, \hat{\gamma}^\star,\hat{\p}_X)$. Under the regularity conditions detailed in Appendix~\ref{app:asym_ate:r2_psi2a}, the $R_2$ remainder admits the bound: 
\vspace{-0.35cm}
\begin{equation}\label{eq:r2_psi2a_propto}
    \begin{aligned}
        \Rem(\hat{\Q}^\star, \Q) &\leq C \Big\{ || \hat{f}_M^r-f_M^r || \! \times \! || \hat{\mu}^{\star} - \mu|| + || \hat{\pi}^{\star} - \pi || \! \times \! \big\{  || \hat{\gamma}^\star- \gamma || +  \textstyle\sum_{a=0}^1 || \hat{\kappa}_a-\kappa_a || \big\} \Big\} \sp ,
    \end{aligned}
\end{equation}

\vspace{-0.35cm}
\noindent for some finite constant $C > 0$. See the detailed form of $\Rem(\hat{\Q}^\star, \Q)$ in Appendix~\ref{app:asym_ate:r2_psi2a}. We have the following theorem establishing the asymptotic linearity of $\psi_{2a}(\hat{\Q}^\star)$. 
\begin{theorem}[Asymptotic linearity of $\psi_{2a}(\hat{\Q}^\star)$] \label{thm:asymp_psi2a} 
Suppose the nuisance estimates in $\hat{\Q}^{\star}$ satisfy the following $L^2(\P)$ convergence rates:
$|| \hat{\pi}^{\star} - \pi || =\smallO(n^{-\frac{1}{k}})$,
$|| \hat{\mu}^{\star} - \mu|| = \smallO(n^{-\frac{1}{q}})$, 
$|| \hat{\gamma}^{\star} - \gamma || =\smallO(n^{-\frac{1}{j}})$, 
$|| \hat{\kappa}_a - \kappa_a || =\smallO(n^{-\frac{1}{\ell}})$, 
$|| \hat{f}_M^r - f_M^r || = \smallO(n^{-\frac{1}{c}})$, and the exponents satisfy:
\begin{enumerate}
    \item[] \textnormal{(A4.2)} \ 
    $\frac{1}{c}+\frac{1}{q} \geq \frac{1}{2}$, \ 
    $\frac{1}{k}+\frac{1}{j} \geq \frac{1}{2}$, and 
    $\frac{1}{\ell}+\frac{1}{k} \geq \frac{1}{2}$.  
\end{enumerate}
Under (A1)-(A3), (A4.2), and regularity conditions (outlined in Appendix~\ref{app:asym_ate:r2_psi2a}), $\psi_{2a}(\hat{\Q}^\star)$ is asymptotically linear: $\psi_{2a}(\hat{\Q}^\star) - \psi(\Q) = \Pn \Phi(\Q) + o_\p(n^{-1/2})$, with influence function $\Phi(\Q)$.
\end{theorem}
$\psi_{2a}(\hat{\Q}^\star)$ also exhibits multiple robustness.
\begin{corollary}[Robustness of $\psi_{2a}(\hat{\Q}^\star)$]  $\psi_{2a}(\hat{\Q}^\star)$ is consistent for $\psi(\Q)$ if at least one of the following conditions hold: 
(i) $||\hat{\pi}^\star - \pi || = o_\p(1)$ and $|| \hat{\mu}^\star - \mu || = o_\p(1)$, \ 
(ii) $|| \hat{\pi}^\star - \pi || = o_\p(1)$ and $|| \hat{f}^r_M - f_M^r || = o_\p(1)$, \ 
(iii) $|| \hat{\mu}^\star -  \mu ||  = o_\p(1)$, $|| \hat{\gamma}^\star - \gamma || = o_\p(1)$, and $|| \hat{\kappa}_a - \kappa_a || = o_\p(1)$, \  
(iv) $||\hat{\gamma}^\star - \gamma|| = o_\p(1)$, $|| \hat{\kappa}_a - \kappa_a || = o_\p(1)$, and $|| \hat{f}_M^r - f_M^r || = o_\p(1)$.
\label{cor:robust_psi2a}
\end{corollary} 
Corollary~\ref{cor:robust_psi2a} highlights that consistency can be achieved either by consistently estimating $(\mu, \pi)$, or by consistently estimating ($\gamma$, $\kappa_a$, and $f_M^r$). In a partially specified scenario where only one of $\hat{\mu}^\star$ or $\hat{\pi}^\star$ is consistent, consistency of the estimator still holds if a subset of components in ($\gamma$, $\kappa_a$, and $f_M^r$) is consistently estimated. 

\subsubsection{$\psi_{2b}(\hat{\Q}^\star)$: {\small TMLE with fully regression-based methods}} \label{subsec:asymp_psi_2b}

Consider the TMLE $\psi_{2b}(\hat{\Q}^\star)$ from Section~\ref{subsec:est_without_density}, where $f_M^r$ is estimated via regression-based components $\pi$ and $\lambda$; thus $\hat{\Q}^\star = (\hat{\mu}^\star, \hat{\kappa}_a, \hat{\lambda}, \hat{\pi}^\star, \hat{\gamma}^\star,\hat{\p}_X)$. Under regularity conditions stated in Appendix~\ref{app:asym_ate:r2_psi2b}, the $R_2(\hat{\Q}^\star, \Q)$ term admits the following upper bound 
\vspace{-0.35cm}
\begin{align}
    C \Big\{ 
    || \hat{\lambda}-\hat{\lambda} || \! \times \! ||\hat{\mu}^\star-\mu||  + 
    || \hat{\pi}^{\star} - \pi || \! \times \! \big\{ ||\hat{\mu}^\star-\mu|| 
    +  || \hat{\gamma}^\star- \gamma || + || (\hat{\kappa}_1-\hat{\kappa}_0)-(\kappa_1-\kappa_0) ||  \big\} \Big\} \sp , 
    \label{eq:r2_psi2b_propto}
\end{align}

\vspace{-0.35cm}
\noindent for some finite constant $C > 0$. The detailed form of $\Rem(\hat{\Q}^\star, \Q)$ is provided in Appendix~\ref{app:asym_ate:r2_psi2b}. 
\begin{theorem}[Asymptotic linearity of $\psi_{2b}(\hat{\Q}^\star)$] \label{thm:asymp_psi2b} 
Suppose the nuisance estimates in $\hat{\Q}^{\star}$ satisfy the following $L^2(\P)$ convergence rates:  
$|| \hat{\pi}^{\star} - \pi || =\smallO(n^{-\frac{1}{k}})$, \ 
$|| \hat{\mu}^{\star} - \mu|| = \smallO(n^{-\frac{1}{q}})$, \ 
$|| \hat{\gamma}^{\star} - \gamma || =\smallO(n^{-\frac{1}{j}})$, \ 
$|| \hat{\kappa}_a - \kappa_a || =\smallO(n^{-\frac{1}{\ell}})$, \
$|| \hat{\lambda} - \lambda || = \smallO(n^{-\frac{1}{d}})$, 
and the exponents satisfy: 
\begin{enumerate}
    \item[] \textnormal{(A4.3)} \ 
    $\frac{1}{q}+\frac{1}{k} \geq \frac{1}{2}$, \ 
    $\frac{1}{d}+\frac{1}{q} \geq \frac{1}{2}$, \ 
    $\frac{1}{k}+\frac{1}{j} \geq \frac{1}{2}$, and 
    $\frac{1}{k}+\frac{1}{\ell} \geq \frac{1}{2}$.  
\end{enumerate}
Under (A1)-(A3), (A4.3), and the regularity conditions (outlined in Appendix~\ref{app:asym_ate:r2_psi2b}), $\psi_{2b}(\hat{\Q}^\star)$ is asymptotically linear $\psi_{2b}(\hat{\Q}^\star) - \psi(\Q) = \Pn \Phi(\Q) + o_\p(n^{-1/2})$, with influence function $\Phi(\Q)$. 
\end{theorem}
We note that for $\psi_{2b}(\hat{\Q}^\star)$, consistency of the estimate $\hat{f}_M^r$ depends on both $\hat{\pi}$ and $\hat{\lambda}$, combining robustness conditions (ii) and (iv) from Corollary~\ref{cor:robust_psi2a}. Robustness properties are formalized below. 
\begin{corollary}[Robustness of $\psi_{2b}(\hat{\Q}^\star)$] 
$\psi_{2b}(\hat{\Q}^\star)$ is consistent for $\psi(\Q)$ if at least one of the following holds: 
(i) $|| \hat{\pi}^\star - \pi || = o_\p(1)$ and $||\hat{\mu}^\star - \mu || = o_\p(1)$, \ 
(ii) $|| \hat{\pi}^\star - \pi || = o_\p(1)$ and $|| \hat{\lambda} - \lambda || = o_\p(1)$, \ 
(iii) $|| \hat{\mu}^\star - \mu || = o_\p(1)$, $|| \hat{\gamma}^\star - \gamma|| = o_\p(1)$, and $|| \hat{\kappa}_a - \kappa_a || = o_\p(1)$. 
\label{cor:robust_psi2b} 
\end{corollary}
Unlike $\psi_1$ and $\psi_{2a}$, where certain components could ensure consistency on their own, $\psi_{2b}$ requires at least one of $\hat{\mu}^\star$ or $\hat{\pi}^\star$ to be consistent even when all auxiliary regressions ($\hat{\lambda}$, $\hat{\gamma}$, $\hat{\kappa}_a$) are consistently estimated. In this sense, $\psi_{2b}$ exhibits a slightly weaker robustness property. Nevertheless, it remains attractive in practice due to its fully regression-based construction.

\subsection{ATT front-door functional estimators}
\label{subsec:asymp_ATT}

We now establish conditions for the asymptotic linearity of our ATT estimators. Following Section~\ref{sec:est_ATT}, we focus on the fully regression-based estimator $\beta_b(\hat{\Q}^\star)$, with $\hat{\Q}^\star = (\hat{\mu}^\star, \hat{\kappa}_{a_1}^\star, \hat{\lambda}, \hat{\pi}, \hat{\p}_A, \hat{\p}_{AX})$. Under the regularity conditions detailed in Appendix~\ref{app:asym_att:r2_betab}, the remainder is bounded by 
\vspace{-0.35cm}
\begin{equation}\label{eq:r2_betab_propto}
\begin{aligned}
\Rem(\hat{\Q}^{\star},\Q) \leq C\Big\{||\hat{\pi}-\pi|| \times ||\hat{\mu}^{\star}-\mu|| + ||\hat{\lambda}-\lambda|| \times ||\hat{\mu}^{\star}-\mu||+ ||\hat{\pi}-\pi||\times ||\hat{\kappa}_{a_1}-{\kappa}_{a_1}||\Big\} \sp ,
\end{aligned}
\end{equation}

\vspace{-0.35cm}
\noindent for some constant $C > 0$. The detailed form is provided in  Appendix~\ref{app:asym_att:r2_betab}. Results for $\beta_1(\hat{\Q}^\star)$ (Appendix~\ref{app:att_density_est}) and $\beta_a(\hat{\Q}^\star)$ (Section~\ref{sec:est_ATT}) are deferred to Appendices~\ref{app:asym_att:r2_beta1} and \ref{app:asym_att:r2_betaa}, respectively.
\begin{theorem}[Asymptotic linearity of $\beta_b(\hat{\Q}^\star)$] \label{thm:asymp_psi2b_att} 
Suppose the nuisance estimates in $\hat{\Q}^{\star}$ satisfy the following $L^2(\P)$ convergence rates: 
$||\hat{\pi} - \pi || =\smallO(n^{-\frac{1}{k}})$, \ 
$||\hat{\mu}^{\star} - \mu|| = \smallO(n^{-\frac{1}{q}})$, \ 
$|| \hat{\kappa}_{a_1} - \kappa_{a_1} || = \smallO(n^{-\frac{1}{\ell}})$, \ 
$|| \hat{\lambda} - \lambda || = \smallO(n^{-\frac{1}{d}})$, 
and the exponents satisfy: 
\begin{enumerate}
    \item[] \textnormal{(A4.4)} \ 
    $\frac{1}{q}+\frac{1}{k} \geq \frac{1}{2}$, \ 
    $\frac{1}{d}+\frac{1}{q} \geq \frac{1}{2}$, and 
    $\frac{1}{k}+\frac{1}{\ell} \geq \frac{1}{2}$. 
\end{enumerate}
Under (A1)-(A3), (A4.4), and the regularity conditions (outlined in Appendix~\ref{app:asym_att:r2_betab}), $\beta_b(\hat{\Q}^\star)$ is asymptotically linear: $\beta_b(\hat{\Q}^\star) - \beta_b(\Q) = \Pn \Phi_{\beta}(\Q) + o_\p(n^{-1/2})$, with influence function $\Phi_{\beta}(\Q)$.
\end{theorem}

Notably, $\beta_b(\hat{\Q}^\star)$ requires weaker conditions than its ATE counterpart $\psi_{2b}(\hat{\Q}^\star)$: it avoids the need to estimate $\gamma$, which simplifies implementation and strengthens robustness, as shown below. 

\begin{corollary}[Robustness of $\beta_b(\hat{\Q}^\star)$] 
$\beta_b(\hat{\Q}^\star)$ is consistent for $\psi(\Q)$ if at least one of the following holds: 
(i) $|| \hat{\pi}^\star - \pi || = o_\p(1)$ and $||\hat{\mu}^\star - \mu || = o_\p(1)$, \ 
(ii) $|| \hat{\pi}^\star - \pi || = o_\p(1)$ and $|| \hat{\lambda} - \lambda || = o_\p(1)$, \ 
(iii) $|| \hat{\mu}^\star - \mu || = o_\p(1)$ and $|| \hat{\kappa}_a - \kappa_a || = o_\p(1)$. 
\label{cor:robust_betab} 
\end{corollary}
These robustness conditions closely resemble those for $\psi_{2b}(\hat{\Q}^\star)$ in Corollary~\ref{cor:robust_psi2b}, with one key distinction: consistency of $\hat{\gamma}$ is no longer required. This relaxation simplifies condition (iii) while retaining the benefits of a fully regression-based approach. 

\subsection{Cross fitting as an alternative to Donsker conditions} \label{subsec:sample_split}

Our various estimators of the ATE and ATT can be made robust to violations of the Donsker condition by using sample splitting for nuisance parameter estimation, yielding what is commonly referred to as cross-validated TMLE \citep{zheng2010asymptotic} or double/debiased machine learning \citep{double17chernozhukov}. 

To implement cross-fitting, the data are partitioned into $K$ approximately equal, non-overlapping folds indexed by $S_i \in \{1, \dots, K\}$. For each fold $k$, nuisance parameters $\Q$ are estimated on the data excluding fold $k$, yielding $\hat{\Q}^{(-k)}$. These estimates are then used to evaluate the EIF and generate a cross-fitted one-step estimator or TMLE.

For example, the $k$-th fold version of the one-step estimator $\psi_1^+$ is:
\vspace{-0.35cm}
\begin{align}
\psi_{1,k}^{+, \text{cf}}(\hat{\Q}^{(-k)})
= \frac{1}{n_k} \sum_{i: S_i=k} &
\frac{\hat{f}^{(-k)}_M(M_i \mid a_0, X_i)}{\hat{f}^{(-k)}_M(M_i \mid A_i, X_i)} \big\{ Y_i - \hat{\mu}^{(-k)}(M_i, A_i, X_i) \big\}  \\
&+ \frac{\mathbb{I}(A_i = a_0)}{\hat{\pi}^{(-k)}(a_0 \mid X_i)} \big\{ \hat{\xi}^{(-k)}(M_i, X_i) - \hat{\theta}^{(-k)}(X_i) \big\}
\hat{\eta}^{(-k)}(A_i, X_i) \sp , \notag 
\end{align}

\vspace{-0.35cm}
\noindent where $\hat{\xi}^{(-k)}$, $\hat{\theta}^{(-k)}$, and $\hat{\eta}^{(-k)}$ are computed as before using the $k$-specific nuisance estimates. The final cross-fitted one-step estimator averages over all folds: $\psi_{1}^{+, \text{cf}}(\hat{\Q}) = \frac{1}{K} \sum_{k=1}^K  \psi_{1,k}^{+, \text{cf}}(\hat{\Q}^{(-k)})$.  

For cross-fitted TMLE, the targeting step is performed using fold-specific submodels that share a common fluctuation parameter. For example, to update $\hat{\pi}$ in $\psi_1(\hat{\Q}^\star)$, we may define for each $k$:  
\vspace{-0.35cm}
\begin{align*}
    \hat{\pi}^{(-k)}\big(\varepsilon_A; \hat{\mu}^{(-k)}, \hat{f}_M^{(-k)}\big)\left(1 \! \mid \! X\right) = \mbox{expit}\Big\{ \mbox{logit}\{ \hat{\pi}^{(-k)}(1 \mid X) \} + \varepsilon_A \big\{ \hat{\eta}^{(-k)}(1, X) - \hat{\eta}^{(-k)}(0, X)\big\} \Big\} \sp ,
\end{align*}

\vspace{-0.35cm}
\noindent and obtain a shared fluctuation parameter $\hat{\varepsilon}_A$ via pooled empirical risk minimization:
\vspace{-0.35cm}
\begin{align*}
    \hat{\varepsilon}_A = \underset{\varepsilon_A \in \mathbb{R}}{\mbox{arg min}} \sum_{k=1}^K \Pnk \ L_A\big(\hat{\pi}^{(-k)}\big(\varepsilon_A; \hat{\mu}^{(-k)}, \hat{f}_M^{(-k)}\big) \big) \sp ,
\end{align*}

\vspace{-0.35cm}
\noindent where $\Pnk$ is the empirical distribution of the $k$-th held-out sample. Analogous submodels can be defined for $\mu$ and $f_M$ (Section \ref{subsec:est_with_density}) to generate a cross-fitted TMLE. 

Cross-fitted estimators retain asymptotic linearity under conditions similar to our earlier theorems, without requiring the Donsker condition (A1). We omit formal statements for brevity.

\section{Model evaluation and semiparametric efficiency gains}
\label{sec:testing}

The assumptions of no unmeasured confounding and no direct effect of $A$ on $Y$ are untestable under the front-door model, which is \textit{nonparametrically saturated} such that it imposes no restrictions on the observed data distribution $\P$. However, \citet{bhattacharya2022testability} proposed methods for evaluating these assumptions when an \textit{anchor} variable $Z$ is present. An anchor variable is a pre-treatment variable associated with $A$ (and possibly $M$), but not a direct cause of $Y$; i.e., it influences $Y$ only through $A$ and $M$. In practice, $Z$ can often be viewed as a baseline analogue of the mediator—e.g., pre-vaccine antibody levels when $M$ denotes post-vaccine immune response. 

The anchor condition (no direct effect of $Z$ on $Y$) induces a \textit{generalized independence constraint}—also known as a \textit{Verma} or \textit{dormant} constraint \citep{verma1990equivalence, shpitser08dormant}—in $\P$ over $O = (X, Z, A, M, Y)$. Such constraints arise as independence relations in truncated or post-intervention distributions. In the anchor-included front-door model, the Verma takes the form $Z \perp Y^m \! \mid \! X$, equivalent to $Z \perp Y$ in the truncated distribution $\P(O)/\P(M \! \mid \! A, Z, X)$; see Appendix~\ref{app:testing_verma} for details. This constraint underlies the parametric tests of \citet{bhattacharya2022testability} for assessing the joint validity of conditional ignorability and the absence of a direct effect of $A$ on $Y$ (see their proof of Theorem~1 and Appendix~B).  

We advance anchor variable testing on three fronts. First, we generalize the prior parametric tests to allow flexible nuisance estimations, e.g., based on modern machine learning, yielding a flexible weighted risk minimization framework (Section~\ref{subsec:weighted_risk_min}). Second, we introduce a novel \textit{doubly robust} test based on a \textit{conditional counterfactual means}, which remains valid under partial model misspecification and is particularly well-suited to settings where the anchor and mediator are discrete or can be discretized (Section~\ref{subsec:dr_CATE_test}). Third, we show that when the Verma constraint holds, it can be leveraged to construct more efficient estimators for causal effects (Section~\ref{subsec:verma}). 

\subsection{Testing via weighted risk minimizations} 
\label{subsec:weighted_risk_min}

The Verma constraint $Z \perp Y^m \mid X$ is equivalently expressed as $Z \perp Y^a \mid X, M^a$ (see Theorem~1 in \citep{bhattacharya2022testability} and Appendix~\ref{app:testing_verma}), which implies that, under the null that the front-door assumptions hold, the conditional distribution $\P(Y^a \mid M^a, Z, X)$ is invariant to $Z$. Here, we test a specific implication of this constraint: that the conditional mean $\E(Y^a \mid M^a, Z, X)$ should be invariant in $Z$. While this implication is weaker than full distributional invariance, it suffices for evaluating identification of the causal effects. Under the null, the following MSE risk minimizers coincide (and for binary $Y$, so do the corresponding distributions): 
\vspace{-0.35cm}
\begin{equation}\label{eq:counterfactual_risk_minimizer}
\begin{aligned}
    \mu^a_\text{primal}(m, z, x) &\coloneqq \argmin_{\tilde{\mu} \in \mathcal{M}_\mu} \int (y - \tilde{\mu}(m, z, x))^2 \sp \diff \P(Y^a=y, M^a=m, z, x) \sp ,  \\ 
    \mu^a_\text{primal}(m, x) &\coloneqq \argmin_{\tilde{\mu} \in \mathcal{M}_\mu} \int (y - \tilde{\mu}(m, x))^2 \sp \diff \P(Y^a=y, M^a=m, x) \sp . 
\end{aligned}
\end{equation}

\vspace{-0.35cm}
The minimizers in \eqref{eq:counterfactual_risk_minimizer} can be re-expressed as weighted risk minimizers under $\P$ (see Appendix~\ref{app:testing_verma} for identification details):
\vspace{-0.35cm}
\begin{equation}\label{eq:primal_risk_minimizer}
\begin{aligned}
\mu^a_\text{primal}(m, z, x) &= \argmin_{\tilde{\mu} \in \mathcal{M}_\mu} \sp \E\big( \q_\text{primal}(A \mid Y, M, Z, X) \sp (Y - \tilde{\mu}(M, Z, X))^2 \big) \sp , \\
\mu^a_\text{primal}(m, x) &= \argmin_{\tilde{\mu} \in \mathcal{M}_\mu} \sp \E\big( \q_\text{primal}(A \mid Y, M, Z, X) \sp (Y - \tilde{\mu}(M, X))^2 \big) \sp , 
\end{aligned}
\end{equation}
where $\q_\text{primal}$ is the \textit{primal weight} \citep{bhattacharya2022semiparametric}, defined as
\vspace{-0.35cm}
\begin{align*}
    \q_\text{primal}(A \mid Y, M, Z, X) = \frac{\sum_{a'} \pi(a' \mid Z, X) \sp f_Y(Y \mid M,a',Z,X)}{\pi(A \mid Z,X) \sp f_Y(Y \mid M, A, Z, X)} \sp .
\end{align*}

\vspace{-0.35cm}
To implement the test via \eqref{eq:primal_risk_minimizer}, we first estimate $\q_\text{primal}$ using models for the propensity score $\pi(A = a \! \mid \! Z, X) \coloneqq \p(A=a \! \mid \! Z, X)$ and the conditional outcome density $f_Y(Y\mid M,A,Z,X)\coloneqq\p(Y\mid M,A,Z,X)$. Notably, the outcome density ratio can be estimated via Bayes’ rule from $\p(A \mid Y, M, Z, X)$ and $\p(A \mid M, Z, X)$. 
Given the estimate $\hat{\q}_\text{primal}$, we fit two primal-weighted regressions of $Y$ on $(M, Z, X)$ and $(M, X)$ to estimate the minimizers in \eqref{eq:primal_risk_minimizer}. 
We define the \textit{primal test statistic} as the difference in empirical MSE risks: 
\vspace{-0.35cm}
\begin{align}
T_{\mathrm{n}, \text{primal}} = \frac{1}{n}\sum_{i = 1}^n \big\{ (Y_i - \hat{\mu}^a_\text{primal}(M_i,X_i))^2 - (Y_i - \hat{\mu}^a_\text{primal}(M_i,Z_i,X_i))^2 \big\} \sp .
\label{eq:primal_test_statistic}
\end{align}

\vspace{-0.35cm}
To approximate the null distribution of $T_{\mathrm{n}, \text{primal}}$, we adopt a permutation approach \citep{paschali2022bridging}. Specifically, we permute the values of $Z$ across observations, refit the two weighted regressions, and recompute the primal test statistic. Repeating this procedure multiple times yields a reference distribution under the null. The one-sided $\p$-value is computed as the proportion of permuted test statistics greater than or equal to the observed value. 

This permutation-based approach remains valid even when regression models are fit using flexible machine learning methods due to the nonparametric nature of the test \citep{paschali2022bridging}. Unlike bootstrap procedures—which may break down in non-Donsker settings or yield unstable results with complex learners—the permutation test relies only on the assumption that, under the null, the primal-weighted distribution of $Y$ is invariant to permutations of $Z$ given $(M, X)$. This form of conditional exchangeability ensures the validity of the test without requiring asymptotic approximations or regularity conditions on the estimators.

The validity of the primal test relies on correct specification of both the treatment and outcome models: $\pi$, $f_Y$. \citet{bhattacharya2022testability} proposed a complementary parametric test based on the following \textit{dual weight}, which re-weights $\P$ using $f_M(M \! \mid \! A, Z, X) \coloneqq \p(M \! \mid \! A, Z, X)$:
\vspace{-0.35cm}
\begin{align*}
\q_\text{dual}(M \mid A, Z, X) = {f_M(M \mid a, Z, X)}/{f_M(M \mid A, Z, X)} \sp .
\end{align*}

\vspace{-0.5cm}
\noindent The counterfactual risk minimizations in \eqref{eq:counterfactual_risk_minimizer} can be implemented via weighted least squares using $\q_\text{dual}$ (see Appendix \ref{app:testing_verma} for a proof.) Consequently, replacing  $\q_\text{primal}$ with $\q_\text{dual}$ in \eqref{eq:primal_risk_minimizer} yields a nonparametric dual test. To implement it, we first estimate $\q_\text{dual}$ (e.g., via density‐ratio estimation or Bayes‐rule decomposition). With the resulting estimate $\hat{\q}_\text{dual}$, we fit two weighted regressions of $Y$ on $(M, Z, X)$ and $(M, X)$, yielding estimates $\hat{\mu}^a_{\text{dual}}(M, Z, X)$ and $\hat{\mu}^a_{\text{dual}}(M, X)$, respectively. The \textit{dual test statistic} is defined analogously to the primal case: 
\vspace{-0.35cm}
\begin{align}
T_{\mathrm{n}, \text{dual}} = \frac{1}{n} \sum_{i=1}^n \big\{ (Y_i - \hat{\mu}^a_{\text{dual}}(M_i,X_i))^2 - (Y_i - \hat{\mu}^a_{\text{dual}}(M_i,Z_i,X_i))^2 \big\} \sp .
\label{eq:dual_test_statistic}
\end{align}

\vspace{-0.35cm}
As in the primal test, we approximate the null distribution of $T_{\mathrm{n}, \text{dual}}$ using a permutation procedure. We repeatedly permute the values of $Z$, refit the weighted regressions, and recalculate the test statistic. The one-sided $\p$-value is defined as the proportion of permuted statistics less than or equal to the observed $T_{\mathrm{n}, \text{dual}}$. This approach supports flexible or nonparametric regressions while maintaining valid inference under the null. 

While the primal and dual tests offer complementary strengths—the former relying on treatment and outcome models, the latter on the mediator model—each requires correct specification of at least one set of nuisance components. In practice, model misspecification can undermine the validity of either test, and conflicting results may be difficult to interpret. This motivates our next \textit{doubly robust} test based on the invariance of a conditional counterfactual mean (CCM). 

\subsection{A doubly robust test} 
\label{subsec:dr_CATE_test}

We assume $M$ and $Z$ are discrete (or discretized), deferring continuous‐valued cases to future work. Under the Verma $Z \perp Y^m \! \mid \! X$, we have $\mu^m(z,x) \coloneqq \E(Y^m \! \mid \! Z = z, X = x)$ constant in $z$, for every $(m, x)$. When $X$ is discrete and low‐dimensional, one can test pointwise invariance by checking $\mu^m(1,x)=\mu^m(0,x)$ within each stratum of $X$ via a Wald‐type test (see Appendix \ref{app:testing_dr}). However, if $X$ is  continuous or high‐dimensional, this approach is not feasible due to the curse-of-dimensionality. In this instance, we suggest that a test could be based on the  marginalized quantity $\mu^m(z) \coloneqq \int \mu^m(z, x) \sp \p(x) \sp \diff x$, and test a weaker null: $\Delta(m) \coloneqq \mu^m(1) - \mu^m(0) = 0$. This test has the advantage of being based on a pathwise differentiable parameter $\Delta(m)$, allowing the utilization of doubly robust methods, as described below. However, depending on the structure of $\mu^m(z,x)$, it may have limited power against some alternatives. Nevertheless, characterizing a robust test based on the marginal parameter $\Delta(m)$ may prove useful in many settings.

Let $\Delta_n$ denote a vector of estimated contrasts $\Delta_n(m)$ for each $m$. Let $\Sigma_n$ denote an estimate of the asymptotic variance-covariance matrix of $n^{1/2} \Delta_n$, which can generally be obtained as the empirical covariance matrix of estimated influence functions. A Wald-style test statistic is defined as $T_{\mathrm{n}, \text{CCM}} \coloneqq \Delta_n^\top \Sigma_n^{-1} \Delta_n / n.$ Under the null and in large samples, $T_{\mathrm{n}, \text{CCM}}$ is approximately Chi-squared distributed with $d$ degrees of freedom, where $d$ is the dimension of $\Delta$. Comparison of the test statistic to relevant quantiles of this distribution allows for appropriate hypothesis tests with correct asymptotic size.

To implement this test, we require robust estimates of both $\Delta$ and the covariance matrix $\Sigma$. Estimators of $\Delta$ are motivated by the identification result that (see Appendix~\ref{app:testing_dr} for proof)
\vspace{-0.35cm}
\begin{align}
\mu^m(z) = \int \sum_{a} \mu(m, a, z, x) \sp \pi(a \mid z, x) \sp \p(x) \sp \diff x \sp .
\label{eq:cate_test_id}
\end{align}

\vspace{-0.35cm}
\noindent Plug-in estimators based on \eqref{eq:cate_test_id} may suffer from the \emph{g-null paradox} \citep{robins1997estimation}, whereby parametric estimation of both $\mu$ and $\pi$ can lead to invalid tests that reject the null even when it holds. This motivates the usage of influence-function-based estimators that remain valid under flexible nonparametric estimation of nuisance components—even when convergence rates fall below root-$n$. For example, a one-step estimator of $\mu^m(z)$ can be computed as follows. We define $f_Z(Z \!\mid \! X) \coloneqq \p(Z \! \mid \! X)$ and propose the estimator (see detailed detivation in Appendix~\ref{app:testing_dr}):

\vspace{-1.1cm}
{\small 
\begin{align}
\hat{\mu}^{+,m}(z) \! &= \! \frac{1}{n} \sum_{i=1}^n \frac{\I(Z_i = z, M_i = m)}{\hat{f}_M(m \mid A_i, z, X_i) \sp \hat{f}_Z(z \mid X_i)}  (Y_i - \hat{\mu}(m, A_i, z, X_i)) \! + \! \sum_a \hat{\mu}(m, a, z, X_i)\sp \hat{\pi}(a \! \mid \! z, X_i) \notag  \\
&\hspace{1.5cm} + \frac{\I(Z_i = z)}{\hat{f}_Z(z \mid X_i)} \big( \hat{\mu}(m, A_i, z, X_i) - \sum_a \hat{\mu}(m, a, z, X_i)\sp  \hat{\pi}(a \mid z, X_i) \big)  \sp . 
\label{eq:one-step_doubly_robuts_test}
\end{align}
}

\vspace{-0.35cm}
\noindent The above estimator, and the TMLE counterpart \citep{gruber2010targeted}, exhibit doubly-robust consistency for $\mu^m(z)$ if either $(\hat{\pi}, \hat{\mu})$ or $(\hat{f}_M, \hat{f}_Z)$ are consistent.

While doubly‐robust estimation of $\mu^m(z)$ (and thereby $\Delta$) is straightforward, ensuring a doubly-robust estimate of the variance‐covariance matrix $\Sigma$ is more challenging. The challenge arises from the fact that under inconsistent estimation of nuisance parameters, the one-step (TMLE) estimate of $\mu^m(z)$, while consistent, will not generally be asymptotically linear, unless it is based on working parametric models. However, as noted above their use in this case is susceptible to the g-null paradox and therefore is not recommended. If flexible regressions with slower-than-parametric convergence rates are adopted, then additional effort is required to ensure doubly robust asymptotic linearity \citep{van2014targeted} and generally this has only been demonstrated to be feasible using TMLE \citep{benkeser2017doubly}.

Thus, we propose to adopt these TMLE-based methods to develop a doubly robust hypothesis test. This involves a careful analysis of the second-order remainder term (see Appendix~\ref{app:testing_dr} for details). We refer to this test as DR-CCM.


The three tests offer flexible validation of the front-door model.  DR-CCM is doubly robust but limited to discrete mediators/anchors; the primal test handles continuous or multivariate settings under correct treatment and outcome models; and the dual test only requires a correct mediator model. The test should be based on which nuisance component can be most reliably estimated.

\subsection{Efficiency gains under the Verma constraint}
\label{subsec:verma}

When the Verma constraint holds (i.e., under the null), it imposes structural restrictions on the observed data distribution, shrinking the statistical model and enabling the construction of more efficient estimation of causal effects. We illustrate this in the context of estimating $\E(Y^{a_0})$. 

Under the front-door model with an anchor variable $Z$, we define a family of identification functionals for $\E(Y^{a_0})$, each indexed by a fixed level $z^*$ in the state space $\mathcal{Z}$ of $Z$: 
\vspace{-0.35cm}
\begin{align}
\psi_{z^*}(\Q) = \iiint \sum_{a=0}^1 \mu(m, a, z^*, x) \sp \pi(a \mid z^*, x) \sp f_M(m \mid A=a_0, z, x) \sp \p(z,x) \sp \diff m \sp \diff z \sp \diff x \sp . 
\label{eq:id_ATE_zstar}
\end{align}

\vspace{-0.35cm}
\noindent See Appendix~\ref{app:testing_eff_gain} for an identification proof. 
Although $\psi_{z^*}(\Q)$ equals $\E(Y^{a_0})$ for all $z^* \in \mathcal{Z}$, the efficiency of plug-in or influence-function-based estimators may vary with the choice of $z^*$. Below, we focus on one-step estimators that avoid density estimation and show how to exploit this structure to improve efficiency, beginning with the case where $Z$ is discrete. 

\vspace{0.15cm}
\noindent \textbf{Estimation under discrete \texorpdfstring{$Z$}{Z}.}
A one-step estimator for \eqref{eq:id_ATE_zstar} can be constructed using this set of nuisance functions: $\p_{ZX}(z,x)\coloneqq \p(Z=z,X=x)$, $f_Z(z \! \mid \! x)$, $\pi(a \! \mid \! z, x)$, $\mu(m, a, z, x)$, $\xi_{z^*}(m,x) \coloneqq \sum_a \mu(m, a, z^*, x) \sp \pi(a \! \mid \! z^*, x)$, $\gamma_{z^*}(z,x)\coloneqq \E(\xi_{z^*}(M,X)\! \mid \! a_0,z,x)$, 
$\kappa_{a, z^*}(z, x) \coloneqq \E(\mu(M, a, z^*, X) \! \mid \! a_0, z, x)$, and
$f_{M, z^*}^r(m, a, z, x) \coloneqq {f_M(m \! \mid \! a_0, z, x)}/{f_M(m \! \mid \! a, z^*, x)}$. 
Let $\Q = \{\mu,\pi,\xi_{z^*}, \gamma_{z^*},\kappa_{a, z^*}, f_{M, z^*}^r,f_Z,\p_{ZX}\}$. Given the nuisance estimates, $\hat{\Q}$, the one-step estimator is given as $\psi^{+}_{z^*}(\hat{\Q}) =  \frac{1}{n}\sum_{i = 1}^n \Phi_{z^*}(\hat{\Q})(O_i) + \hat{\gamma}_{z^*}(Z_i,X_i)$, where $\Phi_{z^*}(\Q)$ denotes the np-EIF of \eqref{eq:id_ATE_zstar} and is given by (see a proof in Appendix~\ref{app:testing_eff_gain}): 
\vspace{-0.35cm}
\begin{align}
\Phi_{z^*}(\Q)(O_i) 
   &=  \frac{\I(Z_i=z^*)}{f_Z(z^*\mid X_i) }\sum_z f^r_{M, z^*}(M_i, A_i, z, X_i) \sp f_Z(z\mid X_i) \sp \big(Y_i-\mu(M_i,A_i,z^*,X_i) \big) \label{eq:verma_if}  \\
   &\hspace{0.5cm}+\frac{\I(Z=z^*)}{f_Z(z^*\mid X_i)}(A_i-\pi(1\mid z^*,X_i))\ \sum_z\big(\kappa_{1, z^*}(z,X_i)-\kappa_{0, z^*}(z,X_i)\big)f_Z(z\mid X_i) 
   \notag \\ 
    &\hspace{0.5cm}+ \frac{\I(A_i=a_0)}{\pi(a_0\mid Z_i,X_i)}\big(\xi_{z^*}(M_i,X_i) - \gamma_{z^*}(Z_i,X_i)\big) 
   +\gamma_{z^*}(Z_i,X_i) - \psi_{z^*}(\Q) \sp .  \notag 
\end{align}

\vspace{-0.35cm}
\noindent Although the estimand in \eqref{eq:id_ATE_zstar} is invariant to the choice of $z^*$, the efficiency of the estimator $\psi^+_{z^*}(\hat{\Q})$ generally is not. To explore this, we define a \textit{class of influence functions} formed by convex combinations of the EIFs corresponding to different anchor levels. Under binary $Z$, this class is 
\vspace{-0.5cm}
\begin{align}
\Lambda_{\alpha} \coloneqq  \big\{ \alpha \sp \Phi_{z^*=1}(\Q) + (1 - \alpha) \sp \Phi_{z^*=0}(\Q),  \quad \text{for } \alpha \in [0,1] \sp  \big\}  \sp . 
\label{eq:class_of_IFs}    
\end{align}

\vspace{-0.5cm}
\noindent For any fixed $\alpha \in [0,1]$ and $\hat{\Q}$, we define the aggregated estimator as $\psi^{+}_{\alpha}(\hat{\Q}) \coloneqq \alpha \sp \psi^{+}_{z^*=1}(\hat{\Q}) + (1 - \alpha) \sp \psi^{+}_{z^* = 0}(\hat{\Q})$. When $\alpha = 0$ or $1$, this reduces to $\psi^{+}_{z^*=0}(\hat{\Q})$ or $\psi^{+}_{z^*=1}(\hat{\Q})$, respectively. To improve efficiency, we derive an \textit{optimal weight} $\alpha^{\mathrm{opt}}$ that minimizes the asymptotic variance of the aggregated estimator, given by the variance of the combined influence functions: 
\vspace{-0.5cm}
\begin{align}
\label{eq:eff_alpha}
\alpha^{\mathrm{opt}} \coloneqq \argmin_{\alpha \in [0,1]} \E\big( \big\{ \alpha \sp \Phi_{z^*=1}(\Q) + (1 - \alpha) \sp \Phi_{z^*=0}(\Q) \big\}^2 \big) \sp . 
\end{align}

\vspace{-0.5cm}
\noindent The minimizer has a closed form: $\alpha^{\mathrm{opt}} = \E(\Phi_{z^*=0}(\Q)(\Phi_{z^*=0}(\Q)-\Phi_{z^*=1}(\Q)))/\E((\Phi_{z^*=1}(\Q)-\Phi_{z^*=0}(\Q))^2)$ (see Appendix~\ref{app:testing_eff_gain} for a proof). An estimator of this weight, denoted as $\hat{\alpha}^{\mathrm{opt}}$, is given by $P_n(\Phi_{z^*=0}(\hat{\Q})(\Phi_{z^*=0}(\hat{\Q})-\Phi_{z^*=1}(\hat{\Q})))/P_n((\Phi_{z^*=1}(\hat{\Q})-\Phi_{z^*=0}(\hat{\Q}))^2)$. The resulting \textit{optimally weighted estimator} is $\psi^{+}_{\alpha^\mathrm{opt}}(\hat{\Q}) = \hat{\alpha}^{\mathrm{opt}} \sp \psi^{+}_{z^*=1}(\hat{\Q}) + (1 - \hat{\alpha}^{\mathrm{opt}}) \sp \psi^{+}_{z^*=0}(\hat{\Q})$. 

\vspace{0.15cm}
\textbf{Extension to continuous \texorpdfstring{$Z$}{Z}.}
When $Z$ is continuous, the functional $\psi_{z^*}(\Q)$ in \eqref{eq:id_ATE_zstar} is not pathwise differentiable, so a von Mises expansion does not apply. One practical solution is to discretize $Z$ using meaningful cutoffs and apply the discrete methods. Alternatively, one can define an integrated functional by averaging $\psi_{z^*}(\Q)$ over a reference distribution $\tilde{\p}(Z)$ with the same support as the true marginal of $Z$ (see Appendix~\ref{app:testing_eff_gain}):  
\vspace{-0.35cm}
\begin{align}
    \psi_{\tilde{\p}}(\Q) = \iiint \! \bigg\{ \int \sum_a \mu(m, a, z, x) \sp \pi(a \mid z, x) \sp \tilde{\p}(z) \sp \diff z \bigg\}  f_M(m \mid a_0, z, x) \sp \p(z,x) \sp \diff m \sp \diff z \sp \diff x \sp . 
    \label{eq:id_ATE_ptilde}
\end{align}

\vspace{-0.35cm}
\noindent As in the discrete case, the estimand remains invariant to the choice of $\tilde{\p}$, though the efficiency of the resulting estimator may depend on it. Details on constructing one-step estimators based on $\psi_{\tilde{\p}}$ and leveraging the Verma constraint in this setting are provided in Appendix~\ref{app:testing_eff_gain}.

\section{Simulation studies}
\label{sec:sims}

All numerical results, including simulation and empirical analyses, are fully reproducible using scripts available in the GitHub repository \href{https://github.com/annaguo-bios/fd-methods}{\texttt{annaguo-bios/fd-methods}}, which provides labeled code for each figure and table, synthetic datasets, and instructions for accessing the real data sets. An accompanying \href{https://github.com/annaguo-bios/fdcausal}{\texttt{fdcausal}} \textsf{R} package implements the proposed methods for broader use.

We conducted six sets of simulation studies, each targeting a distinct methodological question addressed in this paper.
(1) \textit{Theoretical properties}: Assessed the asymptotic behavior of the ATE and ATT estimators under various settings, including both uni- and multivariate mediators. This scenario also compared TMLEs using linear versus nonlinear submodels. 
(2) \textit{Weak overlap:} Examined the potential finite-sample advantages of TMLEs over one-step estimators for both ATE and ATT under weak treatment overlap;  
(3) \textit{Model misspecification:} Evaluated the robustness of the ATE and ATT estimators when nuisance models were correctly specified versus misspecified;
(4) \textit{Cross-fitting:} Investigated whether cross-fitting improves performance for TMLE and one-step estimators of ATE and ATT in settings prone to overfitting; 
(5) \textit{Model evaluation:} Analyzed type I error and power of our three proposed tests for validity of the front-door model assumptions under various null and alternative scenarios; and 
(6) \textit{Efficiency gain:} Demonstrated that incorporating the Verma constraint within a semiparametric model improves the efficiency of ATE estimation. 

ATE was estimated as contrasts of the estimated $\psi_{a_0}(\P)$ for $a_0 \in \{0,1\}$, following Section~\ref{sec:est_ATE}. ATT was estimated by estimating $\beta_{a_0}(\P)$ for $a_0=0$, following Section~\ref{sec:est_ATT}, and subtracting it from the empirical mean of $Y$ among individuals with $A = 1$. With slight abuse of notation, we use the same symbols to represent the corresponding contrasts in the ATE and ATT estimators. 


\vspace{0.15cm}
\noindent \textbf{Simulation 1: Theoretical properties.} 
We assessed asymptotic bias and variance of our ATE and ATT estimators across mediators (binary, univariate to four-dimensional continuous) using parametric and kernel nuisance fits, confirming $\sqrt{n}$‐bias decay and variance convergence to $\P[\Phi(\Q)^2]$ (Appendix~\ref{app:sims:consistency}, ATE: Figs~\eqref{fig:binary}--\eqref{fig:d4}; ATT: Figs~\eqref{fig:att_binary}--\eqref{fig:att_d4}).  We also compared linear versus expit TMLE submodels on bias, standard deviation (SD), mean squared error (MSE), and 95\% confidence interval (CI) coverage and width for select mediators, finding both valid under correct model specification (Appendix~\ref{app:sims:consistency}, ATE: Table~\ref{table:TMLEs_ATE}; ATT: Table~\ref{table:TMLEs_ATT}). 

\vspace{0.15cm}
\noindent \textbf{Simulation 2: Weak overlap.} 
We evaluated TMLE and one-step estimators for ATE and ATT under weak overlap, induced by assigning $A \! \mid \! X = x \sim \text{Bernoulli}(0.001 + 0.998x)$ for $X \sim \text{Uniform}(0,1)$, yielding near-deterministic probabilities. See Appendix~\ref{app:sims:overlap} for details. 

We considered three mediator settings: univariate binary, univariate continuous, and bivariate continuous. For each, we implemented practical ATE estimators. In the binary case, we used $\psi_1^+(\hat{\Q})$ and $\psi_1(\hat{\Q}^\star)$, leveraging the ease of modeling binary mediator densities. For continuous mediators, we included $\psi_1^+(\hat{\Q})$, $\psi_1(\hat{\Q}^\star)$, $\psi_{2a}^+(\hat{\Q})$, $\psi_{2a}(\hat{\Q}^\star)$, $\psi_{2b}^+(\hat{\Q})$, and $\psi_{2b}(\hat{\Q}^\star)$. Mediator-related nuisance functions were estimated using kernel density estimation, density-ratio methods, and Bayes-based regression. ATT estimators were constructed analogously. 

Based on 1000 replicates at sample sizes of 500, 1000, and 2000, we assessed bias, SD, MSE, CI coverage, and width. ATE results, provided in Table~\ref{table:weakoverlap_ATE}, show comparable bias across estimators, but TMLEs had lower SD and narrower CIs, yielding reduced MSE across all mediator types and sample sizes. For the binary mediator case, the TMLE exhibits slight undercoverage at smaller sample sizes of 500 and 1000, while coverage approaches the nominal 95\% level as the sample size increases. ATT results appear in Appendix~\ref{app:sims:overlap}, Table~\ref{table:weakoverlap_ATT}. 

  \providecommand{\huxb}[2]{\arrayrulecolor[RGB]{#1}\global\arrayrulewidth=#2pt}
  \providecommand{\huxvb}[2]{\color[RGB]{#1}\vrule width #2pt}
  \providecommand{\huxtpad}[1]{\rule{0pt}{#1}}
  \providecommand{\huxbpad}[1]{\rule[-#1]{0pt}{#1}}

\begin{table}[t]

\captionsetup{justification=centering,singlelinecheck=off}
\caption{Comparison of ATE TMLE and one-step estimators under weak overlap across mediator types.}
 \setlength{\tabcolsep}{0pt}
 \renewcommand{\arraystretch}{0.6}
\resizebox{\textwidth}{!}{
\begin{tabular}{l l l l l l l l l l l l l l l}

\noalign{\hrule height 1pt} 

\multicolumn{1}{!{\huxvb{0, 0, 0}{0}}l!{\huxvb{0, 0, 0}{0}}}{\huxtpad{-1pt + 1em}\raggedright \hspace{-1pt} \textbf{{\fontsize{6pt}{7.2pt}\selectfont }} \hspace{-1pt}\huxbpad{-1pt}} &
\multicolumn{1}{l!{\huxvb{0, 0, 0}{0}}}{\huxtpad{-1pt + 1em}\raggedright \hspace{-1pt} {\fontsize{6pt}{7.2pt}\selectfont } \hspace{-1pt}\huxbpad{-1pt}} &
\multicolumn{2}{c!{\huxvb{0, 0, 0}{0}}}{\huxtpad{-1pt + 1em}\centering \hspace{-1pt} {\fontsize{6pt}{7.2pt}\selectfont Univariate Binary} \hspace{-1pt}\huxbpad{-1pt}} &
\multicolumn{6}{c!{\huxvb{0, 0, 0}{0}}}{\huxtpad{-1pt + 1em}\centering \hspace{-1pt} {\fontsize{6pt}{7.2pt}\selectfont Univariate Continuous} \hspace{-1pt}\huxbpad{-1pt}} &
\multicolumn{4}{c!{\huxvb{0, 0, 0}{0}}}{\huxtpad{-1pt + 1em}\centering \hspace{-1pt} {\fontsize{6pt}{7.2pt}\selectfont Bivariate Continuous} \hspace{-1pt}\huxbpad{-1pt}} \tabularnewline
\end{tabular}}\label{table:weakoverlap_ATE}

\end{table}

\vspace{0.15cm}
\noindent \textbf{Simulation 3: Model misspecification.} 
We evaluated the sensitivity of the ATE and ATT estimators to model misspecification by comparing misspecified parametric models, arising from the omission of key interaction terms in the DGPs, with flexible nuisance estimation using Super Learner—an ensemble of GLMs, GAMs, random forests, SVMs, BART, and XGBoost \citep{van2007super}. To address potential Donsker violations from complex learners, we also included cross-fitted versions of all estimators; see Appendix~\ref{app:sims:misspecification} for details. 

Simulations used binary and continuous mediators, 1000 replicates, and sample sizes of 500, 1000, and 2000 (details in Appendix~\ref{app:sims:misspecification}). For binary mediators, we used $\psi_1^+(\hat{\Q})$ and $\psi_1(\hat{\Q}^\star)$; for continuous mediators, $\psi_{2a}(\hat{\Q}^\star)$, $\psi_{2b}(\hat{\Q}^\star)$, and their one-step analogues. ATE results, provided in Table~\ref{table:misspecification_ATE}, show that misspecified models led to bias and poor coverage, while Super Learner–based estimators reduced bias and improved coverage with increasing sample size. Some undercoverage persisted for $\psi_1$, and cross-fitting yielded limited additional gains. These results highlight the importance of flexible nuisance estimation. ATT findings (Appendix~\ref{app:sims:misspecification}, Table~\ref{table:misspecification_ATT}) were similar.

  \providecommand{\huxb}[2]{\arrayrulecolor[RGB]{#1}\global\arrayrulewidth=#2pt}
  \providecommand{\huxvb}[2]{\color[RGB]{#1}\vrule width #2pt}
  \providecommand{\huxtpad}[1]{\rule{0pt}{#1}}
  \providecommand{\huxbpad}[1]{\rule[-#1]{0pt}{#1}}

\begin{table}[!t]
\begin{center}
\captionsetup{justification=centering,singlelinecheck=off}
\caption{Performance of ATE estimators under model misspecifications across mediator types.}
 \setlength{\tabcolsep}{0pt}
\resizebox{1\textwidth}{!}{
\begin{tabular}{l l l l l l l l l l l l l l l l l l l l}

\thickline

\multicolumn{1}{!{\huxvb{0, 0, 0}{0}}l!{\huxvb{0, 0, 0}{0}}}{\huxtpad{-1pt + 1em}\raggedright \hspace{-1pt} \textbf{{\fontsize{7pt}{8.4pt}\selectfont }} \hspace{-1pt}\huxbpad{-1pt}} &
\multicolumn{1}{c!{\huxvb{0, 0, 0}{0}}}{\huxtpad{-1pt + 1em}\centering \hspace{-1pt} \textbf{{\fontsize{7pt}{8.4pt}\selectfont }} \hspace{-1pt}\huxbpad{-1pt}} &
\multicolumn{9}{c!{\huxvb{0, 0, 0}{0}}}{\huxtpad{-1pt + 1em}\centering \hspace{-1pt} \textbf{{\fontsize{7pt}{8.4pt}\selectfont TMLEs}} \hspace{-1pt}\huxbpad{-1pt}} &
\multicolumn{9}{c!{\huxvb{0, 0, 0}{0}}}{\huxtpad{-1pt + 1em}\centering \hspace{-1pt} \textbf{{\fontsize{7pt}{8.4pt}\selectfont One-step estimators}} \hspace{-1pt}\huxbpad{-1pt}} \tabularnewline
\end{tabular}}\label{table:misspecification_ATE}
\par\end{center}

\end{table}

\vspace{0.15cm}
\noindent \textbf{Simulation 4: Cross-fitting.} 
We examined the role of cross-fitting by focusing on random forests, which are known to perform poorly without sample splitting in high-dimensional settings \citep{double17chernozhukov, biau2012analysis}. Details are provided in Appendix~\ref{app:sims:cross-fit} (see Tables~\ref{table:crossfitting_ATE}-\ref{apptable:crossfitting_att}). 



\vspace{0.15cm}
\noindent \textbf{Simulation 5: Model evaluation.} 
We evaluated the performance of the proposed tests from Section~\ref{sec:testing} using simulations designed to assess type I error and power. Each scenario involved 200 replicates per sample size, with the rejection rate interpreted as type I error when the data-generating process satisfied front-door assumptions, and as power when it did not. We used four data-generating models: in DAG1, $Z$ has direct effects on both $A$ and $M$; in DAG2, $Z$ affects $A$ and shares unmeasured confounding with $M$—both satisfying the front-door conditions. Violations were introduced in DAG3, which includes unmeasured confounding between $A$–$M$ and $M$–$Y$, and in DAG4, which includes a direct effect of $A$ on $Y$. See Appendix~\ref{app:sims:tests} for details.  

We conducted three sets of simulations. The \textit{first} confirmed that all tests controlled type I error and gained power with increasing sample size under correctly specified models across various variable-type configurations (deferred to Appendix Table~\ref{table:test_three}). The \textit{second} examined model misspecification, highlighting the double-robustness of the DR-CCM test, shown in Table~\ref{table:test_DR}. The \textit{third} evaluated the dual and primal tests in continuous-variable settings, with and without Super Learner; while Super Learner mitigated type I error inflation under complex DGPs, it reduced power—likely due to increased estimator variance (deferred to Appendix Table~\ref{table:test_SL}). 

\providecommand{\huxb}[2]{\arrayrulecolor[RGB]{#1}\global\arrayrulewidth=#2pt}
  \providecommand{\huxvb}[2]{\color[RGB]{#1}\vrule width #2pt}
  \providecommand{\huxtpad}[1]{\rule{0pt}{#1}}
  \providecommand{\huxbpad}[1]{\rule[-#1]{0pt}{#1}}

\begin{table}[!t]
\begin{center}
\captionsetup{justification=centering,singlelinecheck=off}
\caption{Comparative analysis of DR-CCM, dual, and primal tests under model misspecifications.}
 \setlength{\tabcolsep}{0pt}
\resizebox{0.8\textwidth}{!}{\begin{tabular}{l l l l l l l l l l l l l}

\hhline{>{\huxb{0, 0, 0}{1}}->{\huxb{0, 0, 0}{1}}->{\huxb{0, 0, 0}{1}}->{\huxb{0, 0, 0}{1}}->{\huxb{0, 0, 0}{1}}->{\huxb{0, 0, 0}{1}}->{\huxb{0, 0, 0}{1}}->{\huxb{0, 0, 0}{1}}->{\huxb{0, 0, 0}{1}}->{\huxb{0, 0, 0}{1}}->{\huxb{0, 0, 0}{1}}->{\huxb{0, 0, 0}{1}}->{\huxb{0, 0, 0}{1}}-}
\arrayrulecolor{black}

\multicolumn{1}{!{\huxvb{0, 0, 0}{0}}l!{\huxvb{0, 0, 0}{0}}}{\huxtpad{-1pt + 1em}\raggedright \hspace{-1pt} \textbf{{\fontsize{6pt}{7.2pt}\selectfont }} \hspace{-1pt}\huxbpad{-1pt}} &
\multicolumn{4}{c!{\huxvb{0, 0, 0}{0}}}{\huxtpad{-1pt + 1em}\centering \hspace{-1pt} \textbf{{\fontsize{6pt}{7.2pt}\selectfont DR-CCM test}} \hspace{-1pt}\huxbpad{-1pt}} &
\multicolumn{4}{c!{\huxvb{0, 0, 0}{0}}}{\huxtpad{-1pt + 1em}\centering \hspace{-1pt} \textbf{{\fontsize{6pt}{7.2pt}\selectfont Dual test}} \hspace{-1pt}\huxbpad{-1pt}} &
\multicolumn{4}{c!{\huxvb{0, 0, 0}{0}}}{\huxtpad{-1pt + 1em}\centering \hspace{-1pt} \textbf{{\fontsize{6pt}{7.2pt}\selectfont Primal test}} \hspace{-1pt}\huxbpad{-1pt}} \tabularnewline[-0.5pt]

\hhline{>{\huxb{255, 255, 255}{0.4}}->{\huxb{0, 0, 0}{0.4}}->{\huxb{0, 0, 0}{0.4}}->{\huxb{0, 0, 0}{0.4}}->{\huxb{0, 0, 0}{0.4}}->{\huxb{0, 0, 0}{0.4}}->{\huxb{0, 0, 0}{0.4}}->{\huxb{0, 0, 0}{0.4}}->{\huxb{0, 0, 0}{0.4}}->{\huxb{0, 0, 0}{0.4}}->{\huxb{0, 0, 0}{0.4}}->{\huxb{0, 0, 0}{0.4}}->{\huxb{0, 0, 0}{0.4}}-}
\arrayrulecolor{black}

\multicolumn{1}{!{\huxvb{0, 0, 0}{0}}l!{\huxvb{0, 0, 0}{0}}}{\huxtpad{-1pt + 1em}\raggedright \hspace{-1pt} \textit{{\fontsize{6pt}{7.2pt}\selectfont }} \hspace{-1pt}\huxbpad{-1pt}} &
\multicolumn{2}{c!{\huxvb{0, 0, 0}{0}}}{\huxtpad{-1pt + 1em}\centering \hspace{-1pt} \textit{{\fontsize{6pt}{7.2pt}\selectfont Type I error}} \hspace{-1pt}\huxbpad{-1pt}} &
\multicolumn{2}{c!{\huxvb{0, 0, 0}{0}}}{\huxtpad{-1pt + 1em}\centering \hspace{-1pt} \textit{{\fontsize{6pt}{7.2pt}\selectfont Power}} \hspace{-1pt}\huxbpad{-1pt}} &
\multicolumn{2}{c!{\huxvb{0, 0, 0}{0}}}{\huxtpad{-1pt + 1em}\centering \hspace{-1pt} \textit{{\fontsize{6pt}{7.2pt}\selectfont Type I error}} \hspace{-1pt}\huxbpad{-1pt}} &
\multicolumn{2}{c!{\huxvb{0, 0, 0}{0}}}{\huxtpad{-1pt + 1em}\centering \hspace{-1pt} \textit{{\fontsize{6pt}{7.2pt}\selectfont Power}} \hspace{-1pt}\huxbpad{-1pt}} &
\multicolumn{2}{c!{\huxvb{0, 0, 0}{0}}}{\huxtpad{-1pt + 1em}\centering \hspace{-1pt} \textit{{\fontsize{6pt}{7.2pt}\selectfont Type I error}} \hspace{-1pt}\huxbpad{-1pt}} &
\multicolumn{2}{c!{\huxvb{0, 0, 0}{0}}}{\huxtpad{-1pt + 1em}\centering \hspace{-1pt} \textit{{\fontsize{6pt}{7.2pt}\selectfont Power}} \hspace{-1pt}\huxbpad{-1pt}} \tabularnewline[-0.5pt]

\hhline{>{\huxb{255, 255, 255}{0.4}}->{\huxb{0, 0, 0}{0.4}}->{\huxb{0, 0, 0}{0.4}}->{\huxb{0, 0, 0}{0.4}}->{\huxb{0, 0, 0}{0.4}}->{\huxb{0, 0, 0}{0.4}}->{\huxb{0, 0, 0}{0.4}}->{\huxb{0, 0, 0}{0.4}}->{\huxb{0, 0, 0}{0.4}}->{\huxb{0, 0, 0}{0.4}}->{\huxb{0, 0, 0}{0.4}}->{\huxb{0, 0, 0}{0.4}}->{\huxb{0, 0, 0}{0.4}}-}
\arrayrulecolor{black}

\multicolumn{1}{!{\huxvb{0, 0, 0}{0}}l!{\huxvb{0, 0, 0}{0}}}{\huxtpad{-1pt + 1em}\raggedright \hspace{-1pt} {\fontsize{6pt}{7.2pt}\selectfont n} \hspace{-1pt}\huxbpad{-1pt}} &
\multicolumn{1}{c!{\huxvb{0, 0, 0}{0}}}{\huxtpad{-1pt + 1em}\centering \hspace{-1pt} {\fontsize{6pt}{7.2pt}\selectfont DAG1} \hspace{-1pt}\huxbpad{-1pt}} &
\multicolumn{1}{c!{\huxvb{0, 0, 0}{0}}}{\huxtpad{-1pt + 1em}\centering \hspace{-1pt} {\fontsize{6pt}{7.2pt}\selectfont DAG2} \hspace{-1pt}\huxbpad{-1pt}} &
\multicolumn{1}{c!{\huxvb{0, 0, 0}{0}}}{\huxtpad{-1pt + 1em}\centering \hspace{-1pt} {\fontsize{6pt}{7.2pt}\selectfont DAG3} \hspace{-1pt}\huxbpad{-1pt}} &
\multicolumn{1}{c!{\huxvb{0, 0, 0}{0}}}{\huxtpad{-1pt + 1em}\centering \hspace{-1pt} {\fontsize{6pt}{7.2pt}\selectfont DAG4} \hspace{-1pt}\huxbpad{-1pt}} &
\multicolumn{1}{c!{\huxvb{0, 0, 0}{0}}}{\huxtpad{-1pt + 1em}\centering \hspace{-1pt} {\fontsize{6pt}{7.2pt}\selectfont DAG1} \hspace{-1pt}\huxbpad{-1pt}} &
\multicolumn{1}{c!{\huxvb{0, 0, 0}{0}}}{\huxtpad{-1pt + 1em}\centering \hspace{-1pt} {\fontsize{6pt}{7.2pt}\selectfont DAG2} \hspace{-1pt}\huxbpad{-1pt}} &
\multicolumn{1}{c!{\huxvb{0, 0, 0}{0}}}{\huxtpad{-1pt + 1em}\centering \hspace{-1pt} {\fontsize{6pt}{7.2pt}\selectfont DAG3} \hspace{-1pt}\huxbpad{-1pt}} &
\multicolumn{1}{c!{\huxvb{0, 0, 0}{0}}}{\huxtpad{-1pt + 1em}\centering \hspace{-1pt} {\fontsize{6pt}{7.2pt}\selectfont DAG4} \hspace{-1pt}\huxbpad{-1pt}} &
\multicolumn{1}{c!{\huxvb{0, 0, 0}{0}}}{\huxtpad{-1pt + 1em}\centering \hspace{-1pt} {\fontsize{6pt}{7.2pt}\selectfont DAG1} \hspace{-1pt}\huxbpad{-1pt}} &
\multicolumn{1}{c!{\huxvb{0, 0, 0}{0}}}{\huxtpad{-1pt + 1em}\centering \hspace{-1pt} {\fontsize{6pt}{7.2pt}\selectfont DAG2} \hspace{-1pt}\huxbpad{-1pt}} &
\multicolumn{1}{c!{\huxvb{0, 0, 0}{0}}}{\huxtpad{-1pt + 1em}\centering \hspace{-1pt} {\fontsize{6pt}{7.2pt}\selectfont DAG3} \hspace{-1pt}\huxbpad{-1pt}} &
\multicolumn{1}{c!{\huxvb{0, 0, 0}{0}}}{\huxtpad{-1pt + 1em}\centering \hspace{-1pt} {\fontsize{6pt}{7.2pt}\selectfont DAG4} \hspace{-1pt}\huxbpad{-1pt}} \tabularnewline[-0.5pt]

\hhline{>{\huxb{255, 255, 255}{0.4}}->{\huxb{0, 0, 0}{0.4}}->{\huxb{0, 0, 0}{0.4}}->{\huxb{0, 0, 0}{0.4}}->{\huxb{0, 0, 0}{0.4}}->{\huxb{0, 0, 0}{0.4}}->{\huxb{0, 0, 0}{0.4}}->{\huxb{0, 0, 0}{0.4}}->{\huxb{0, 0, 0}{0.4}}->{\huxb{0, 0, 0}{0.4}}->{\huxb{0, 0, 0}{0.4}}->{\huxb{0, 0, 0}{0.4}}->{\huxb{0, 0, 0}{0.4}}-}
\arrayrulecolor{black}

\multicolumn{1}{!{\huxvb{0, 0, 0}{0}}l!{\huxvb{0, 0, 0}{0}}}{\huxtpad{-1pt + 1em}\raggedright \hspace{-1pt} {\fontsize{6pt}{7.2pt}\selectfont 500} \hspace{-1pt}\huxbpad{-1pt}} &
\multicolumn{1}{c!{\huxvb{0, 0, 0}{0}}}{\huxtpad{-1pt + 1em}\centering \hspace{-1pt} {\fontsize{6pt}{7.2pt}\selectfont 0.06} \hspace{-1pt}\huxbpad{-1pt}} &
\multicolumn{1}{c!{\huxvb{0, 0, 0}{0.4}}}{\huxtpad{-1pt + 1em}\centering \hspace{-1pt} {\fontsize{6pt}{7.2pt}\selectfont 0.055} \hspace{-1pt}\huxbpad{-1pt}} &
\multicolumn{1}{c!{\huxvb{0, 0, 0}{0}}}{\huxtpad{-1pt + 1em}\centering \hspace{-1pt} {\fontsize{6pt}{7.2pt}\selectfont 0.09} \hspace{-1pt}\huxbpad{-1pt}} &
\multicolumn{1}{c!{\huxvb{0, 0, 0}{0.6}}!{\huxvb{0, 0, 0}{0.6}}}{\huxtpad{-1pt + 1em}\centering \hspace{-1pt} {\fontsize{6pt}{7.2pt}\selectfont 0.525} \hspace{-1pt}\huxbpad{-1pt}} &
\multicolumn{1}{c!{\huxvb{0, 0, 0}{0}}}{\huxtpad{-1pt + 1em}\centering \hspace{-1pt} {\fontsize{6pt}{7.2pt}\selectfont 0.76} \hspace{-1pt}\huxbpad{-1pt}} &
\multicolumn{1}{c!{\huxvb{0, 0, 0}{0.4}}}{\huxtpad{-1pt + 1em}\centering \hspace{-1pt} {\fontsize{6pt}{7.2pt}\selectfont 0.145} \hspace{-1pt}\huxbpad{-1pt}} &
\multicolumn{1}{c!{\huxvb{0, 0, 0}{0}}}{\huxtpad{-1pt + 1em}\centering \hspace{-1pt} {\fontsize{6pt}{7.2pt}\selectfont 0.57} \hspace{-1pt}\huxbpad{-1pt}} &
\multicolumn{1}{c!{\huxvb{0, 0, 0}{0.6}}!{\huxvb{0, 0, 0}{0.6}}}{\huxtpad{-1pt + 1em}\centering \hspace{-1pt} {\fontsize{6pt}{7.2pt}\selectfont 0.865} \hspace{-1pt}\huxbpad{-1pt}} &
\multicolumn{1}{c!{\huxvb{0, 0, 0}{0}}}{\huxtpad{-1pt + 1em}\centering \hspace{-1pt} {\fontsize{6pt}{7.2pt}\selectfont 0.31} \hspace{-1pt}\huxbpad{-1pt}} &
\multicolumn{1}{c!{\huxvb{0, 0, 0}{0.4}}}{\huxtpad{-1pt + 1em}\centering \hspace{-1pt} {\fontsize{6pt}{7.2pt}\selectfont 0.125} \hspace{-1pt}\huxbpad{-1pt}} &
\multicolumn{1}{c!{\huxvb{0, 0, 0}{0}}}{\huxtpad{-1pt + 1em}\centering \hspace{-1pt} {\fontsize{6pt}{7.2pt}\selectfont 0.12} \hspace{-1pt}\huxbpad{-1pt}} &
\multicolumn{1}{c!{\huxvb{0, 0, 0}{0}}}{\huxtpad{-1pt + 1em}\centering \hspace{-1pt} {\fontsize{6pt}{7.2pt}\selectfont 0.33} \hspace{-1pt}\huxbpad{-1pt}} \tabularnewline[-0.5pt]

\hhline{>{\huxb{0, 0, 0}{0.4}}|>{\huxb{0, 0, 0}{0.6}}||>{\huxb{0, 0, 0}{0.4}}|>{\huxb{0, 0, 0}{0.6}}||>{\huxb{0, 0, 0}{0.4}}|}
\arrayrulecolor{black}

\multicolumn{1}{!{\huxvb{0, 0, 0}{0}}l!{\huxvb{0, 0, 0}{0}}}{\huxtpad{-1pt + 1em}\raggedright \hspace{-1pt} {\fontsize{6pt}{7.2pt}\selectfont 1000} \hspace{-1pt}\huxbpad{-1pt}} &
\multicolumn{1}{c!{\huxvb{0, 0, 0}{0}}}{\huxtpad{-1pt + 1em}\centering \hspace{-1pt} {\fontsize{6pt}{7.2pt}\selectfont 0.055} \hspace{-1pt}\huxbpad{-1pt}} &
\multicolumn{1}{c!{\huxvb{0, 0, 0}{0.4}}}{\huxtpad{-1pt + 1em}\centering \hspace{-1pt} {\fontsize{6pt}{7.2pt}\selectfont 0.04} \hspace{-1pt}\huxbpad{-1pt}} &
\multicolumn{1}{c!{\huxvb{0, 0, 0}{0}}}{\huxtpad{-1pt + 1em}\centering \hspace{-1pt} {\fontsize{6pt}{7.2pt}\selectfont 0.185} \hspace{-1pt}\huxbpad{-1pt}} &
\multicolumn{1}{c!{\huxvb{0, 0, 0}{0.6}}!{\huxvb{0, 0, 0}{0.6}}}{\huxtpad{-1pt + 1em}\centering \hspace{-1pt} {\fontsize{6pt}{7.2pt}\selectfont 0.725} \hspace{-1pt}\huxbpad{-1pt}} &
\multicolumn{1}{c!{\huxvb{0, 0, 0}{0}}}{\huxtpad{-1pt + 1em}\centering \hspace{-1pt} {\fontsize{6pt}{7.2pt}\selectfont 0.86} \hspace{-1pt}\huxbpad{-1pt}} &
\multicolumn{1}{c!{\huxvb{0, 0, 0}{0.4}}}{\huxtpad{-1pt + 1em}\centering \hspace{-1pt} {\fontsize{6pt}{7.2pt}\selectfont 0.225} \hspace{-1pt}\huxbpad{-1pt}} &
\multicolumn{1}{c!{\huxvb{0, 0, 0}{0}}}{\huxtpad{-1pt + 1em}\centering \hspace{-1pt} {\fontsize{6pt}{7.2pt}\selectfont 0.795} \hspace{-1pt}\huxbpad{-1pt}} &
\multicolumn{1}{c!{\huxvb{0, 0, 0}{0.6}}!{\huxvb{0, 0, 0}{0.6}}}{\huxtpad{-1pt + 1em}\centering \hspace{-1pt} {\fontsize{6pt}{7.2pt}\selectfont 0.995} \hspace{-1pt}\huxbpad{-1pt}} &
\multicolumn{1}{c!{\huxvb{0, 0, 0}{0}}}{\huxtpad{-1pt + 1em}\centering \hspace{-1pt} {\fontsize{6pt}{7.2pt}\selectfont 0.255} \hspace{-1pt}\huxbpad{-1pt}} &
\multicolumn{1}{c!{\huxvb{0, 0, 0}{0.4}}}{\huxtpad{-1pt + 1em}\centering \hspace{-1pt} {\fontsize{6pt}{7.2pt}\selectfont 0.13} \hspace{-1pt}\huxbpad{-1pt}} &
\multicolumn{1}{c!{\huxvb{0, 0, 0}{0}}}{\huxtpad{-1pt + 1em}\centering \hspace{-1pt} {\fontsize{6pt}{7.2pt}\selectfont 0.06} \hspace{-1pt}\huxbpad{-1pt}} &
\multicolumn{1}{c!{\huxvb{0, 0, 0}{0}}}{\huxtpad{-1pt + 1em}\centering \hspace{-1pt} {\fontsize{6pt}{7.2pt}\selectfont 0.3} \hspace{-1pt}\huxbpad{-1pt}} \tabularnewline[-0.5pt]

\hhline{>{\huxb{0, 0, 0}{0.4}}|>{\huxb{0, 0, 0}{0.6}}||>{\huxb{0, 0, 0}{0.4}}|>{\huxb{0, 0, 0}{0.6}}||>{\huxb{0, 0, 0}{0.4}}|}
\arrayrulecolor{black}

\multicolumn{1}{!{\huxvb{0, 0, 0}{0}}l!{\huxvb{0, 0, 0}{0}}}{\huxtpad{-1pt + 1em}\raggedright \hspace{-1pt} {\fontsize{6pt}{7.2pt}\selectfont 2000} \hspace{-1pt}\huxbpad{-1pt}} &
\multicolumn{1}{c!{\huxvb{0, 0, 0}{0}}}{\huxtpad{-1pt + 1em}\centering \hspace{-1pt} {\fontsize{6pt}{7.2pt}\selectfont 0.07} \hspace{-1pt}\huxbpad{-1pt}} &
\multicolumn{1}{c!{\huxvb{0, 0, 0}{0.4}}}{\huxtpad{-1pt + 1em}\centering \hspace{-1pt} {\fontsize{6pt}{7.2pt}\selectfont 0.04} \hspace{-1pt}\huxbpad{-1pt}} &
\multicolumn{1}{c!{\huxvb{0, 0, 0}{0}}}{\huxtpad{-1pt + 1em}\centering \hspace{-1pt} {\fontsize{6pt}{7.2pt}\selectfont 0.32} \hspace{-1pt}\huxbpad{-1pt}} &
\multicolumn{1}{c!{\huxvb{0, 0, 0}{0.6}}!{\huxvb{0, 0, 0}{0.6}}}{\huxtpad{-1pt + 1em}\centering \hspace{-1pt} {\fontsize{6pt}{7.2pt}\selectfont 0.915} \hspace{-1pt}\huxbpad{-1pt}} &
\multicolumn{1}{c!{\huxvb{0, 0, 0}{0}}}{\huxtpad{-1pt + 1em}\centering \hspace{-1pt} {\fontsize{6pt}{7.2pt}\selectfont 0.995} \hspace{-1pt}\huxbpad{-1pt}} &
\multicolumn{1}{c!{\huxvb{0, 0, 0}{0.4}}}{\huxtpad{-1pt + 1em}\centering \hspace{-1pt} {\fontsize{6pt}{7.2pt}\selectfont 0.42} \hspace{-1pt}\huxbpad{-1pt}} &
\multicolumn{1}{c!{\huxvb{0, 0, 0}{0}}}{\huxtpad{-1pt + 1em}\centering \hspace{-1pt} {\fontsize{6pt}{7.2pt}\selectfont 0.945} \hspace{-1pt}\huxbpad{-1pt}} &
\multicolumn{1}{c!{\huxvb{0, 0, 0}{0.6}}!{\huxvb{0, 0, 0}{0.6}}}{\huxtpad{-1pt + 1em}\centering \hspace{-1pt} {\fontsize{6pt}{7.2pt}\selectfont 1} \hspace{-1pt}\huxbpad{-1pt}} &
\multicolumn{1}{c!{\huxvb{0, 0, 0}{0}}}{\huxtpad{-1pt + 1em}\centering \hspace{-1pt} {\fontsize{6pt}{7.2pt}\selectfont 0.19} \hspace{-1pt}\huxbpad{-1pt}} &
\multicolumn{1}{c!{\huxvb{0, 0, 0}{0.4}}}{\huxtpad{-1pt + 1em}\centering \hspace{-1pt} {\fontsize{6pt}{7.2pt}\selectfont 0.095} \hspace{-1pt}\huxbpad{-1pt}} &
\multicolumn{1}{c!{\huxvb{0, 0, 0}{0}}}{\huxtpad{-1pt + 1em}\centering \hspace{-1pt} {\fontsize{6pt}{7.2pt}\selectfont 0.075} \hspace{-1pt}\huxbpad{-1pt}} &
\multicolumn{1}{c!{\huxvb{0, 0, 0}{0}}}{\huxtpad{-1pt + 1em}\centering \hspace{-1pt} {\fontsize{6pt}{7.2pt}\selectfont 0.26} \hspace{-1pt}\huxbpad{-1pt}} \tabularnewline[-0.5pt]

\hhline{>{\huxb{0, 0, 0}{0.4}}|>{\huxb{0, 0, 0}{0.6}}||>{\huxb{0, 0, 0}{0.4}}|>{\huxb{0, 0, 0}{0.6}}||>{\huxb{0, 0, 0}{0.4}}|}
\arrayrulecolor{black}

\multicolumn{1}{!{\huxvb{0, 0, 0}{0}}l!{\huxvb{0, 0, 0}{0}}}{\huxtpad{-1pt + 1em}\raggedright \hspace{-1pt} {\fontsize{6pt}{7.2pt}\selectfont 4000} \hspace{-1pt}\huxbpad{-1pt}} &
\multicolumn{1}{c!{\huxvb{0, 0, 0}{0}}}{\huxtpad{-1pt + 1em}\centering \hspace{-1pt} {\fontsize{6pt}{7.2pt}\selectfont 0.05} \hspace{-1pt}\huxbpad{-1pt}} &
\multicolumn{1}{c!{\huxvb{0, 0, 0}{0.4}}}{\huxtpad{-1pt + 1em}\centering \hspace{-1pt} {\fontsize{6pt}{7.2pt}\selectfont 0.02} \hspace{-1pt}\huxbpad{-1pt}} &
\multicolumn{1}{c!{\huxvb{0, 0, 0}{0}}}{\huxtpad{-1pt + 1em}\centering \hspace{-1pt} {\fontsize{6pt}{7.2pt}\selectfont 0.48} \hspace{-1pt}\huxbpad{-1pt}} &
\multicolumn{1}{c!{\huxvb{0, 0, 0}{0.6}}!{\huxvb{0, 0, 0}{0.6}}}{\huxtpad{-1pt + 1em}\centering \hspace{-1pt} {\fontsize{6pt}{7.2pt}\selectfont 1} \hspace{-1pt}\huxbpad{-1pt}} &
\multicolumn{1}{c!{\huxvb{0, 0, 0}{0}}}{\huxtpad{-1pt + 1em}\centering \hspace{-1pt} {\fontsize{6pt}{7.2pt}\selectfont 0.99} \hspace{-1pt}\huxbpad{-1pt}} &
\multicolumn{1}{c!{\huxvb{0, 0, 0}{0.4}}}{\huxtpad{-1pt + 1em}\centering \hspace{-1pt} {\fontsize{6pt}{7.2pt}\selectfont 0.685} \hspace{-1pt}\huxbpad{-1pt}} &
\multicolumn{1}{c!{\huxvb{0, 0, 0}{0}}}{\huxtpad{-1pt + 1em}\centering \hspace{-1pt} {\fontsize{6pt}{7.2pt}\selectfont 0.98} \hspace{-1pt}\huxbpad{-1pt}} &
\multicolumn{1}{c!{\huxvb{0, 0, 0}{0.6}}!{\huxvb{0, 0, 0}{0.6}}}{\huxtpad{-1pt + 1em}\centering \hspace{-1pt} {\fontsize{6pt}{7.2pt}\selectfont 1} \hspace{-1pt}\huxbpad{-1pt}} &
\multicolumn{1}{c!{\huxvb{0, 0, 0}{0}}}{\huxtpad{-1pt + 1em}\centering \hspace{-1pt} {\fontsize{6pt}{7.2pt}\selectfont 0.18} \hspace{-1pt}\huxbpad{-1pt}} &
\multicolumn{1}{c!{\huxvb{0, 0, 0}{0.4}}}{\huxtpad{-1pt + 1em}\centering \hspace{-1pt} {\fontsize{6pt}{7.2pt}\selectfont 0.1} \hspace{-1pt}\huxbpad{-1pt}} &
\multicolumn{1}{c!{\huxvb{0, 0, 0}{0}}}{\huxtpad{-1pt + 1em}\centering \hspace{-1pt} {\fontsize{6pt}{7.2pt}\selectfont 0.085} \hspace{-1pt}\huxbpad{-1pt}} &
\multicolumn{1}{c!{\huxvb{0, 0, 0}{0}}}{\huxtpad{-1pt + 1em}\centering \hspace{-1pt} {\fontsize{6pt}{7.2pt}\selectfont 0.3} \hspace{-1pt}\huxbpad{-1pt}} \tabularnewline[-0.5pt]

\hhline{>{\huxb{0, 0, 0}{0.4}}|>{\huxb{0, 0, 0}{0.6}}||>{\huxb{0, 0, 0}{0.4}}|>{\huxb{0, 0, 0}{0.6}}||>{\huxb{0, 0, 0}{0.4}}|}
\arrayrulecolor{black}

\multicolumn{1}{!{\huxvb{0, 0, 0}{0}}l!{\huxvb{0, 0, 0}{0}}}{\huxtpad{-1pt + 1em}\raggedright \hspace{-1pt} {\fontsize{6pt}{7.2pt}\selectfont 10000} \hspace{-1pt}\huxbpad{-1pt}} &
\multicolumn{1}{c!{\huxvb{0, 0, 0}{0}}}{\huxtpad{-1pt + 1em}\centering \hspace{-1pt} {\fontsize{6pt}{7.2pt}\selectfont 0.065} \hspace{-1pt}\huxbpad{-1pt}} &
\multicolumn{1}{c!{\huxvb{0, 0, 0}{0.4}}}{\huxtpad{-1pt + 1em}\centering \hspace{-1pt} {\fontsize{6pt}{7.2pt}\selectfont 0.03} \hspace{-1pt}\huxbpad{-1pt}} &
\multicolumn{1}{c!{\huxvb{0, 0, 0}{0}}}{\huxtpad{-1pt + 1em}\centering \hspace{-1pt} {\fontsize{6pt}{7.2pt}\selectfont 0.805} \hspace{-1pt}\huxbpad{-1pt}} &
\multicolumn{1}{c!{\huxvb{0, 0, 0}{0.6}}!{\huxvb{0, 0, 0}{0.6}}}{\huxtpad{-1pt + 1em}\centering \hspace{-1pt} {\fontsize{6pt}{7.2pt}\selectfont 1} \hspace{-1pt}\huxbpad{-1pt}} &
\multicolumn{1}{c!{\huxvb{0, 0, 0}{0}}}{\huxtpad{-1pt + 1em}\centering \hspace{-1pt} {\fontsize{6pt}{7.2pt}\selectfont 1} \hspace{-1pt}\huxbpad{-1pt}} &
\multicolumn{1}{c!{\huxvb{0, 0, 0}{0.4}}}{\huxtpad{-1pt + 1em}\centering \hspace{-1pt} {\fontsize{6pt}{7.2pt}\selectfont 0.975} \hspace{-1pt}\huxbpad{-1pt}} &
\multicolumn{1}{c!{\huxvb{0, 0, 0}{0}}}{\huxtpad{-1pt + 1em}\centering \hspace{-1pt} {\fontsize{6pt}{7.2pt}\selectfont 0.995} \hspace{-1pt}\huxbpad{-1pt}} &
\multicolumn{1}{c!{\huxvb{0, 0, 0}{0.6}}!{\huxvb{0, 0, 0}{0.6}}}{\huxtpad{-1pt + 1em}\centering \hspace{-1pt} {\fontsize{6pt}{7.2pt}\selectfont 1} \hspace{-1pt}\huxbpad{-1pt}} &
\multicolumn{1}{c!{\huxvb{0, 0, 0}{0}}}{\huxtpad{-1pt + 1em}\centering \hspace{-1pt} {\fontsize{6pt}{7.2pt}\selectfont 0.14} \hspace{-1pt}\huxbpad{-1pt}} &
\multicolumn{1}{c!{\huxvb{0, 0, 0}{0.4}}}{\huxtpad{-1pt + 1em}\centering \hspace{-1pt} {\fontsize{6pt}{7.2pt}\selectfont 0.095} \hspace{-1pt}\huxbpad{-1pt}} &
\multicolumn{1}{c!{\huxvb{0, 0, 0}{0}}}{\huxtpad{-1pt + 1em}\centering \hspace{-1pt} {\fontsize{6pt}{7.2pt}\selectfont 0.115} \hspace{-1pt}\huxbpad{-1pt}} &
\multicolumn{1}{c!{\huxvb{0, 0, 0}{0}}}{\huxtpad{-1pt + 1em}\centering \hspace{-1pt} {\fontsize{6pt}{7.2pt}\selectfont 0.355} \hspace{-1pt}\huxbpad{-1pt}} \tabularnewline[-0.5pt]

\hhline{>{\huxb{0, 0, 0}{1}}->{\huxb{0, 0, 0}{1}}->{\huxb{0, 0, 0}{1}}->{\huxb{0, 0, 0}{1}}->{\huxb{0, 0, 0}{1}}->{\huxb{0, 0, 0}{1}}->{\huxb{0, 0, 0}{1}}->{\huxb{0, 0, 0}{1}}->{\huxb{0, 0, 0}{1}}->{\huxb{0, 0, 0}{1}}->{\huxb{0, 0, 0}{1}}->{\huxb{0, 0, 0}{1}}->{\huxb{0, 0, 0}{1}}-}
\arrayrulecolor{black}

\end{tabular}}\label{table:test_DR}
\par\end{center}

\end{table}

\vspace{0.15cm}
\noindent \textbf{Simulation 6: Efficiency gain.} 
This simulation evaluated the efficiency of ATE one-step estimators leveraging the Verma constraint via an anchor variable $Z$ (Section~\ref{subsec:verma}). We considered two scenarios: (1) Binary $Z$, comparing $\psi_{z^*=1}^+$, $\psi_{z^*=0}^+$, and $\psi_{\alpha^\mathrm{opt}}^+$; and (2) Continuous $Z$, evaluating $\psi_{\tilde{\p}}^+$ under three choices of $\tilde{\p}(Z)$: the true density $p(Z)$, $\operatorname{Uniform}(-1,1)$, and a Truncated Normal ($\operatorname{TN}$) with mean $0.6$ and standard deviation $0.7$, truncated at the $0.1\%$ and $99.9\%$ quantiles of $p(Z)$. We compare these to the ``non-Verma'' one-step estimators $\psi_{1}^+$ from Section~\ref{sec:est_ATE} where $Z$ is considered as part of baseline covariates. Each setting was replicated 1000 times at sample sizes from 500 to 8000. Full data-generating details are provided in Appendix~\ref{app:sims:eff}. 

Under binary $Z$, $\psi_{\alpha^\mathrm{opt}}^+$ achieved lower variance than either fixed-level estimator, with larger gains relative to $\psi_{z^*=1}^+$, yielding \~4-fold variance reductions. It also reduced variance by \~1.7-fold relative to the non-Verma estimator (Appendix Fig.~\ref{fig:sim6_binary_interAZ}). For continuous $Z$, $\tilde{\p}(Z)=\operatorname{TN}(0.6,0.7)$ yielded the lowest variance, corresponding to a \~7-fold reduction compared to the non-Verma estimator (Appendix Fig.~\ref{fig:sim6_continuous_interAZ}). See Appendix~\ref{app:sims:eff} for details. These results suggest that leveraging the Verma constraint can yield efficiency gains, with the magnitude of improvement depending on the underlying data-generating process. 

\section{Real data application} \label{sec:real_data}

We applied our front-door estimation framework to two real-world data sets: a longitudinal Finnish cohort examining the effect of early academic performance on future income \citep{fsd} (results in Appendix~\ref{app:real_data_academic}) and an observational study evaluating the impact of mobile stroke unit (MSU) dispatch on post-stroke outcomes in the Berlin prehospital stroke care trial, known as B\_PROUD \citep{ebinger2017berlin}. We focus on the latter as our primary application below.  

The B\_PROUD study is a nonrandomized investigation of MSU care conducted in Berlin between February 2017 and May 2019 \citep{ebinger2017berlin}. This dataset was previously analyzed by \citet{piccininni2023effect} using a front-door approach to estimate the causal effect of MSU dispatch on 3-month functional outcomes. To enable estimation with continuous mediators, their analysis discretized the time from ambulance dispatch to thrombolysis into coarse categories—an approach that, while practical, can lead to information loss and sensitivity to bin definitions. In contrast, our framework accommodates mixed-type mediators without discretization, leveraging flexible machine learning tools to preserve the full resolution of the data. 

We applied our method to 768 patients eligible for reperfusion therapy in the B\_PROUD cohort, of whom 588 (77\%) received MSU care ($A = 1$) and 180 (23\%) received conventional emergency services ($A = 0$). The outcome of interest, $Y$, is the 3-month modified Rankin Scale (mRS) score, an ordinal measure ranging from 0 (no symptoms) to 6 (death). The assumed causal pathway from $A$ to $Y$ is fully mediated through two variables: (i) $M_1$, a binary indicator of thrombolysis receipt, and (ii) $M_2$, the time from ambulance dispatch to thrombolysis (set to 0 if thrombolysis was not received). This assumption is plausible since reducing time to thrombolysis is the intended effect of the treatment. We adjust for baseline covariates, systolic blood pressure ($X_1$) and stroke severity ($X_2$), and assume conditional ignorability given these variables, justified by the highly standardized emergency stroke care protocol \citep{piccininni2023effect}.

To handle the ordinal outcome, we constructed binary indicators $Y_k := \I(Y \leq k)$ for $k = 0, \dots, 5$, applied our estimators to each, and recovered the marginal probability mass function $\p(Y^a = k)$ by differencing cumulative probabilities. This allowed us to estimate the full distribution of potential outcomes under each treatment level. For comparability with \citet{piccininni2023effect}, we also replicated their discretization of $M_2$ using the first quartile and median as cutoffs. Results from this secondary analysis are provided in Appendix~\ref{app:real_data_msu_dispatch}. 

We estimated the ATE using both the one-step estimator $\psi_{2b}^{+}(\hat{\Q})$ and its TMLE counterpart $\psi_{2b}(\hat{\Q}^\star)$, and estimated the ATT using estimators described in Section~\ref{sec:est_ATT}. Details and results for the ATT are provided in Appendix~\ref{app:real_data_msu_dispatch}. To flexibly capture potential nonlinearities and interactions, we used Super Learner with five-fold cross-fitting. The ensemble library included intercept-only models, GLMs, multivariate adaptive regression splines, and random forests. 

The one-step estimate of ATE was $-0.079$ (95\% CI: $(-0.468,0.311)$), while TMLE yielded $-0.074$ (95\% CI: $(-0.464,0.315)$). Although not statistically significant, both estimates suggest a shift toward improved outcomes with MSU care. To further characterize this effect, we estimated the full potential outcome distributions. Under MSU care, TMLE estimated the following mRS distribution: 0(29\%), 1(20\%), 2(11\%), 3(15\%), 4(13\%), 5(3\%), 6(9\%). These estimates are generally consistent with those reported in the original analysis by \citet{piccininni2023effect}, which found corresponding values of 0(30\%), 1(19\%), 2(12\%), 3(15\%), 4(12\%), 5(4\%), 6(9\%).


\section{Discussions}
\label{sec:conc}

While the front-door model provides a powerful framework for causal inference in the presence of treatment-outcome unmeasured confounding, its practical utility depends on both robust estimation strategies and the validity of its identifying assumptions. In this paper, we developed a suite of influence function-based estimators for both the ATE and ATT that accommodate complex, multivariate mediators without relying on  parametric assumptions. Our estimators incorporate modern machine learning methods and use sample-splitting to avoid reliance on Donsker conditions, enabling valid inference in flexible settings. Beyond estimation, we also addressed the testability of key identification assumptions by leveraging a generalized equality constraint involving an anchor variable, which we incorporate into a semiparametric model under the null to both test these assumptions and construct more efficient estimators in this setting. 

Despite these advances, several important directions remain. One is to extend our estimation strategies to more complex causal structures, such as hidden variable DAGs represented by acyclic directed mixed graphs. While identification theory in these models is well developed, efficient and flexible estimation remains an open challenge more broadly \citep{guo2024average, guo2025causal}. Expanding one-step and TMLE methods to this setting would improve applicability when mediators only partially explain the treatment effect or unmeasured confounding extends beyond the treatment–outcome link. Another direction is to refine our doubly robust evaluation tools—such as test statistics and confidence intervals—for settings with multiple mediators of mixed types. Finally, extending these methods to longitudinal data with time-varying treatments and mediators would support more realistic analyses where mediation mechanisms and front-door structure evolve over time.


\section*{Competing interests}
The authors declare no competing interests.



\section*{Acknowledgments} 
This work was supported by the National Science Foundation under Grant No. 2502299. 

\section*{Data availability}
Due to data sharing restrictions, the raw datasets are not publicly available. The B\_PROUND dataset may be requested from the authors of \cite{piccininni2023effect}, and the Finnish cohort dataset is available upon application through the Finnish Social Science Data Archive. The implementation of all numerical experiments is available in the GitHub repository: \href{https://github.com/annaguo-bios/fd-methods}{\texttt{annaguo-bios/fd-methods}}. We have also developed the \href{https://github.com/annaguo-bios/fdcausal}{\texttt{fdcausal}} \textsf{R} package for causal inference under the front-door model. 


\vspace{0.75cm}
\begingroup
\renewcommand{\baselinestretch}{0.95}
\selectfont  
\setlength{\bibsep}{11pt}    
\bibliographystyle{abbrvnat}
\bibliography{references}
\endgroup




\section*{Figure Legends} 

\noindent\textbf{Figure 1:}
(a) Example of a DAG with measured confounders $X$ and unmeasured
confounders $U$; 
(b) The front-door DAG with unmeasured confounders $U$ between $A$ and $Y$ (dashed edges indicate assumptions);
(c) The front-door DAG with the inclusion of measured confounders $X$.

\newpage
\appendix

\setcounter{page}{1}

\setcounter{section}{0}
\renewcommand{\thesection}{S\arabic{section}}
\renewcommand{\thesubsection}{S\arabic{section}.\arabic{subsection}}
\renewcommand{\thesubsubsection}{S\arabic{section}.\arabic{subsubsection}}

\setcounter{figure}{0}
\renewcommand{\thefigure}{S\arabic{figure}}

\setcounter{table}{0}
\renewcommand{\thetable}{S\arabic{table}}

\setcounter{equation}{0}
\renewcommand{\theequation}{S\arabic{equation}}

\begin{center}
\noindent {\bf \LARGE Supplementary Materials for: Flexible Nonparametric Inference for Causal Effects under the Front-Door Model}
\end{center}

\vspace{0.5cm}

\begin{spacing}{1.6}

\begin{description}
    \item[GitHub repository:] The GitHub repository \href{https://github.com/annaguo-bios/fd-methods}{\texttt{annaguo-bios/fd-methods}} contains the complete codebase required to reproduce all numerical studies reported in this paper, including simulation experiments and empirical analyses.
    
    \vspace{0.25cm}
    \item[R package:] The R package \href{https://github.com/annaguo-bios/fdcausal}{\texttt{fdcausal}} provides a general-purpose implementation of the estimators proposed in this work for the causal effects under the front-door model, and is intended for broader use beyond the specific settings considered in this paper.
    
    \vspace{0.25cm}
    \item[Supplementary appendix:] This appendix contains additional methodological details, theoretical results, and supporting analyses. It is organized as follows.
\end{description}

\noindent 
Appendix~\ref{app:notation} summarizes the notations used throughout the manuscript for ease of reference.


\vspace{0.2cm}
\noindent 
Appendix~\ref{app:model_front-door} presents identification proofs for the ATE and ATT under the front-door model, derives the corresponding efficient influence functions, and briefly outlines the model’s geometric structure and tangent space decomposition. 

\vspace{0.2cm}
\noindent 
Appendices~\ref{app:tmle_details} and~\ref{app:att_details} provide additional technical details on one-step estimators and TMLE procedures for the ATE and ATT, including loss–submodel validations, binary outcome adjustments, and algorithmic summaries of the TMLE steps. 

\vspace{0.2cm}
\noindent 
Appendix \ref{app:asym} presents the proofs underlying inference results for the ATE and ATT estimators, including second-order remainder terms, regularity conditions, and robustness properties. It also includes the formal asymptotic theorems for the ATT estimators. 

\vspace{0.2cm}
\noindent 
Appendix \ref{app:testing} details the three testing procedures for front‐door assumptions, more efficient ATE estimators under a semiparametric front‐door model, and the Verma constraint, with all relevant identification and estimation proofs.

\vspace{0.2cm}
\noindent 
Appendix~\ref{app:sims} presents details of the simulation studies, along with additional simulation results. 

\vspace{0.2cm} 
\noindent 
Appendix~\ref{app:real} expands the main real data application and includes a second study on the effect of academic performance on future income under the front-door model.

\vspace{0.4cm}
\noindent We use the following integration notations interchangeably in the supplementary material: $\int(.) \diff \P(x)=\int(.) \p(x)\sp \diff x$, $\int(.) \diff \P(x,y)=\iint(.) \p(x,y)\sp \diff x\sp \diff y$, for any random variables $X$ and $Y$. 

\vspace{0.4cm}
\noindent For ease of reference, a table of contents for the supplementary materials is provided below.

\end{spacing}

\pagebreak

\startcontents[supp]
{\small \printcontents[supp]{}{1}{}}

\pagebreak
\section{Glossary of terms and notations} 
\label{app:notation} 

To ease navigation of the notations, we provide a comprehensive list in Table~\ref{tab:notations}. 

\begin{table}[H]
\begin{center}
\caption{\centering Glossary of terms and notations}
\label{tab:notations}
\addtolength{\tabcolsep}{8pt}
{\small
\renewcommand{\arraystretch}{0.65} 
\resizebox{0.9\textwidth}{!}{\begin{tabular}{ ll | ll} 
    \hline  
    \textbf{Symbol}     & \textbf{Definition}  &  \textbf{Symbol}     & \textbf{Definition}  
    \\ \hline 
    $A, a_0$   & Treatment, fixed assignment  &  $\pi(A\mid X)$ & propensity score 
    \\
    $Y, Y^a$  & Outcome, potential outcome   &  $\mu(M, A, X)$  &  Outcome regression 
    \\
    $X$  & Observed confounders & $f_M(M \mid A, X)$ &  Mediator density  
    \\
    $M$ & Mediator(s)  &    $\xi(M, X)$ & $\sum_{a \in \{0, 1\}} \mu(M,a,X) \pi(a\mid X)$ 
    \\
    $U$    & Unmeasured variables  & $\eta(A, X)$ & $\int \mu(m, A, X) f_M(m \mid a_0, X) \sp \diff m$  
    \\ 
    $O$    & Observed data $(X, A, M, Y)$  & $\theta(X)$  &  $\int \xi(m, X) f_M(m \mid a_0, X) \diff m$ 
     \\ 
    $\P$   & Observed data distribution  & $\gamma(X)$ & $\E\big(  \xi(M, X)  \sp \big| \sp a_0, X \big) \equiv \theta(X)$ 
    \\
    $\Q$ &  Collection of nuisances & $f^r_M(M, A, X)$ & $f_M(M\mid a_0,X)/f_M(M \mid A,X)$ 
     \\
    $\psi(\P)$   & Target parameter for ATE ($\equiv \psi(\Q)$) & $\lambda(A \mid M, X)$ & $\p(A \mid M, X)$ 
    \\
    $\beta(\P)$   & Target parameter for ATT ($\equiv \beta(\Q)$) & $\p_A(A)$ & $\p(A)$ 
    \\ 
    $\Phi(\Q)$ & Efficient influence function for $\psi(\P)$ &  $\kappa_a(X)$ & $\E\big( \mu(M, a, X)  \sp \big| \sp a_0, X  \big)$ 
    \\
    $\Phi_{\beta}(\Q)$ & Efficient influence function for $\beta(\P)$ &  $\p_{AX}(A,X)$ & $\p(A,X)$ 
    \\ 
    $\hat{\Q}$   & Initial estimate of $\Q$ & $\tau(A, X)$ & $\E\big( \sp f_M^{r}(M, A, X) \sp \mu(M, A, X)  \sp \big| \sp A, X  \big)$
    \\ 
    $\hat{\Q}^\star$ & TMLE estimate of $\Q$  & $H_A(X)$ & Clever covariate in treatment model
    \\ 
    $\p_X$ & Covariates distribution & $H_M(X)$ & Clever covariate in mediator model  
    \\ 
    $\Pn$ &  Empirical distribution   &  $\mathcal{M}, \mathcal{X}$ & Domains for variables $M, X$
    \\
    $L_{\Q_j}$ &  Loss function for nuisance $\Q_j \in \Q$ &   $\mathcal{M}_{\Q_j}, \mathcal{M}_\mathcal{\Q}$ & Model space for nuisance $\Q_j$ and $\Q$
    \\
    $\psi^+_{\cdot}(\hat{\Q})$, $\beta^+_{\cdot}(\hat{\Q})$ & One-step estimators & $\psi_{\cdot}(\hat{\Q}^\star)$, $\beta_{\cdot}(\hat{\Q}^\star)$ & TMLEs
    \\
    $\psi^{\cdot}_{1}(\cdot)$, $\beta^{\cdot}_{1}(\cdot)$ & Estimators with density estimation & $\psi^{\cdot}_{2a}(\cdot)$, $\beta^{\cdot}_{a}(\cdot)$ & Estimators with density ratio estimation
    \\
    $\psi^{\cdot}_{2b}(\cdot)$, $\beta^{\cdot}_{b}(\cdot)$ & Estimators with Bayes' rule & $R_2$ & Second-order remainder
    \\
    $\psi_{1, k}^{+, \mathrm{cf}}\left(\hat{\mathrm{Q}}^{(-k)}\right)$ & k-th fold cross-fitted estimator of $\psi^+_{1}$ & Z & Anchor variable 
    \\
    $\q_{\text{primal}}$ & Primal weight & $\q_{\text{dual}}$ & Dual weight
    \\
    $\mu^a_\text{primal}$ & MSE risk minimizer with primal weight & $\mu^a_\text{dual}$ & MSE risk minimizer with dual weight
    \\
    $T_{\mathrm{n}, \text { primal }}$ & Primal test statistic & $T_{\mathrm{n}, \text { dual }}$ & Dual test statistic
    \\
    $\mu^m(z, x)$ & $\E(Y^m \mid Z = z, X = x)$ & $T_{\mathrm{n}, \mathrm{CCM}}$ & Conditional counterfactual mean test statistic
    \\
    $\psi_{z^*}(\Q)$ & $\psi(\Q)$ at $Z=z^*$ & $\Phi_{z^*}$ & Efficient influence function for $\psi_{z^*}(\Q)$
    \\
    $\Lambda_{\alpha}$ & Class of IFs defined by weight $\alpha$ & $\alpha^{\mathrm{opt}}$ & Optimal weight
\end{tabular}
}
}
\end{center}
\end{table}

\section{Details on causal front-door model}
\label{app:model_front-door}

\subsection{Nonparametric identification}
\label{app:model_id_proof}

\noindent \textbf{\underline{\large Identification of $\E(Y^{a_0})$}}

\noindent 
Given the stated identification assumptions, $\p(Y^{a_0} = y)$ can be identified as follows:

\vspace{-1.5cm}
\begin{align*}
&\hspace{-0.5cm} \p(Y^{a_0} = y) \\
&= \iint \p(Y^{a_0} = y, M^{a_0} = m, X=x) \sp \diff m\ \diff x \\
&= \iint \p(Y^{m} = y \mid M^{a_0} = m, x) \sp \p(M^{a_0} = m \mid x) \sp \p(x) \sp \diff m\ \diff x \\
&= \iint \Big\{\sum_{a=0}^1 \p(Y^{m} = y, A = a\mid M^{a_0}=m, x) \Big\}  \sp \p(M^{a_0}=m \mid x)  \sp \p(x) \sp \diff m\ \diff x\\
&= \iint \Big\{\sum_{a=0}^1 \p(Y^{m} = y \mid A = a, x) \sp \p(A=a \mid x)\Big\}  \sp \p(M=m \mid A = a_0, x)  \sp \p(x) \sp \diff m\ \diff x\\
&= \iint \Big\{\sum_{a=0}^1 \p(Y = y \mid M=m, A = a, x) \sp \p(A=a \mid x) \Big\}  \sp \p(M=m \mid A = a_0, x)  \sp \p(x) \sp \diff m\ \diff x \sp ,  
\end{align*}

\vspace{-.3cm} \noindent 
where the first equality holds by probability rules, second by factorization rules, and a combination of consistency and no direct effect assumptions, the third holds by probability rules, the fourth holds by factorization rules, consistency, positivity, and conditional ignorability, and the fifth  holds by conditional ignorability, consistency, and positivity. Thus, the target parameter $\E(Y^{a_0})$ is identified via the following functional:

\vspace{-1.25cm}
\begin{align*}
\psi_{a_0}(\P) =\iint \sum_{a = 0}^1 y \sp \p(y \mid m, a, x) \sp \p(a \mid x) \sp \p(m\mid a_0,x) \sp \p(x)\ \diff y \sp \diff m \sp \diff x \sp . 
\end{align*}

\noindent \textbf{\underline{\large Identification of $\E(Y^{a_0} \mid A = a_1)$}}

\noindent 
Similarly, given the stated identification assumptions, $\p(Y^{a_0}\mid A=a_1)$ can be identified as follows:

\vspace{-1.5cm}
\begin{align*}
&\p(Y^{a_0} = y\mid A=a_1)
\\
&= \iint \p(Y^{a_0} = y, M^{a_0} = m, X=x\mid A=a_1) \sp \diff m\ \diff x \\
&= \iint \p(Y^{m} = y \mid M^{a_0} = m, x, A=a_1) \sp \p(M^{a_0} = m \mid x, A=a_1) \sp \p(x\mid A=a_1) \sp \diff m\ \diff x\\
&= \iint \p(Y^{m} = y \mid  x, A=a_1) \sp \p(M^{a_0} = m \mid x, A=a_0) \sp \p(x\mid A=a_1) \sp \diff m\ \diff x\\
&= \iint \p(Y = y \mid M = m, x, A=a_1) \sp \p(M = m \mid x,A=a_0) \sp \p(x\mid A=a_1) \sp \diff m\ \diff x \sp ,
\end{align*}

\vspace{-.3cm} \noindent 
where the first and second qualities hold by probability rules, the third holds by ignorability, and the last equality holds by consistency and positivity. Thus, the target parameter $\E(Y^{a_0} \mid A = a_1)$ is identified via the following functional:

\vspace{-1.25cm}
\begin{align*}
    \beta_{a_0}(\P)=\int y \sp \p( y \mid m, A=a_1, x) \sp \p(m \mid A=a_0,x) \sp \p(x\mid A=a_1) \sp \diff y\sp \diff m\sp \diff x \sp.
\end{align*}

\noindent \textbf{\underline{\large Identification of $\E(Y^{a_1, M^{a_0}} \mid A=a_1)$}}

\noindent 
In addition to (i) consistency, (ii) conditional ignorability, and (iii) positivity, identification of $\E(Y^{a_1, M^{a_0}}\mid A=a_1)$ requires an additional assumption: (iv) \textit{cross-world independence} stating that $M^{a_0} \perp Y^{a_1, m} \mid A=a_1,X$. Under these assumptions, $\E(Y^{a_1, M^{a_0}} \mid A=a_1)$ is identified as:

\vspace{-1.25cm}
\begin{align*}
    &\E(Y^{a_1, M^{a_0}} \mid A=a_1) 
    \\
    &=\iiint y\ \p(Y^{a_1,m}=y\mid M^{a_0}=m, A=a_1,x)\ \p(M^{a_0}=m\mid  A=a_1,x)\ \p(x\mid A=a_1)\ \diff y \ \diff m \ \diff x
    \\
    &=\iiint y\ \p(Y^{a_1,m}=y\mid A=a_1,x)\ \p(M^{a_0}=m\mid  A=a_1,x)\ \p(x\mid A=a_1)\ \diff y \ \diff m \ \diff x
    \\
    &=\iiint y\ \p(Y=y\mid M=m, A=a_1,x)\ \p(M^{a_0}=m\mid  A=a_1,x)\ \p(x\mid A=a_1)\ \diff y \ \diff m \ \diff x
    \\
    &=\iint \E(Y=y\mid M=m, A=a_1,x)\ \p(M=m\mid  A=a_0,x)\ \p(x\mid A=a_1) \ \diff m \ \diff x \sp ,
\end{align*}

\vspace{-.3cm} \noindent 
where the second equality holds by (iv), and the third and fourth holds by (i) and (ii). Thus, the target parameter $\E(Y^{a_1, M^{a_0}} \mid A=a_1)$ is identifiable via the same functional as $\beta_{a_0}(\P)$.

\subsection{Alternative interpretations of the front-door functionals}
\label{app:alternative_interpretations}

The ATE front-door functional in \eqref{eq:id_ATE} corresponds to the \textit{population intervention indirect effect} (PIIE) introduced by \citet{fulcher19robust}. The PIIE, indexed by a fixed treatment level $a_0$, captures the mean difference between $Y$ (the observed outcome) and $Y^{A, M^{a_0}}$ (the counterfactual outcome) under an intervention that shifts the mediator to the value it would have taken had treatment been set to $a_0$; i.e., $\text{PIIE}(a_0) \coloneqq \E(Y - Y^{A, M^{a_0}})$. Instead of assuming no direct effect of treatment on the outcome—as required by the front-door model—\citet{fulcher19robust} identify the PIIE by replacing this condition with a cross-world independence assumption: $M^{a_0} \perp Y^{a_1, m} \! \mid \! A=a_1, X$. Under this alternative assumption, the counterfactual mean $\E(Y^{A, M^{a_0}})$ remains identified by the front-door functional $\psi_{a_0}(\P)$ in \eqref{eq:id_ATE}. This connection implies that our proposed estimators, outlined in the next section, retain some meaningful interpretation even when the full mediation assumption fails, thereby broadening their applicability to settings where treatment has both direct and indirect effects.

A closely related interpretation applies to the ATT front-door functional in \eqref{eq:id_ATT}, which corresponds to a PIIE among the treated (PIIE-T) or among the controls (PIIE-C), depending on the conditioning group, $\text{PIIE-T} \coloneqq \E(Y - Y^{1, M^0} \! \mid \! A = 1)$ and $\text{PIIE-C} \coloneqq \E(Y - Y^{0, M^1} \! \mid \! A = 0).$ The counterfactual parameter $\E(Y^{a_1, M^{a_0}} \! \mid \! A = a_1)$ captures the expected outcome for individuals who received treatment level $a_1$, had they retained their treatment assignment but experienced mediator values as if they had received $A = a_0$. This quantity is directly identified by the conditional front-door functional $\beta_{a_0}(\P)$ under the same cross-world assumption of \citet{fulcher19robust}; see Appendix~\ref{app:model_id_proof} for a proof. These interpretations imply that our ATT and ATC estimators also recover subgroup-specific PIIEs, capturing the component of the treatment effect that operates through shifting the values of $M$ under specific interventions within each subpopulation, under alternative assumptions to those required by the standard front-door model.

\citet{wen2024causal} provide another interpretation of the front-door functional, viewing it as the \textit{average causal effect on an intervening variable}, defined as $\E(Y^{a_M = 1} - Y^{a_M = 0})$. Here, $A_M$ represents an intervenable component of the treatment, distinct from the original variable $A$, which may not correspond to a well-defined or manipulable intervention. In one of their motivating examples, $A$ reflects chronic pain—an inherently non-manipulable construct—that influences a doctor’s perception of the patient’s pain status, captured by $A_M$. This perceived status in turn affects opioid use ($M$) and mortality ($Y$) in their data application. Under identification assumptions, they show that $\E(Y^{a_M = a_0})$ is identified by the same front-door functional $\psi_{a_0}(\P)$. This reinforces the relevance of the functional in \eqref{eq:id_ATE} for policy settings in which direct intervention on $A$ is infeasible, but meaningful action can still be taken on modifiable components such as $A_M$. Our estimators thus support not only classical mediation analysis, but also modern frameworks that emphasize intervenable causal mechanisms. 

These connections substantially broaden the scope of our estimation framework, which remains valid in settings where the effect of $A$ on $Y$ is only partially mediated by $M$. They also underscore the policy relevance of front-door estimands in scenarios where interventions must target modifiable components of treatment pathways, rather than treatment itself.

\subsection{Statistical model and EIF derivations}
\label{app:model_eif}

Let $\mathcal{H}$ denote the \emph{Hilbert space} defined as the space of all mean-zero, square-integrable scalar functions of observed data $O=(X, A, M, Y)$, equipped with the inner product $\E(h_1(O) \times h_2(O)), \forall h_1, h_2 \in \mathcal{H}.$ Let $\M$ denote the front-door statistical model, which consists of distributions defined over observed data $O$. By chain rule of probability, we can write down this joint distribution as $\p(o) = \p(y \mid m, a, x)\ \p(m \mid a, x) \sp \p(a \mid x) \sp \p(x).$ Given this factorization,  we can write down the joint score as $S(o) = S(y \mid m, a, x) + S(m \mid a, x) + S(a \mid x) + S(x)$. 

The \textit{tangent space} of $\M$, denoted by $\mathscr{T}$, is defined as the mean-square closure of all linear combinations of scores in corresponding parametric submodels for $\M$. We can partition $\mathscr{T}$ into a \textit{direct sum} of four orthogonal subspaces, $\mathscr{T} = \mathscr{T}_Y \oplus \mathscr{T}_M \oplus \mathscr{T}_A \oplus \mathscr{T}_X $, defined as follows: 
\begin{align*}
    &\mathscr{T}_{Y}= \Big\{h_Y(Y,M,A,X)\in\mathcal{H} \ ,  \ \sp \text{s.t.}  \ \sp  \E\big(h_Y(Y,M,A,X) \sp \big| \sp M, A, X\big)=0 \Big\} \sp ,  \\
    &\mathscr{T}_{M}= \Big\{h_M(M,A,X)\in\mathcal{H} \ ,  \ \sp \text{s.t.}  \ \sp \E\big(h_M(M,A,X) \sp \big| \sp A, X\big)=0 \Big\} \sp , \\
    &\mathscr{T}_{A}= \Big\{h_A(A,X)\in\mathcal{H} \ ,  \ \sp \text{s.t.}  \ \sp  \E\big(h_A(A,X) \sp \big| \sp X\big)=0 \Big\} \sp , \\
    &\mathscr{T}_{X}= \big\{h_X(X)\in\mathcal{H} \ ,  \ \sp \text{s.t.}  \ \sp \E(h_X(X))=0\big\} \sp . 
\end{align*}  

\vspace{-.3cm}  
Demonstrating the mutual orthogonality of these tangent spaces is straightforward. For instance, the inner product of any $h_Y(Y,M,A,X) \in \mathscr{T}_Y$ and $h_M(M,A,X) \in \mathscr{T}_M$ is zero, since: 

\vspace{-1.25cm} 
\begin{align*}
\E\big(h_Y(Y,M,A,X) \times h_M(M,A,X)\big) = \E\big(h_M(M,A,X) \times \E(h_Y(Y,M,A,X) \mid M,A,X)\big) = 0 \sp,
\end{align*}

\vspace{-.3cm} \noindent 
which confirms the orthogonality of $\mathscr{T}_Y$ and $\mathscr{T}_M$. Similar arguments can be applied to prove orthogonality between other pairs of tangent spaces. In the context of the front-door model, where there is no independence restriction among any sets of variables, the tangent space encompasses the entire Hilbert space. Broadly speaking, any statistical model in which $\mathscr{T}$ is equivalent to $\mathcal{H}$ is classified as \textit{nonparametric saturated}.

Any function $h(O)$ within the Hilbert space $\mathcal{H}$ can be \textit{uniquely} decomposed into orthogonal components, expressed as $h = h_Y + h_M + h_A + h_X$. Here, $h_V$ represents the projection of $h$ onto $\mathscr{T}_V$ for each $V$ in the set $\{Y, M, A, X\}$. A prime example of this decomposition is observed in the nonparametric EIF, which is an element in $\mathcal{H}$. An EIF, say denoted by $\Phi(\Q)(O)$, can be broken down into four distinct components, each corresponding to the unique projection of $\Phi(\Q)(O)$ onto one of the four mutually orthogonal tangent spaces. The projection $\Phi_Y(\Q)(O)$ is specifically shown as a unique projection of $\Phi(\Q)(O))$ onto $\mathscr{T}_Y$. Similar proofs for $\Phi_M(\Q)(O)$, $\Phi_A(\Q)(O)$, and $\Phi_X(\Q)(O)$ as projections onto $\mathscr{T}_M$, $\mathscr{T}_A$, and $\mathscr{T}_X$, respectively, can be readily formulated. Demonstrating that $\Phi_Y(\Q)(O)$ is a projection of $\Phi(O)$ onto $\mathscr{T}_Y$ is equivalent to showing that for any $h_Y(Y,M,A,X) \in \mathscr{T}_Y$, the equation $\E\big((\Phi(\Q)(O) - \Phi_Y(\Q)(O)) h_Y(Y,M,A,X)\big) = 0$ holds true. Note that $\Phi(\Q)(O) - \Phi_Y(\Q)(O)$ is only a function of $M, A, X$. Thus, via the tower rule, we have: 
$\E\big((\Phi(\Q)(O) - \Phi_Y(\Q)(O))h_Y(Y,M,A,X)) = \E\Big((\Phi(\Q)(O) - \Phi_Y(\Q)(O))\E\big(h_Y(Y,M,A,X) \sp \big| \sp M,A,X\big) \Big) = 0.$


In the following, we let $o=(x,a,m,y)$ denote realizations of $O=(X,A,M,Y)$. 

\vspace{0.5cm}
\noindent \textbf{\underline{\large EIF for the identification functional of $\E(Y^{a_0})$}}

\noindent 
The EIF for the ID functional of $\E(Y^{a_0})$, denoted by $\psi(\Q)$ ($\equiv \psi(\P)$), is derived as follows:

\vspace{-1.5cm}
\begin{align*}
\frac{\partial}{\partial \varepsilon} \psi\left(\P_{\varepsilon}\right) \Big|_{\varepsilon=0} 
&=\frac{\partial}{\partial \varepsilon} \int y \sp \diff \P_{\varepsilon}\left(y \mid m, a, x\right) \diff \P_{\varepsilon}(m \mid a_0, x) \diff \P_{\varepsilon}\left(a \mid x\right) \diff \P_{\varepsilon}(x)\Big|_{\varepsilon=0}  \\
& =\int y S\left(y \mid m, a, x\right) \diff \P\left(y \mid m, a, x\right) \diff\P(m \mid a_0, x) \diff \P\left(a \mid x\right) \diff\P(x) \quad(1) \\
& \hspace{0.25cm}+\int y S(m \mid a_0, x) \diff \P\left(y \mid m, a, x\right) \diff\P(m \mid a_0, x) \diff \P\left(a \mid x\right) \diff\P(x) \quad(2) \\
& \hspace{0.25cm}+\int y S\left(a, x\right) \diff \P\left(y \mid m, a, x\right) \diff\P(m \mid a_0, x) \diff \P\left(a \mid x\right) \diff\P(x) \sp . \quad(3)
\end{align*}

\vspace{-0.3cm} \noindent 
Given our notations, line (1) simplifies to: 

\vspace{-1.25cm} 
\begin{align*}
&\int y S\left(y \mid m, a, x\right) \diff \P\left(y \mid m, a, x\right) \diff\P(m \mid a_0, x) \diff \P\left(a \mid x\right) \diff\P(x)\\
&\hspace{1cm}=\int f_M^{r}(m,a,x)\left[y-\mu(m, a,x)\right] S\left(y \mid m, a, x\right) \diff \P\left(y, m, a, x\right) \\
&\hspace{1cm}=  \int f_M^{r}(m,a,x)\left[y-\mu(m,a,x)\right] S\left(o\right) \diff\P(o) \sp .
\end{align*}
Line (2) simplifies to: 

\vspace{-1.5cm} 
\begin{align*}
&\int y S(m \mid a_0, x) \diff \P\left(y \mid m, a, x\right) \diff\P(m \mid a_0, x) \diff \P\left(a \mid x\right) \diff\P(x) \\
&\hspace{1cm}=\int \sum_{a} \mu(m, a,x) \pi(a\mid x) S(m \mid a_0, x) \diff\P(m \mid x, a_0) \diff\P(x)\\
&\hspace{1cm}=\int \frac{\I\left(a=a_0\right)}{\pi(a\mid x)}\xi(m,x)S(m \mid a_0, x) \diff\P(o)\\
&\hspace{1cm}=\int \frac{\I\left(a=a_0\right)}{\pi(a\mid x)}\Big[\xi(m, x)-\theta(x)\Big] S(m \mid a, x) \diff\P(o)\\
&\hspace{1cm}=\int \frac{\I\left(a=a_0\right)}{\pi(a\mid x)}\left[\xi(m, x)-\theta(x)\right] S(o) \diff\P(o) \sp .
\end{align*}

Line (3) simplifies to: 

\vspace{-1.5cm} 
\begin{align*}
& \int y S\left(a, x\right) \diff \P\left(y \mid m, a, x\right) \diff\P(m \mid a_0, x) \diff \P(a, x) \\
&\hspace{1cm}= \int\left(\eta(a, x)-\psi\right) S\left(a, x\right) \diff \P(a, x) \\
&\hspace{1cm}= \int\left(\eta(a, x)-\psi\right) S(o) \diff\P(o) \sp .
\end{align*}
Therefore, the EIF for $\psi(\Q)$, denoted by $\Phi(\Q)(O)$, is: 

\vspace{-1.25cm} 
\begin{align*}
    \Phi(\Q)(O) 
    &= \sp \underbrace{\frac{f_M(M \mid a_0, X)}{f_M(M \mid A, X)} \left\{ Y - \mu(M, A, X) \right\}}_{\Phi_Y(\Q)(O)}
    \sp + \sp \underbrace{\frac{\mathbb{I}(A = a_0)}{\pi(a_0 \mid X)} \left\{\xi(M,X) - \theta(X) \right\}}_{\Phi_M(\Q)(O)}  \\ 
    &\hspace{0.5cm} + \underbrace{\eta(A, X) - \theta(X)}_{\Phi_A(\Q)(O)}  
   \sp + \sp \underbrace{\theta(X) - \psi(\Q)}_{\Phi_X(\Q)(O)} \sp .
\end{align*}
When $A$ is binary, $\Phi_A(\Q)$ can be simplified as: 

\vspace{-1.25cm} 
\begin{align*}
    \eta(A, X) - \theta(X)&= \sum_{a=0}^{1} \left[ \I(A = a) \sp \eta(a, X) - \eta(a,X) \sp \pi(a\mid X) \right] \\
    &=\sum_{a' = 0}^1 \eta(a, X) \{ \I(A = a) - \pi(a\mid X) \} \\
    & = \{\eta(1, X) - \eta(0, X)\} \{A - \pi(1\mid X) \} \sp . 
\end{align*}
Similarly, when $M$ is binary, $\Phi_M(\Q)$ can be simplified as:

\vspace{-1.15cm} 
\begin{align*}
\frac{\mathbb{I}(A = a_0)}{\pi(a_0\mid X)}  \sp \{ \xi(M,X) - \theta(X)\} &=\frac{\mathbb{I}(A = a_0)}{\pi(a_0\mid X)} \sum_{m=0}^{1} \{\I(M = m) \xi(m,X) - \xi(m,X) \sp f_M(m\mid a_0, X) \}\\
&= \frac{\I(A = a_0)}{\pi(a_0\mid X)} \sum_{m=0}^1 \xi(m,X) \sp \{\I(M = m)  -  f_M(m\mid a_0, X) \} \\
&= \frac{\I(A = a_0)}{\pi(a_0\mid X)} \{\xi(1, X) - \xi(0,X) \} \{M - f_{M}(1\mid a_0, X) \} \sp . 
\end{align*}

\noindent \textbf{\underline{\large EIF for the identification functional of $\E(Y^{a_0} \! \mid \! A=a_1)$}}

\noindent 
The EIF for  the ID functional of $\E(Y^{a_0} \! \mid \! A=a_1)$, denoted by $\beta(\P)$ ($\equiv \beta(\Q)$), is derived as follows:

\vspace{-1.25cm}
\begin{align*}
\frac{\partial}{\partial \varepsilon} \beta\left(\P_{\varepsilon}\right) \Big|_{\varepsilon=0} 
&=\frac{\partial}{\partial \varepsilon} \int y \sp \diff \P_{\varepsilon}\left(y \mid m, a_1, x\right) \diff \P_{\varepsilon}(m \mid a_0, x) \diff \P_{\varepsilon}\left(x \mid a_1\right) \Big|_{\varepsilon=0}  \\
& =\int y S\left(y \mid m, a_1, x\right) \diff \P\left(y \mid m, a_1, x\right) \diff\P(m \mid a_0, x) \diff \P\left(x \mid a_1\right) \quad(4) \\
& \hspace{0.25cm}+\int y S(m \mid a_0, x) \diff \P\left(y \mid m, a_1, x\right) \diff\P(m \mid a_0, x) \diff \P\left(x \mid a_1\right) \quad(5) \\
& \hspace{0.25cm}+\int y S\left(x\mid a_1\right) \diff \P\left(y \mid m, a_1, x\right) \diff\P(m \mid a_0, x) \diff \P\left(x \mid a_1\right)\sp. \quad(6)
\end{align*}

\noindent
Given our notations, line (4) simplifies to:

\vspace{-1.25cm} 
\begin{align*}
    &\int y S\left(y \mid m, a_1, x\right) \diff \P\left(y \mid m, a_1, x\right) \diff\P(m \mid a_0, x) \diff \P\left(x \mid a_1\right)
    \\
    &\hspace{1cm}=\int \frac{\I(a=a_1)}{\p_A(a_1)}\frac{f_M(m\mid a_0,x)}{f_M(m\mid a_1,x)}[y-\mu(m,a_1,x)]\ S(y\mid m,a,x)\ \diff\P(y,m,a,x)
    \\
    &\hspace{1cm}=\int \frac{\I(a=a_1)}{\p_A(a_1)}\frac{f_M(m\mid a_0,x)}{f_M(m\mid a,x)}[y-\mu(m,a_1,x)] \sp S(o)\ \diff\P(o) \sp .
\end{align*}

Line (5) simplifies to:

\vspace{-1.5cm} 
\begin{align*}
    &\int y S(m \mid a_0, x) \diff \P\left(y \mid m, a_1, x\right) \diff \P(m \mid a_0, x) \diff \P\left(x \mid a_1\right)
    \\
    &\hspace{1cm}=\int \frac{\I(a=a_0)}{\p_A(a_1)}\frac{\pi(a_1\mid x)}{\pi(a_0\mid x)} [\mu(m,a_1,x) - \kappa_{a_1}(x)] \sp S(m\mid a,x) \sp \diff\P(m,a,x)
    \\
    &\hspace{1cm}=\int \frac{\I(a=a_0)}{\p_A(a_1)}\frac{\pi(a_1\mid x)}{\pi(a_0\mid x)} [\mu(m,a_1,x) - \kappa_{a_1}(x)] \sp S(o) \sp \diff\P(o) \sp .
\end{align*}

Line (6) simplifies to:

\vspace{-1.5cm} 
\begin{align*}
    &\int y S\left(x\mid a_1\right) \diff \P\left(y \mid m, a_1, x\right) \diff \P(m \mid a_0, x) \diff \P\left(x \mid a_1\right)
    \\
    &\hspace{1cm}=\int \frac{\I(a=a_1)}{\p_A(a_1)}[\kappa_{a_1}(x) - \beta] \sp S(x\mid a) \sp \diff\P(x,a)
    \\
    &\hspace{1cm}=\int \frac{\I(a=a_1)}{\p_A(a_1)}[\kappa_{a_1}(x) - \beta] \sp S(o) \sp \diff\P(o) \sp .
\end{align*}

Therefore, the EIF for $\beta(\Q)$, denoted by $\Phi_{\beta}(\Q)(O)$ is given by:

\vspace{-1.25cm} 
\begin{align*}
    \Phi_{\beta}(\Q)(O)  
    &=\underbrace{\frac{\I(a=a_1)}{\p_A(a_1)}\frac{f_M(m\mid a_0,x)}{f_M(m\mid a,x)}[y-\mu(m,a_1,x)]}_{\Phi_{\beta,Y}(\Q)(O)} \\
    &+ \underbrace{\frac{\I(a=a_0)}{\p_A(a_1)}\frac{\pi(a_1\mid x)}{\pi(a_0\mid x)} [\mu(m,a_1,x) - \kappa_{a_1}(x)]}_{\Phi_{\beta,M}(\Q)(O)}
    \\
    &+\underbrace{\frac{\I(a=a_1)}{\p_A(a_1)}[\kappa_{a_1}(x) - \beta]}_{\Phi_{\beta,AX}(\Q)(O)} \sp .
\end{align*}

\subsection{Overview of one-step corrected plug-in estimation} 
\label{app:one-step_overview}

The stochastic behavior of a plug-in estimator $\psi(\hat{\Q})$ can be studied using a linear expansion of the parameter. Given an $\P$-integrable function $f$ of the observed data $O$, let $\P f \coloneqq \int \! f(o) \p(o) \diff o$ and $\Pn f \coloneqq \frac{1}{n} \sum_{i = 1}^n f(O_i)$. A linear expansion of $\psi(\hat{\Q})$ yields $\psi(\hat{\Q}) = \psi(\Q) - \P \Phi(\hat{\Q}) + \Rem(\hat{\Q}, \Q)$, where $\Phi$ is a gradient of $\psi$ satisfying $\P \Phi(\Q) = 0$, and $\Rem(\hat{\Q}, \Q)$ denotes a second-order remainder term. While multiple gradients may satisfy the expansion in general, the tangent space of our model is saturated such that there is only a single, unique gradient. This gradient is also known as the efficient influence function (EIF) due to its foundational link to the theory of regular, asymptotically linear estimators \citep{bickel1993efficient}. 

To better characterize the stochastic behavior of  $\psi(\hat{\Q})$,  we rewrite its linear expansion as
\begin{align}
    \psi(\hat{\Q}) - \psi(\Q) = \Pn \Phi(\Q) - \Pn \Phi(\hat{\Q}) + (\Pn - \P) \big\{ \Phi(\hat{\Q}) - \Phi(\Q) \big\} + \Rem(\hat{\Q}, \Q) \sp . 
    \label{eq:expansion_plus}
\end{align}

\vspace{-0.35cm}
\noindent The \textit{first} term in \eqref{eq:expansion_plus} is a sample average of mean-zero i.i.d. terms and thus enjoys standard root-$n$ asymptotic behavior. The \textit{third} term is an empirical process term, which can be shown to be $o_\p(n^{-1/2})$ if $\Phi(\hat{\Q}) - \Phi(\Q)$ falls in a $\P$-Donsker class with probability tending to 1 and $\P \{ \Phi(\hat{\Q}) - \Phi(\Q)\}^2 = o_\p(1)$ \citep{vaart2023empirical}. In Section~\ref{sec:asymptotic}, we use sample-splitting procedure to assure that the third term is $o_\p(n^{-1/2})$, even if Donsker conditions are not met \citep{kennedy2024semiparametric, double17chernozhukov}. The \textit{fourth} term is the second-order remainder, which can generally be bounded by the convergence rates of respective components of $\hat{\Q}$ to their true counterparts. To precisely bound the second-order remainder, we must consider its explicit form. We characterize this remainder in Section~\ref{sec:asymptotic}. For the time being, it suffices to state that if the rates of convergence of nuisance estimators are sufficiently fast, then we generally expect $\Rem(\hat{\Q}, \Q) = o_\p(n^{-1/2})$. Finally, the \textit{second} term in \eqref{eq:expansion_plus} is the first-order bias of the plug-in estimator. When flexible nuisance estimation strategies are used (e.g., based on machine learning), this term may not have standard root-$n$ asymptotic behavior. This  motivates the one-step corrected plug-in estimator, denoted by $ \psi^+_1(\hat{\Q})$, to be $ \psi(\hat{\Q}) + \Pn \Phi(\hat{\Q})$.

\subsection{Overview of the TMLE framework}
\label{app:tmle_overview} 

Given a plug-in estimator $\psi(\hat{\Q})$ of the parameter of interest $\psi(\Q)$, the core idea of a TMLE procedure is to find a replacement for $\hat{\Q}$, say $\hat{\Q}^\star$, such that the following two aims hold: 
\begin{itemize}
    \item[]\textbf{(I)} \sp $\hat{\Q}^\star$ is at least as good  an estimate of $\Q$ as $\hat{\Q}$, w.r.t. a valid measure of empirical risk,  

    \item[]\textbf{(II)} \sp $\Pn \Phi(\hat{\Q}^\star) = o_\p(n^{-1/2})$, so that the first-order bias of $\psi(\hat{\Q}^\star)$ would be negligible. 
\end{itemize}
Consider the general setting where $\psi(\Q)$ is the parameter of interest and $\Q$ is parameterized as $(\Q_1, \Q_2, \dots, \Q_J)$, i.e.,  there are $J$ key nuisance parameters needed to evaluate $\psi$ and its EIF. We assume $\Q$ belongs in a functional space $\calQ$, defined as $\M_{\Q_1} \times \M_{\Q_2} \times \cdots \times \M_{\Q_J}$, i.e., the Cartesian product of the functional spaces of each nuisance functional, denoted by $\M_{\Q_j}$. Suppose also that the EIF can be written as $\Phi = \sum_{j=1}^J \Phi_j$, where $\Phi_j$ is the component of $\Phi$ that belongs to the tangent space associated with $\Q_j$. For example, for $\psi(\Q)$ in \eqref{eq:id_ATE}, we can set $\Q = (\mu, f_M, \pi, \p_X)$, and according to the EIF in \eqref{eq:eif_tangent_space} $\Phi_1 = \Phi_Y, \Phi_2 = \Phi_M, \Phi_3 = \Phi_A, \Phi_4=\Phi_X$. 

To achieve both aims (I)-(II), the TMLE procedure comprises two main steps: the \textit{initialization} step, where the initial estimate $\hat{\Q}$ is obtained, and the subsequent \textit{targeting} step, where $\hat{\Q}$ is updated to a new estimate $\hat{\Q}^\star$. In the \textit{initialization} step, we obtain an initial estimate of $\Q$ based on a collection of estimates for each nuisance parameter individually, $\hat{\Q} = (\hat{\Q}_1, \ldots, \hat{\Q}_J)$. In the \textit{targeting} step, we require (i) a \emph{submodel} and (ii) a \emph{loss function} for each component $\Q_j$ of $\Q$. For requirement (i), with an estimate $\hat{\Q}$ of $\Q$, we define a submodel $\{ \hat{\Q}_{j}(\varepsilon_j; \hat{\Q}_{-j}), \sp \varepsilon_j \in \R \}$ within $\mathcal{M}_{\Q_j}$. This submodel is indexed by a univariate real-valued parameter $\varepsilon_j$ and may also depend on $\hat{\Q}_{-j}$ (the components of $\hat{\Q}$ excluding component $j$) or a subset of $\hat{\Q}_{-j}$ (including the possibility of an empty subset). For requirement (ii), with a given $\tilde{\Q} \in \calQ$, we denote a loss function for $\tilde{\Q}_j$ by $L(\tilde{\Q}_j; \tilde{\Q}_{-j}): \mathcal{O} \rightarrow \R$, where $\mathcal{O}$ denotes the state space of the observed data. 
Note that the loss function for $\tilde{\Q}_j$ can also be indexed by $\tilde{\Q}_{-j}$, or possibly by a subset of $\tilde{\Q}_{-j}$, which may sometimes be an empty set. The submodel and loss function must be chosen to satisfy:
\begin{itemize}
    \item[]\textbf{(C1)} \sp $\hat{\Q}_j(0; \hat{\Q}_{-j}) = \hat{\Q}_j \sp ,$ 
    \item[] \textbf{(C2)} \sp $\Q_j = \argmin_{\tilde{\Q}_j \in \M_{\Q_j}} \int L(\tilde{\Q}_j; \Q_{-j})(o) \sp \p(o) \sp \diff o \sp ,$
    \item[] \textbf{(C3)} \sp $\frac{\partial
    }{\partial\varepsilon_j} L\big(\hat{\Q}_j(\varepsilon_j; \hat{\Q}_{-j}); \hat{\Q}_{-j}\big)\big|_{\varepsilon_j=0} = \Phi_j(\hat{\Q}) \sp .$ 
\end{itemize}
(C1) implies that the submodel aligns with the given estimate $\hat{\Q}_j$ at $\varepsilon_j=0$; (C2) indicates that the expectation of the loss function under the true distribution $\P$ is minimized at $\Q_j$; and (C3) ensures that the evaluation of the derivative of the loss function with respect to $\varepsilon_j$ at $0$ is equivalent to evaluation of the corresponding component of the EIF at $\hat{\Q}$. 

Given appropriate choices of submodels and loss functions, we proceed to update $\hat{\Q}$ via an iterative risk minimization process. Given current estimates at iteration $t$, say $\hat{\Q}^{(t)}$, we update $\hat{\Q}_j^{(t)}$ via empirical risk minimization along the selected submodel using the selected loss function. That is, we define $\hat{\varepsilon}_j = \argmin_{\varepsilon_j \in \R} \Pn L(\hat{\Q}_j(\varepsilon_j; \hat{\Q}_{-j}^{(t)}); \hat{\Q}_{-j}^{(t)})$ to be the value of $\varepsilon_j$ that minimizes empirical risk  given current estimates $\hat{\Q}_{-j}^{(t)}$. Condition (C2) suggests that the updated estimate $\hat{\Q}_j^{(t+1)} = \hat{\Q}_j(\hat{\varepsilon}_j; \hat{\Q}_{-j}^{(t)})$ should satisfy (I), as $\hat{\Q}_j^{(t+1)}$ will have lower empirical risk than $\hat{\Q}_j^{(t)}$. This process is repeated for each of the $J$ components of $\Q$ resulting in an updated estimate $\hat{\Q}^{(t+1)}$. Condition (C3) suggests that if during this updating process we have found that $\hat{\varepsilon}_j \approx 0$ for each $j$, then we might expect $\Pn \Phi_j(\hat{\Q}^{(t+1)}) \approx 0$ for each $j$ and thus (II) may be satisfied. If after iteration $t$, we find that (II) is not approximately satisfied, we would repeat the  updating process. The process is repeated until $\Pn \Phi(\hat{\Q}^{(t)}) < C_n$, where $C_n=o_\p(n^{-1/2})$, e.g., $C_n = \{n^{1/2} \mbox{log}(n) \}^{-1}$. The final estimate of $\Q$ is denoted as $\hat{\Q}^\star$ and the TMLE is defined as the plug-in estimator $\psi(\hat{\Q}^\star)$.

We derive TMLEs for all three representations of the ATE and ATT front-door functionals. These estimators differ in both stages of the TMLE procedure: (i) they use different parameterizations of the nuisance functions comprising $\Q$, requiring distinct estimation strategies, and (ii) they employ different techniques to achieve (II), the TMLE approximate-equation-solving property where $\Pn \Phi(\hat{\Q}^\star) = o_\p(n^{-1/2})$. Further methodological details are provided in Sections~\ref{sec:est_ATE} and \ref{sec:est_ATT}.  For a general overview of the TMLE framework, see \citet{van2011targeted}. 

\pagebreak
\section{Details on estimators for the ATE front-door functional}
\label{app:tmle_details}

\subsection{Validity of loss function and submodel combinations}
\label{app:tmle_loss+submodel}

We establish the validity of the loss function and submodel combinations (discussed in Appendix~\ref{app:tmle_overview}) for the binary mediator case, as detailed in Algorithm~\ref{appalg:binary} (Appendix~\ref{app:tmle_alg}) and discussed in Section~\ref{subsec:est_with_density}. Similar proofs for the remaining TMLE procedures follow analogously.

Since the proof for the $f_M$ update closely mirrors that for the propensity score $\pi$, we focus here on verifying conditions (C1)–(C3) for the updates to $\pi$ and $\mu$.

\vspace{0.15cm}
\noindent \underline{Loss function and submodel combination used for updating $\pi$}:  

\vspace{-1.25cm}
\begin{align*}
\hat{\pi}\left(\varepsilon_A; \hat{\mu}^{(0)}, \hat{f}_m^{(t)}\right)(1 \mid X) &= \operatorname{expit} \left\{\operatorname{logit}\{ \hat{\pi}^{(t)}(1 \mid X) \}+ \varepsilon_A\left\{\hat{\eta}^{(t)}(1, X)-\hat{\eta}^{(t)}(0, X)\right\} \right\}, \sp \varepsilon_A \in \mathbb{R} \sp ,\\
    L_{A}(\tilde{\pi})(O) &= - \log \tilde{\pi}(A \mid X) \sp . 
\end{align*}

\vspace{-0.3cm}
\noindent \textit{Proof of (C1):} 
$\hat{\pi}\big(\varepsilon_A=0; \hat{\mu}^{(0)}, \hat{f}_m^{(t)}\big)(1 \mid X) = \operatorname{expit} \left\{\operatorname{logit}\{ \hat{\pi}^{(t)}(1 \mid X) \}\right\}=\hat{\pi}^{(t)}(1 \mid X).$ 

\noindent \textit{Proof of (C2):}
$\E(L_A(\tilde{\pi})(O))
= \E(-\log \tilde{\pi}(A \mid X))
= \int \left\{ - \sum_{a} \pi(a\mid x) \log \tilde{\pi}(a \mid x) \right\} \diff\P(x) 
$ is minimized if $- \sum_{a} \pi(a\mid x) \log \tilde{\pi}(a \mid x)$ is minimized for any $x \in \mathcal{X}$. Since 

\vspace{-1.cm} 
\begin{align*}
    - \sum_{a} \pi(a\mid x) \log \tilde{\pi}(a \mid x)
    &= - \sum_{a} \pi(a\mid x) \log \left( \frac{\tilde{\pi}(a \mid x)}{\pi(a\mid x)} \times \pi(a\mid x) \right) \\
    &= - \sum_{a} \pi(a\mid x) \log  \frac{\tilde{\pi}(a \mid x)}{\pi(a\mid x)} - \sum_{a} \pi(a\mid x) \log \pi(a\mid x) \sp ,  
\end{align*}
we only need to focus on the minimization of $- \sum_{a} \pi(a\mid x) \log  \frac{\tilde{\pi}(a \mid x)}{\pi(a\mid x)}$, which corresponds to the Kullback-Leibler (KL) divergence  from $\pi(a\mid x)$ to $\tilde{\pi}(a \mid x)$, denoted by $D_\text{KL}(\pi \sp || \sp  \tilde{\pi})$. This KL-divergence is minimized if $\tilde{\pi}(A \mid X=x) = \pi(A \mid X=x)$, for all $x \in \mathcal{X}.$ 

\noindent\textit{Proof of (C3):}

\vspace{-1.25cm} 
{\small\begin{align*}
 \frac{\partial}{\partial\varepsilon_A}L_{A}(&\hat{\pi}(\varepsilon_A; \hat{\mu}^{(0)}, \hat{f}_m^{(0)}))\Bigg|_{\varepsilon_A=0} \\
 &= - \frac{\partial}{\partial\varepsilon_A} \bigg\{A \log \hat{\pi}(\varepsilon_A; \hat{\mu}^{(0)}, \hat{f}_m^{(0)}) +(1-A) \log \left\{1 -\hat{\pi}(\varepsilon_A; \hat{\mu}^{(0)}, \hat{f}_m^{(t)})\right\} \bigg\}\Bigg|_{\varepsilon_A=0}\\
  &= - \bigg\{A \frac{\frac{\partial}{\partial\varepsilon_A}\hat{\pi}(\varepsilon_A; \hat{\mu}^{(0)}, \hat{f}_m^{(0)})}{\hat{\pi}(\varepsilon_A; \hat{\mu}^{(0)}, \hat{f}_m^{(0)})} +(1-A) \frac{-\frac{\partial}{\partial\varepsilon}\hat{\pi}(\varepsilon_A; \hat{\mu}^{(0)}, \hat{f}_m^{(0)})}{1-\hat{\pi}(\varepsilon_A; \hat{\mu}^{(0)}, \hat{f}_m^{(0)})}\bigg\}\Bigg|_{\varepsilon_A=0}\\
  &=\left\{\hat{\eta}^{(t)}(1, X)-\hat{\eta}^{(t)}(0, X)\right\} \left\{\hat{\pi}^{(t)}(1\mid X)-A\right\} \sp \propto \sp \Phi_A(\hat{\Q}^{(t)}) \sp . 
\end{align*}}

\noindent \underline{Loss function and submodel combination used for updating $\mu$}: 

\vspace{-1.25cm} 
\begin{align*}
    \hat{\mu}(\varepsilon_Y)(M,A,X) &= \hat{\mu}^{(t)}(M,A,X) + \varepsilon_Y \sp , \sp \varepsilon_Y \in \mathbb{R} \sp , \\
    L_Y\left(\tilde{\mu}; \hat{f}_M^{(t)}\right)(O) 
    &= \frac{\hat{f}_M^{(t)}(M \mid a_0, X)}{\hat{f}_M^{(t)}(M \mid A, X)} \{ Y - \tilde{\mu}(M, A, X) \}^2 \sp .
\end{align*}

\noindent \textit{Proof of (C1):} 
    $\hat{\mu}(\varepsilon_Y=0)(M,A,X) = \hat{\mu}^{(t)}(M,A,X)$. 

\noindent \textit{Proof of (C2):}

\vspace{-1.25cm} 
\begin{align*}
    \E(L_Y(&\tilde{\mu}; \hat{f}_M^{(t)})(O)) \\ 
    &=\E\bigg(\frac{\hat{f}_M^{(t)}(M \mid a_0, X)}{\hat{f}_M^{(t)}(M \mid A, X)} \{ Y-\tilde{\mu}(M,A,X) \}^2\bigg)\\
    &=\E\bigg(\frac{\hat{f}_M^{(t)}(M \mid a_0, X)}{\hat{f}_M^{(t)}(M \mid A, X)} \{ Y-\mu(M,A,X) \}^2 + \frac{\hat{f}_M^{(t)}(M \mid a_0, X)}{\hat{f}_M^{(t)}(M \mid A, X)} \{ \mu(M,A,X) - \tilde{\mu}(M, A, X) \}^2\bigg) \sp ,
\end{align*}
which is minimized when $\tilde{\mu}(M,A,X)=\mu(M,A,X)$. 

\noindent \textit{Proof of (C3):}

\vspace{-1.25cm} 
\begin{align*}
    \frac{\partial}{\partial\varepsilon}L_Y(\hat{\mu}(\varepsilon_Y; \hat{f}_M^{(t)}))\Bigg|_{\varepsilon=0}=2\ \frac{\hat{f}_M^{(t)}(M \mid a_0, X)}{\hat{f}_M^{(t)}(M \mid A, X)}(Y-\hat{\mu}^{(t)}(M,A,X)) \propto \Phi_Y(\hat{\Q}^{(t)}) \sp .
\end{align*}

\subsection{TMLE considerations for binary outcome}
\label{app:tmle_binaryY}

For binary outcomes, the TMLE procedure for computing $\psi_{1}(\hat{\Q}^\star)$—originally described in Section~\ref{subsec:est_with_density} for continuous outcomes—requires the following modifications.

We adopt a new loss function and submodel for updating $\hat{\mu}$:

\vspace{-1.cm} 
\begin{equation}\label{app:eq:submodels_binaryY}
\begin{aligned} 
    &\hat{\mu}(\varepsilon_Y;\hat{f}_M^{(t)})(M,A,X) = \operatorname{expit} \Big\{\operatorname{logit}\hat{\mu}^{(t)}(M,A,X)+\varepsilon_Y \frac{\hat{f}_M^{(t)}(M\mid a_0,X)}{\hat{f}_M^{(t)}(M\mid A,X)} \Big\} \sp , \sp \varepsilon_Y\in \mathbb{R} \sp , \\
    &L_Y(\tilde{\mu}) = - \log \tilde{\mu}(M,A,X) \sp .
\end{aligned}
\end{equation}

Due to the nonlinear nature of the parametric submodel in \eqref{app:eq:submodels_binaryY} with respect to $\varepsilon_Y$,  computations of $ \hat{\eta}^{(t)}(1,X) - \hat{\eta}^{(t)}(0,X)$ and $ \hat{\xi}^{(t)}(1,X) - \hat{\xi}^{(t)}(0,X)$ would depend on updated estimate of $\hat{\mu}^{(t)}$. Therefore, unlike the continuous outcome case, the dependence of submodels $\hat{\pi}\big(\varepsilon_A; \hat{\mu}^{(t)}, \hat{f}_m^{(t)}\big)$ and $\hat{f}_M\big(\varepsilon_M; \hat{\mu}^{(t)}, \hat{\pi}^{(t)}\big)$ on $\hat{\mu}^{(t)}$ would be through the updated estimate $\hat{\mu}^{(t)}$. This implies that once the estimate of $\mu$ is updated, the estimates for $f_M$ and $\pi$ must be updated accordingly. Given $\hat{\Q}^{(t)} = (\hat{\mu}^{(t)}, \hat{f}^{(t)}_M, \hat{\pi}^{(t)}, \hat{\p}_X)$, we modify Step 2 of the continuous outcome case, discussed in Section~\ref{subsec:est_with_density}, as follows. 

\vspace{0.15cm}
\noindent \emph{Step 2a: Update $\hat{\pi}$,} by following the exact same procedure as the one discussed in Section~\ref{subsec:est_with_density}, modula the fact that $\hat{\mu}$ is replaced with $\hat{\mu}^{(t)}$. After performing the empirical risk minimization and obtaining $\hat{\varepsilon}_A$, we update $\hat{\pi}^{(t+1)} = \pi(\hat{\varepsilon}_A; \hat{\mu}^{(t)}, \hat{f}_M^{(t)})$ and define $\hat{\Q}^{(\text{temp}_1)} = (\hat{\mu}^{(t)}, \hat{\pi}^{(t+1)}, \hat{f}_M^{(t)}, \hat{\p}_X)$. Condition (C3) implies that $\Pn \Phi_A(\hat{\Q}^{(\text{temp}_1)}) = o_\p(n^{-1/2})$.

\vspace{0.15cm}
\noindent \emph{Step 2b: Update $\hat{f}_M$,} by following the exact same procedure as the one discussed in Section~\ref{subsec:est_with_density}, modula the fact that $\hat{\mu}$ is replaced with $\hat{\mu}^{(t)}$. After performing the empirical risk minimization and obtaining $\hat{\varepsilon}_M$, we update $\hat{f}^{(t+1)}_M = \hat{f}_M(\hat{\varepsilon}_M; \hat{\mu}^{(t)}, \hat{\pi}^{(t+1)})$ and define $\hat{\Q}^{(\text{temp}_2)} = (\hat{\mu}^{(t)}, \hat{\pi}^{(t+1)}, \hat{f}_M^{(t+1)}, \hat{\p}_X)$. Condition (C3) implies that $\Pn \Phi_M(\hat{\Q}^{(\text{temp}_2)}) = o_\p(n^{-1/2})$.

\vspace{0.15cm}
\noindent \emph{Step 2c: Update $\hat{\mu}$,} by performing an empirical risk minimization to find 

\vspace{-1.25cm} 
\begin{align}
    \hat{\varepsilon}_Y = \argmin_{\varepsilon_Y \in \mathbb{R}} \Pn L_Y(\hat{\mu}(\varepsilon_Y; \hat{f}_M^{(t+1)})) \sp .
    \label{appeq:eps_mu}
\end{align}

\vspace{-.3cm} \noindent 
This corresponds to fitting a logistic regression without an intercept term:

\vspace{-1.25cm} 
\begin{align*}
    Y\sim\mathrm{offset}\big(\mathrm{logit}\ \hat{\mu}^{(t)})+\hat{H}_Y^{(t)}(M,A,X) \sp , \quad \text{where } \sp \hat{H}_Y^{(t)}(M,A,X)\coloneqq \frac{\hat{f}_M^{(t+1)}(M\mid a_0,X)}{\hat{f}_M^{(t+1)}(M\mid A,X)} \sp .
\end{align*}

\vspace{-.3cm} \noindent 
The coefficient of $\hat{H}_Y^{(t)}(M,A,X)$ corresponds to $\hat{\varepsilon}_Y$ as a solution to \eqref{appeq:eps_mu}. We update $\hat{\mu}^{(t+1)} = \hat{\mu}(\hat{\varepsilon}_Y; \hat{f}_M^{(t+1)})$, and define $\hat{\Q}^{(t+1)} = (\hat{\mu}^{(t+1)}, \hat{\pi}^{(t+1)}, \hat{f}_M^{(t+1)}, \hat{\p}_X)$. Condition (C3) implies that $\Pn\Phi_{Y}(\hat{\Q}^{(t+1)})=o_\p(n^{-1/2})$. We increment $t$ and repeat \textit{Step 2} until convergence. 

\vspace{0.15cm}
Assume convergence at iteration $t^\star$. Let $\hat{\pi}^\star=\hat{\pi}^{(t^\star)}$, $\hat{f}_M^\star=\hat{f}_M^{(t^\star)}$, $\hat{\mu}^\star=\hat{\mu}^{(t^\star)}$, and define $\hat{\Q}^\star=(\hat{\mu}^\star,\hat{\pi}^\star,\hat{f}_M^\star,\hat{\p}_X)$.  
The TMLE plug-in is then given by $\psi_1(\hat{\Q}^\star)$, as described in \eqref{eq:tmle_1}.  

The TMLE procedure for computing $\psi_{2}(\hat{\Q}^\star)$—originally described in Section~\ref{subsec:est_without_density} for continuous outcomes—remains largely unchanged for binary outcomes, with the submodel–loss function pair in \eqref{app:eq:submodels_binaryY} used for updating $\hat{\mu}$.

\subsection{An alternative submodel for targeting $\hat{\mu}$ under continuous $Y$}
\label{app:tmle_nonlinear}
The TMLEs proposed for continuous $Y$ in the main manuscript rely on linear parametric submodels for targeting $\hat{\mu}$. However, such models may exhibit instability in sparse data settings with low Fisher information, as demonstrated in simulations by \cite{zhou2015coarsened}. To address this issue, \cite{gruber2010targeted} showed that TMLEs using parametric submodels constrained to remain within the semiparametric model of the observed data distribution tend to be more robust than those based on linear submodels. Motivated by this, we introduce an alternative TMLE that employs a nonlinear submodel for targeting the outcome regression $\mu$,when $Y$ is continuous. We outline the key ideas for this construction below, noting that the procedure closely parallels that described in Appendix~\ref{app:tmle_binaryY}.

Let $a$ and $b$ denote the minimum and maximum observed values of $Y$, respectively. To enable the use of nonlinear submodels designed for binary outcomes, we rescale $Y$ to the unit interval by defining $Y^* = (Y - a)/(b - a)$, so that $Y^* \in [0, 1]$. Targeting is then performed using $Y^*$ in place of $Y$, applying the nonlinear parametric submodels defined in \eqref{app:eq:submodels_binaryY}. All remaining steps of the TMLE procedure follow exactly as described in Appendix~\ref{app:tmle_binaryY}. Finally, to return to the original scale, we multiply the point estimate by $(b - a)$ and add $a$, and rescale the estimated EIF by multiplying it by $(b - a)$.

\subsection{Valid submodels for conditional density of a continuous mediator}
\label{app:tmle_valid_submodel}

To ensure that the submodel in \eqref{eq:submodel_cont} is a valid submodel of $\mathcal{M}_{f_M}$, the range of $\varepsilon_M$ must be restricted so that the submodel defines a valid probability density function; that is, $\hat{f}_M(\varepsilon_M; \hat{\mu}^{(0)}, \hat{\pi}^{(t)})(M \mid a_0, X) \geq 0$ for all $\varepsilon_M \in (-\delta, \delta)$. Recall that $\hat{\xi}^{(t)}(M,X) = \sum_{a=0}^{1}\hat{\mu}^{(0)}(M,a,X)\;\hat{\pi}^{(t)}(a\mid X)$ and $\hat{\theta}^{(t)}(X) = \int \hat{\xi}^{(t)}(m,X) \sp \hat{f}^{(t)}_M(m \mid a_0, X) \sp \diff m$. 

\noindent Let $S^{(t)}_{\text{pos}}$ denote the set of indices for observations with
\begin{align*}
    \frac{\hat{\xi}^{(t)}(M_i,X_i)-\hat{\theta}^{(t)}(X_i)}{\hat{\pi}^{(t)}(a_0\mid X_i)}>0 \sp . 
\end{align*}
For $i\in S^{(t)}_{\text{pos}}$, $\hat{f}_M(\varepsilon_M, \hat{\Q}^{(t)})(M \mid a_0, X)\geq 0$ implies that
$\varepsilon_M\geq L^{(t)}_i$, where 
$$L^{(t)}_i\coloneqq-\frac{\hat{\pi}^{(t)}(a_0\mid X_i)}{\hat{\xi}^{(t)}(M_i,X_i)-\hat{\theta}^{(t)}(X_i)} \sp .$$

\noindent Similarly, define $S^{(t)}_{\text{neg}}$ to be the set of indices for observations with 
\begin{align*}
\frac{\hat{\xi}^{(t)}(M_i,X_i)-\hat{\theta}^{(t)}(X_i)}{\hat{\pi}^{(t)}(a_0\mid X_i)}<0 \sp .
\end{align*} 
For $i\in S^{(t)}_{\text{neg}}$, $\hat{f}_M(\varepsilon_M, \hat{\Q}^{(t)})(M \mid a_0, X)\geq 0$ implies that $\varepsilon_M\leq R^{(t)}_i,$ where 
$$R^{(t)}_i\coloneqq-\frac{\hat{\pi}^{(t)}(a_0\mid X_i)}{\hat{\xi}^{(t)}(M_i,X_i)-\hat{\theta}^{(t)}(X_i)}\sp .$$

Let $L^{(t)}=\argmax_{i\in S^{(t)}_{\text{pos}}}L^{(t)}_i$ and $R^{(t)}=\argmin_{i\in S^{(t)}_{\text{neg}}}R^{(t)}_i$. For the given dataset, $(L, R)$ constitutes a valid domain for $\varepsilon_M$. For any $\varepsilon_M\in (L, R)$, we have $\hat{f}_M(\varepsilon_M; \hat{\mu},\hat{\pi}^{(t)})(M \mid a_0, X)\geq 0$. Any selection of $\delta$ ensuring $(-\delta, \delta) \subseteq (L, R)$ would be applicable for carrying out the TMLE procedure. Note that the valid domain for $\varepsilon_M$ changes over iteration alongside the iterative updates of estimates for $f_M$ and $\pi$. Consequently, the choice of $\delta$ should be relatively small to guarantee the submodel defined in \eqref{eq:submodel_cont} is a valid submodel over all iterations. 

Alternatively, we may use the following submodel where $\varepsilon_M$ can span the entire real line, 
\begin{align}
    \hat{f}_M(\varepsilon_M; \hat{\mu}^{(0)},  \hat{\pi}^{(t)})(M \mid a_0, X) \! = \! \frac{\hat{f}_M^{(t)}(M \! \mid \! a_0,X)\exp\bigg[ \displaystyle \frac{\varepsilon_M}{\hat{\pi}^{(t)}(a_0 \! \mid \! X)}  \Big( \hat{\xi}^{(t)}(M,X) - \hat{\theta}^{(t)}(X)  \Big)\bigg]}{\displaystyle \iint \hat{f}_M^{(t)}(m \! \mid \! a_0,x)\exp\bigg[ \displaystyle \frac{\varepsilon_M}{\hat{\pi}^{(t)}(a_0 \! \mid \! x)} \Big( \hat{\xi}^{(t)}(m,x) - \hat{\theta}^{(t)}(x)  \Big)\bigg] \diff m \sp \diff x} \sp .  
    \label{app:eq:alternative_submodel_fM_continuous}
\end{align}
This alternative submodel increases computational complexity, as the denominator must be numerically approximated at each iteration.

\subsection{TMLEs that avoid mediator density estimation}
\label{app:tmle_multi}
Given initial estimates $\hat{\Q}$, a TMLE version of $\psi_2^+(\hat{\Q})$ can be formulated as follows.

\vspace{0.2cm}
\noindent \emph{Step 1: Define loss functions and submodels} through $\hat{\mu}, \hat{\pi}$, $\hat{\gamma}$. Given $\hat{\Q} \in \calQ$, $\varepsilon_Y, \varepsilon_A, \varepsilon_\gamma \in \mathbb{R}$, define 

\vspace{-0.5cm} 
\begin{equation} \label{eq:submodels_multivariate}
\begin{aligned} 
    \hat{\mu}(\varepsilon_Y)(M,A,X) &= \hat{\mu}(M,A,X) + \varepsilon_Y \sp ,  \\
    \hat{\pi}(\varepsilon_A; \hat{\kappa})(1 \mid X) 
    &= \operatorname{expit} \Big[\operatorname{logit}\big\{ \hat{\pi}(1 \mid X) \big\}+ \varepsilon_A \big\{\hat{\kappa}_1(X)-\hat{\kappa}_0(X)\big\} \Big] \sp , \\
    \hat{\gamma}(\varepsilon_\gamma)(X) &= \hat{\gamma}(X)+\varepsilon_\gamma \sp . 
\end{aligned}
\end{equation}
For a given $\tilde{\mu}\in\M_{\mu}$, $\tilde{\pi}\in\M_{\pi}$, and $\tilde{\gamma}\in\M_{\gamma}$, define the following loss functions:

\vspace{-1.cm}  
\begin{equation} \label{eq:loss_functions_multivariate}
\begin{aligned} 
    &L_Y(\tilde{\mu};\hat{f}_M^r)(O) = \hat{f}_M^r(M,A,X) \{ Y - \tilde{\mu}(M, A, X) \}^2 \sp , \quad 
    L_{A}(\tilde{\pi})(O) = - \log \tilde{\pi}(A \mid X) \sp , \\
    &\hspace{2cm} L_{\gamma}(\tilde{\gamma};\hat{\pi},\hat{\xi})(O) = \frac{\I(A=a_0)}{\hat{\pi}(a_0\mid X)}\left(\hat{\xi}(M,X)-\tilde{\gamma}(X)\right)^2 \sp .
\end{aligned}
\end{equation}
See Appendix~\ref{app:tmle_loss+submodel} for a proof of validity of these submodel–loss function pairs under (C1)–(C3). 

Note that the submodel $\hat{\pi}(\varepsilon_A;\hat{\kappa})$ is indexed by $\hat{\kappa}$, which in turn depends on $\hat{\mu}$. However, this submodel remains invariant to updates of $\hat{\mu}$ due to the linearity of the $\mu$ submodel in $\varepsilon_Y$, which makes $\hat{\kappa}_1(X) - \hat{\kappa}_0(X)$ effectively fixed by the initial $\hat{\mu}$. Moreover, since the submodels and loss functions for $\hat{\pi}$ and $\hat{\mu}$ are independent of each other’s updates, their targeting steps can be performed simultaneously in a single step. In contrast, the submodel and loss function for $\hat{\gamma}$ depend on the targeted versions of $\hat{\pi}$ and $\hat{\mu}$. Thus, targeting $\hat{\gamma}$ must follow the updates of $\hat{\pi}$ and $\hat{\mu}$, using $\hat{\xi}$ and $\hat{\gamma}$ computed from those updated estimates. This sequencing ensures that $\hat{\gamma}$ is targeted using the most recent nuisance values.

\vspace{0.15cm}
\noindent \emph{Step 2: Perform empirical risk minimizations using submodels and loss functions for $\mu$ and $\pi$.} 

\vspace{0.15cm}
\noindent \emph{Step 2a: Update an estimate of $\mu$} by performing an empirical risk minimization to find $\hat{\varepsilon}_Y = \argmin_{\varepsilon_Y \in \mathbb{R}} \Pn L_Y(\hat{\mu}(\varepsilon_Y);\hat{f}_M^r)$. This minimization problem can be solved by fitting $Y\sim\mathrm{offset}(\hat{\mu})+1$ with weight $\hat{f}_M^r(M,A,X)$. The intercept coefficient corresponds to $\hat{\varepsilon}_Y$ as the minimizer of the empirical risk. Define $\hat{\mu}^\star = \hat{\mu}(\hat{\varepsilon}_Y;\hat{f}_M^r )$ and let $\hat{\Q}^{(1)}=(\hat{\mu}^\star, \hat{\gamma}, \hat{f}^{r}_M, \hat{\kappa}, \hat{\pi}, \hat{\p}_X)$. Condition (C3) implies that $\Pn\Phi_{Y}(\hat{\Q}^{(1)})=0$. 

\vspace{0.15cm}
\noindent \emph{Step 2b: Update an estimate of $\pi$} by performing an empirical risk minimization to find $\hat{\varepsilon}_A = \argmin_{\varepsilon_A \in \mathbb{R}} \Pn L_A(\hat{\pi}(\varepsilon_A; \hat{\kappa}))$. The solution  is obtained by fitting the following logistic regression without an intercept term:

\vspace{-1.5cm} 
\begin{align*}
    A\sim\mathrm{offset}(\operatorname{logit}\ \hat{\pi}(1\mid X))+\hat{H}_{A}(X) \sp ,  \sp \text{where } \sp \hat{H}_A(X)=\hat{\kappa}_1(X) - \hat{\kappa}_0(X) \sp .
\end{align*}%

\vspace{-.3cm} \noindent 
The coefficient in front of the clever covariate $\hat{H}_{A}(X)$ corresponds to $\hat{\varepsilon}_A$ as the minimizer of the empirical risk. Define $\hat{\pi}^\star = \pi(\hat{\varepsilon}_A; \hat{\mu})$ and let $\hat{\Q}^{(2)}=(\hat{\mu}^\star, \hat{\gamma}, \hat{f}^{r}_M, \hat{\kappa}, \hat{\pi}^\star, \hat{\p}_X)$.  Condition (C3) implies that $\Pn\Phi_{A}(\hat{\Q}^{(2)}) = 0$. Compute $\hat{\gamma}(X)$ by fitting the following linear regression using only data points where $A_i=a_0$ and making prediction using all the data points of $X$:

\vspace{-1.25cm} 
\begin{align*}
    \hat{\xi}^\star(M, X) \sim X \sp , \sp \text{where } \hat{\xi}^\star(M, X) = \sum_{a = 0}^1 \hat{\mu}^\star(M, a, X) \sp \hat{\pi}^\star(a \mid X) \sp . 
\end{align*}

\noindent \emph{Step 3: Perform one-step risk minimization using pre-defined submodel and loss function for $\gamma$.} Update $\gamma$ by performing an empirical risk minimization to find 

\vspace{-1.25cm}  
\begin{align}
    \hat{\varepsilon}_\gamma \sp = \sp \argmin_{\varepsilon_\gamma \in \mathbb{R}} \sp \Pn L_{\gamma}\left(\hat{\gamma}(\varepsilon_\gamma); \hat{\pi}^\star,\hat{\xi}^\star\right) \sp .
    \label{eq:eps_mu_multi}
\end{align}%

\vspace{-.3cm} \noindent 
The solution can be obtained by fitting $\hat{\xi}^\star\sim\mathrm{offset}(\hat{\gamma})+1$ with weight $\frac{\I(A=a_0)}{\hat{\pi}^\star(a_0\mid X)}$.
The intercept coefficient corresponds to $\hat{\varepsilon}_\gamma$ as a solution to the optimization problem in \eqref{eq:eps_mu_multi}. Define $\hat{\gamma}^\star = \hat{\gamma}(\hat{\varepsilon}_\gamma)$ and let $\hat{\Q}^\star =(\hat{\mu}^\star, \hat{\gamma}^\star, \hat{f}^{r}_M, \hat{\kappa}, \hat{\pi}^\star,  \hat{\p}_X)$. Condition (C3) implies that $\Pn\Phi(\hat{\Q}^\star) = 0$. 

\vspace{0.15cm}
\noindent \emph{Step 4: Evaluate the plug-in estimator} in \eqref{eq:plugin_2} based on updated estimate $\hat{\gamma}^{\star}$, 

\vspace{-1.25cm} 
\begin{align}
    \psi_2(\hat{\Q}^\star) = \frac{1}{n} \sum_{i=1}^n \hat{\gamma}^\star(X_i) \sp . 
    \label{appeq:tmle_2} 
\end{align}%

\noindent \textbf{\large Remark.} \ 
One can also adopt an alternative sequential regression for $\theta(X)$, redefined as $\sum_{a=0}^1 \eta(a, X) \pi(a \! \mid \! X)$, with $\eta(a, X) = a \kappa_1(X) + (1-a) \kappa_0(X)$. This reverses the integration order in \eqref{eq:id_ATE}, marginalizing over $M$ first to derive $\eta(A, X)$, rather than over $A$ to obtain $\xi(M, X)$. The resulting plug-in estimator, $\psi_3(\hat{\Q})$, is given by $\frac{1}{n} \sum_{i = 1}^n \hat{\kappa}_1(X_i) \hat{\pi}(1 \! \mid \! X_i) + \hat{\kappa}_0(X_i) \hat{\pi}(0 \! \mid \! X_i).$ For TMLE based on this formulation, targeting $\hat{\kappa}$ is necessary, unlike in $\psi_{2}(\hat{\Q}^\star)$ where $\hat{\gamma}$ was targeted. This also includes targeting $\hat{\mu}$ and $\hat{\pi}$. The goal of targeting  $\hat{\kappa}$ is to satisfy $\Pn \Phi_M(\Q) = o_\p(n^{-1/2})$, where $\Phi_M(\Q)(O_i)$ is rewritten in terms of $\kappa_a(X)$ as: 
\begin{align*} 
\Phi_M(\Q)(O_i)
    \! = \!  \frac{\mathbb{I}(A_i = a_0)}{\pi(a_0  \! \mid \! X_i)} \Big\{ \pi(1 \! \mid \! X_i) \big\{\mu(M_i, 1, X_i) \! - \! \kappa_1(X_i) \big\} + \pi(0 \! \mid \! X_i) \big\{\mu(M_i, 0, X_i) \! - \! \kappa_0(X_i) \big\}  \Big\}  \sp . 
\end{align*}%
Implementing the TMLE $\psi_3(\hat{\Q}^\star)$ requires iterative updates of $(\hat{\mu}, \hat{\pi}, \hat{\kappa})$, making it more complex than $\psi_2(\hat{\Q}^\star)$. For practical use, we therefore recommend $\psi_2(\hat{\Q}^\star)$ due to its simpler implementation. 

\subsection{TMLE algorithms for estimating the ATE front-door functional}
\label{app:tmle_alg}

The detailed procedures of constructing a TMLE-based plug-in estimator for $\psi(\Q)$ in \eqref{eq:id_ATE}, when $M$ is binary, continuous, or multivariate are shown in Algorithms~\ref{appalg:binary}, \ref{appalg:continuous}, and \ref{appalg:multi}, respectively. 

\begin{spacing}{1.45}
\begin{algorithm}[H]
	\caption{\textproc{TMLE based on mediator density estimation with binary $M$ $(\psi_1(\hat{\Q}^\star)$)}}  
    \label{appalg:binary}
    
    \begin{algorithmic}[1] 
		
		\State \textbf{Obtain initial nuisance estimates}: $\hat{\mu}^{(0)}$, $\hat{f}^{(0)}_{M}$, $\hat{\pi}^{(0)}$, and $\hat{\p}_X$.

        \noindent {\small Estimate of $\Q_j$ at the $t^\text{th}$ iteration is denoted by $\hat{\Q}_j^{(t)}$.} 

		\vspace{0.1cm}
        \State \textbf{Define loss functions \& submodels} indexed by $\varepsilon_A, \varepsilon_M, \varepsilon_{Y} \in \mathbb{R}$. {\small Given $\hat{\Q}^{(t)} = (\hat{\mu}^{(0)},\hat{f}_M^{(t)},\hat{\pi}^{(t)},\hat{\p}_X)$: }
        {\small 
        \begin{itemize}
            \item Define the parametric submodels at iteration $t$ as follows:\par\vspace{-3em}
            \begin{align*}
                \hat{\pi}\big(\varepsilon_A; \hat{\mu}^{(0)}, \hat{f}_M^{(t)}\big)(1 \mid X) 
                &= \operatorname{expit}\{\operatorname{logit}\{ \hat{\pi}^{(t)}(1 \mid X) \}+ \varepsilon_A\{\hat{\eta}^{(t)}(1, X)-\hat{\eta}^{(t)}(0, X)\} \} \sp ,
                \\
                \hat{f}_M(\varepsilon_M; \hat{\mu}^{(0)}, \hat{\pi}^{(t)})(1 \mid A, X) 
                &= \operatorname{expit}\{\operatorname{logit}\{\hat{f}_M^{(t)}(1 \mid A, X) \} + \varepsilon_M \sp \frac{\hat{\xi}^{(t)}(1, X)-\hat{\xi}^{(t)}(0, X)}{\hat{\pi}^{(t)}(A \mid X)} \} \sp , \\
                \hat{\mu}(\varepsilon_Y) 
                &= \hat{\mu}^{(t)} + \varepsilon_Y \sp , 
            \end{align*} \par\vspace{-1.5em}
            where $\hat{\eta}^{(t)}(a,X) = \int\hat{\mu}^{(0)}(m,a,X) \sp \hat{f}^{(t)}_M(m | a_0,X) \sp \diff m$, $\hat{\xi}^{(t)}(m,X) = \sum_{a=0}^1 \hat{\mu}^{(0)}(m,a,X) \sp \hat{\pi}^{(t)}(a | X)$. 
            \item Define the loss functions at iteration $t$ as follows:\par\vspace{-3em}
            \begin{align*} 
            &L_{A}(\tilde{\pi})(O) = - \log \tilde{\pi}(A \mid X) \sp , \qquad 
            L_M(\tilde{f}_M)(O) = - \I(A = a_0) \log \tilde{f}_M(M \mid A, X) \sp , \\
            &\hspace{1.25cm} L_Y(\tilde{\mu}; \hat{f}_M^{(t)})(O) 
            = \{\hat{f}_M^{(t)}(M \mid a_0, X) / \hat{f}_M^{(t)}(M \mid A, X)\} \{ Y - \tilde{\mu}(M, A, X) \}^2 \sp .
        \end{align*}
        \end{itemize}        
        }
	\par\vspace{-1em}
        \State \textbf{Update $\hat{\pi}^{(0)}$ and $\hat{f}^{(0)}_M$ iteratively.} {\small We begin by updating $\hat{\pi}$, though updates can start with either $\hat{\pi}$ or $\hat{f}_M$. At the $t^\text{th}$ iteration: }
        
        \vspace{0.15cm}
        {\small 
        \begin{itemize}
        \item Given $\hat{\Q}^{(t)} = (\hat{\mu}^{(0)},\hat{f}_M^{(t)},\hat{\pi}^{(t)},\hat{\p}_X)$, fit the following logistic regression without an intercept:
        \par\vspace{-1em}
        \[
            A\sim\mathrm{offset}\big(\operatorname{logit}\ \hat{\pi}^{(t)}(1\mid X)\big)+ \hat{H}_{A}^{(t)}\big(X\big), \text{ where } \hat{H}_{A}^{(t)}(X) \coloneqq \hat{\eta}^{(t)}(1,X)-\hat{\eta}^{(t)}(0,X) \sp .
        \]
        The coefficient in front of $\hat{H}_{A}^{(t)}(X)$ is the minimizer $\hat{\varepsilon}_A$. Update $\hat{\pi}^{(t)}$ to $ \hat{\pi}^{(t+1)} = \hat{\pi}(\hat{\varepsilon}_A; \hat{\mu}^{(0)}, \hat{f}_M^{(t)})$.
        %
        \item Given $\hat{\Q}^{(\mathrm{temp})} = (\hat{\mu}^{(0)},\hat{f}_M^{(t)},\hat{\pi}^{(t+1)},\hat{\p}_X)$, fit the following logistic regression without an intercept:
         \par\vspace{-1em}
        \[
        M\sim\mathrm{offset}\big(\mathrm{logit}\sp \hat{f}_M^{(t)}(1\mid a_{0},X)\big)+\hat{H}_{M}^{(t)}\big(X\big) , \text{ where } \hat{H}_{M}^{(t)}\big(X\big) \coloneqq \frac{\hat{\xi}^{(t)}(1,X)-\hat{\xi}^{(t)}(0,X)}{\hat{\pi}^{(t+1)}(a_0\mid X)}.
        \]

        Note that $\hat{\xi}^{(t)}$ is computed using $\hat{\mu}^{(0)}$ and $\hat{\pi}^{(t+1)}$. 

        \vspace{0.15cm}
        \item[] The coefficient of $\hat{H}_{M}^{(t)}(X)$ is the minimizer $\hat{\varepsilon}_M$.
        Update $\hat{f}_M^{(t)}$ to $\hat{f}^{(t+1)}_M = \hat{f}_M(\hat{\varepsilon}_M; \hat{\mu}, \hat{\pi}^{(t+1)})$. 
        \vspace{0.15cm}
        \item Let $\hat{\Q}^{(t+1)} = (\hat{\mu}^{(0)},\hat{f}_M^{(t+1)},\hat{\pi}^{(t+1)},\hat{\p}_X)$. Iterate over this step while $|\Pn \Phi(\hat{\Q}^{(t+1)})| > C_n = o_\p(n^{-1/2})$. 
        
        \item[] Assume convergence is achieved at iteration $t=t^\star$. Let $\hat{\pi}^\star=\hat{\pi}^{(t^\star)}$ and $\hat{f}_M^\star=\hat{f}_M^{(t^\star)}$.
        \end{itemize}
        }

		\vspace{0.3cm}
        \State \textbf{Update $\hat{\mu}^{(0)}$ in one step}. 

        \vspace{0.15cm}
        {\small
        \begin{itemize}
            \item Given $\hat{\Q}^{(t^*)} = (\hat{\mu}^{(0)},\hat{f}_M^{\star},\hat{\pi}^{\star},\hat{\p}_X)$, fit the weighted following regression: 
            \par\vspace{-1em}
            \[
            Y\sim\operatorname{offset}(\hat{\mu}^{(0)}(M,A,X))+1, \sp \text{with weight }=\hat{f}^\star_M(M\mid a_0,X)/\hat{f}^\star_M(M\mid A,X).
            \]
            
            \item[] The intercept is the minimizer $\hat{\varepsilon}_Y$.
            Update $\hat{\mu}^{(0)}(M,A,X)$ as $\hat{\mu}^\star(M,A,X)=\hat{\mu}^{(0)}(M,A,X)+\hat{\varepsilon}_Y$.

            \item Let $\hat{\Q}^{\star} = (\hat{\mu}^{\star},\hat{f}_M^{\star},\hat{\pi}^{\star},\hat{\p}_X)$.
        \end{itemize}}

		\vspace{0.3cm}
		\State \textbf{Return} $\psi_1(\hat{\Q}^\star) = \frac{1}{n} \sum_{i=1}^n \hat{\theta}^\star(X_i)$ as the TMLE estimator, where \par\vspace{-1em}
  {\small\[
      \hat{\theta}^\star(x) = \sum_{m=0}^1 \hat{\xi}^\star(m, x) \sp \hat{f}_M^\star(m \mid a_0, x) \sp , \sp \text{and } \hat{\xi}^\star(m, x) = \sum_{a=0}^1 \hat{\mu}^\star(m, a, x) \sp \hat{\pi}^\star(a \mid x).
  \]}
	\end{algorithmic}
\end{algorithm}

\end{spacing}

\begin{spacing}{1.5}
\begin{algorithm}[H]
	\caption{\textproc{TMLE based on mediator density estimation with continuous $M$  ($\psi_1(\hat{\Q}^\star)$)}}  
    \label{appalg:continuous}
    
    \begin{algorithmic}[1] 
		
		\State \textbf{Obtain initial nuisance estimates}: $\hat{\mu}^{(0)}$, $\hat{f}^{(0)}_{M}$, $\hat{\pi}^{(0)}$,  and $\hat{\p}_X$.
  
        \noindent {\small Estimate of $\Q_j$ at the $t^\text{th}$ iteration is denoted by $\hat{\Q}_j^{(t)}$.} 

		\vspace{0.1cm}
        \State \textbf{Define loss functions \& submodels} indexed by $\varepsilon_A, \varepsilon_M, \varepsilon_{Y}$. {\small Given $\hat{\Q}^{(t)} = (\hat{\mu}^{(0)},\hat{f}_M^{(t)},\hat{\pi}^{(t)},\hat{\p}_X)$:}
        {\small 
        \begin{itemize}
            \item Define the parametric submodels at iteration $t$ as follows: ($\varepsilon_A, \varepsilon_Y \in \mathbb{R}$, and $-\delta < \varepsilon_M < \delta$)
            \vspace{-0.2cm}
            \begin{align*}
                \hat{\pi}\left(\varepsilon_A; \hat{\mu}^{(t)}, \hat{f}_M^{(t)}\right)(1 \mid X) 
                &= \operatorname{expit}\big\{\operatorname{logit}\{ \hat{\pi}^{(t)}(1 \mid X) \}+ \varepsilon_A\{\hat{\eta}^{(t)}(1, X)-\hat{\eta}^{(t)}(0, X)\} \big\} \sp ,
                \\
                \hat{f}_M(\varepsilon_M; \hat{\mu}^{(t)},  \hat{\pi}^{(t)})(M \mid a_0, X) &= \hat{f}_M^{(t)}(M\mid a_0,X) \big\{ 1 + \varepsilon_M \sp \frac{\hat{\xi}^{(t)}(M,X)-\hat{\theta}^{(t)}(X)}{\hat{\pi}^{(t)}(a_0\mid X)}  \big\} \sp ,  \\
                \hat{\mu}(\varepsilon_Y) 
                &= \hat{\mu}^{(t)} + \varepsilon_Y \sp ,
            \end{align*}\par\vspace{-1.5em}
            where $\hat{\eta}^{(t)}(a,X) = \int\hat{\mu}^{(t)}(m,a,X) \hat{f}^{(t)}_M(m | a_0,X) \sp \diff m$, $\hat{\xi}^{(t)}(m,X) = \sum_{a=0}^1 \hat{\mu}^{(t)}(m,a,X) \sp \hat{\pi}^{(t)}(a | X)$. 

            The parametric submodel for $\hat{f}_M$ can also be chosen to be \eqref{app:eq:alternative_submodel_fM_continuous} with $\varepsilon_M \in \R$. 
            \item Define the loss functions at iteration $t$ as follows:\par\vspace{-3em}
            \begin{align*} 
            &L_{A}(\tilde{\pi})(O) 
            = - \log \tilde{\pi}(A \mid X) \sp , \qquad 
            L_M(\tilde{f}_M)(O) 
            = - \I(A = a_0) \log \tilde{f}_M(M \mid A, X) \sp , \\
            &\hspace{1.25cm} L_Y(\tilde{\mu}; \hat{f}_M^{(t)})(O) 
            = \{\hat{f}_M^{(t)}(M \mid a_0, X) / \hat{f}_M^{(t)}(M \mid A, X) \} \{ Y - \tilde{\mu}(M, A, X) \}^2 \sp .
        \end{align*}\par\vspace{-1em}
        \end{itemize}        
        }
        
		\vspace{0.1cm}
        \State \textbf{Update $\hat{\pi}^{(0)}$ and $\hat{f}^{(0)}_M$ iteratively.} {\small We begin by updating $\hat{\pi}$, though updates can start with either $\hat{\pi}$ or $\hat{f}_M$.  At the $t^\text{th}$ iteration: }
        
        \vspace{0.15cm}
        {\small 
        \begin{itemize}
        \item Given $\hat{\Q}^{(t)} = (\hat{\mu}^{(0)},\hat{f}_M^{(t)},\hat{\pi}^{(t)},\hat{\p}_X)$, fit the following logistic regression without an intercept:\par\vspace{-1em}
        \[
            A\sim\mathrm{offset}\big(\operatorname{logit}\ \hat{\pi}^{(t)}(1\mid X)\big)+ \hat{H}_{A}^{(t)}\big(X\big), \text{ where } \hat{H}_{A}^{(t)}(X) \coloneqq \hat{\eta}^{(t)}(1,X)-\hat{\eta}^{(t)}(0,X) \sp .  
        \]
        The coefficient of $\hat{H}_{A}^{(t)}(X)$ is the minimizer $\hat{\varepsilon}_A$. Update $\hat{\pi}^{(t)}$ to $ \hat{\pi}^{(t+1)} = \hat{\pi}(\hat{\varepsilon}_A; \hat{\mu}, \hat{f}_M^{(t)})$. 
        \item Given $\hat{\Q}^{(\mathrm{temp})} = (\hat{\mu}^{(0)},\hat{f}_M^{(t)},\hat{\pi}^{(t+1)},\hat{\p}_X)$, obtain $\hat{\varepsilon}_M$ by numerically solving this optimization problem: \par\vspace{-1em}
        \[\hat{\varepsilon}_M \sp = \sp \argmin_{\varepsilon_M \in \R} \sp \Pn L_M\Big(\hat{f}_M\big(\varepsilon_M; \hat{\mu}^{(0)}, \hat{\pi}^{(t+1)}\big)\Big).\]
        \par\vspace{-1em}
        \item[] Update $\hat{f}_M^{(t)}$ to $\hat{f}^{(t+1)}_M = \hat{f}_M(\hat{\varepsilon}_M; \hat{\mu}, \hat{\pi}^{(t+1)})$. 
        \vspace{0.15cm}
        \item Let $\hat{\Q}^{(t+1)} = (\hat{\mu}^{(0)},\hat{f}_M^{(t+1)},\hat{\pi}^{(t+1)},\hat{\p}_X)$. Iterate over this step while $|\Pn \Phi(\hat{\Q}^{(t+1)})| > C_n = o_\p(n^{-1/2})$. 
        
        \item[] Assume convergence is achieved at iteration $t=t^\star$. Let $\hat{\pi}^\star=\hat{\pi}^{(t^\star)}$ and $\hat{f}_M^\star=\hat{f}_M^{(t^\star)}$.
        \end{itemize}
        }

		\vspace{0.1cm}
        \State \textbf{Update $\hat{\mu}^{(0)}$ in one step}. 

        \vspace{0.15cm}
        {\small
        \begin{itemize}
            \item Given $\hat{\Q}^{(t^*)} = (\hat{\mu}^{(0)},\hat{f}_M^{\star},\hat{\pi}^{\star},\hat{\p}_X)$, fit the following weighted regression: \par\vspace{-1em}
            \[Y\sim\operatorname{offset}(\hat{\mu}^{(0)}(M,A,X))+1, \sp \text{with weight }=\hat{f}^\star_M(M\mid a_0,X)/\hat{f}^\star_M(M\mid A,X).\]\par\vspace{-1em}
            
            \item[] The intercept is the minimizer $\hat{\varepsilon}_Y$. Update $\hat{\mu}^{(0)}(M,A,X)$ as $\hat{\mu}^\star(M,A,X)=\hat{\mu}^{(0)}(M,A,X)+\hat{\varepsilon}_Y$.

            \item Let $\hat{\Q}^{\star} = (\hat{\mu}^{\star},\hat{f}_M^{(t)},\hat{\pi}^{(t)},\hat{\p}_X)$.
        \end{itemize}}

		\vspace{0.1cm}
		\State \textbf{Return} $\psi_1(\hat{\Q}^\star) = \frac{1}{n} \sum_{i=1}^n \hat{\theta}^\star(X_i)$ as the TMLE estimator, where \par\vspace{-1em}
  {\small\[
      \hat{\theta}^\star(x) = \int \hat{\xi}^\star(m, x) \sp \hat{f}_M^\star(m \mid a_0, x) \sp \diff m \sp , \ \text{and } \ \hat{\xi}^\star(m, x) = \sum_{a=0}^1 \hat{\mu}^\star(m, a, x)\sp \hat{\pi}^\star(a \mid x).
  \]}\par 
	\end{algorithmic}
\end{algorithm}
\end{spacing}

\begin{spacing}{1.5}
\begin{algorithm}[H]
	\caption{\textproc{TMLE that avoids mediator density estimation ($\psi_2(\hat{\Q}^\star)$)}}  
    \label{appalg:multi}
    
    \begin{algorithmic}[1] 
		
		\State \textbf{Obtain initial nuisance estimates}: $\hat{\mu}, \hat{\kappa}_a, \hat{f}_M^r, \hat{\pi}, \hat{\gamma}$, and $\hat{\p}_X$. 
        {\small\begin{itemize}
            \item $f^{r}_{M}(M,a_1,X)$ can be estimated either via direct estimation of the density ratio, or by applying the Bayes' rule to reparameterize the ratio in terms of $\hat{\pi}(A\mid X)$ and $\hat{\lambda}(A \mid M, X)$, as in \eqref{eq:bayes}.
            \item $\hat{\kappa}_{a_1}(X)$ is obtained via a regression of $\hat{\mu}(M, a, X)$ on $X$ using only rows with $A=a_0.$  
        \end{itemize}}
		\vspace{0.2cm}
        \State \textbf{Define loss functions and parametric fluctuations} indexed by $\varepsilon_A$, $\varepsilon_{\gamma}, \varepsilon_Y \in \R$.
        {\small
        \begin{itemize}
            \item Define the parametric submodels as follows:
            \par\vspace{-3em}
            \begin{align*} 
                \hat{\mu}(\varepsilon_Y) &= \hat{\mu} + \varepsilon_Y \sp ,  \\
                \hat{\pi}(\varepsilon_A; \hat{\kappa})(1 \mid X) 
                &= \operatorname{expit}\Big\{\operatorname{logit}\big\{ \hat{\pi}(1 \mid X) \big\}+ \varepsilon_A \big\{\hat{\kappa}_1(X)-\hat{\kappa}_0(X)\big\} \Big\} \sp , \\
                \hat{\gamma}(\varepsilon_\gamma)(X) &= \hat{\gamma}(X)+\varepsilon_\gamma \sp . 
            \end{align*}\par\vspace{-1.5em}
            \item Define the loss functions as follows:\par\vspace{-3em}
            \begin{align*} 
                L_Y(\tilde{\mu};\hat{f}_M^r)(O) &= \hat{f}_M^r(M,A,X) \{ Y - \tilde{\mu}(M, A, X) \}^2 \sp , \\
                L_{A}(\tilde{\pi})(O) &= - \log \tilde{\pi}(A \mid X) \sp ,\\
                L_{\gamma}(\tilde{\gamma};\hat{\pi},\hat{\xi})(O) &= \frac{\I(A=a_0)}{\hat{\pi}(a_0\mid X)}\left(\hat{\xi}(M,X)-\tilde{\gamma}(X)\right)^2 \sp .
            \end{align*}
        \end{itemize}
        }

        \State \textbf{Update $\hat{\mu}$ and $\hat{\pi}$ in one step} by solving the followings optimization problem:\par\vspace{-1.5em}
        {\small 
            \[
                \hat{\varepsilon}_Y  = \argmin_{\varepsilon_Y \in \mathbb{R}} \Pn L_Y(\hat{\mu}(\varepsilon_Y);\hat{f}_M^r) \sp , \quad
                \hat{\varepsilon}_A = \argmin_{\varepsilon_A \in \mathbb{R}} \Pn L_A(\hat{\pi}(\varepsilon_A)).
            \]\par\vspace{-1em}
        
        \begin{itemize}
            \item Fit the following weighted regression and logistic regression without intercept term
            \par\vspace{-3em}
            \begin{align*}
                &Y\sim\operatorname{offset}(\hat{\mu}(M,A,X))+1, \operatorname{weight}=\hat{f}_M^r \sp ; 
                \\
                &A\sim \operatorname{offset}(\operatorname{logit} \hat{\pi}(1\mid X))+\hat{H}_{A}(X) \sp ,  \quad \text{where } \sp \hat{H}_A(X)=\hat{\kappa}_1(X) - \hat{\kappa}_0(X) \sp .
            \end{align*} 
            \par\vspace{-1.5em}
            \item $\hat{\varepsilon}_Y$ and $\hat{\varepsilon}_A$ equal the coefficients of the intercept and in front of $\hat{H}_A(X)$, respectively.
            \item Update $\hat{\mu}$ and $\hat{\pi}$ as follows\par\vspace{-1.5em}
            \[
                \hat{\mu}^\star = \hat{\mu}(\hat{\varepsilon}_Y;\hat{f}_M^r ) \sp , \quad \hat{\pi}^\star = \pi(\hat{\varepsilon}_A; \hat{\mu}).
            \]\par\vspace{-1em}
            \item Define $\hat{\xi}^\star(m, x) = \sum_{a = 0}^1 \hat{\mu}^\star(m, a, x) \sp \hat{\pi}^\star(a \mid x)$. Estimate $\hat{\gamma}(X)$ by fitting the following linear regression using only data points with $A=a_0$:\par\vspace{-2em}
            \[
                \hat{\xi}^\star(m, x) \sim X.
            \]\par\vspace{-1.5em}
        \end{itemize}}
    \State \textbf{Update $\hat{\gamma}$ in one step} by solving the followings optimization problem:\par\vspace{-1.5em}
    \[
        \hat{\varepsilon}_\gamma \sp = \sp \argmin_{\varepsilon_\gamma \in \mathbb{R}} \sp \Pn L_{\gamma}\left(\hat{\gamma}(\varepsilon_\gamma); \hat{\pi}^\star,\hat{\xi}^\star\right) \sp .
    \]\par\vspace{-1em}
    \begin{itemize}
        \item Fit the following weighted linear regression\par\vspace{-1.5em}
        \[
            \hat{\xi}^\star\sim\mathrm{offset}(\hat{\gamma})+1 \sp , \quad \text{with weight}=\frac{\I(A=a_0)}{\hat{\pi}^\star(a_0\mid X)} \sp . 
        \]\par\vspace{-1em}
        \item The coefficient of the intercept is $\hat{\varepsilon}_{\gamma}$, which minimize the empirical risk. 
        \item Update $\hat{\gamma}(X)$ as $\hat{\gamma}^\star = \hat{\gamma}(\hat{\varepsilon}_\gamma)$.
    \end{itemize}
		\vspace{0.1cm}
		\State \textbf{Return}  $\psi_2(\hat{\Q}^\star) = \frac{1}{n} \sum_{i=1}^n \hat{\gamma}^\star(X_i)$ as the TMLE estimator.  
	\end{algorithmic}
\end{algorithm}

\end{spacing}

\pagebreak
\section{Details on estimators for the ATT front-door functional}
\label{app:att_details}


\citet{fulcher19robust} discuss estimation of the ATT under the front-door model by leveraging estimators developed for the population intervention indirect effect (PIIE). Under the no direct effect assumption, the PIIE coincides with the population intervention effect (PIE) \citep{hubbard2008population}, and is related to the ATT through the identity 

\vspace{-1.25cm}
\begin{align*}
    \underbrace{\mathbb{E}(Y-Y^{A, M^{0}})}_{\text{PIE}} & = \E(Y^1-Y^0 \mid A=1)\ p(A=1).
\end{align*}

\vspace{-0.3cm} \noindent 
This relationship implies that the ATT can be obtained by appropriately rescaling an estimator of the PIE by the marginal treatment probability. 

Leveraging the same idea, the ATT estimand in \eqref{eq:id_ATT} can be re-expressed as: $\beta(\Q) = \psi(\Q)/\p(A = a_1) - \E(Y \mid A = a_0) \sp \p(A = a_0)$, where $\psi(\Q)$ is defined in  \eqref{eq:id_ATE}. 
This reparameterization enables the use of any ATE estimator from Section~\ref{sec:est_ATE}, together with empirical estimates of $\E(Y \mid A = a_0)$ and $\p(A)$, to construct one-step or TMLE estimators for the ATT. While straightforward, this approach introduces unnecessary complexity by estimating nuisance components tailored to the ATE—such as $\hat{\xi}(M,X)$ and $\hat{\theta}(X)$ in $\psi_1(\hat{\Q})$, or pseudo-outcome regressions like $\hat{\gamma}(X)$ in $\psi_{2a}(\hat{\Q})$ and $\psi_{2b}(\hat{\Q})$—that are irrelevant for ATT estimation. This increases the risk of model misspecification and computational burden. In contrast, the EIF-based estimators proposed in Section~\ref{sec:est_ATT} target the ATT directly and avoid such extraneous steps.

\subsection{Plug-in and one-step ATT estimators under standard factorization}
\label{app:att_density_est}
Let $\hat{\Q} = (\hat{\mu}, \hat{f}_M, \hat{\pi}, \hat{\p}_A, \hat{\p}_{AX})$ denote the collection of nuisance estimates. When $M$ is discrete, $\hat{f}_M$ can be obtained via regression-based methods; for univariate continuous or mixed-type multivariate mediators, it may be estimated using parametric models, kernel-based approaches, or flexible density estimation techniques \citep{hayfield2008nonparametric,benkeser2016highly}. The quantities $\hat{\p}A$ and $\hat{\p}{AX}$ denote the empirical estimates of the marginal and joint distributions of $A$ and $(A, X)$, respectively. A plug-in estimator of $\beta(\Q)$, denoted by $\beta_1(\hat{\Q})$, is given by: 

\vspace{-1.cm}
\begin{align*}
    \beta_1(\hat{\Q}) = \frac{1}{n} \sum_{i=1}^{n} \Big\{ \frac{\I(A_i=a_1)}{\hat{\p}_A(a_1)} \int \hat{\mu}(m,a_1,X_i)\ \hat{f}_M(m\mid a_0,X_i)\sp \diff m\Big\} \sp ,
\end{align*}

\vspace{-.2cm} \noindent 
where the integral over $m$ simplifies to a summation $\sum_m \hat{\mu}(m, a_1, X_i)\ \hat{f}_M(m \mid a_0, X_i)$ when $M$ is discrete, and requires numerical evaluation when $M$ is continuous or of mixed variable types.

The corresponding one-step corrected plug-in estimator, denoted by $\beta_1^{+}(\hat{\Q})$, is given by:

\vspace{-1.15cm}
\begin{align*}
    \beta_1^+(\hat{\Q}) &= \beta_1(\hat{\Q}) + \frac{1}{n} \sum_{i=1}^n \bigg\{\frac{\I(A_i = a_1)}{\hat{\p}_A(a_1)} \frac{\hat{f}_M(M_i,a_0,X_i)}{\hat{f}_M(M_i,A_i,X_i)} \sp \big\{Y_i - \hat{\mu}(M_i, a_1, X_i)\big\} 
    \\
    &\hspace{0.5cm} + \frac{\I(A_i  = a_0)}{\hat{\p}_A(a_1)} \frac{\hat{\pi}(a_1 \! \mid \! X_i)}{\hat{\pi}(a_0 \! \mid \! X_i)} \Big\{\hat{\mu}(M_i, a_1, X_i) \! - \! \int \hat{\mu}(m,a_1,X_i)\ \hat{f}_M(m\mid a_0,X_i)\sp \diff m\Big\} 
    \\
    &\hspace{0.5cm}+ \frac{\I(A_i = a_1)}{\hat{\p}_A(a_1)} \Big\{\int \hat{\mu}(m,a_1,X_i)\ \hat{f}_M(m\mid a_0,X_i)\sp \diff m \! - \! \beta_1(\hat{\Q})\Big\}  \bigg\} \sp . \notag 
\end{align*}

To construct the corresponding TMLE, $\beta_1(\hat{\Q}^\star)$, we follow the same general approach as for the ATE, with details varying by whether $M$ is binary or continuous. The procedures are summarized in Algorithms~\ref{appalg:binary_ATT} and \ref{appalg:continuous_ATT} in Appendix~\ref{app:att_tmle_alg}.

\subsection{TMLE algorithms for estimating the ATT front-door functional} 
\label{app:att_tmle_alg}

The detailed procedures of constructing a TMLE-based plug-in estimator for $\beta(\Q)$ in \eqref{eq:id_ATT}, when $M$ is binary, continuous, or multivariate are shown in Algorithms~\ref{appalg:binary_ATT}, \ref{appalg:continuous_ATT}, and \ref{appalg:multi_ATT}, respectively. 

\begin{spacing}{1.6}
\begin{algorithm}[H]
	\caption{\textproc{TMLE based on mediator density estimation with binary $M$ $(\beta_1(\hat{\Q}^\star)$)}}  
    \label{appalg:binary_ATT}
    
    \begin{algorithmic}[1] 
		
		\State \textbf{Obtain initial nuisance estimates}: $\hat{\mu}, \hat{f}_{M}, \hat{\pi}, \hat{p}_A$ and $\hat{\p}_{AX}$.

        \State \textbf{Define loss functions \& submodels} indexed by $\varepsilon_M, \varepsilon_{Y}$. {\small Given $\hat{\Q} = (\hat{\mu}, \hat{f}_M, \hat{\pi}, \hat{\p}_A, \hat{\p}_{AX})$: }
        {\small 
        \begin{itemize}
            \item Define the parametric submodels as follows: ($\varepsilon_M, \varepsilon_Y \in \mathbb{R}$)
            \begin{align*}
                \hat{f}_M\left(\varepsilon_M; \hat{\mu}\right)(1 \mid a_0, X) 
                &= \operatorname{expit}\left\{\operatorname{logit}\left\{ \hat{f}_M(1 \mid a_0, X) \right\} + \varepsilon_M \sp \frac{\hat{\pi}(a_1\mid X)}{\hat{\pi}(a_0\mid X)}\frac{\hat{\mu}(1,a_1,X)-\hat{\mu}(0,a_1,X)}{\hat{\p}_A(a_1)} \right\} \sp , 
                \\
                \hat{\mu}(\varepsilon_Y) (M,a_1,X)
                &= \hat{\mu}(M,a_1,X) + \varepsilon_Y \sp . 
            \end{align*}
            \item Define the loss functions as follows:
            \begin{align*} 
            L_Y\left(\tilde{\mu}; \hat{f}_M\right)(O) 
            &= \frac{\I(A=a_1)}{\hat{\p}_A(a_1)}\frac{\hat{f}_M(M \mid a_0, X)}{\hat{f}_M(M \mid a_1, X)} \{ Y - \tilde{\mu}(M, a_1, X) \}^2,
            \\
            L_M(\tilde{f}_M)(O) 
            &= - \I(A = a_0) \log \tilde{f}_M(M \mid a_0, X) \sp .
        \end{align*}
        \end{itemize}        
        }
        \State \textbf{Update $\hat{f}_M$.} 
        
        \vspace{0.15cm}
        {\small 
        Given $\hat{\Q} = (\hat{\mu}, \hat{f}_M, \hat{\pi}, \hat{\p}_A, \hat{\p}_{AX})$, fit the following logistic regression without an intercept:
        \begin{align*}
            M\sim\mathrm{offset}\big(\mathrm{logit}\sp \hat{f}_M(1\mid a_{0},X)\big)+\hat{H}_{M}\big(X\big) , \text{ where } \hat{H}_{M}\big(X\big) \coloneqq \frac{\hat{\pi}(a_1\mid X)}{\hat{\pi}(a_0\mid X)}\frac{\hat{\mu}(1,a_1,X)-\hat{\mu}(0,a_1,X)}{\hat{p}_A(a_1)} \sp . 
        \end{align*}
        
        \begin{itemize}
        \item The coefficient in front of $\hat{H}_{M}(X)$ is the minimizer $\hat{\varepsilon}_M \sp \coloneqq \sp \argmin_{\varepsilon_M \in \R} \sp \Pn L_M\Big(\hat{f}_M\big(\varepsilon_M; \hat{\mu}\big)\Big)$. 
        \item Update $\hat{f}_M(M|a_0,X)$ to $\hat{f}^{\star}_M(M|a_0,X) = \hat{f}_M(\hat{\varepsilon}_M; \hat{\mu})$. 
        \end{itemize}
        }

    \State \textbf{Update $\hat{\mu}$.} 
        
        \vspace{0.15cm}
        {\small 
        Given $\hat{\Q} = (\hat{\mu}, \hat{f}^\star_M, \hat{\pi}, \hat{\p}_A, \hat{\p}_{AX})$, fit the following weighted regression:
         $$Y\sim\operatorname{offset}(\hat{\mu}(M,a_1,X))+1, \sp \text{with weight }=\frac{\I(A=a_1)}{\hat{p}_A(a_1)}\frac{\hat{f}^\star_M(M \mid a_0, X)}{\hat{f}_M(M \mid a_1, X)}\sp .$$ 
        
        \begin{itemize}
        \item[] The intercept is the minimizer to $\hat{\varepsilon}_Y \sp \coloneqq \sp \argmin_{\varepsilon_Y \in \mathbb{R}} \sp \Pn L_Y\Big(\hat{\mu}(\varepsilon_Y); \hat{f}^\star_M \Big) \sp .$  
        \item[] Update $\hat{\mu}(M,a_1,X)$ as $\hat{\mu}^\star(M,a_1,X)=\hat{\mu}(M,a_1,X)+\hat{\varepsilon}_Y$.
        \end{itemize}
        }

		\vspace{0.3cm}
		\State \textbf{Return} the TMLE estimator $\beta_1(\hat{\Q}^\star)$ as
  {\small\begin{align*}
      \beta_1(\hat{\Q}^\star) = \frac{1}{n}\sum_{i=1}^{n}\frac{\I(A_i=a_1)}{\hat{\p}_A(a_1)} \ \sum_{m\in\{0,1\}}\hat{\mu}^{\star}(m,a_1,X_i)\hat{f}_M^{\star}(m\mid a_0,X_i) \sp .
  \end{align*}}
	\end{algorithmic}
\end{algorithm}

\end{spacing}

\begin{spacing}{1.6}
\begin{algorithm}[H]
	\caption{\textproc{TMLE based on mediator density estimation with continuous $M$  ($\beta_1(\hat{\Q}^\star)$)}}  
    \label{appalg:continuous_ATT}
    
    \begin{algorithmic}[1] 

    \State \textbf{Obtain initial nuisance estimates}: $\hat{\mu}, \hat{f}_{M}, \hat{\pi}, \hat{p}_A$ and $\hat{\p}_{AX}$.

		\vspace{0.1cm}
        \State \textbf{Define loss functions \& submodels} indexed by $\varepsilon_M, \varepsilon_{Y}$. {\small Given $\hat{\Q} = (\hat{\mu}, \hat{f}_M, \hat{\pi}, \hat{\p}_A, \hat{\p}_{AX})$: }
        {\small 
        \begin{itemize}
            \item Define the parametric submodels as follows: ($\varepsilon_Y \in \mathbb{R}$, $\varepsilon_M  \in -\delta<\varepsilon_M < \delta$)
            \begin{align*}
                \hat{f}_M\left(\varepsilon_M; \hat{\mu}\right)(1 \mid a_0, X) 
                &=  \hat{f}_M(1 \mid a_0, X) \left[ 1 + \varepsilon_M \left\{\frac{\hat{\pi}(a_1\mid X)}{\hat{\pi}(a_0\mid X)}\frac{\hat{\mu}(M,a_1,X)-\hat{\kappa}_{a_1}(X)}{\hat{\p}_A(a_1)}\right\} \right] \sp , 
                \\
                \hat{\mu}(\varepsilon_Y) (M,a_1,X)
                &= \hat{\mu}(M,a_1,X) + \varepsilon_Y \sp .
            \end{align*}
            The parametric submodel for $\hat{f}_M$ can also be chosen to be \eqref{app:eq:alternative_submodel_fM_continuous} with $\varepsilon_M \in \R$. 
            
            \item Define the loss functions as follows:
            \begin{align*} 
            L_Y\left(\tilde{\mu}; \hat{f}_M\right)(O) 
            &= \frac{\I(A=a_1)}{\hat{\p}_A(a_1)}\frac{\hat{f}_M(M \mid a_0, X)}{\hat{f}_M(M \mid a_1, X)} \{ Y - \tilde{\mu}(M, a_1, X) \}^2,
            \\ L_M(\tilde{f}_M)(O) 
            &= - \I(A = a_0) \log \tilde{f}_M(M \mid a_0, X) \sp .
        \end{align*}
        \end{itemize}        
        }
        
		\vspace{0.1cm}
        \State \textbf{Update $\hat{f}_M(M\mid A,X)$ in one step}. 
        
        \vspace{0.15cm}
        {\small 
        Given $\hat{\Q} = (\hat{\mu}, \hat{f}_M, \hat{\pi}, \hat{\p}_A, \hat{\p}_{AX})$, obtain $\hat{\varepsilon}_M$ by numerically solving this optimization problem: 
        $$\hat{\varepsilon}_M \sp = \sp \argmin_{\varepsilon_M \in \R} \sp \Pn L_M\Big(\hat{f}_M\big(\varepsilon_M; \hat{\mu}\big)\Big)\sp .$$
        \begin{itemize}        
        \item Update $\hat{f}_M(M\mid a_0,X)$ to $\hat{f}^\star_M(M\mid a_0,X) = \hat{f}_M(\hat{\varepsilon}_M; \hat{\mu})$. 
        \end{itemize}
        }

		\vspace{0.3cm}
        \State \textbf{Update $\hat{\mu}(M, A, X)$ in one step}. 

        \vspace{0.15cm}
        {\small
        \begin{itemize}
            \item Given $\hat{\Q} = (\hat{\mu}, \hat{f}^\star_M, \hat{\pi}, \hat{\p}_A, \hat{\p}_{AX})$, fit the following weighted regression:
         $$Y\sim\operatorname{offset}(\hat{\mu}(M,a_1,X))+1, \sp \text{with weight }=\frac{\I(A=a_1)}{\hat{\p}_A(a_1)}\frac{\hat{f}^\star_M(M \mid a_0, X)}{\hat{f}_M(M \mid a_1, X)} \sp .$$ 
        
        \item[] The intercept is the minimizer to $\hat{\varepsilon}_Y \sp = \sp \argmin_{\varepsilon_Y \in \mathbb{R}} \sp \Pn L_Y\Big(\hat{\mu}(\varepsilon_Y); \hat{f}^\star_M \Big) \sp .$  
        \item[] Update $\hat{\mu}(M,a_1,X)$ as $\hat{\mu}^\star(M,a_1,X)=\hat{\mu}(M,a_1,X)+\hat{\varepsilon}_Y$.
        \end{itemize}}

		\vspace{0.3cm}
		\State \textbf{Return} the TMLE estimator $\beta_1(\hat{\Q}^\star)$ as
  {\small\begin{align*}
      \beta_1(\hat{\Q}^\star) = \frac{1}{n}\sum_{i=1}^{n}\frac{\I(A_i=a_1)}{\hat{\p}_A(a_1)}\int\hat{\mu}^{\star}(m,a_1,X_i)\sp \hat{f}_M^{\star}(m\mid a_0,X_i)\sp \diff m \sp .
  \end{align*}}
	\end{algorithmic}
\end{algorithm} 

\end{spacing}

\begin{spacing}{1.5}
\begin{algorithm}[H]
	\caption{\textproc{TMLE that avoids mediator density estimation ($\beta(\hat{\Q}^\star)$)}}  
    \label{appalg:multi_ATT}
    
    \begin{algorithmic}[1] 
		
		\State \textbf{Obtain initial nuisance estimates}: $\hat{\mu}, \hat{\pi}, \hat{f}^r_M, \hat{\kappa}_{a_1}, \hat{\p}_A$, and $\hat{\p}_{AX}$.
        \begin{itemize}
            \item $f^{r}_{M}(M,a_1,X)$ can be estimated either via direct estimation of the density ratio, or by applying the Bayes' rule to reparameterize the ratio in terms of $\hat{\pi}(A\mid X)$ and $\hat{\lambda}(a_1 \mid M, X)$, as in \eqref{eq:bayes}.
            \item $\hat{\kappa}_{a_1}(X)$ is obtained via a regression of $\hat{\mu}(M, a_1, X)$ on $X$ using only rows with $A=a_0.$  
        \end{itemize}
		\vspace{0.2cm}
        \State \textbf{Define loss functions and parametric fluctuations} indexed by $\varepsilon_{\kappa}$ and $\varepsilon_Y$.
        {\small
        \begin{itemize}
            \item Define the parametric submodels as follows: ($\varepsilon_Y, \varepsilon_\kappa  \in \mathbb{R}$)
            \begin{align*} 
                \hat{\mu}(\varepsilon_Y) &= \hat{\mu} + \varepsilon_Y \sp , \quad
                \hat{\kappa}_{a_1}(\varepsilon_\kappa)(X) = \hat{\kappa}_{a_1}(X)+\varepsilon_\kappa \sp . 
            \end{align*}
            \item Define the loss functions as follows:
            \begin{align*} 
                L_Y(\tilde{\mu};\hat{f}_M^r)(O) &= \frac{\I(A=a_1)}{\hat{\p}_A(a_1)}\hat{f}_M^r(M,a_1,X) \{ Y - \tilde{\mu}(M, a_1, X) \}^2 \sp , \\
                L_{\kappa}(\tilde{\kappa}_{a_1};\hat{\pi},\hat{\mu})(O) &= \frac{\I(A=a_0)}{\hat{\p}_A(a_1)}\left(\hat{\mu}(M,a_1,X)-\tilde{\kappa}_{a_1}(X)\right)^2 \sp .
            \end{align*}
        \end{itemize}
        }

    \State \textbf{Update $\hat{\mu}$ in one step.} 
        
        \vspace{0.15cm}
        {\small 
        Given $\hat{\Q} = (\hat{\mu}, \hat{\pi}, \hat{f}^r_M, \hat{\kappa}_{a_1}, \hat{\p}_A, \hat{\p}_{AX})$, fit the following weighted regression:
         $$Y\sim\operatorname{offset}(\hat{\mu}(M,a_1,X))+1, \sp \text{with weight }=\frac{\I(A=a_1)}{\hat{\p}_A(a_1)}\hat{f}^r_M(M,a_1,X).$$ 
        
        \begin{itemize}
        \item The intercept is the minimizer to $\hat{\varepsilon}_Y \sp = \sp \argmin_{\varepsilon_Y \in \mathbb{R}} \sp \Pn L_Y\Big(\hat{\mu}(\varepsilon_Y); \hat{f}^r_M \Big) \sp .$  
        \item Update $\hat{\mu}(M,a_1,X)$ as $\hat{\mu}^\star(M,a_1,X)=\hat{\mu}(M,a_1,X)+\hat{\varepsilon}_Y$.
        \item Estimate $\hat{\kappa}_{a_1}$ by fitting the following linear regression using only data points with $A=a_0$:
        \begin{align*}
            \hat{\mu}^\star(M,a_1,X) \sim X \sp.
        \end{align*}
        \end{itemize}
        }

    \vspace{-0.5cm}
    \State \textbf{Update $\hat{\kappa}_{a_1}$ in one step.} 
    
    \vspace{0.15cm}
        {\small 
        Given $\hat{\Q} = (\hat{\mu}^\star, \hat{\pi}, \hat{f}^r_M, \hat{\kappa}_{a_1}, \hat{\p}_A, \hat{\p}_{AX})$, fit the following weighted regression:
         $$\hat{\mu}^\star(M,a_1,X)\sim\operatorname{offset}(\hat{\kappa}_{a_1}(X))+1, \sp \text{with weight }=\frac{\I(A=a_0)}{\hat{\p}_A(a_1)}\frac{\hat{\pi}(a_1\mid X)}{\hat{\pi}(a_0\mid X)}.$$ 
        
        \begin{itemize}
        \item The intercept is the minimizer to $\hat{\varepsilon}_\kappa \sp = \sp \argmin_{\varepsilon_\kappa \in \mathbb{R}} \sp \Pn L_\kappa\Big(\hat{\kappa}_{a_1}(\varepsilon_\kappa); \hat{\pi},\hat{\mu}^\star \Big) \sp .$  
        \item Update $\hat{\kappa}_{a_1}(X)$ as $\hat{\kappa}^\star(X)=\hat{\kappa}(X)+\hat{\varepsilon}_\kappa$.
        \end{itemize}}

		\vspace{0.1cm}
		\State \textbf{Return} the TMLE estimator $\beta(\hat{\Q}^\star)$ as 
        \begin{align*}
            \beta(\hat{\Q}^\star) = \frac{1}{n} \sum_{i=1}^n \frac{\I(A_i=a_1)}{\hat{\p}_A(a_1)}\hat{\kappa}_{a_1}^\star(X_i) \sp . 
        \end{align*}
	\end{algorithmic}
\end{algorithm}

\end{spacing}

\newpage
\section{Details on inference and asymptotic properties}
\label{app:asym}

We assume the following convergence rates for our nuisance estimates: 

\vspace{-.5cm}
\begin{equation} \label{eq:convergence_rates}
\begin{aligned}
    &|| \hat{\pi}^{\star}- \pi || = \smallO(n^{-\frac{1}{k}}) \sp , 
    \
    &|| \hat{f}_{M}^\star - f_{M} || = \smallO(n^{-\frac{1}{b}}) \sp , 
    \\
    &|| \hat{\mu}^{\star} - \mu || = \smallO(n^{-\frac{1}{q}}) \sp , 
    \
    &|| \hat{\gamma}^\star- \gamma || = o_\p(n^{-\frac{1}{j}}) \sp , 
    \\
    &|| \hat{\kappa}_a-\kappa_a ||  = o_\p(n^{-\frac{1}{\ell}}) \sp , 
    \
    &|| \hat{f}_M^r - f_M^r  || = o_\p(n^{-\frac{1}{c}}) \sp , 
    \\
    &|| \hat{\lambda}- \lambda ||  = \smallO(n^{-\frac{1}{d}}) \sp . \\
\end{aligned}
\end{equation}

\subsection{ATE front-door functional estimators}
\label{app:asym_ate}

\subsubsection{The remainder term, asymptotic linearity, and robustness for $\psi_1(\hat{\Q}^\star)$} \label{app:asym_ate:r2_psi1}  \vspace{0.1cm}

\noindent\underline{\textbf{$\Rem(\hat{\Q}^\star, \Q)$ derivation}}
\newline
Given the von Mises expansion, we can write: 

\vspace{-1.5cm}
{\small\begin{align*}
    &\Rem(\hat{\Q}^\star, \Q)
    \\
    &= \psi(\hat{\Q}^\star) - \psi(\Q) + \int \Phi(\hat{\Q}^\star) \sp \diff\P(o)  \\
    &= \iiint \left\{ \mu(m, a, x) - \hat{\mu}^\star(m, a, x) \right\} \sp \left\{ \frac{\hat{f}^\star_M(m\mid a_0, x)}{\hat{f}^\star_M(m\mid a,x)} \sp f_M(m \mid a,x)\right\} \pi(a\mid x) \sp \p(x) \sp \diff x\ \diff a\ \diff m\\
    &\hspace{0.05cm} + \iiint \hat{\mu}^\star(m,a,x) \sp \{ f_M(m\mid a_0, x) - \hat{f}^\star_M(m\mid a_0, x)\} \sp \big\{ \frac{\pi(a_0\mid x)}{\hat{\pi}^\star(a_0\mid x)} \sp \hat{\pi}^\star(a\mid x) \big\} \sp \p(x) \sp  \diff x\ \diff a\ \diff m\\
    &\hspace{0.05cm} + \iiint \left\{ \hat{\mu}^\star(m,a,x) \sp \hat{f}^\star_M(m\mid a_0,x) - \mu(m,a,x) f_M(m\mid a_0, x) \right\} \sp \pi(a\mid x) \sp \p(x) \sp  \diff x\ \diff a\ \diff m \sp . 
\end{align*}}
For a clear formulation, we introduce a term that is equal to zero into the above expression: 

\vspace{-1.25cm}
\begin{align*}
    0&=\iiint \frac{f_{M}(m\mid a_0, x)}{f_{M}(m\mid a, x)} \sp [\mu(m, a, x)-\hat{\mu}(m, a, x)] \sp f_{M}(m\mid a, x) \sp \pi(a\mid x)\  \p(x) \sp \diff x\ \diff a\ \diff m\\
    &\hspace{0.25cm}+\iiint\left[\hat{\mu}^\star(m,a,x) f_{M}(m\mid a_0, x)-\mu(m,a,x) f_{M}(m\mid a_0, x)\right]\ \pi(a\mid x) \sp \p(x)\ \diff x\ \diff a\ \diff m \sp .  
\end{align*}

\vspace{-.3cm} \noindent 
For a more clear derivation of the convergence behavior, we can further decompose $\Rem(\hat{\Q}^\star, \Q)$ as:

\vspace{-1.5cm}
{\small
\begin{align}
    &\Rem(\hat{\Q}^\star, \Q)
   \notag  \\
    &=\int \bigg[ \frac{\hat{f}^\star_{M}(m\mid a_0, x)}{\hat{f}^\star_{M}(m\mid a, x)f_{M}(m\mid a, x)}(f_{M}(m\mid a, x) -\hat{f}^\star_{M}(m\mid a, x))\ (\mu(m, a, x)-\hat{\mu}^\star(m, a, x)) \notag  \\
    &\hspace{0.25cm}+ \frac{1}{f_M(m\mid a,x)}(\hat{f}^\star_{M}(m\mid a_0, x) -f_{M}(m\mid a_0, x))\ (\mu(m, a, x)-\hat{\mu}^\star(m, a, x)) \notag  \\
    &\hspace{0.25cm}+ \frac{\hat{\mu}^\star(m, a, x)}{f_M(m\mid a,x)}\frac{\pi(a_0\mid x)}{\hat{\pi}^\star(a_0\mid x)\pi(a\mid x)}\ (\hat{\pi}^\star(a\mid x)-\pi(a\mid x))\ (f_{M}(m\mid a_0, x)-\hat{f}^\star_{M}(m\mid a_0, x)) \notag \\
    &+ \frac{\hat{\mu}^\star(m, a, x)}{f_M(m\mid a,x)\ \hat{\pi}^\star(a_0\mid x)}(\pi(a_0\mid x)-\hat{\pi}^\star(a_0\mid x))[f_{M}(m\mid a_0, x)-\hat{f}^\star_{M}(m\mid a_0, x)] \bigg] \sp \diff\P(x, a, m) \sp . 
    \label{appeq:r2_est1}
\end{align}
}

\noindent\underline{\textbf{Regularity discussions}}

\noindent 
In the following, we discuss two sets of regularity conditions.

\noindent 
\underline{[First set of regularity conditions.]} Let $\mathcal{X}$ and $\mathcal{M}$ denote the domain of $X$ and $M$. Assume

\vspace{-1cm}
\begin{equation}\label{app:eq:psi1_regularity1_a}
\begin{aligned}
    &\sup_{x \in \mathcal{X}, a \in \{0, 1\}, m \in \mathcal{M}} \hat{f}^\star_M(m \mid a, x)/\hat{f}^\star_M(m \mid 1-a, x) < +\infty \sp ,\quad
    &&\inf_{x \in \mathcal{X}, a \in \{0,1\}} \hat{\pi}^\star(a \mid x) > 0 \sp ,
    \\ 
    &\inf_{x \in \mathcal{X}, a \in \{0, 1\}, m \in \mathcal{M}} f_M(m \mid a, x) > 0,\quad
    &&\hspace{-2.2cm}\sup_{x \in \mathcal{X}, a \in \{0,1\}} \pi(a \mid x)/\pi(1-a \mid x) < +\infty \sp .\\
\end{aligned}
\end{equation}

Under the boundedness conditions of \eqref{app:eq:psi1_regularity1_a}, we apply the Cauchy–Schwarz inequality to each term in \eqref{appeq:r2_est1}, leading to the following inequality:

\vspace{-1.25cm} 
\begin{align*}
    \Rem(\hat{\Q}^\star, \Q) & \leq C\bigg[ ||\hat{f}^\star_{M}-f_{M} || \times ||\hat{\mu}^\star-\mu|| +||\hat{f}^\star_{M}-f_{M} || \times || \hat{\pi}^{\star} - \pi || \bigg] \sp ,
\end{align*}

\vspace{-.3cm} \noindent 
where $C$ is a finite positive constant. Given the nuisance convergence rates in \eqref{eq:convergence_rates}, we obtain
\begin{equation}\label{appeq:r2_cvg_est1}
    \begin{aligned}
    \Rem(\hat{\Q}^\star, \Q) \leq \smallO\left(n^{\max \left\{-\left(\frac{1}{b}+\frac{1}{q}\right),-\left(\frac{1}{b}+\frac{1}{k}\right)\right\}}\right)\sp.
\end{aligned}
\end{equation}

\noindent\underline{[Second set of regularity conditions.]} Let $|| f ||_4 = (\P f^4)^{1/4}$ denote the $L^4(\P)$ norm of the function $f$. Assume there exists finite constant $C > 0$ such that
\begin{equation}\label{appeq:r2_est1_condtion2}
\begin{aligned}
        &\bigg|\bigg|\frac{\hat{f}^\star_{M}(. \mid a_0, .)}{\hat{f}^\star_{M}f_{M}}\bigg|\bigg|_4 \leq C \sp , \ 
        \bigg|\bigg|\frac{1}{f_M}\bigg|\bigg|_4\leq C \sp , \
        \bigg|\bigg| \frac{1}{f_M}\frac{\pi(a_0\mid .)}{\hat{\pi}^\star(a_0\mid .)\pi}\bigg|\bigg|_4 \leq C \sp , \ \bigg|\bigg|\frac{1}{f_M}\frac{1}{\hat{\pi}^\star(a_0\mid .)}\bigg|\bigg|_4\leq C \sp .
    \end{aligned}
\end{equation}

Given that the boundedness conditions in \eqref{appeq:r2_est1_condtion2} hold, we apply the Cauchy–Schwarz inequality to each term in \eqref{appeq:r2_est1}, resulting in the following inequality: 

\vspace{-1.25cm} 
\begin{align*}
    \Rem(\hat{\Q}^\star, \Q) & \leq C\bigg[ ||\hat{f}^\star_{M}-f_{M} ||_4 \times ||\hat{\mu}^\star-\mu|| 
    +||\hat{f}^\star_{M}-f_{M} || \times || \hat{\pi}^{\star} - \pi ||_4\bigg] \sp .
\end{align*}

\vspace{-.3cm} \noindent 
We can arrive at the same result as in \eqref{appeq:r2_cvg_est1} by modifying the convergence behaviors of $\hat{f}_M^\star$ and $\hat{\pi}^\star$ in \eqref{eq:convergence_rates} to reflect a stronger $L^4(\P)$-consistency. This can be expressed as follows:  

\vspace{-.75cm}
\begin{equation}
\begin{aligned}
    || \hat{\pi}^{\star} - \pi ||_4 &= \smallO(n^{-\frac{1}{k}}) \sp , \quad 
    \sp || \hat{f}_{M}^\star - f_{M} ||_4 = \smallO(n^{-\frac{1}{b}}) \sp .
\end{aligned}
\end{equation}

\subsubsection{The remainder term, asymptotic linearity, and robustness for $\psi_{2a}(\hat{\Q}^\star)$} \label{app:asym_ate:r2_psi2a} \vspace{0.1cm}

\noindent\underline{\textbf{$\Rem(\hat{\Q}^\star, \Q)$ derivation}}

\noindent
Given the von Mises expansion, we can write: 

\vspace{-1.25cm}
\begin{align*}
\Rem(\hat{\Q}^\star, \Q)
&= \psi(\hat{\Q}^\star) - \psi(\Q) + \int \Phi(\hat{\Q}^\star) \sp \diff\P(o)  \\
&= \iiint \hat{f}_M^{r}(m,a,x)[\mu(m, a, x)-\hat{\mu}^\star(m, a, x)] f_M(m\mid a,x) \pi(a\mid x) \p(x) \sp \diff x\sp \diff a\sp \diff m\\
&\hspace{0.25cm}+\iint \frac{\pi(a_0\mid x)}{\hat{\pi}^\star(a_0\mid x)}\ (\hat{\xi}^\star(m, x)-\hat{\gamma}^\star(x))\ f_M(m\mid a_0,x)\ \p(x) \diff x\ \diff m\\
&\hspace{0.25cm}+\int [\hat{\kappa}_1(x)-\hat{\kappa}_0(x)] \left(\pi(1\mid x)-\hat{\pi}^\star(1\mid x)\right)\ \p(x)\ \diff  x\\
&\hspace{0.25cm}+\int \hat{\gamma}^\star(x)\ \p(x)\ dx-\int \E\left(\xi(m, x) \mid a_0, x\right) \sp \p(x) \sp \diff  x\\
&=\iiint \left(\hat{f}_M^{r}(m,a,x)-f_M^{r}(m,a,x)\right)\ [\mu(m, a, x)-\hat{\mu}^\star(m, a, x)]
\\
&\hspace{6cm}\times f_M(m\mid a,x) \sp \pi(a\mid x) \sp \p(x) \sp \diff x\ \diff a\ \diff m\\
&\hspace{0.25cm}+\iiint f_M^{r}(m,a,x)\ [\mu(m, a, x)-\hat{\mu}^\star(m, a, x)] \sp f_M(m\mid a,x) \sp \pi(a\mid x)\ \p(x) \sp \diff x\ \diff a\\
&\hspace{0.25cm}+\iint \left(\frac{\pi(a_0\mid x)}{\hat{\pi}^\star(a_0\mid x)}-1\right)\ (\hat{\xi}^\star(m, x)-\hat{\gamma}^\star(x))\ f_M(m\mid a_0,x)\ \p(x)\  \diff x\ \diff m\\
&\hspace{0.25cm}+\iint (\hat{\xi}^\star(m, x)-\hat{\gamma}^\star(x))\ f_M(m\mid a_0,x)\ \p(x) \sp \diff x\ \diff m\\
&\hspace{0.25cm}+\int \left[\left(\hat{\kappa}_1(x)-\hat{\kappa}_0(x)\right)-\left(\kappa_1(x)-\kappa_0(x)\right)\right] \sp \left(\pi(1\mid x)-\hat{\pi}^\star(1\mid x)\right)\ \p(x)\ \diff  x\\
&\hspace{0.25cm}+ \int \left(\kappa_1(x)-\kappa_0(x)\right)\ \left(\pi(1\mid x)-\hat{\pi}^\star(1\mid x)\right)\ \p(x)\ \diff  x \\
&\hspace{0.25cm}+\int \hat{\gamma}^\star(x)\ \p(x)\ \diff x-\int \E\left(\xi(m, x) \mid a_0, x\right) \sp \p(x) \sp \diff  x\\
&=\int \left(\hat{f}_M^{r}(m,a,x)-f_M^{r}(m,a,x)\right)\ [\mu(m, a, x)-\hat{\mu}^\star(m, a, x)] \sp \diff\P(m,a,x)\\
&\hspace{0.25cm}+\int \left(\frac{\pi(a_0\mid x)}{\hat{\pi}^\star(a_0\mid x)}-1\right)\ (\gamma(x))-\hat{\gamma}^\star(x))\ \diff\P(x)\\
&\hspace{0.25cm}+\int \bigg[\left(\hat{\kappa}_1(x)-\hat{\kappa}_0(x)\right)-\left(\kappa_1(x)-\kappa_0(x)\right)\bigg] \left(\pi(1\mid x)-\hat{\pi}^\star(1\mid x)\right) \sp \diff\P(x) \sp .
\end{align*}

\noindent\underline{\textbf{Regularity discussions}}

\noindent 
In the following, we discuss two regularity conditions. 

\noindent\underline{[First regularity condition.]} Let $\mathcal{X}$ denote the domain of $X$. Assume 

\vspace{-1.25cm}
\begin{align}
    &\inf_{x \in \mathcal{X}, a \in \{0,1\}} \hat{\pi}^\star(a \mid x) > 0 \sp . 
    \label{appeq:r2_est2a_condtion1}
\end{align}

\vspace{-0.3cm} \noindent 
If the condition in \eqref{appeq:r2_est2a_condtion1} holds, then by applying Cauchy–Schwarz inequality, we arrive at the following inequality:

\vspace{-1.5cm}
\begin{align*}
\Rem(\hat{\Q}^\star, \Q) &\leq || \hat{f}_M^r -f_M^r|| \sp \times || \hat{\mu}^{\star} - \mu|| 
+ C \sp || \hat{\pi}^{\star} - \pi || \sp \times || \hat{\gamma}^\star - \gamma || \\
&\hspace{0.25cm}+ || \hat{\kappa}_1 -\kappa_1 || \sp \times || \hat{\pi}^{\star} - \pi || +|| \hat{\kappa}_0 -\kappa_0 || \sp \times || \hat{\pi}^{\star} - \pi || \sp ,
\end{align*}

\vspace{-0.3cm} \noindent 
where $C$ is a finite positive constant. Given the nuisance convergence rates in \eqref{eq:convergence_rates}, we have
\begin{align}\label{appeq:r2_cvg_est2a}
    \Rem(\hat{\Q}^\star, \Q) \leq \smallO\left[n^{\max \left\{-\left(\frac{1}{c}+\frac{1}{q}\right),-\left(\frac{1}{k}+\frac{1}{j}\right),-\left(\frac{1}{k}+\frac{1}{\ell}\right)\right\}}\right] \sp.
\end{align}
\noindent\underline{[Second set of regularity conditions.]} Let $|| f ||_4 = (\P f^4)^{1/4}$ denote the $L^4(\P)$ norm of the function $f$. Assume there exists a finite positive constant $C$ such that

\vspace{-0.5cm}
{\small
\begin{equation}\label{appeq:r2_est2a_condtion2}
    \begin{aligned}
        &\bigg|\bigg|\frac{1}{\hat{\pi}^\star}\bigg|\bigg|_4 \leq C \sp .
    \end{aligned}
\end{equation}}

Given the boundedness conditions in \eqref{appeq:r2_est2a_condtion2} hold, we apply the Cauchy-Schwarz inequality to each term in \eqref{appeq:r2_est2b}, resulting in the following inequality: 

\vspace{-1.25cm}
\begin{align*}
\Rem(\hat{\Q}^\star, \Q) &\leq || \hat{f}_M^r-f_M^r || \sp \times || \hat{\mu}^{\star} - \mu|| +C\ || \hat{\pi}^{\star} - \pi ||_4 \sp \times || \hat{\gamma}^\star- \gamma ||\\
&\hspace{0.25cm}+ || \hat{\kappa}_1-\kappa_1 || \sp \times || \hat{\pi}^{\star} - \pi || +|| \hat{\kappa}_0-\kappa_0 || \sp \times || \hat{\pi}^{\star} - \pi || \sp .
\end{align*}

\vspace{-0.3cm}
We can arrive at the same result as in \eqref{appeq:r2_cvg_est2a} by modifying the convergence behaviors of $\hat{\pi}^\star(A\mid X)$ in \eqref{eq:convergence_rates} to reflect a stronger $L^4(\P)$-consistency. This can be expressed as follows:

\vspace{-1.25cm}
\begin{align*}
    || \hat{\pi}^{\star} - \pi ||_4 = \smallO(n^{-\frac{1}{k}}) \sp .
\end{align*}

\begin{remark}
    It is important to note that the nuisance estimates $ \hat{\gamma}^\star $ and $ \hat{\kappa}_a $ depend on the estimates of $ \hat{\xi}^\star $ and $ \hat{\mu}^\star $, respectively. Moreover, $ \hat{\xi}^\star $ itself relies on the estimates of $ \hat{\mu}^\star $ and $ \hat{\pi}^\star $. However, the $L^2(\P)$ convergence conditions $ || \hat{\gamma}^\star - \gamma || = o_\p(n^{-\frac{1}{j}}) $ and $ || \hat{\kappa}_a-\kappa_a ||  = o_\p(n^{-\frac{1}{\ell}})$, from display~\ref{eq:convergence_rates}, indicate the convergence of the sequential regressions for any choice of $ \tilde{\pi} \in \M_\pi $ and $ \tilde{\mu} \in \M_\mu $, irrespective of the correctness of these nuisance estimates. To make this dependence more explicit, the respective convergence rates can be restated as follows: 

    \vspace{-1.25cm}
    \begin{align}
        || \hat{\gamma}^\star(.; \hat{\mu}^\star, \hat{\pi}^\star)- \gamma(.; \hat{\mu}^\star, \hat{\pi}^\star) || &= o_\p(n^{-\frac{1}{j}}) \sp , \quad 
        || \hat{\kappa}_a(.; \hat{\mu}^\star)-\kappa_a(.; \hat{\mu}^\star) ||  = o_\p(n^{-\frac{1}{\ell}}) \sp . 
    \end{align}
\end{remark}


\subsubsection{The remainder term, asymptotic linearity, and robustness for $\psi_{2b}(\hat{\Q}^\star)$} \label{app:asym_ate:r2_psi2b} \vspace{0.1cm}

\underline{\textbf{$\Rem(\hat{\Q}^\star, \Q)$ derivation}}

\noindent 
Given the von Mises expansion, we can write:

\vspace{-1.25cm}
{\small\begin{align*}
&\Rem(\hat{\Q}^\star, \Q)
\\
&= \psi(\hat{\Q}^\star) - \psi(\Q) + \int \Phi(\hat{\Q}^\star) \sp \diff\P(o)  \\
&= \iiint \frac{\hat{\lambda}(a_0\mid m,x)}{\hat{\lambda}(a\mid m,x)}\frac{\hat{\pi}^\star(a\mid x)}{\hat{\pi}^\star(a_0\mid x)}[\mu(m, a, x)-\hat{\mu}^\star(m, a, x)] \sp f_M(m\mid a,x)\ \pi(a\mid x)\ \p(x) \sp \diff x\ \diff a\ \diff m\\
&\hspace{0.25cm}+\iint \frac{\pi(a_0\mid x)}{\hat{\pi}^\star(a_0\mid x)}(\hat{\xi}^\star(m, x)-\hat{\gamma}^\star(x))\ f_M(m\mid a_0,x)\ \p(x) \sp \diff x\ \diff m\\
&\hspace{0.25cm}+\int [\hat{\kappa}_1(x)-\hat{\kappa}_0(x)] \sp \left(\pi(1\mid x)-\hat{\pi}^\star(1\mid x)\right)\ \p(x)\ \diff  x\\
&\hspace{0.25cm}+\int \hat{\gamma}^\star(x)\ \p(x)\ \diff x-\int \E\left(\xi(m, x) \mid a_0, x\right) \sp \p(x) \sp \diff  x\\
&=\iiint \frac{\hat{\lambda}(a_0\mid m,x)}{\hat{\lambda}(a\mid m,x)}\big(\frac{\hat{\pi}(a\mid x)}{\hat{\pi}^\star(a_0\mid x)}-\frac{\pi(a\mid x)}{\pi(a_0\mid x)}\big) [\mu(m, a, x)-\hat{\mu}^\star(m, a, x)] \\
&\hspace{8cm}\times f_M(m\mid a,x)\pi(a\mid x)\ \p(x) \sp \diff x\ \diff a\ \diff m\\
&\hspace{0.25cm}+\iiint \frac{\pi(a\mid x)}{\pi(a_0\mid x)}\big(\frac{\hat{\lambda}(a_0\mid m,x)}{\hat{\lambda}(a\mid m,x)}-\frac{\lambda(a_0\mid m,x)}{\lambda(a\mid m,x)}\big)\ [\mu(m, a, x)-\hat{\mu}^\star(m, a, x)] \\
&\hspace{8cm}\times f_M(m\mid a,x)\  \pi(a\mid x)\ \p(x) \sp \diff x\ \diff a\ \diff m\\
&\hspace{0.25cm}+\iiint f_M^{r}(m,a,x)[\mu(m, a, x)-\hat{\mu}^\star(m, a, x)]\ f_M(m\mid a,x)\ \pi(a\mid x)\ \p(x) \sp \diff x\ \diff a\\
&\hspace{0.25cm}+\iint \left(\frac{\pi(a_0\mid x)}{\hat{\pi}^\star(a_0\mid x)}-1\right)(\hat{\xi}^\star(m, x)-\hat{\gamma}^\star(x))\ f_M(m\mid a_0,x)\ \p(x) \sp \diff x\ \diff m\\
&\hspace{0.25cm}+\iint (\hat{\xi}^\star(m, x)-\hat{\gamma}^\star(x))\ f_M(m\mid a_0,x)\ \p(x) \sp \diff x\ \diff m\\
&\hspace{0.25cm}+\int \left[\left(\hat{\kappa}_1(x)-\hat{\kappa}_0(x)\right)-\left(\kappa_1(x)-\kappa_0(x)\right)\right]\ \left(\pi(1\mid x)-\hat{\pi}(1\mid x)\right)\ \p(x)\ \diff  x\\
&\hspace{0.25cm}+ \int \left(\kappa_1(x)-\kappa_0(x)\right)\ \left(\pi(1\mid x)-\hat{\pi}^\star(1\mid x)\right)\ \p(x)\ \diff  x \\
&=\int \frac{\hat{\lambda}(a_0\mid m,x)}{\hat{\lambda}(a\mid m,x)}\left(\frac{\hat{\pi}^\star(a\mid x)}{\hat{\pi}^\star(a_0\mid x)}-\frac{\pi(a\mid x)}{\pi(a_0\mid x)}\right)\ [\mu(m, a, x)-\hat{\mu}(m, a, x)]\ \diff\P(m,a,x)\\
&\hspace{0.25cm}+\int \frac{\pi(a\mid x)}{\pi(a_0\mid x)}\left(\frac{\hat{\lambda}(a_0\mid m,x)}{\hat{\lambda}(a\mid m,x)}-\frac{\lambda(a_0\mid m,x)}{\lambda(a\mid m,x)}\right)\ [\mu(m, a, x)-\hat{\mu}^\star(m, a, x)]\ \diff\P(m,a,x)\\
&\hspace{0.25cm}+\int \left(\frac{\pi(a_0\mid x)}{\hat{\pi}^\star(a_0\mid x)}-1\right)(\gamma(x))-\hat{\gamma}^\star(x)) \sp \diff\P(x)\\
&\hspace{0.25cm}+\int \left[\left(\hat{\kappa}_1(x)-\hat{\kappa}_0(x)\right)-\left(\kappa_1(x)-\kappa_0(x)\right)\right]\ \left(\pi(1\mid x)-\hat{\pi}^\star(1\mid x)\right)\ \diff\P(x) \sp .
\end{align*}}
For a clearer derivation of the convergence behavior, we can further decompose $\Rem(\hat{\Q}^\star, \Q)$ as:

\vspace{-0.5cm}
\begin{equation}\label{appeq:r2_est2b}
{\small\begin{aligned}
&\Rem(\hat{\Q}^\star, \Q)\\
&=\int \frac{\hat{\lambda}(a_0\mid m,x)}{\hat{\lambda}(a\mid m,x)\hat{\pi}^\star(a_0\mid x)}\ \left(\hat{\pi}^\star(a\mid x)-\pi(a\mid x)\right)[\mu(m, a, x)-\hat{\mu}(m, a, x)] \sp \diff \P(m,a,x)\\
&\hspace{0.25cm}+\int \frac{\hat{\lambda}(a_0\mid m,x)}{\hat{\lambda}(a\mid m,x)\hat{\pi}^\star(a_0\mid x)}\frac{\pi(a\mid x)}{\hat{\pi}^\star(a_0\mid x)\pi(a_0\mid x)}\\
&\hspace{5cm}\times (\pi(a_0\mid x)-\hat{\pi}^\star(a_0\mid x)) [\mu(m, a, x)-\hat{\mu}^\star(m, a, x)] \sp \diff \P(m,a,x)\\
&\hspace{0.25cm}+\int \frac{\pi(a\mid x)}{\pi(a_0\mid x)\hat{\lambda}(a_0\mid m,x)}(\hat{\lambda}(a\mid m,x)-\lambda(a\mid m,x))[\mu(m, a, x)-\hat{\mu}^\star(m, a, x)] \sp \diff \P(m,a,x)\\
&\hspace{0.25cm}+\int \frac{\pi(a\mid x)}{\pi(a_0\mid x)}\frac{\lambda(a\mid m,x)}{\hat{\lambda}(a_0\mid m,x)\lambda(a_0\mid m,x)}(\lambda(a_0\mid m,x)-\hat{\lambda}(a_0\mid m,x))\\
&\hspace{8cm}\times[\mu(m, a, x)-\hat{\mu}^\star(m, a, x)] \sp \diff \P(m,a,x)\\
&\hspace{0.25cm}+\int \left(\frac{\pi(a_0\mid x)}{\hat{\pi}^\star(a_0\mid x)}-1\right)\ (\gamma(x))-\hat{\gamma}^\star(x))\ \diff \P(x)\\
&\hspace{0.25cm}+\int \left[\left(\hat{\kappa}_1(x)-\hat{\kappa}_0(x)\right)-\left(\kappa_1(x)-\kappa_0(x)\right)\right] \left(\pi(1\mid x)-\hat{\pi}^\star(1\mid x)\right) \sp \diff \P(x) \sp .
\end{aligned}}
\end{equation}

\noindent\underline{\textbf{Regularity discussions}} \newline
In the following, we discuss two sets of regularity conditions. 

\noindent\underline{[First set of regularity conditions.]} Let $\mathcal{X}$ and $\mathcal{M}$ denote the domain of $X$ and $M$. Assume 

\vspace{-0.75cm}
\begin{equation}\label{appeq:r2_est2b_condtion1}
    \begin{aligned}
        &\inf_{a \in \{0,1\}, x\in\mathcal{X}} \hat{\pi}^\star(a\mid x) >0 \sp , \quad 
        &&\hspace{-0.2cm}\sup_{x \in \mathcal{X}, a \in \{0,1\}, m \in \mathcal{M}} \hat{\lambda}(a \mid x, m)/\hat{\lambda}(1-a \mid x, m) < +\infty \sp ,
        \\
        &\sup_{x \in \mathcal{X}, a \in \{0,1\}} \pi(a \mid x)/\pi(1-a \mid x) < +\infty \sp , \quad 
        &&\inf_{x \in \mathcal{X}, a \in \{0,1\}, m \in \mathcal{M}} \hat{\lambda}(a\mid m, x) > 0 \sp .
    \end{aligned}
\end{equation}

Under the boundedness conditions of \eqref{appeq:r2_est2b_condtion1}, we apply the Cauchy–Schwarz inequality to each term in \eqref{appeq:r2_est2b}, leading to the following inequality:
\begin{align*}
    \Rem(\hat{\Q}^\star, \Q) & \leq C\bigg[ || \hat{\pi}^{\star} - \pi || \sp \times ||\hat{\mu}^\star-\mu|| +|| \hat{\lambda}-\hat{\lambda} || \sp \times ||\hat{\mu}^\star-\mu|| \bigg]\\
    &\hspace{0.75cm}+|| \hat{\pi}^{\star} - \pi|| \sp \times || \hat{\gamma}^\star- \gamma || +|| (\hat{\kappa}_1-\hat{\kappa}_0)-(\kappa_1-\kappa_0) || \sp \times || \hat{\pi}^{\star} - \pi || \bigg] \sp , 
\end{align*}%
where $C$ is a finite positive constant. Given the nuisance convergence rates in \eqref{eq:convergence_rates}, we obtain
\begin{equation}\label{appeq:r2_cvg_est2b}
    \begin{aligned}
    \Rem(\hat{\Q}^\star, \Q) \leq \smallO\left[n^{\max \left\{-\left(\frac{1}{q}+\frac{1}{k}\right),-\left(\frac{1}{d}+\frac{1}{q}\right),-\left(\frac{1}{k}+\frac{1}{j}\right),-\left(\frac{1}{k}+\frac{1}{\ell}\right)\right\}}\right] \sp.
\end{aligned}
\end{equation}

\noindent\underline{[Second set of regularity conditions.]} 
Let $|| f ||_4 = (\P f^4)^{1/4}$ denote the $L^4(\P)$ norm of the function $f$. Assume there exists a finite positive constant $C$ such that
{\small\begin{equation}\label{appeq:r2_est2b_condtion2}
    \begin{aligned}
        &\bigg|\bigg|\frac{\hat{\lambda}(a_0 \mid.)}{\hat{\lambda} \sp \hat{\pi}^\star(a_0\mid .)}\bigg|\bigg|_4 \leq C \sp , \qquad \bigg|\bigg|\frac{\hat{\lambda}(a_0\mid .)}{\hat{\lambda} \sp \hat{\pi}^\star(a_0\mid .)}\frac{\pi}{\hat{\pi}^\star(a_0\mid X)\ \pi(a_0\mid .)}\bigg|\bigg|_4\leq C \sp ,\\
        &\bigg|\bigg|\frac{\pi}{\pi(a_0\mid .) \sp \hat{\lambda}(a_0\mid .)}\bigg|\bigg|_4 \leq C \sp , \qquad \bigg|\bigg|\left(\frac{\pi(a_0\mid .)}{\hat{\pi}^\star(a_0\mid .)}-1\right)\bigg|\bigg|_4\leq C \sp .
    \end{aligned}
\end{equation}}

Given that the boundedness conditions in \eqref{appeq:r2_est2b_condtion2} hold, we apply the Cauchy-Schwarz inequality to each term in \eqref{appeq:r2_est2b}, resulting in the following inequality: 
\begin{align*}
    \Rem(\hat{\Q}^\star, \Q) & \leq C\bigg[ || \hat{\pi}^{\star} - \pi ||_4 \times ||\hat{\mu} -\mu|| +|| \hat{\lambda} -\hat{\lambda} ||_4 \times ||\hat{\mu}^\star-\mu||\bigg]\\
    &\hspace{0.75cm}+|| \hat{\pi}^{\star} - \pi ||_4 \times || \hat{\gamma}^\star- \gamma || +|| (\hat{\kappa}_1-\hat{\kappa}_0)-(\kappa_1-\kappa_0) || \times || \hat{\pi}^{\star} - \pi || \bigg] \sp . 
\end{align*}

We can arrive at the same result as in \eqref{appeq:r2_cvg_est2b} by modifying the convergence behaviors of $\hat{\lambda}(A\mid M,X)$ and $\hat{\pi}^\star(1\mid X)$ in \eqref{eq:convergence_rates} to reflect a stronger $L^4(\P)$-consistency, expressed as:

\vspace{-1.25cm}
\begin{align*}
    || \hat{\pi}^{\star} - \pi ||_4 = \smallO(n^{-\frac{1}{k}}) \sp , 
    \quad
    || \hat{\lambda}-\lambda ||_4  = \smallO(n^{-\frac{1}{d}}) \sp .
\end{align*}

\subsection{ATT front-door functional estimators}\label{app:asym_att}
\subsubsection{The remainder term, asymptotic linearity, and robustness for $\beta_{1}(\hat{\Q}^\star)$} \label{app:asym_att:r2_beta1} \vspace{0.1cm}

\noindent\underline{\textbf{$\Rem(\hat{\Q}^\star, \Q)$ derivation}}

\noindent Given the von Mises expansion, we can write: 

\vspace{-1.25cm}
\begin{align*}
\Rem(\hat{\Q}^\star, \Q)
&= \beta_1(\hat{\Q}^\star) - \beta_1(\Q) + \int \Phi_{\beta}(\hat{\Q}^\star) \sp \diff \P(o)  \\
&=\beta_1(\hat{\Q}^\star) - \beta_1(\Q) + \int \bigg\{\frac{\I(a=a_1)}{\hat{\p}_A(a_1)} \frac{\hat{f}_M(m,a_0,x)}{\hat{f}_M(m,a_1,x)} \sp \big\{y-\hat{\mu}^{\star}(m,a_1,x)\big\}
    \\
    &\hspace{0.25cm}+\frac{\I(a=a_0)}{\hat{p}_A(a_1)}\frac{\hat{\pi}^{\star}(a_1\mid x)}{\hat{\pi}^{\star}(a_0\mid x)}\big\{\hat{\mu}(m,a_1,x)-\hat{\kappa}_{a_1}(x)\big\}
    +\frac{\I(a=a_1)}{\hat{\p}_A(a_1)}\big\{\hat{\kappa}_{a_1}(X)-\beta(\hat{\Q}^{\star})\big\}\bigg\} \sp \diff \P(o)
    \\
    &=\int \bigg\{\frac{\I(a=a_1)}{\hat{\p}_A(a_1)} \big(\frac{\hat{f}_M(m,a_0,x)}{\hat{f}_M(m,a_1,x)}-\frac{{f}_M(m,a_0,x)}{{f}_M(m,a_1,x)}\big) \sp \big\{\mu(m,a_1,x)-\hat{\mu}^{\star}(m,a_1,x)\big\}  
    \\
    &\hspace{0.25cm}+ \frac{\I(a=a_1)}{\hat{\p}_A(a_1)}\frac{{f}_M(m,a_0,x)}{{f}_M(m,a_1,x)}\big\{\mu(m,a_1,x)-\hat{\mu}^{\star}(m,a_1,x)\big\} 
    \\
    &\hspace{0.25cm}+\frac{\I(a=a_0)}{\hat{\p}_A(a_1)}\big(\frac{\hat{\pi}^{\star}(a_1\mid x)}{\hat{\pi}^{\star}(a_0\mid x)}-\frac{{\pi}(a_1\mid x)}{{\pi}(a_0\mid x)}\big)\ \hat{\mu}^\star(m,a_1,x)\ \big\{f_M(m,a_0,x)-\hat{f}_M(m,a_0,x)\big\} 
    \\
    &\hspace{0.25cm}+ \frac{\I(a=a_0)}{\hat{\p}_A(a_1)}\frac{{\pi}(a_1\mid x)}{{\pi}(a_0\mid x)}\ \hat{\mu}^\star(m,a_1,x)\ \big\{f_M(m,a_0,x)-\hat{f}_M(m,a_0,x)\big\} 
    \\
    &\hspace{0.25cm}+\frac{\I(a=a_1)}{\hat{\p}_A(a_1)}\big\{\hat{\kappa}_{a_1}(X)-\beta(\hat{\Q}^{\star})\big\}\bigg\} \sp \diff\P(o) 
    \\
    &\hspace{0.25cm}+\beta_1(\hat{\Q}^\star) - \beta_1(\Q) \sp . 
\end{align*}

\vspace{-0.3cm} \noindent 
Note that the second, fourth, fifth, and sixth lines sum to a term with $o_\p(n^{-\frac{1}{2}})$ rate of convergence:

\vspace{-1.25cm}
\begin{align*}
    &(2)+(4)+(5-6) \\
    &\hspace{0.25cm} = \beta_1(\hat{\Q}^\star) - \beta_1(\Q)
    \\
    &\hspace{0.25cm}+\int\bigg\{\underbrace{\frac{\I(a=a_1)}{\hat{p}_A(a_1)}\frac{{f}_M(m,a_0,x)}{{f}_M(m,a_1,x)}\mu(m,a_1,x)}_{\textcircled{1}}\underbrace{-\frac{\I(a=a_1)}{\hat{\p}_A(a_1)}\frac{{f}_M(m,a_0,x)}{{f}_M(m,a_1,x)}\hat{\mu}^{\star}(m,a_1,x)}_{\textcircled{2}}
    \\
    &\hspace{0.25cm}+\underbrace{\frac{\I(a=a_0)}{\hat{p}_A(a_1)}\frac{{\pi}(a_1\mid x)}{{\pi}(a_0\mid x)}\ \hat{\mu}^\star(m,a_1,x)\ f_M(m,a_0,x)}_{\textcircled{3}}\underbrace{-\frac{\I(a=a_0)}{\hat{\p}_A(a_1)}\frac{{\pi}(a_1\mid x)}{{\pi}(a_0\mid x)}\ \hat{\mu}^\star(m,a_1,x)\ \hat{f}_M(m,a_0,x)}_{\textcircled{4}}
    \\
    &\hspace{0.25cm}+\underbrace{\frac{\I(a=a_1)}{\hat{\p}_A(a_1)}\ \hat{\kappa}_{a_1}(X)}_{\textcircled{5}}\underbrace{-\frac{\I(a=a_1)}{\hat{\p}_A(a_1)}\ \beta(\hat{\Q}^{\star})}_{\textcircled{6}}\bigg\}\ \diff \P(o) \sp ,
\end{align*}
where

\vspace{-1.25cm}
\begin{align*}
    &\int \textcircled{2}+\textcircled{3}\ \diff\P(o)=\int \textcircled{4}+\textcircled{5}\ \diff \P(o)=0 \sp , 
    \\
    &\int \textcircled{1}\ \diff\P(o)-\beta(\Q) = \int \I\left(a=a_1\right)\left(\frac{1}{\hat{\p}_A\left(a_1\right)}-\frac{1} {\p_A\left(a_1\right)}\right)\ f_M(m,a_0,x)\ \mu(m,a_1,x)\ \diff \P(o) \sp , 
    \\
    &\beta(\hat{\Q}^\star)-\int \textcircled{6}\ \diff\P(o) = (1-\frac{\p_A(a_1)}{\hat{\p}_A(a_1)}) \sp \beta(\hat{\Q}^\star) \sp .
\end{align*}

Therefore, we have the second-order remainder term to have the final form:
\begin{equation}\label{appeq:r2_beta1}
    \begin{aligned}
    \Rem(\hat{\Q}^\star, \Q) &=\int \bigg\{\frac{\I(a=a_1)}{\hat{\p}_A(a_1)} \big(\frac{\hat{f}_M(m,a_0,x)}{\hat{f}_M(m,a_1,x)}-\frac{{f}_M(m,a_0,x)}{{f}_M(m,a_1,x)}\big) \sp \big\{\mu(m,a_1,x)-\hat{\mu}^{\star}(m,a_1,x)\big\}
    \\
    &\hspace{-0.3cm}+\frac{\I(a=a_0)}{\hat{\p}_A(a_1)}\big(\frac{\hat{\pi}^{\star}(a_1\mid x)}{\hat{\pi}^{\star}(a_0\mid x)}-\frac{{\pi}(a_1\mid x)}{{\pi}(a_0\mid x)}\big)\ \hat{\mu}^\star(m,a_1,x)\ \big\{f_M(m,a_0,x)-\hat{f}_M(m,a_0,x)\big\}\bigg\} \sp \diff\P(o)
    \\
    &\hspace{0.25cm}+\int \I\left(a=a_1\right)\left(\frac{1}{\hat{\p}_A\left(a_1\right)}-\frac{1} {\p_A\left(a_1\right)}\right)\ f_M(m,a_0,x)\ \mu(m,a_1,x)\ \diff\P(o)
    \\
    &\hspace{0.25cm}+(1-\frac{\p_A(a_1)}{\hat{p}_A(a_1)}) \sp \beta_1(\hat{\Q}^\star) \sp .
\end{aligned}
\end{equation}

\noindent\underline{\textbf{Regularity discussions}}

\noindent 
Let $\mathcal{X}$ and $\mathcal{M}$ denote the domain of $X$ and $M$. Assume 
\begin{equation}\label{appeq:r2_beta1_condtion1}
    \begin{aligned}
        &\inf_{x\in\mathcal{X},\sp m \in\mathcal{M}} \hat{f}_M(m,a_1,x) >0 \sp , \quad 
        && \inf_{x\in\mathcal{X},\sp m \in\mathcal{M}} f_M(m,a_1,x) >0  \sp , 
        \\
        &\inf_{x\in\mathcal{X}} \hat{\pi}(a_0\mid x) >0 \sp , \quad 
        && \inf_{x\in\mathcal{X}} {\pi}(a_0\mid x) >0 \sp , 
        \\
        &\sup_{x \in \mathcal{X}, m \in \mathcal{M}} \hat{\mu}^{\star}(m,a_1,x) < \infty \sp .
        &&
    \end{aligned}
\end{equation}

Under the boundedness conditions of \eqref{appeq:r2_beta1_condtion1}, we apply the Cauchy–Schwarz inequality to each term in \eqref{appeq:r2_beta1}, leading to the following inequality:

\vspace{-1.25cm}
\begin{align*}
\Rem(\hat{\Q}^{\star},\Q) \leq C\bigg[||\hat{f}_M-f_M|| \times ||\hat{\mu}^{\star}-\mu|| + ||\hat{f}_M-f_M||\times ||\hat{\pi}^{\star}-\pi||\bigg] \sp ,
\end{align*}

\vspace{-0.3cm} \noindent 
where $C$ is a finite positive constant. Given the nuisance convergence rates in \eqref{eq:convergence_rates}, we obtain
\begin{equation}\label{appeq:r2_cvg_beta1}
    \begin{aligned}
    \Rem(\hat{\Q}^\star, \Q) \leq \smallO\left[n^{\max \left\{-\left(\frac{1}{b}+\frac{1}{k}\right),-\left(\frac{1}{k}+\frac{1}{q}\right)\right\}}\right] \sp.
\end{aligned}
\end{equation}

\noindent\underline{\textbf{Asymptotic linearity}}
\begin{theorem}[Asymptotic linearity of $\beta_1(\hat{\Q}^\star)$] 
\label{thm:asymp_psi1_att} 
In addition to (A1)-(A3) and the boundedness condition \eqref{appeq:r2_beta1_condtion1} , we assume that the nuisance estimates $\hat{\Q}^{\star} = (\hat{\mu}^\star, \hat{f}_M^\star, \hat{\pi},\hat{p}_A,\hat{\p}_{AX})$ satisfy:
\begin{enumerate}
    \item[] (A5.4) \emph{$L^2(\P)$ convergence of nuisance regressions}: Let $|| \hat{\pi}^{\star} - \pi || =\smallO(n^{-\frac{1}{k}})$, $ 
    || \hat{f}_{M}^\star - f_{M} || =\smallO(n^{-\frac{1}{b}})$, $ 
    || \hat{\mu}^{\star} - \mu|| = \smallO(n^{-\frac{1}{q}})$, and assume that both $\frac{1}{b} + \frac{1}{q} \geq \frac{1}{2}$ and $\frac{1}{k} + \frac{1}{b} \geq \frac{1}{2}$.
\end{enumerate}
Under these conditions, $\beta_1(\hat{\Q}^\star) - \beta_1(\Q) = \Pn \Phi_{\beta}(\Q) + o_\p(n^{-1/2})$ implying that the TMLE $\beta_1(\hat{\Q}^\star)$ is asymptotically linear and with influence function equal to $\Phi_{\beta}(\Q)$.
\end{theorem}
Note that $\hat{\p}_A$ is estimated nonparametrically as the sample mean. Therefore, it converges to the true mean at a rate of $o_\p(n^{-1/2})$, ensuring the last two lines in $\Rem(\hat{\Q}^\star, \Q)$ also converge to the truth at a rate of $o_\p(n^{-1/2})$.

An immediate corollary of Theorem \ref{thm:asymp_psi1_att} is that $\beta_1(\hat{\Q}^\star)$ shares the same multiple robustness properties as its corresponding ATE estimator, as stated in the Corollary~\ref{cor:robust_psi1}. To avoid redundancy, we omit a restatement of these properties here.

\subsubsection{The remainder term, asymptotic linearity, and robustness for $\beta_{a}(\hat{\Q}^\star)$} \label{app:asym_att:r2_betaa} \vspace{0.1cm}

\noindent\underline{\textbf{$\Rem(\hat{\Q}^\star, \Q)$ derivation}}

Given the $\Rem(\hat{\Q}^\star, \Q)$ term of $\beta_1(\hat{\Q}^\star)$, it immediately follows that $\beta_a(\hat{\Q}^\star)$ has an $\Rem(\hat{\Q}^\star, \Q)$ term as follows:

\vspace{-1.25cm}
\begin{equation}\label{appeq:r2_betaa}
    \begin{aligned}
    \Rem(\hat{\Q}^\star, \Q) &=\int \bigg\{\frac{\I(a=a_1)}{\hat{\p}_A(a_1)} \big(\hat{f}_M^r(m,a,x)-f_M^r(m,a,x)\big) \sp \big\{\mu(m,a_1,x)-\hat{\mu}^{\star}(m,a_1,x)\big\}
    \\
    &\hspace{0.25cm}+\frac{\I(a=a_0)}{\hat{\p}_A(a_1)}\big(\frac{\hat{\pi}^{\star}(a_1\mid x)}{\hat{\pi}^{\star}(a_0\mid x)}-\frac{{\pi}(a_1\mid x)}{{\pi}(a_0\mid x)}\big)\ \big\{\kappa_{a_1}(x;\hat{\mu}^{\star})-\hat{\kappa}_{a_1}(x;\hat{\mu}^{\star})\big\}\bigg\} \sp \diff\P(o)
    \\
    &\hspace{0.25cm}+\int \I\left(a=a_1\right)\left(\frac{1}{\hat{\p}_A\left(a_1\right)}-\frac{1} {\p_A\left(a_1\right)}\right)\ f_M(m,a_0,x)\ \mu(m,a_1,x)\ \diff\P(o)
    \\
    &\hspace{0.25cm}+(1-\frac{\p_A(a_1)}{\hat{\p}_A(a_1)}) \sp \beta_a(\hat{\Q}^\star) \sp .
\end{aligned}
\end{equation}

\noindent\underline{\textbf{Regularity discussions}}
\newline
Let $\mathcal{X}$ and $\mathcal{M}$ denote the domain of $X$ and $M$. Assume 

\vspace{-.75cm}
\begin{equation}\label{appeq:r2_betaa_condtion1}
    \begin{aligned}
        &\inf_{x\in\mathcal{X}} \hat{\pi}(a_0\mid x) >0 \sp , \quad 
        && \inf_{x\in\mathcal{X}} {\pi}(a_0\mid x) >0 \sp . 
    \end{aligned}
\end{equation}

Under the boundedness conditions of \eqref{appeq:r2_betaa_condtion1}, we apply the Cauchy–Schwarz inequality to each term in \eqref{appeq:r2_betaa}, leading to the following inequality:
\begin{align*}
\Rem(\hat{\Q}^{\star},\Q) \leq C\bigg[||\hat{f}_M^r-f_M|| \times ||\hat{\mu}^{\star}-\mu|| + ||\hat{\pi}-\pi||\times ||\hat{\kappa}_{a_1}-{\kappa}_{a_1}||\bigg] \sp ,
\end{align*}
where $C$ is a finite positive constant. Given the nuisance convergence rates in \eqref{eq:convergence_rates}, we obtain
\begin{equation}\label{appeq:r2_cvg_beta_a}
    \begin{aligned}
    \Rem(\hat{\Q}^\star, \Q) \leq \smallO\left[n^{\max \left\{-\left(\frac{1}{c}+\frac{1}{q}\right),-\left(\frac{1}{k}+\frac{1}{\ell}\right)\right\}}\right] \sp .
\end{aligned}
\end{equation}

\noindent\underline{\textbf{Asymptotic linearity}}

\begin{theorem}[Asymptotic linearity of $\beta_{a}(\hat{\Q}^\star)$] \label{thm:asymp_betaa} 
In addition to (A1)-(A3) and the regularity conditions \eqref{appeq:r2_betaa_condtion1}, we assume the nuisance estimates $\hat{\Q}^{\star} = (\hat{\mu}^\star, \hat{\kappa}_{a_1}, \hat{f}_M^r, \hat{\pi}^\star,\hat{\p}_A,\hat{\p}_{AX})$ satisfy: 
\begin{enumerate}
    \item[] (A5.5) \sp \emph{$L^2(\P)$-rates of nuisance estimates}: Let $|| \hat{\pi}^{\star} - \pi || =\smallO(n^{-\frac{1}{k}})$, $
    || \hat{\mu}^{\star} - \mu|| = \smallO(n^{-\frac{1}{q}})$,
    $|| \hat{\kappa}_{a_1} - \kappa_{a_1} || =\smallO(n^{-\frac{1}{\ell}})$, $ || \hat{f}_M^r - f_M^r || = \smallO(n^{-\frac{1}{c}})$, and assume that 
    $\frac{1}{c}+\frac{1}{q} \geq \frac{1}{2}$, and $\frac{1}{\ell}+\frac{1}{k} \geq \frac{1}{2}$. 
\end{enumerate}
Under these conditions, $\beta_a(\hat{\Q}^\star) - \beta_a(\Q) = \Pn \Phi_{\beta}(\Q) + o_\p(n^{-1/2})$ implying that the TMLE $\beta_{a}(\hat{\Q}^\star)$ is asymptotically linear and with influence function equal to $\Phi_{\beta}(\Q)$.
\end{theorem}
Comparing the asymptotic behavior of $\beta_{a}(\hat{\Q}^\star)$ with its ATE counterpart, we find that $\beta_{a}(\hat{\Q}^\star)$ demonstrates greater robustness. Specifically, the convergence of $\hat{\gamma}^\star$ to its truth at a certain rate—required for $\psi_{2a}(\hat{\Q}^\star)$ to achieve asymptotic linearity—is no longer necessary for $\beta_{a}(\hat{\Q}^\star)$. For the same reason, $\beta_{a}(\hat{\Q}^\star)$ achieves consistency under weaker conditions, as illustrated below.
\begin{corollary}[Robustness of $\beta_{a}(\hat{\Q}^\star)$]  $\beta_{a}(\hat{\Q}^\star)$ is consistent for $\beta(\Q)$ if at least one of the following conditions hold: 

\vspace{-1.5cm}
\begin{align*}
    &(i) \sp  ||\hat{\pi}^\star - \pi || = o_\p(1) \sp \text{and} \sp || \hat{\mu}^\star - \mu || = o_\p(1) \sp , 
    \\
&(ii) \sp || \hat{\pi}^\star - \pi || = o_\p(1) \sp \text{and} \sp  || \hat{f}^r_M - f_M^r || = o_\p(1) \sp , \\
    &(iii) \sp  || \hat{\mu}^\star -  \mu ||  = o_\p(1) \sp , \sp  \text{and} \sp  || \hat{\kappa}_1 - \kappa_1 || = o_\p(1) \sp , 
    \\
        &(iv) \sp || \hat{\kappa}_{a_1} - \kappa_{a_1} || = o_\p(1) \sp , \sp \text{and} \sp  || \hat{f}_M^r - f_M^r || = o_\p(1) \sp .
\end{align*} \label{cor:robust_betaa}
\end{corollary} 

\vspace{-1.2cm} 
Corollary \ref{cor:robust_betaa} suggests that either the nuisance estimates $\hat{\mu}^\star$ and $\hat{\pi}^\star$ need to converge to their respective truths (conditions (i)-(iii)) or the estimates introduced to circumvent density estimation, $\hat{\kappa}_{a_1}, \hat{f}_M^r$, should converge to their true values (condition (iv)).

\subsubsection{The remainder term, asymptotic linearity, and robustness for $\beta_{b}(\hat{\Q}^\star)$} \label{app:asym_att:r2_betab} \vspace{0.1cm}

\noindent\underline{\textbf{$\Rem(\hat{\Q}^\star, \Q)$ derivation}}

Given the $\Rem(\hat{\Q}^\star, \Q)$ term for $\beta_1(\hat{\Q}^\star)$, it immediately follows that $\beta_b(\hat{\Q}^\star)$ has an $\Rem(\hat{\Q}^\star, \Q)$ term as follows:

\vspace{-1.25cm}
\begin{equation}\label{appeq:r2_betab}
    \begin{aligned}
    &\Rem(\hat{\Q}^\star, \Q)\\
    &=\int \bigg\{\frac{\I(a=a_1)}{\hat{\p}_A(a_1)} \big(\frac{\hat{\lambda}(a_0\mid m,x)}{\hat{\lambda}(a_1\mid m,x)}/\frac{\hat{\pi}(a_0\mid x)}{\hat{\pi}(a_1\mid x)} - \frac{\lambda(a_0\mid m,x)}{\lambda(a_1\mid m,x)}/\frac{{\pi}(a_0\mid x)}{{\pi}(a_1\mid x)}\big) \sp \big\{\mu(m,a_1,x)-\hat{\mu}^{\star}(m,a_1,x)\big\}
    \\
    &\hspace{0.25cm}+\frac{\I(a=a_0)}{\hat{\p}_A(a_1)}\big(\frac{\hat{\pi}^{\star}(a_1\mid x)}{\hat{\pi}^{\star}(a_0\mid x)}-\frac{{\pi}(a_1\mid x)}{{\pi}(a_0\mid x)}\big)\ \big\{\kappa_{a_1}(x;\hat{\mu}^{\star})-\hat{\kappa}_{a_1}(x;\hat{\mu}^{\star})\big\}\bigg\} \sp \diff\P(o)
    \\
    &\hspace{0.25cm}+\int \I\left(a=a_1\right)\left(\frac{1}{\hat{\p}_A\left(a_1\right)}-\frac{1} {\p_A\left(a_1\right)}\right)\ f_M(m,a_0,x)\ \mu(m,a_1,x)\ \diff\P(o)
    \\
    &\hspace{0.25cm}+(1-\frac{\p_A(a_1)}{\hat{\p}_A(a_1)}) \sp \beta_a(\hat{\Q}^\star) \sp ,
\end{aligned}
\end{equation}
where the first line can be further simplified as
\begin{align*}
    (1) &= \int \bigg\{\frac{\I(a=a_1)}{\hat{\p}_A(a_1)}\frac{\pi(a_1\mid x)}{\pi(a_0\mid x)} \big(\frac{\hat{\lambda}(a_0\mid m,x)}{\hat{\lambda}(a_1\mid m,x)} - \frac{\lambda(a_0\mid m,x)}{\lambda(a_1\mid m,x)}\big) \sp \big\{\mu(m,a_1,x)-\hat{\mu}^{\star}(m,a_1,x)\big\}
    \\
    &\hspace{0.25cm}+\frac{\I(a=a_1)}{\hat{\p}_A(a_1)}\frac{\hat{\lambda}(a_0\mid m, x)}{\hat{\lambda}(a_1\mid m,x)} \big(\frac{\hat{\pi}(a_0\mid x)}{\hat{\pi}(a_1\mid x)} - \frac{{\pi}(a_0\mid x)}{{\pi}(a_1\mid x)}\big) \sp \big\{\mu(m,a_1,x)-\hat{\mu}^{\star}(m,a_1,x)\big\}\bigg\}\ \diff\P(o) \sp .
\end{align*}

\noindent\underline{\textbf{Regularity discussions}}
\newline
Let $\mathcal{X}$ and $\mathcal{M}$ denote the domain of $X$ and $M$. Assume 
\begin{equation}\label{appeq:r2_betab_condtion1}
    \begin{aligned}
        &\inf_{x\in\mathcal{X}} \hat{\pi}(a_0\mid x) >0 \sp , \quad 
        && \inf_{x\in\mathcal{X}} {\pi}(a_0\mid x) >0 \sp , 
        \\
        &\inf_{x\in\mathcal{X},\sp m\in\mathcal{M}} \hat{\lambda}(a_1\mid m,x) >0 \sp , \quad 
        && \inf_{x\in\mathcal{X},\sp m\in\mathcal{M}} {\lambda}(a_1\mid m,x) >0 \sp . 
    \end{aligned}
\end{equation}

Under the boundedness conditions of \eqref{appeq:r2_betab_condtion1}, we apply the Cauchy–Schwarz inequality to each term in \eqref{appeq:r2_betab}, leading to the following inequality:
\begin{align*}
\Rem(\hat{\Q}^{\star},\Q) \leq C\bigg[||\hat{\pi}-\pi|| \times ||\hat{\mu}^{\star}-\mu|| + ||\hat{\lambda}-\lambda|| \times ||\hat{\mu}^{\star}-\mu||+ ||\hat{\pi}-\pi||\times ||\hat{\kappa}_{a_1}-{\kappa}_{a_1}||\bigg] \sp,
\end{align*}
where $C$ is a finite positive constant. Given the nuisance convergence rates in \eqref{eq:convergence_rates}, we obtain
\begin{equation}\label{appeq:r2_cvg_beta_b}
    \begin{aligned}
    \Rem(\hat{\Q}^\star, \Q) \leq \smallO\left[n^{\max \left\{-\left(\frac{1}{k}+\frac{1}{q}\right),-\left(\frac{1}{d}+\frac{1}{q}\right),-\left(\frac{1}{k}+\frac{1}{\ell}\right)\right\}}\right] \sp .
\end{aligned}
\end{equation}

\pagebreak
\section{Details on model evaluation and efficiency gains}
\label{app:testing}

To complement the front-door identification and estimation strategies, we expand here on how the presence of an anchor variable $Z$ can support both model evaluation and efficiency gains. This extends our discussion in Section~\ref{sec:testing} by formalizing the statistical tests and empirical criteria that can be applied when an anchor is available.

\subsection{Details on Verma constraint}\label{app:testing_verma}  

\cite{bhattacharya2022testability} introduced the concept of an anchor variable $Z$ to facilitate empirical evaluation of the front-door assumptions. An anchor satisfies the \textit{relevance} criterion, meaning it must be associated with the treatment $A$, either via a direct effect, through unmeasured confounding, or both. The extended front-door model incorporating such an anchor is shown in Fig.~\ref{fig:front-door_anchor}, where bidirected arrows indicate the presence of unmeasured confounding between endpoint variables. The anchor variable may also exhibit marginal dependence with the mediator $M$, either through a direct causal link ($Z \blue{\rightarrow} M$), as in Fig.~\ref{fig:front-door_anchor}(a), or via unmeasured confounding ($Z \red{\leftrightarrow} M$), as in Fig.~\ref{fig:front-door_anchor}(b). 

In the anchor-included front-door models shown in Fig.~\ref{fig:front-door_anchor}, $Z$ is marginally associated with $Y$, even though these variables are not adjacent; that is, $Z \not\perp Y \mid X, A, M$, or conditional on any subset of ${X, A, M}$. According to d-separation rules \citep{pearl09causality}, this dependence arises due to an open path from $Z$ to $Y$ through $A$ and $M$, which, when blocked (e.g., by conditioning on either $A$ or $M$), opens up a collider path via the unmeasured confounder between treatment and outcome. 

Despite the lack of ordinary conditional independence between $Z$ and $Y$, their non-adjacency induces a constraint on the observed distribution $\P(O)$ where $O=(X, Z, A, M, Y)$, formalized by the nested Markov model for DAGs with hidden variables \citep{richardson2023nested}. For concreteness and without loss of generality, we focus on an anchor configuration consistent with the structure in Fig.~\ref{fig:graphs_truncation}(a).

\begin{figure}[t] 
	\begin{center}
    \scalebox{0.65}{
    \begin{tikzpicture}[>=stealth, node distance=1.7cm]
        \tikzstyle{format} = [thick, circle, minimum size=1.0mm, inner sep=2pt]
        \tikzstyle{square} = [draw, thick, minimum size=4.5mm, inner sep=2pt]

    \begin{scope}[xshift=0cm, yshift=0cm]
		\path[->, thick]

        node[] (a) {$A$}
		node[left of=a, xshift=-0.75cm] (z) {$Z$}
		node[right of=a, xshift=0.75cm] (m) {$M$}
		node[above right of=m, yshift=-0.25cm] (u) {}
		node[above left of=m, yshift=-0.25cm] (x) {$X$}
		node[right of=m, xshift=0.75cm] (y) {$Y$}
		 
        (z) edge[blue, dashed] (a)
        (z) edge[red, <->, bend left, dashed] (a)
        (z) edge[blue, bend right=20] (m)
        
        (a) edge[blue] (m)
        (m) edge[blue] (y)

        (a) edge[red, <->, bend left=20] (y)
        (x) edge[blue, bend right=25] (z)
        (x) edge[blue, bend left=0] (a)
        (x) edge[blue, bend left=0] (m)
        (x) edge[blue, bend left=20] (y)

        node[below right of=a, xshift=-0.25cm, yshift=-0.2cm] (t1) {(a)}
        ;
	\end{scope}
    \begin{scope}[xshift=11cm, yshift=0cm]
		\path[->, thick]

        node[] (a) {$A$}
		node[left of=a, xshift=-0.75cm] (z) {$Z$}
		node[right of=a, xshift=0.75cm] (m) {$M$}
		node[above right of=m, yshift=-0.25cm] (u) {}
		node[above left of=m, yshift=-0.25cm] (x) {$X$}
		node[right of=m, xshift=0.75cm] (y) {$Y$}
		 
        (z) edge[blue, dashed] (a)
        (z) edge[red, <->, bend left, dashed] (a)
        (z) edge[red, <->, bend right=20] (m)
        
        (a) edge[blue] (m)
        (m) edge[blue] (y)

        (a) edge[red, <->, bend left=20] (y)
        (x) edge[blue, bend right=25] (z)
        (x) edge[blue, bend left=0] (a)
        (x) edge[blue, bend left=0] (m)
        (x) edge[blue, bend left=20] (y)

        node[below right of=a, xshift=-0.25cm, yshift=-0.2cm] (t2) {(b)}
        ;
	\end{scope}
	\end{tikzpicture}
	}
	\caption{Two variations of the front-door graph incorporating an anchor variable $Z$.  At least one of the dashed edges between $Z$ and $A$ must be present to satisfy the \textit{relevance} criterion.  This means $Z$ may influence $A$ either directly ($Z \blue{\rightarrow} A$), through unmeasured confounding ($Z \red{\leftrightarrow} A$), or both. Additionally, (a) $Z$ may directly affect the mediator $M$ ($Z \blue{\rightarrow} M$); or (b) $Z$ may be confounded with $M$ via unmeasured factors ($Z \red{\leftrightarrow} M$).} 
	\label{fig:front-door_anchor}
	\end{center}
\end{figure}

\begin{figure}[t] 
	\begin{center}
    \scalebox{0.65}{
    \begin{tikzpicture}[>=stealth, node distance=1.7cm]
        \tikzstyle{format} = [thick, circle, minimum size=1.0mm, inner sep=2pt]
        \tikzstyle{square} = [draw, thick, minimum size=4.5mm, inner sep=2pt]

    \begin{scope}[xshift=0cm, yshift=0cm]
		\path[->, thick]

        node[] (a) {$A$}
		node[left of=a, xshift=-0.75cm] (z) {$Z$}
		node[right of=a, xshift=0.75cm] (m) {$M$}
		node[above right of=m, yshift=-0.25cm] (u) {}
		node[above left of=m, yshift=-0.25cm] (x) {$X$}
		node[right of=m, xshift=0.75cm] (y) {$Y$}
		 
        (z) edge[blue] (a)
        (z) edge[blue, bend right=20] (m)
        
        (a) edge[blue] (m)
        (m) edge[blue] (y)

        (a) edge[red, <->, bend left=20] (y)
        (x) edge[blue, bend right=0] (z)
        (x) edge[blue, bend left=0] (a)
        (x) edge[blue, bend left=0] (m)
        (x) edge[blue, bend left=20] (y)

        node[below right of=a, xshift=-0.25cm, yshift=-0.2cm] (t1) {(a)}
        ;
	\end{scope}
    \begin{scope}[xshift=11cm, yshift=0cm]
		\path[->, thick]
		
		node[] (a) {$A$}
		node[square, left of=a, xshift=-0.75cm] (z) {$Z$}
		node[square, right of=a, xshift=0.75cm] (m) {$M$}
		node[above right of=m, yshift=-0.25cm] (u) {}
		node[square, above left of=m, yshift=-0.25cm] (x) {$X$}
		node[right of=m, xshift=0.75cm] (y) {$Y$}
		 
        (z) edge[blue] (a)
        
        (m) edge[blue] (y)

        (a) edge[red, <->, bend right=25] (y)
        (x) edge[blue, bend left=0] (a)
        (x) edge[blue, bend left=0] (y)

        node[below right of=a, xshift=-0.25cm, yshift=-0.2cm] (t2) {(b)}
        ;
	\end{scope}
	\end{tikzpicture}
	}
	\caption{(a) An example of an anchor-included front-door graph; (b) The conditional graph corresponding to the kernel $q_{AY}(A, Y \mid X, Z, M)$, where $\{X, Z, M\}$ are fixed by intervention (indicated by square nodes). } 
	\label{fig:graphs_truncation}
	\end{center}
\end{figure}

According to the nested Markov factorization \citep{richardson2023nested}, the observed data distribution $\P(O)$ for the graph in Fig.~\ref{fig:graphs_truncation}(a) factorizes as follows:

\vspace{-1.25cm}
\begin{align*}
    \P(X, Z, A, M, Y) = \q_{X}(X) \times \q_{Z}(Z \mid X) \times \q_{M}(M \mid X, Z, A) \times \q_{AY}(A, Y \mid X, Z, M) \ , 
\end{align*}
where $q_{D}(D \mid \pa_{\mathcal{G}}(D))$ denotes a Markov kernel. Here, $D$ is a \textit{district} which is a set of variables connected by bidirected edges, and the kernel corresponds to a post-intervention distribution in which all variables in $O \! \setminus \! D$ are fixed by intervention. Each kernel is identifiable from $\P(O)$ via sequential application of the g-formula. In this example, we have $\q_{X}(X) \equiv \P(X)$, $\q_{Z}(Z \mid X) \equiv \P(Z \mid X)$, $\q_{M}(M \mid X, Z, A) \equiv \P(M \mid X, Z, A)$, and $\q_{AY}(A, Y \mid X, Z, M) \equiv \P(A \mid X, Z) \times \P(Y \mid X, Z, A, M)$. 

The kernel $q_{AY}(A, Y \mid X, Z, M)$ corresponds to the conditional graph in Fig.~\ref{fig:graphs_truncation}(b), where the variables $X$, $Z$, and $M$ are treated as fixed (i.e., all incoming edges into these nodes are removed), as indicated by the square boxes around them. In this conditional graph, $Y$ is d-separated from $Z$ given $\{X, M\}$, implying the independence $Y \perp Z \mid X, M$. This independence is encoded in the marginal kernel $q_{AY}(Y \mid X, Z, M)$, which therefore must not depend on $Z$. Applying the rules of marginalization to $q_{AY}(A, Y \mid X, Z, M)$ yields the following constraint:

\vspace{-1.25cm}
\begin{align}
   \q_{AY}(Y \! \mid \! X, Z, M) \coloneqq \sum_{a'} \P(A=a' \! \mid \! X, Z) \sp \P(Y \! \mid \! X, Z, A=a', M) \ \text{is not a function of} \ Z \ . 
   \label{eq:verma}
\end{align}

\vspace{-0.3cm} \noindent 
This restriction is an example of a \textit{generalized} independence, also known as a \textit{Verma}, constraint. 

Per the results of \cite{tian02general}, the post-intervention distributions $\P(X, Z, A, Y^m)$ and $\P(X, Z, M^a, Y^a)$ are both identifiable, since $M$ and $A$ satisfy the graphical condition of primal fixability \citep{bhattacharya2022semiparametric}. A variable $O_i \in O$ is said to be primal fixable if it does not have a path to any of its children that passes only through unmeasured variables. The identified forms of these distributions are as follows:
\begin{equation}\label{eq:mse_minimizer_dual_id}
\begin{aligned}
    \P(X,Z,A,Y^m) 
    &\coloneqq \frac{\P(O)}{\q_{M}(M \! \mid \! X, Z, A)}\Big|_{M=m} 
    = \frac{\P(X,Z,A,M = m, Y)}{\P(M = m \! \mid \! A, Z, X)} \\
    &= \P(X, Z, A) \times \P(Y  \mid X, Z, A, M=m) \ , 
\end{aligned}
\end{equation}
which is Markov relative to the graph in Fig.~\ref{fig:interventions}(a), where $Z \perp Y^m \mid X$.
Similarly,
\begin{equation}\label{eq:mse_minimizer_primal_id}
\begin{aligned}
    \P(X,Z,M^a, Y^a) 
    &\coloneqq \frac{\P(O)}{\q_{AY}(A \mid X, Z, M, Y)}\Big|_{A=a}
    = \frac{\P(X,Z,A=a,M, Y)}{ \frac{\P(A=a \mid X, Z) \ \P(Y \mid X, Z, A=a, M)}{\sum_{a'} \P(A=a' \mid X, Z) \ \P(Y  \mid X, Z, A=a', M)}}  \\
    &\hspace{-0.75cm}= \P(X, Z) \times \P(M \mid X, Z, A = a) \times \sum_{a'} \P(A = a' \mid X, Z) \times \P(Y \mid X, Z, A = a', M) \ ,  
\end{aligned}
\end{equation}
which is Markov relative to the graph in Fig.~\ref{fig:interventions}(b), where $Y^a \perp Z \mid X, M^a$. 

\begin{figure}[t] 
	\begin{center}
    \scalebox{0.6}{
    \begin{tikzpicture}[>=stealth, node distance=1.5cm]
        \tikzstyle{format} = [thick, circle, minimum size=1.0mm, inner sep=2pt]
        \tikzstyle{square} = [draw, thick, minimum size=4.5mm, inner sep=2pt]

    \begin{scope}[xshift=0cm, yshift=0cm]
		\path[->, thick]
		
		node[] (a) {$A$}
		node[left of=a, xshift=-0.75cm] (z) {$Z$}
		node[square, right of=a, xshift=0.75cm] (m) {$m$}
		node[above right of=m, yshift=-0.25cm] (u) {}
		node[above left of=m, yshift=-0.25cm] (x) {$X$}
		node[right of=m, xshift=0.75cm] (y) {$Y^m$}
		 
        (z) edge[blue] (a)
        
        (m) edge[blue] (y)

        (a) edge[red, <->, bend right=20] (y)
        (x) edge[blue, bend right=0] (z)
        (x) edge[blue, bend left=0] (a)
        (x) edge[blue, bend left=0] (y)

        node[below right of=a, xshift=-0.25cm, yshift=-0.2cm] (t1) {(a)}
        ;
	\end{scope}
    \begin{scope}[xshift=9cm, yshift=0cm]
		\path[->, thick]
		
		node[square] (a) {$a$}
		node[left of=a, xshift=-0.75cm] (z) {$Z$}
		node[right of=a, xshift=0.75cm] (m) {$M^a$}
		node[above right of=m, yshift=-0.25cm] (u) {}
		node[above left of=m, yshift=-0.25cm] (x) {$X$}
		node[right of=m, xshift=0.75cm] (y) {$Y^a$}
		 
        (z) edge[blue, bend right=20] (m)
        
        (a) edge[blue] (m)
        (m) edge[blue] (y)

        (x) edge[blue, bend right=0] (z)
        (x) edge[blue, bend left=0] (m)
        (x) edge[blue, bend left=0] (y)

        node[below right of=a, xshift=-0.25cm, yshift=-0.2cm] (t2) {(b)}
        ;
	\end{scope}
    \begin{scope}[xshift=18cm, yshift=0cm]
		\path[->, thick]
		
		node[square] (a0) {$a$}
        node[left of=a0, xshift=0.5cm] (a) {$A$}
		node[left of=a, xshift=-0.25cm] (z) {$Z$}
		node[right of=a0, xshift=0.25cm] (m) {$M^a$}
		node[above right of=m, yshift=-0.25cm] (u) {}
		node[above left of=m, yshift=-0.25cm] (x) {$X$}
		node[right of=m, xshift=0.75cm] (y) {$Y^a$}
		 
        (z) edge[blue] (a)
        (a) edge[red, <->, bend right=25] (y)
        (z) edge[blue, bend right=20] (m)
        
        (a0) edge[blue] (m)
        (m) edge[blue] (y)

        (x) edge[blue, bend right=0] (z)
        (x) edge[blue, bend left=0] (m)
        (x) edge[blue, bend left=0] (y)

        node[below right of=a, xshift=-0.25cm, yshift=-0.2cm] (t3) {(c)}
        ;
	\end{scope}
	\end{tikzpicture}
	}
	\caption{(a) Fixing $M = m$ induces the independence $Z \perp Y^m \mid X$ in $\P(X, Z, A, Y^m)$; 
    (b) Fixing $A = a$ induces the independence $Z \perp Y^a \mid X, M^a$ in $\P(X, Z, M^a, Y^a)$; 
    (c) The graph corresponding to $\P(X, Z, A, M^a, Y^a)$. } 
	\label{fig:interventions}
	\end{center}
\end{figure}

The independencies between counterfactual and factual variables shown in Fig.~\ref{fig:interventions} represent two equivalent forms of the Verma constraint in \eqref{eq:verma}, which underlie our proposed testing procedures. Our weighted risk minimization tests are designed to assess these equivalent independence conditions: the dual test targets the independence in Fig.~\ref{fig:interventions}(a), while the primal test targets the one in Fig.~\ref{fig:interventions}(b).  

Given the identification of $\P(X, Z, M^a, Y^a)$ in \eqref{eq:mse_minimizer_primal_id}, we can thus write the risk minimizer in \eqref{eq:counterfactual_risk_minimizer} in terms of observed data via \eqref{eq:primal_risk_minimizer}, where $\q_\text{primal}(A \!\mid\! Y, M, Z, X)$ is simply $1/\q_{AY}(A \!\mid\! X, Z, M, Y)$.

The minimizers in the dual test are formally defined as: 

\vspace{-0.25cm}
\begin{equation}\label{eq:counterfactual_dual_risk_minimizer}
\begin{aligned}
    \mu^a_\text{dual}(m, z, x) &\coloneqq \argmin_{\tilde{\mu} \in \mathcal{M}_\mu} \int (y - \tilde{\mu}(m, z, x))^2 \sp \diff \P(Y^m=y, a, z, x) \sp ,  \\ 
    \mu^a_\text{dual}(m, x) &\coloneqq \argmin_{\tilde{\mu} \in \mathcal{M}_\mu} \int (y - \tilde{\mu}(m, x))^2 \sp \diff \P(Y^m=y, a, x) \sp . 
\end{aligned}
\end{equation}

According to the identification of $\P(X, Z, A, Y^m)$ in \eqref{eq:mse_minimizer_dual_id}, the minimizers in \eqref{eq:counterfactual_dual_risk_minimizer} can be re-expressed as weighted risk minimizers under $\P$ via: 
\begin{equation}\label{eq:dual_risk_minimizer}
\begin{aligned}
\mu^a_\text{dual}(m, z, x) &= \argmin_{\tilde{\mu} \in \mathcal{M}_\mu} \sp \E\left( \frac{1}{f_M(M \mid X, Z, A)} \sp (Y - \tilde{\mu}(M, Z, X))^2 \right) \sp , \\
\mu^a_\text{dual}(m, x) &= \argmin_{\tilde{\mu} \in \mathcal{M}_\mu} \sp \E\left(\frac{1}{f_M(M \mid X, Z, A)} \sp (Y - \tilde{\mu}(M, X))^2 \right) \sp . 
\end{aligned}
\end{equation}
A more stabilized version of the weighted risk minimizers in \eqref{eq:dual_risk_minimizer} can be obtained by replacing the inverse weight $1/f_M(M \!\mid\! X, Z, A)$ with the dual weight defined as 

\vspace{-1.25cm}
\begin{align*}
\q_\text{dual}(M \mid A, Z, X) = {f_M(M \mid a, Z, X)}/{f_M(M \mid A, Z, X)} \sp ,
\end{align*}

\vspace{-0.3cm} \noindent 
motivated by the equivalence between the distribution $\P(X, Z, A, M^a, Y^a)$ (which is Markov relative to the graph in Fig.~\ref{fig:interventions}(c)) and the following weighted version of the observed distribution: 

\vspace{-1.25cm}
\begin{align*}
    \P(X, Z, A, M^a, Y^a) 
    &\coloneqq \P(X, Z, M^a, Y^a) \times \q_{AY}(A \mid X, Z, M, Y) \\
    &= \q_\text{dual}(M \mid A, Z, X) \times \P(O) \ .  
\end{align*}

\subsection{Details on a doubly robust test}
\label{app:testing_dr}

\subsubsection{Identification proofs}\label{app:testing_id}
The parameters used in the test based on conditional counterfactual mean (CCM) are identified under three assumptions: 
\begin{itemize}
    \item[(i)] \textit{Consistency:} $Y^m = Y$ when $M=m$, for all $m$ in the state space of $M$.
    \item[(ii)] \textit{Conditional ignorability:} $Y^m\perp M \mid A,Z,X$, for all $m$ within its domain.
    \item[(iii)] \textit{Positivity:} $\p(M=m\mid A=a,Z=z,X=x)>0$ for any $(a,z,x)$ with $\p(A=a,Z=z,X=x)>0$, and $\p(A=a\mid Z=z,X=x)>0$ for any $(z,x)$ with $\p(Z=z,X=x)>0$.
\end{itemize}
Given these identification assumptions, $\mu^m(z, x) \coloneqq \E(Y^m \mid Z = z, X = x)$ is identified as:
\begin{align*}
    \mu^m(z, x)  
    &= \sum_a\E(Y^m \mid A=a, Z = z, X = x) \sp \p(A=a\mid Z=z, X=x)
    \\
    &= \sum_a\E(Y \mid M=m, A=a, Z = z, X = x) \sp \p(A=a\mid Z=z, X=x) \sp,
\end{align*}
where the first equality follows from probability rules and the second equality follows from the consistency and conditional ignorability assumptions.\par

Given the identification of $\mu^m(z,x)$, the identification of $\mu^m(z)$ immediately follows by integrating $\mu^m(z,x)$ over the observed distribution of $X$.

\subsubsection{Derivations of influence functions and one-step estimators}\label{app:testing_if}

The one-step estimators of $\mu^m(z,x)$ and $\mu^m(z)$ can be obtained via deriving the corresponding influence functions. 

\noindent \underline{\bf Influence function for $\mu^m(z,x)$} 

\vspace{-1.25cm}
\begin{align*}
&\frac{\partial}{\partial \varepsilon} \mu^{m}(z,x)\left(\P_{\varepsilon}\right) \Big|_{\varepsilon=0} \\
&=\frac{\partial}{\partial \varepsilon} \int y' \sp \diff \P_{\varepsilon}\left(y' \mid m, a', z, x\right) \sp \diff \P_{\varepsilon}(a' \mid z, x) \sp \Big|_{\varepsilon=0}
\\
&=\int \frac{\I(m'=m,z'=z,x'=x)}{\p(m\mid a',z,x)\ \p(z,x)} \big(y'-\E(Y\mid m,a',z,x)\big) \sp S(y'\mid m', a',z',x') \sp \diff\P(y', m', a', z', x') 
\\
&\hspace{0.25cm}+\int \frac{\I(z'=z,x'=x)}{\p(z,x)}\big(\E(Y \mid m,a',z,x) - \mu^{m}(z,x)\big) \ S(a'\mid z',x')  \sp \diff\P( a', z', x') \\ 
&=\int \frac{\I(m'=m,z'=z,x'=x)}{\p(m\mid a',z,x)\ \p(z,x)} \big(y'-\E(Y\mid m,a',z,x)\big) \sp S(y', m', a',z',x') \sp \diff\P(y', m', a', z', x') 
\\
&\hspace{0.25cm}+\int \frac{\I(z'=z,x'=x)}{\p(z,x)}\big(\E(Y \mid m,a',z,x) - \mu^{m}(z,x)\big) \ S(a', z',x')  \sp \diff\P( a', z', x')\sp. 
\end{align*}%
Given our notations, the np-EIF for $\mu^m(z,x)$ is given by: 

\vspace{-0.25cm}
\begin{equation}\label{eq:app:EIF_mu_mzx}
\begin{aligned}
\Phi_{m,z,x}(\Q)(O) 
&= \frac{\I(M=m, Z=z, X=x)}{f_M(m \mid A,z, x) \ \p(z, x)} (Y - \mu(m,A, z, x)) \\
&\hspace{0.5cm} + \frac{\I(Z=z, X=x)}{\p(z,x)} \big(\mu(m,A, z, x) - \mu^m(z,x)\big) \sp . 
\end{aligned}
\end{equation}

\noindent\underline{\bf Influence function for $\mu^m(z)$}

\vspace{-1.25cm}
{\small
\begin{align*}
&\frac{\partial}{\partial \varepsilon} \mu^{m}(z)\left(\P_{\varepsilon}\right) \Big|_{\varepsilon=0} \\
&=\frac{\partial}{\partial \varepsilon} \int y' \sp \diff\P_{\varepsilon}\left(y' \mid m, a', z, x'\right) \sp  \diff \P_{\varepsilon}(a' \mid z, x')\ \diff \P_{\varepsilon}(x')\Big|_{\varepsilon=0}
\\
&=\int \frac{\I(m'=m,z'=z)}{\p(m\mid a',z,x')\ \p(z\mid x')}\big(y'-\E(Y\mid m,a',z,x')\big)\ S(y'\mid m',a',z',x')\ \diff\P(y', m', a', z', x')
\\
&\hspace{0.25cm}+\int \frac{\I(z'=z)}{\p(z\mid x')}\big(\E(Y\mid m,a',z,x')- \sum_{a^*} \E(Y\mid m,a^*,z,x')\ \p(a^*\mid z,x')\big)\ S(a'\mid z',x') \sp \diff\P(a', z', x')
\\
&\hspace{0.25cm}+\int \big(\sum_{a^*} \E(Y\mid m,a^*,z,x')\ \p(a^*\mid z,x') - \mu^{m}(z)\big) \sp S(x')\ \diff\P(x') \sp .
\end{align*}}
Given our notations, the np-EIF for $\mu^m(z)$ is given by:

\vspace{-0.25cm}
\begin{equation}\label{eq:app:EIF_mu_mz}
\begin{aligned}
   \Phi_{m,z}(\Q)(O)
   &= \frac{\I(M=m,Z=z)}{f_M(m\mid A,z,X)\ f_Z(z\mid X)}\big(Y-\mu(m,A,z,X)\big) \\
   &\hspace{0.5cm}+ \frac{\I(Z=z)}{f_Z(z\mid X)}\big(\mu(m,A,z,X)- \sum_a \mu( m,a,z,X)\ \pi(a\mid z,X)\big)
   \\
   &\hspace{0.5cm}+\sum_a \mu(m,a,z,X)\ \pi(a\mid z,X) - \mu^m(z) \sp .
\end{aligned}
\end{equation}

\subsubsection{DR-CCM test for continuous $X$} 
\label{app:dr_ccm_cont}

Our proposed DR-CCM test is based on a TMLE for $\mu^{m}(z)$ in \eqref{eq:cate_test_id} that satisfies doubly robust asymptotic linearity. This ensures that both the test statistic and its confidence interval are consistently estimated if either $(\hat{\pi}, \hat{\mu})$ or $(\hat{f}_M, \hat{f}_Z)$ is correctly specified. Achieving this property requires quantifying the first-order bias of the initial one-step estimator $\hat{\mu}^{+, m}(z)$ in \eqref{eq:one-step_doubly_robuts_test} (and its TMLE counterpart), and updating the nuisance estimates to approximately solve both the score equation induced by the influence function of $\mu^{m}(z)$ and the estimating equations that render the first-order remainder bias negligible. Following the framework of \citet{van2014targeted}, \citet{benkeser2015data}, and \citet{benkeser2017doubly}, we summarize the construction procedure below and refer readers to those works for further details. 

Throughout this section, we assume that either $(\hat{\pi}, \hat{\mu})$ or $(\hat{f}_M, \hat{f}_Z)$ is correctly specified, though we do not assume knowledge of which. Define $\xi(m, z, x; \mu) = \sum_a \pi(a \mid z, x)\ \mu(m, a, z, x)$. Given the influence function $\Phi_{m,z}(\Q)$ in \eqref{eq:app:EIF_mu_mz}, the $R_2$ term for $\mu^{+,m}(z)$ is derived as follows: 

\vspace{-1.25cm}
\begin{align}
    \Rem(\hat{\Q},\Q)&= \hat{\mu}^{+,m}(z) - \mu^{m}(z) + \int \Phi_{m,z}(\hat{\Q})(o) \sp \diff \P(o) \notag \\
    &=\int \Big\{\frac{f_M(m \mid a',z, x') \ \I(z'=z)}{\hat{f}_M(m \mid a',z, x') \ \hat{f}_Z(z\mid x')} (\mu(m,a',z,x') - \hat{\mu}(m,a', z, x')) \notag \\
   &\hspace{0.5cm}+ \frac{f_Z(z\mid x')}{\hat{f}_Z(z\mid x')} \sum_{a^*} \hat{\mu}(m,a^*, z, x') ( \pi(a^*\mid z,x')- \hat{\pi}(a^*\mid z,x'))
   \notag \\
   &\hspace{0.5cm}+\sum_{a^*} \hat{\mu}(m,a^*,z,x')\ \hat{\pi}(a^*\mid z,x') - \mu^{m}(z)\Big\}\ \diff\P(a', x')
   \notag \\
   &=\int \Big\{ \frac{(f_M(m \mid a',z, x')-\hat{f}_M(m \mid a',z, x')) \ \I(z'=z)}{\hat{f}_M(m \mid a',z, x') \ \hat{f}_Z(z\mid x')} (\mu(m,a',z,x') - \hat{\mu}(m,a', z, x'))
   \notag \\
   &\hspace{0.5cm}+\frac{f_Z(z\mid x')-\hat{f}_Z(z\mid x')}{\hat{f}_Z(z\mid x')}\sum_{a^*} \hat{\mu}(m,a^*, z, x')(\pi(a^*\mid z,x') - \hat{\pi}(a^*\mid z,x'))
   \notag \\
   &\hspace{0.5cm}+\frac{f_Z(z\mid x')-\hat{f}_Z(z\mid x')}{\hat{f}_Z(z\mid x')}\sum_{a^*} \pi(a^*\mid z,x')(\mu(m,a^*,z,x') - \hat{\mu}(m,a^*, z, x'))\Big\} \ \diff\P(a', x')\sp.
   \label{appeq:drccm_r2}
\end{align}

The $R_2$ term can be decomposed as $\Rem(\hat{\Q},\Q)=\Rem^1(\hat{\Q},\Q)+\Rem^2(\hat{\Q},\Q)$ where
\begin{align*}
    \Rem^1(\hat{\Q},\Q)&=\E\left(\frac{f_Z(z\mid X)-\hat{f}_Z(z\mid X)}{\hat{f}_Z(z\mid X)}(\xi(m,z,X;\mu)-\hat{\xi}(m,z,X;\hat{\mu}))\right), \text{ and}
    \\
    \Rem^2(\hat{\Q},\Q)&=\E\left(\frac{\I(Z=z)}{\hat{f}_Z(z\mid X)}\frac{f_M(m \mid A,z, X)-\hat{f}_M(m \mid A,z, X)}{\hat{f}_M(m \mid A,z, X)}\left(\mu(m,A,z,X) - \hat{\mu}(m,A, z, X) \right) \right)\sp .
\end{align*}

To analyze their behavior under model misspecification, let $(f_{M,+},f_{Z,+})$ and $(\mu_{+},\xi_{+})$ denote the probability limits of the possibly misspecified nuisance estimates $(\hat{f}_M,\hat{f}_Z)$  and $(\hat{\mu},\hat{\xi})$, respectively. We further define the following two mapping functions:
\begin{align*}
\Phi_{1}(\tilde{\xi})=\E\left(\frac{f_{Z,+}-f_{Z}}{f_{Z,+}} \ \tilde{\xi}\right), \quad \Gamma_{0}(\tilde{f}_Z) =\E\left(\frac{\xi_{+}-\xi}{f_{Z,+}} \tilde{f}_Z\right) \sp .
\end{align*}
These mappings can be used to describe the first-order behavior of $\Rem^1(\hat{\Q},\Q)$ as:

\vspace{-1.25cm}
\begin{align*}
    \Rem^1(\hat{\Q},\Q)=\{\Gamma_{0}(\hat{f}_Z)-\Gamma_{0}(f_Z)\} + \{\Phi_{1}(\hat{\xi})-\Phi_{1}(\xi)\} + \smallO(n^{-1/2}) \sp ,
\end{align*}

\vspace{-0.3cm} \noindent 
where the $\smallO(n^{-1/2})$ term captures the second-order terms that are asymptotically negligible, given that the model for either $f_Z$ or $\xi$ is correctly specified.

A similar expansion applies to $\Rem^2(\hat{\Q}, \Q)$. To derive it, we first define:
\begin{align*}
\Phi_{2}(\tilde{\mu}) =\E\left( \frac{\I(Z=z)}{\hat{f}_Z}\ \frac{f_{M,+}-f_M}{f_{M,+}} \ \tilde{\mu} \right), \quad \Gamma_{1}(\tilde{f}_M) = \E\left(\frac{\I(Z=z)}{\hat{f}_Z} \ \frac{\mu_{+}-\mu}{f_M} \ \tilde{f}_M\right) \sp .
\end{align*}
The first-order behavior of $\Rem^2(\hat{\Q}, \Q)$ can be characterized as:

\vspace{-1.25cm}
\begin{align*}
    \Rem^2(\hat{\Q},\Q)=\{\Phi_{2}(\hat{\mu})-\Phi_{2}(\mu_{+})\}+\{\Gamma_{1}(\hat{f}_M)-\Gamma_{1}(f_M)\}+ \smallO(n^{-1/2}) \sp ,
\end{align*}

\vspace{-0.3cm} \noindent 
where $\smallO(n^{-1/2})$ captures the second-order terms that are asymptotically negligible, provided that the model for either $f_M$ or $\mu$ is correctly specified.

To achieve doubly-robust inference, the key lies in quantifying and correcting the first-order bias in the remainder term $\Rem$, defined in \eqref{appeq:drccm_r2}, using additional nuisance parameters that can be consistently estimated through nonparametric smoothing techniques at desired convergence rates. The general strategy for addressing model misspecification involves four main steps. We illustrate this below using the case where $(\mu_{+}, \xi_{+}) = (\mu, \xi)$ and $(f_{M,+}, f_{Z,+}) \neq (f_M, f_Z)$, in which case both $\Gamma_{0}(\hat{f}_Z) - \Gamma_{0}(f_Z)$ and $\Gamma_{1}(\hat{f}_M) - \Gamma_{1}(f_M)$ are zero:
\begin{enumerate}
    \item Characterize the first-order behavior of each remainder term, $\Rem^1=\Phi_{1}(\hat{\xi})-\Phi_{1}(\xi)+\smallO(n^{-1/2})$, $\Rem^2=\Phi_{2}(\hat{\mu})-\Phi_{2}(\mu)+\smallO(n^{-1/2})$.
    \item Approximate the first-order behavior of $\Phi_1$ and $\Phi_2$ by constructing mappings $\Phi_{1,n}$ and $\Phi_{2,n}$ that are estimable from the observed data. Specifically, we aim to represent $\Phi_1(\hat{\xi}) - \Phi_1(\xi) = \Phi_{1,n}(\hat{\xi}) - \Phi_{1,n}(\xi) + \smallO(n^{-1/2})$, with an analogous expression holding for $\Phi_2$.
    \item Perform linear expansions of $\Phi_{1,n}$ and $\Phi_{2,n}$ around $\xi$ and $\mu$, respectively. Taking $\Phi_{1,n}$ as an example, we have: 

    \vspace{-1.5cm}
    \begin{align*}
        \Phi_{1,n}(\hat{\xi})-\Phi_{1,n}(\xi)=\P_n D_{\xi}(\P) - \P_n D_{\xi,n}(\hat{\P}) + \smallO(n^{-1/2}) \sp ,
    \end{align*}

    \vspace{-0.5cm} \noindent 
    where $D_{\xi}(\P)$ denotes the canonical gradient of $\Phi_{1,n}$ evaluated at the true distribution $\P$, expressed in terms of several nuisance parameters to be defined later. The term $\smallO(n^{-1/2})$ captures the empirical process term and second-order remainder term that are negligible. The empirical quantity $\P_n D_{\xi,n}(\hat{\P})$ represents the first-order bias of the $\Rem$ term, which we seek to correct.
    \item Construct an estimator $\hat{\mu}^{\star,m}(z)$ that account for the first-order bias of each remainder term, such as $\P_n D_{\xi,n}(\hat{\P})$, thus establishing the asymptotic linearity.
\end{enumerate}

Below, we illustrate Steps 1–4 using the case where $(\mu_{+},\xi_{+})\neq(\mu,\xi)$ and $(f_{M,+},f_{Z,+}) = (f_M,f_Z)$, focusing on the expansion of $\Rem^1$. The approach for the complementary case, where $(\mu_{+},\xi_{+}) = (\mu,\xi)$ and $(f_{M,+},f_{Z,+}) \neq (f_M,f_Z)$, as well as for $\Rem^2$, follows analogously. A more detailed discussion of each scenario can be found in \citet{benkeser2015data}.

Under the discussed case, both $\Phi_{1}(\hat{\xi})-\Phi_{1}(\xi)$ and $\Phi_{2}(\hat{\mu})-\Phi_{2}(\mu)$ are zero, and we have

\vspace{-1.25cm}
\begin{align*}
    \Gamma_0(\hat{f}_Z)-\Gamma_0(f_Z)&=\E\Big(\frac{\xi_{+}-\xi}{\xi}(\hat{f}_Z-f_Z)\Big)
    \\
    &=-\E\Big(\frac{\I(M=m,Z=z)}{f_M\ f_Z}\ \frac{Y-\xi_{+}}{f_Z} \ (\hat{f}_Z-f_Z)\Big)
    \\
    &=-\E\Big(\frac{\xi^r}{f^2_Z}(\hat{f}_Z-f_Z)\Big) \sp ,
\end{align*}

\vspace{-0.3cm} \noindent 
where $\xi^r$ is defined as:

\vspace{-1.5cm}
\begin{align*}
    \xi^r(X)=\E\Big(\frac{\I(M=m,Z=z)}{f_M}\big(Y-\xi_{+}\big)\mid \hat{f}_Z,{f}_Z\Big) \sp .
\end{align*}
 We note, however, that we cannot directly estimate $\xi^r(X)$ in practice because
it involves unknown quantities. Thus, we proceed by approximating the first-order behavior of this quantity using a mapping that can be computed based only on the data. We proceed as
\begin{align*}
    -\E\Big(\frac{\xi^r}{f^2_Z}(\hat{f}_Z-f_Z)\Big)&=-\E\Big(\frac{\hat{\xi}^r}{\hat{f}^2_Z}(\hat{f}_Z-f_Z)\Big)+\smallO(n^{-1/2}) \sp ,
\end{align*}
where $\hat{\xi}^r$ denotes an estimate of $\xi^r$, obtained by substituting the true nuisance parameters in its definition with their estimated counterparts and using nonparametric methods to estimate the conditional expectation through a univariate regression. These methods are assumed to yield consistent estimates at a sufficiently fast convergence rate. The term $\smallO(n^{-1/2})$ captures second-order contributions that are negligible given that $(\mu_{+},\xi_{+}) \neq (\mu,\xi)$ and $(f_{M,+},f_{Z,+}) = (f_M,f_Z)$.

We conclude Step 2 with the following approximation, where $\Gamma_{0,n}(\tilde{f}_Z)=\E\big(\{\hat{\xi}^r/\hat{f}^2_Z \}\tilde{f}_Z\big)$: 

\vspace{-1.25cm}
\begin{align*} 
    &\Gamma_0(\hat{f}_Z) -\Gamma_0(f_Z) \\
    &\hspace{0.5cm}=-\I(\mu_{+}\neq \mu, \xi_{+}\neq\xi, f_{M,+}=f_M,f_{Z,+}=f_Z)\big(\Gamma_{0,n}(\hat{f}_Z)-\Gamma_{0,n}(f_Z)+\smallO(n^{-1/2})\big),
\end{align*}

\vspace{-0.3cm} 
We then proceed to Step 3, where we perform a first-order expansion of $\Gamma_{0,n}$ around $f_Z$:

\vspace{-1.25cm}
\begin{align*}
    \Gamma_{0,n}(\hat{f}_Z)-\Gamma_{0,n}(f_Z)&=(\P_n-\P)D_{Z}(\xi^r,f_Z) + \P_n\ D_{Z}(\hat{\xi}^r,\hat{f}_Z)+\smallO(n^{-1/2}) \sp ,
\end{align*}

\vspace{-0.3cm} \noindent 
where $D_{Z}(\xi^r,f_Z)$ is the canonical gradient of $\Gamma_{0,n}$, defined as $D_{Z}(\xi^r,f_Z)=\frac{{\xi}^r}{{f}^2_Z}(\I(Z=z)-f_Z),$ and $\smallO(n^{-1/2})$ involves the empirical process term that is negligible.

This derivation isolates the first-order bias term $\P_n\, D_{Z}(\hat{\xi}^r,\hat{f}_Z)$, which forms the basis for constructing the TMLE of interest. Recall that TMLE is a general framework for refining initial estimates of nuisance parameters through a targeting procedure. These updated nuisance estimates are designed to solve user-defined estimating equations, most commonly, the one associated with the efficient influence function of the estimand. To achieve doubly robust asymptotic linearity, the TMLE procedure is extended to solve additional estimating equations that correct for the first-order bias in the remainder term $\Rem$, such as ensuring that $\P_n\, D_{Z}(\hat{\xi}^r,\hat{f}_Z)$ is negligible.

\subsubsection{CCM test for discrete $X$} 

When $X$ is discrete, we test pointwise invariance of $\mu^m(z, x)$ in $z$ by defining $\Delta(m, x) \coloneqq \mu^m(1, x) - \mu^m(0, x)$, where (see Appendix~\ref{app:testing_id})

\vspace{-1.25cm} 
\begin{align}
\mu^m(z,x) = \sum_{a} \mu(m, a, z, x) \sp \pi(a \mid z, x) \sp . 
\label{eq:app_mu_m_zx_id}
\end{align}

\vspace{-0.3cm} 
Let $\Delta$ denote the full collection of contrasts $\Delta(m,x)$ across all observed $(m,x)$ pairs.  Let $\Sigma$ denote the variance-covariance matrix of $\Delta$. Given estimates of $\mu^m(z,x)$, we obtain plug-in estimates $\Delta_n$ and $\Sigma_n$, and define the test statistic as $T_{\mathrm{n}, \text{CCM}} \coloneqq \Delta_n^\top \Sigma_n^{-1} \Delta_n.$ Under the null, $T_{\mathrm{n}, \text{CCM}}$ asymptotically follows a chi-squared distribution with $d$ degrees of freedom, where $d$ is the dimension of $\Delta$ (e.g., $d = 4$ when both $M$ and $X$ are binary). Alternatively, one may perform univariate Wald tests for each $(m, x)$, adjusting for multiple comparisons via Bonferroni or Benjamini–Hochberg. 

We next describe how to estimate $\mu^m(z, x)$, to ensure \textit{doubly robust inference} for both (i) the \textit{test statistic} and (ii) its \textit{confidence intervals.}  

Given that $X, Z, M$ are discrete, we can achieve a root-$n$ consistent estimator for $\mu^m(z,x)$ based on a simple plug-in estimate of \eqref{eq:app_mu_m_zx_id}. Alternatively, we can use a one-step estimator, which takes the following form, with corresponding EIF derived in Appendix~\ref{app:testing_if}:

\vspace{-1.25cm}
{\small 
\begin{align}
\hat{\mu}^{+,m}(z,x) &= \frac{1}{n} \sum_{i=1}^n \frac{\I(Z_i = z, X_i = x)}{\hat{\p}(z, x)} \bigg\{ \frac{\I(M_i = m)}{\hat{f}_M(m \mid A_i, z, x)} (Y_i - \hat{\mu}(m, A_i, z, x)) +  \hat{\mu}(m, A_i, z, x) \bigg\} \sp . 
\label{eq:one-step_DR_test_dicreteX}
\end{align}
} 

\vspace{-0.5cm} 
To construct a doubly robust confidence interval for $\hat{\mu}^{+,m}(z,x)$, we follow the approach in Appendix~\ref{app:dr_ccm_cont}, which requires the $R_2$ remainder term. This is derived below, using the EIF $\Phi_{m,z,x}(\Q)$ given in \eqref{eq:app:EIF_mu_mzx}.

\vspace{-1.25cm}
{\small
\begin{align*}
\Rem(\hat{\Q},\Q)&\coloneqq  \hat{\mu}^{+,m}(z,x) - \mu^{m}(z,x) + \int \Phi_{m,z,x}(\hat{\Q})(o') \sp \diff \P(o')\\
&=\int \Big\{ \frac{\I(z'=z,x'=x)}{\hat{\p}(z, x)} \ \frac{f_M(m \mid a',z, x)}{\hat{f}_M(m \mid a',z, x)} (\mu(m,a',z,x) - \hat{\mu}(m,a', z, x))\Big\} \ \diff\P(z', x', a') \\
&\hspace{0.5cm}+ \frac{\p(z, x)}{\hat{\p}(z,x)} \sum_{a^*} \hat{\mu}(m,a^*, z, x) ( \pi(a^*\mid z,x)- \hat{\pi}(a^*\mid z,x))+\hat{\mu}^{+,m}(z,x) - \mu^{m}(z,x)
\\
&= \int  \frac{\I(z'=z,x'=x)}{\hat{\p}(z,x)}  \Bigg\{ \sum_{a^*} \Big\{ \frac{\pi(a^* \mid z, x)}{\hat{f}_M(m \mid a^*,z, x)} \big(f_M(m \mid a^*,z, x)-\hat{f}_M(m \mid a^*,z, x) \big)
\\
&\hspace{6.5cm}\times \big(\mu(m,a^*,z,x) - \hat{\mu}(m,a^*, z, x) \big)\Big\} \Bigg\}  \ \diff\P(z', x') 
\\
&\hspace{0.5cm}+\frac{\p(z, x)-\hat{\p}(z,x)}{\hat{\p}(z,x)}\sum_{a^*}  \Big\{ \hat{\mu}(m,a^*, z, x) \big(\pi(a^*\mid z,x) - \hat{\pi}(a^*\mid z,x) \big)
\\
&\hspace{4cm}+  \pi(a^*\mid z,x) \big(\mu(m,a^*,z,x) - \hat{\mu}(m,a^*, z, x) \big) \Big\} \sp .
\end{align*}
}

\subsection{Details on efficiency gains under the Verma constraint}
\label{app:testing_eff_gain}

\subsubsection{Identification proofs} 

\underline{Binary $Z$.} \ 
To illustrate this, we first rewrite the ATE front-door functional in \eqref{eq:id_ATE} to incorporate the anchor variable $Z$:

\vspace{-1.cm}
\begin{equation}\label{eq:id_ATE_z}
    \begin{aligned}
        \psi(\Q)&=\iiint \sum_{a=0}^1 \mu(m, a, z, x) \ \pi(a \mid z, x) \ f_M(m \mid A=a_0, z, x) \ \p(z,x) \ \diff m \sp \diff z \sp \diff x \sp.
    \end{aligned}
\end{equation}
According to the Verma constraint in \eqref{eq:verma}, the term $\sum_{a=0}^1 \mu(m, a, z, x) \sp \pi(a  \! \mid \! z, x)$ is invariant to the value $z$. Leveraging this invariance, we can fix $z$ in the outcome and treatment models to a pre-specified level $z^*\in\mathcal{Z}$, which results in:

\vspace{-1.cm}
\begin{equation}\label{eq:id_ATE_z*}
    \begin{aligned}
        \psi_{z^*}(\Q)&=\iiint \sum_{a=0}^1 \mu(m, a, z^*, x) \ \pi(a \mid z^*, x) \ f_M(m \mid A=a_0, z, x) \ \p(z,x) \  \diff m \sp \diff z \sp \diff x.
    \end{aligned}
\end{equation}

\noindent \underline{Continuous $Z$.} \ 
When $Z$ is continuous, we can construct a pathwise differentiable functional by leveraging the Verma constraint. Specifically, we equate $\sum_{a=0}^1 \mu(m, a, z^*, x) \ \pi(a \mid z^*, x)$ with the integral $\int \sum_{a=0}^1 \mu(m, a, z, x) \ p(a \! \mid \! z, x)\sp \tilde{\p}(z)\sp \diff z$, where $\tilde{\p}(Z)$ is a user-specified valid reference distribution for $Z$ that need not match the true $\p(Z)$. This yields the following:
\begin{equation}\label{eq:id_ATE_tildepz}
    \begin{aligned}
        \psi_{\tilde{\p}}(\Q)&=\iiint \Big\{\int \sum_{a=0}^1 \mu(m, a, z, x) \ p(a \mid z, x)\ \tilde{\p}(z)\ dz\Big\} \ \p(m \mid A=a_0, z, x) \ \p(z,x) \ \diff m \ \diff z \ \diff x \sp.
    \end{aligned}
\end{equation}
We denote this functional by $\psi_{\tilde{\p}}$ to emphasize that the outcome regression and propensity score are integrated over the reference distribution $\tilde{\p}(Z)$.

\subsubsection{Nonparametric EIF derivation} 

\underline{Binary $Z$.} \ 
The np-EIF for $\psi_{z^*}(\Q)$ in \eqref{eq:id_ATE_z*} is derived as follows: 

\begin{align*}
&\frac{\partial}{\partial \varepsilon} \psi_{z^*}(\P_{\varepsilon})\Big|_{\varepsilon=0} 
\\
&=\frac{\partial}{\partial \varepsilon} \int y \ \p_{\varepsilon}(y \mid m, a, z^*,x) \ \p_{\varepsilon}\left(m \mid a_0, z,x\right) \ \p_{\varepsilon}(a \mid z^*,x) \ \p_{\varepsilon}(z\mid x)\ \p_{\varepsilon}(x) \ \diff y \ \diff  m \ \diff a \ \diff z\ \diff x\Big|_{\varepsilon=0}
\\
&=\int \I(z=z^*)\frac{\sum_{z'} \p(m\mid a_0,z',x)\ \p(z'\mid x)}{\p(m\mid a,z^*,x)\ \p(z^*\mid x)} [y-\E(Y\mid m,a,z^*,x)] \ S(y\mid m,a,z,x) \ \diff\P(o)
\\
&\hspace{0.25cm}+\int \frac{\I(a=a_0)}{\p(a\mid z,x)}\big[\xi_{z^*}(m,x) - \gamma_{z^*}(z,x)\big] \ S(m\mid a,z,x) \ \diff\P(o)
\\
&\hspace{0.25cm}+\int \frac{\I(z=z^*)}{\p(z^*\mid x)} (a-\pi(1\mid z^*,x))\sum_{z'} [\kappa_{1,z^*}(z',x) - \kappa_{0,z^*}(z',x)]\p(z'\mid x) \ S(z,x) \ \diff\P(o)
\\
&\hspace{0.25cm}+\int \big[\gamma_{z^*}(z,x) - \psi_{z^*}(\Q)\big] \ S(z,x) \ \diff\P(o)
\\
&=\int \I(z=z^*)\frac{\sum_{z'} \p(m\mid a_0,z',x)\ \p(z'\mid x)}{\p(m\mid a,z^*,x)\ \p(z^*\mid x)} [y-\E(Y\mid m,a,z^*,x)] \ S(o) \ \diff\P(o)
\\
&\hspace{0.25cm}+\int \frac{\I(a=a_0)}{\p(a\mid z,x)}\big[\xi_{z^*}(m,x) - \gamma_{z^*}(z,x)\big] \ S(o) \ \diff\P(o)
\\
&\hspace{0.25cm}+\int \frac{\I(z=z^*)}{\p(z^*\mid x)}(a-\pi(1\mid z^*,x))\sum_{z'} [\kappa_{1,z^*}(z',x) - \kappa_{0,z^*}(z',x)]\p(z'\mid x) \ S(z,x) \ \diff\P(o)
\\
&\hspace{0.25cm}+\int \big[\gamma_{z^*}(z,x) - \psi_{z^*}(\Q)\big] \ S(z,x) \ \diff\P(o) \sp.
\end{align*}

\noindent Therefore, the np-EIF for $\psi_{z^*}(\Q)$ is:
\begin{align*}
\Phi_{z^*}(\Q)(O_i) 
   &=  \frac{\I(Z_i=z^*)}{f_Z(z^*\mid X_i) }\sum_z f^r_{M,z^*}(M_i, A_i, z, X_i) \sp f_Z(z\mid X_i) \sp \big(Y_i-\mu(M_i,A_i,z^*,X_i) \big)
   \\
   &\hspace{0.5cm}+\frac{\I(Z=z^*)}{f_Z(z^*\mid X_i)}(A_i-\pi(a\mid z^*,X_i))\ \sum_z\big(\kappa_{1,z^*}(z,X_i)-\kappa_{0,z^*}(z,X_i)\big)f_Z(z\mid X_i) 
   \notag \\ 
    &\hspace{0.5cm}+ \frac{\I(A_i=a_0)}{\pi(a_0\mid Z_i,X_i)}\big(\xi_{z^*}(M_i,X_i) - \gamma_{z^*}(Z_i,X_i)\big) 
   +\gamma_{z^*}(Z_i,X_i) - \psi_{z^*}(\Q) \sp .
\end{align*}

\noindent \underline{Continuous $Z$.} \ 
The np-EIF for $\psi_{\tilde{\p}}(\Q)$ in \eqref{eq:id_ATE_tildepz} is derived as follows: 
\begin{align*}
&\frac{\partial}{\partial \varepsilon} \psi_{\tilde{\p}}(\P_{\varepsilon})\Big|_{\varepsilon=0}
\\
&=\int \frac{\partial}{\partial \varepsilon} \psi_{z^*}(\P_{\varepsilon})\Big|_{\varepsilon=0} \ \tilde{\p}(z^*)\ \diff z^*
\\
&=\frac{\partial}{\partial \varepsilon} \int y \ \big[\int \sum_a \ \p_{\varepsilon}(y \mid m, a, z^*,x) \ \p_{\varepsilon}(a \mid z^*,x)\ \tilde{\p}(z^*) \ \diff z^*\big]
\\
&\hspace{5cm}\times\p_{\varepsilon}(m \mid a_0, z,x) \ \p_{\varepsilon}(z\mid x) \p_{\varepsilon}(x) \diff y \ \diff  m \ \diff z\ \diff x\Big|_{\varepsilon=0}
\\
&=\int \tilde{\p}(z^*)\frac{\int \p(m\mid a_0,z,x)\ \p(z\mid x)\ \diff z}{\p(m\mid a,z^*,x)\ \p(z^*\mid x)} [y-\E(Y\mid m,a,z^*,x)] \ S(y,m,a,z^*,x) \ \P(y,m,a,z^*,x)
\\
&\hspace{0.25cm}+\int \frac{\I(a=a_0)}{\p(a\mid z,x)}\big[\xi_{\tilde{\p}}(m,x) - \gamma_{\tilde{\p}}(z,x)\big] \ S(m, a,z,x) \ \diff\P(m,a,z,x)
\\
&\hspace{0.25cm}+\int \frac{\tilde{\p}(z^*)}{\p(z^*\mid x)}(a-\p(a=1\mid z^*,x))[\int (\kappa_{1}(z,x)
\\
&\hspace{6cm}-\kappa_{0}(z,x))\ \p(z\mid x)\ \diff z] S(a, z^*,x) \ \diff\P(a,z^*,x)
\\
&\hspace{0.25cm}+\int \big[\gamma_{\tilde{\p}}(z,x) - \psi_{\tilde{\p}}(\Q)\big] \ S(z,x) \ \diff\P(z,x) \sp, 
\end{align*}
where $\xi_{\tilde{\p}}(m,x)=\int \sum_a \ \p(y \mid m, a, z^*,x) \ \p(a \mid z^*,x)\ \tilde{\p}(z^*) \ dz^*$, and $\gamma_{\tilde{\p}}(z,x)=\int \xi_{\tilde{\p}}(m,x) \ p(m\mid a_0,z,x) \ \diff m = \E(\xi_{\tilde{\p}}(M,X)\mid a_0,z,x)$.
\newline
Let $\Q = \{\mu,\gamma_{\tilde{\p}},\kappa_{a}, \pi,\xi_{\tilde{\p}}, f_Z,\p_{ZX}\}$. The np-EIF for $\psi_{\tilde{\p}}(\Q)$ is:
\begin{align*}
    \Phi_{\tilde{\p}}(\Q)(O) 
   &=  \tilde{\p}(Z)\frac{\int f_M(M\mid a_0,z,X)\ f_Z(z\mid X)\ \diff z}{f_M(M\mid a,Z,X)\ f_Z(Z\mid X)} [Y-\mu(M,A,Z,X)] \\
   &\hspace{0.5cm}+ \frac{\I(A=a_0)}{\pi(A\mid Z,X)}\big[\xi_{\tilde{\p}}(M,X) - \gamma_{\tilde{\p}}(Z,X)\big]
   \\
   &\hspace{0.5cm}+\frac{\tilde{\p}(Z)}{f_Z(Z\mid X)}(A-\pi(A=1\mid Z,X))\int [\kappa_1(z,X)-\kappa_0(z,X)]\ f_Z(z\mid X) \ \diff z
   \\
   &\hspace{0.5cm}+\gamma_{\tilde{\p}}(Z,X) - \psi_{\tilde{\p}}(\P)) \sp .
\end{align*}

\noindent \underline{Estimators of $\psi_{\tilde{\p}}(\Q)$ under univariate continuous $Z$.} \ 
Constructing IF-based estimators now requires estimating the conditional densities $f_M$ and $f_Z$. Given nuisance estimates $\hat{\Q}$, the one-step estimator is given by: 
\begin{align*}
    \psi^{+}_{\tilde{\p}}(\hat{\Q}) &=\frac{1}{n}\sum_{i=1}^{n}\Big[\tilde{\p}(Z_i)\frac{\int \hat{f}_M(M_i\mid a_0,z,X_i)\ \hat{f}_Z(z\mid X_i)\ \diff z}{\hat{f}_M(M_i\mid a,Z_i,X_i)\ \hat{f}_Z(Z_i\mid X_i)} [Y-\hat{\mu}(M_i,A_i,Z_i,X_i)]
    \\
    &\hspace{0.25cm}+\frac{\I(A_i=a_0)}{\hat{\pi}(A_i\mid Z_i,X_i)}\big[\hat{\xi}_{\tilde{\p}}(M_i,X_i;) - \hat{\gamma}_{\tilde{\p}}(Z_i,X_i)\big]
    \\
    &\hspace{0.25cm}+\frac{\tilde{\p}(Z_i)}{\hat{f}_Z(Z_i\mid X_i)}(A_i-\hat{\pi}(1\mid Z_i,X_i))\int [\hat{\kappa}_1(z,X_i)-\hat{\kappa}_0(z,X_i)]\ \hat{f}_Z(z\mid X_i) \ \diff z
    \\
    &\hspace{0.25cm}+\hat{\gamma}_{\tilde{\p}}(Z_i,X_i)
    \Big] \sp ,
\end{align*}
where $\hat{\xi}_{\tilde{\p}}(m,x)$ and $\hat{\gamma}_{\tilde{\p}}(z,x)$ are nuisance estimates obtained via numerical integration with respect to the corresponding estimated conditional densities.

The Verma constraint enables the construction of a family of one-step estimators indexed by $\tilde{\p}(Z)$. The choice of $\tilde{\p}(Z)$ impacts the efficiency of the corresponding estimator. When $Z$ is continuous, there are infinitely many valid choices of $\tilde{\p}(Z)$ that respect the support of $Z$. As a result, identifying the optimal $\tilde{\p}(Z)$, the one that minimizes asymptotic variance, becomes a more complex task. In such settings, a closed-form expression for the optimal $\tilde{\p}(Z)$ is no longer available. Instead, we recommend that practitioners explore multiple choices of $\tilde{\p}(Z)$, construct the corresponding one-step estimators, and compare their estimated variances to guide selection.

\subsubsection{Semiparametric gains: The optimal choice of $\alpha$}

The optimal weight $\alpha^{\mathrm{opt}}$ aims to minimize the variance of $\psi^+_{\alpha}(\Q)$, quantified by the IF as
\begin{align*}
    &\E\big(\big\{\alpha \Phi_{z^*=1}(\Q)+(1-\alpha) \Phi_{z^*=0}(\Q)\big\}^2\big)
    \\
    &=\alpha^2\ \E(\Phi^2_{z^*=1}(\Q)) + (1-\alpha)^2\ \E(\Phi^2_{z^*=0}(\Q)) + 2\alpha(1-\alpha)\E(\Phi_{z^*=1}(\Q)\Phi_{z^*=0}(\Q)) \sp.
\end{align*}

An optimizer is derived by differentiating the variance function with respect to $\alpha$ and setting the derivative to zero:
\begin{align*}
    &\partial\E\big(\big\{\alpha \Phi_{z^*=1}(\Q)+(1-\alpha) \Phi_{z^*=0}(\Q)\big\}^2\big)/\partial\alpha
    \\
    &=2\alpha \E(\Phi^2_{z^*=1}(\Q)) - 2(1-\alpha)\ \E(\Phi^2_{z^*=0}(\Q)) + 2(1-2\alpha)\E(\Phi_{z^*=1}(\Q)\Phi_{z^*=0}(\Q))=0
    \\
    &\Longrightarrow \alpha_{\mathrm{opt}}={\E(\Phi_{z^*=0}(\Q)(\Phi_{z^*=0}(\Q)-\Phi_{z^*=1}(\Q)))}/\E((\Phi_{z^*=1}(\Q)-\Phi_{z^*=0}(\Q))^2) \sp.
\end{align*}

To prove $\alpha_{\mathrm{opt}}$ minimizes the variance of $\psi^+{\alpha}(\Q)$, we take the second derivative and show that it is greater than 0:
\begin{align*}
    &\partial^2\E\big(\big\{\alpha \Phi_{z^*=1}(\Q)+(1-\alpha) \Phi_{z^*=0}(\Q)\big\}^2\big)/\partial\alpha^2
    \\
    &=2\E(\Phi^2_{z^*=1}(\Q))  +2\ \E(\Phi^2_{z^*=0}(\Q)) -4\E(\Phi_{z^*=1}(\Q)\Phi_{z^*=0}(\Q))
    \\
    &=2\E((\Phi_{z^*=1}(\Q)-\Phi^2_{z^*=0}(\Q))^2)\geq 0 \sp .
\end{align*}

\pagebreak
\section{Details on simulation studies}
\label{app:sims}

\subsection{Simulation 1: Theoretical properties}\label{app:sims:consistency} 

\begin{figure}[H]
    \centering
    \includegraphics[width=1\textwidth]{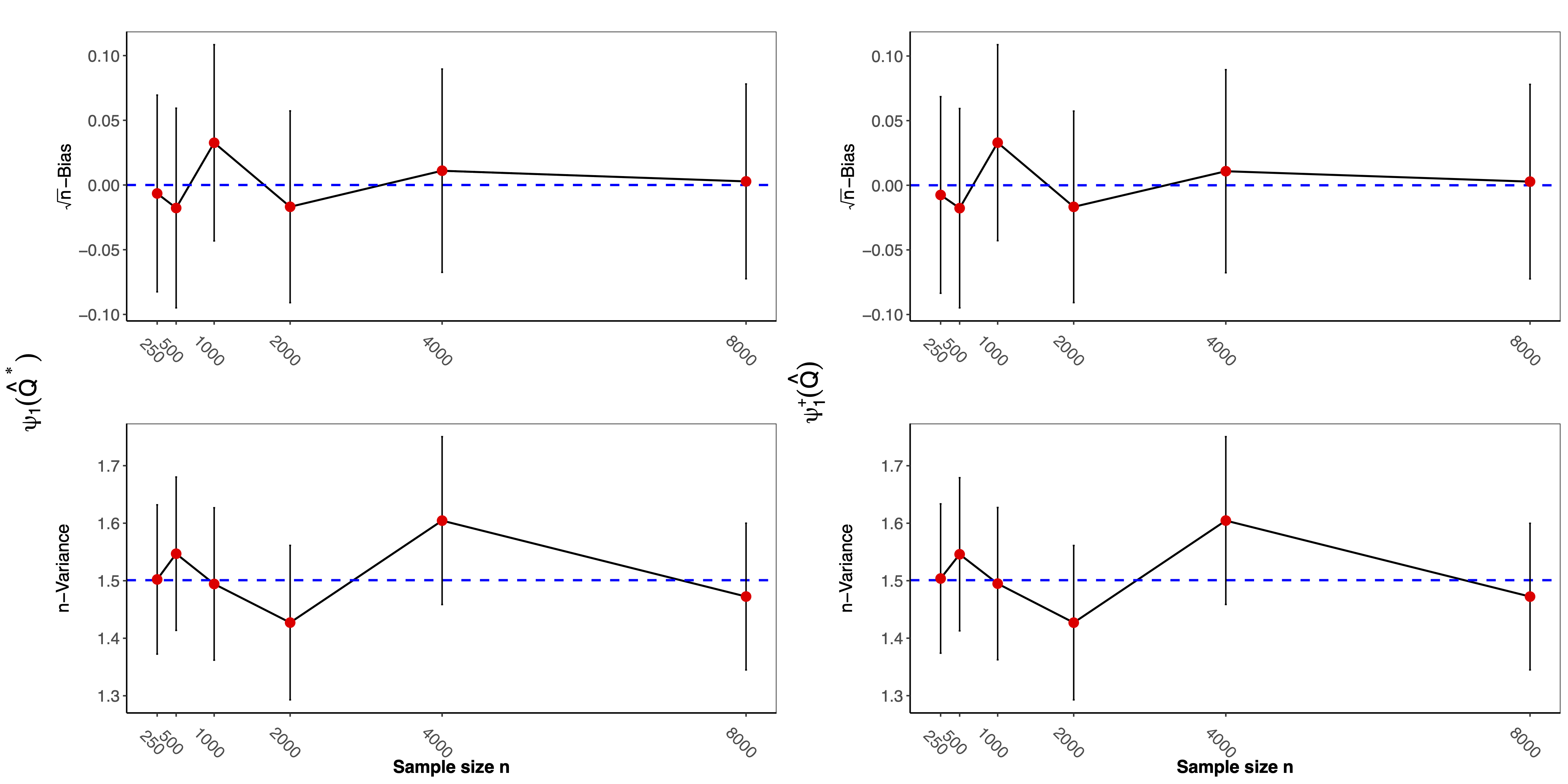}
    \caption{Simulation results validating the $\sqrt{n}$-consistency behaviors of the ATE estimators, under \textbf{univariate binary mediator}: (left) TMLE; (right) one-step estimator. 
    }
    \label{fig:binary}
\end{figure}

\begin{figure}[H]
    \centering
    \includegraphics[width=1\textwidth]{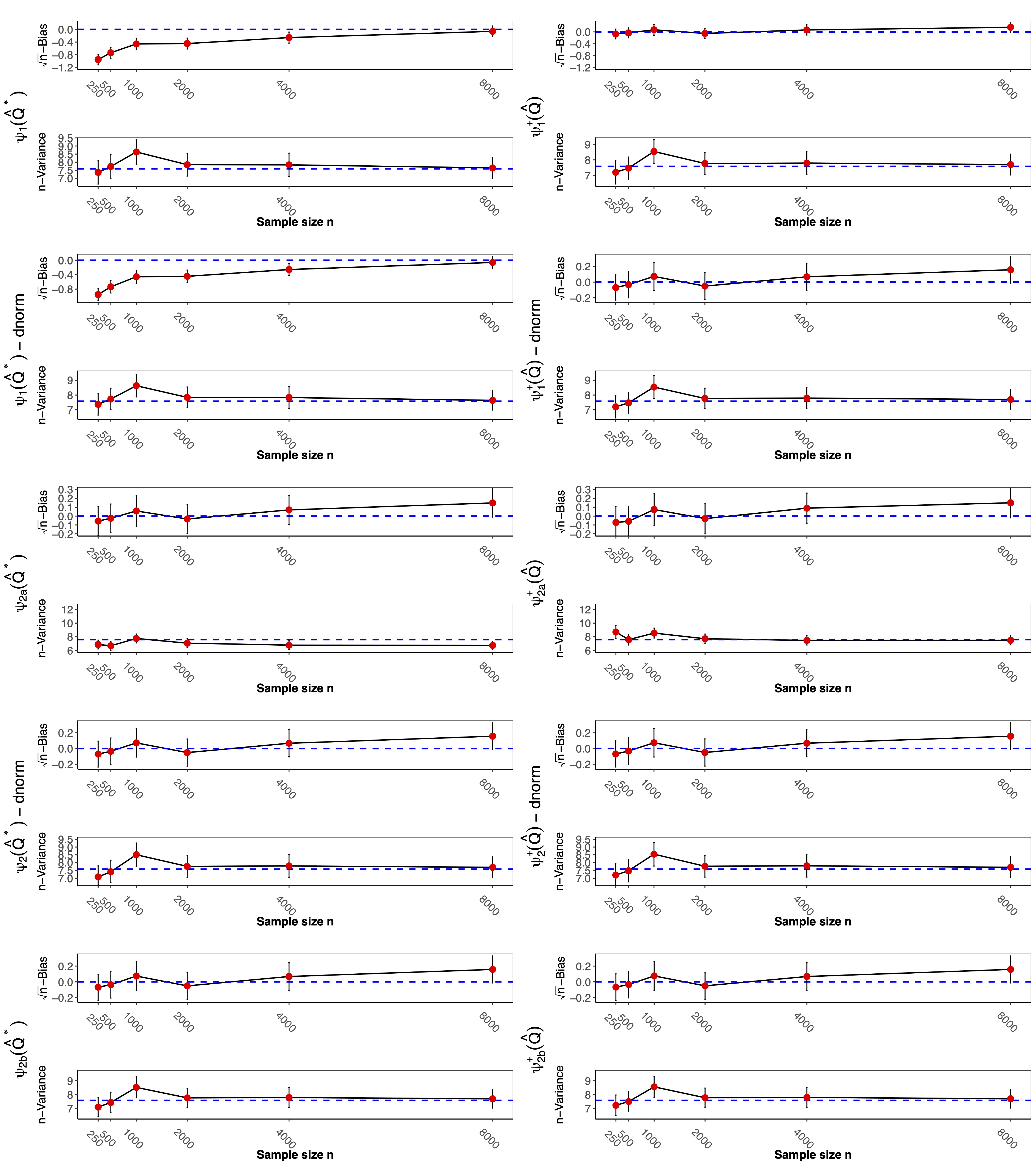}
    \caption{Simulation results validating the $\sqrt{n}$-consistency behaviors of the ATE estimators, under \textbf{univariate continuous mediator}: (left) TMLEs; (right) one-step estimators. 
    }
    \label{fig:continuous}
\end{figure}

\begin{figure}[H]
    \centering
    \includegraphics[width=1\textwidth]{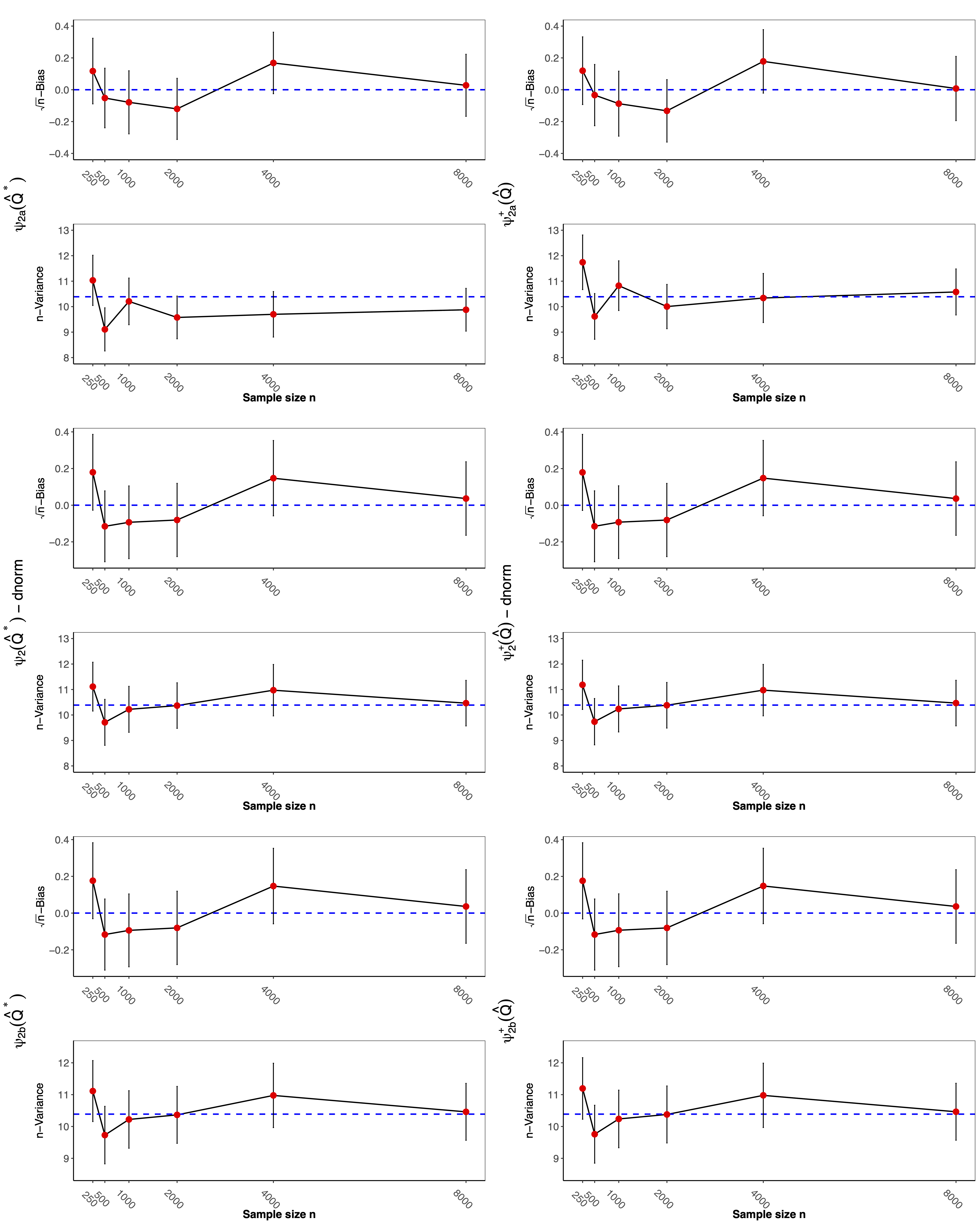}
    \caption{Simulation results validating the $\sqrt{n}$-consistency behaviors of the ATE estimators, under \textbf{bivariate continuous mediators}: (left) TMLEs; (right) one-step estimators. 
    }
    \label{fig:d2}
\end{figure}

\begin{figure}[H]
    \centering
    \includegraphics[width=1\textwidth]{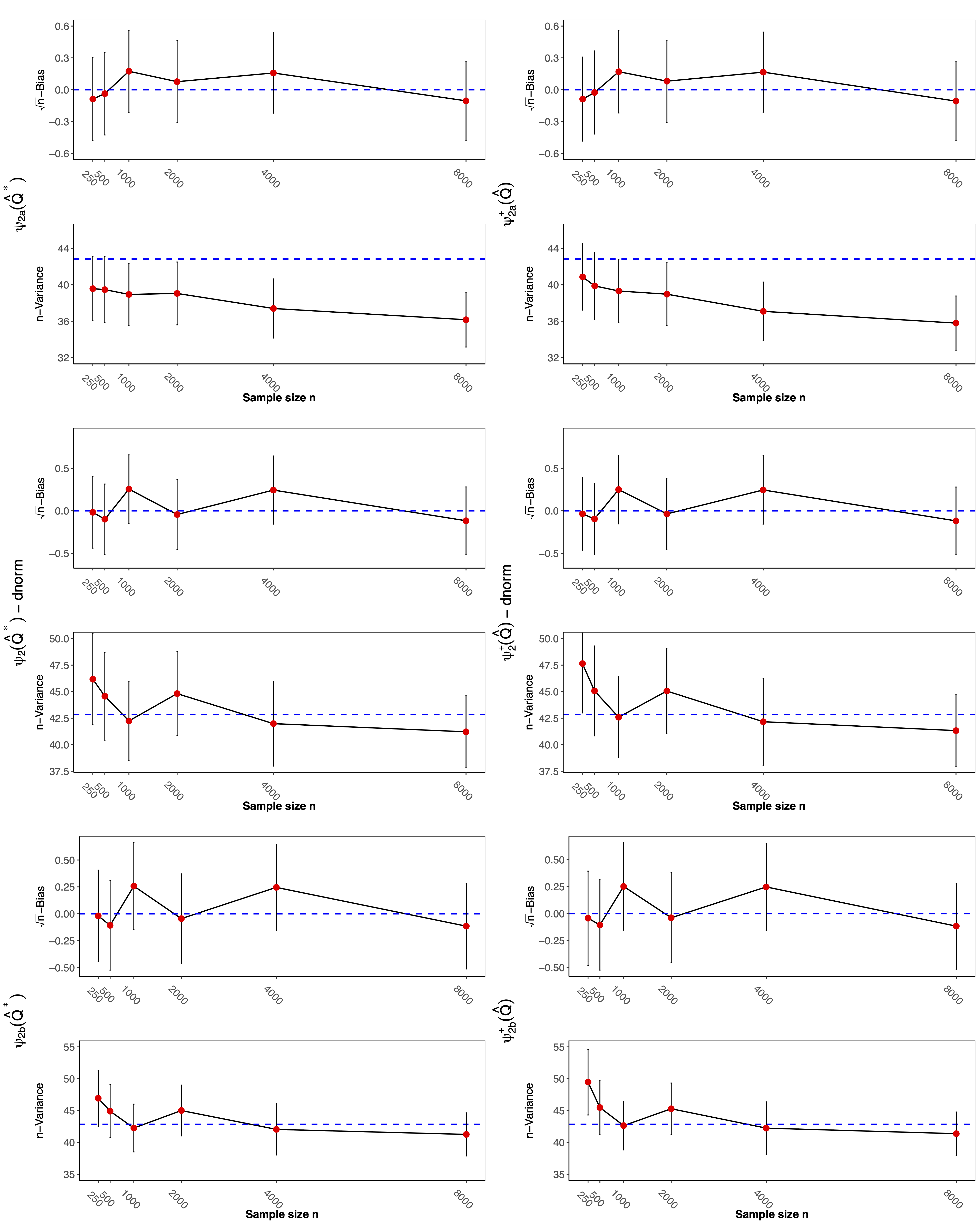}
    \caption{Simulation results validating the $\sqrt{n}$-consistency behaviors of the ATE estimators, under \textbf{quadrivariate continuous mediators}: (left) TMLEs; (right) one-step estimators. 
    }
    \label{fig:d4}
\end{figure}

\begin{figure}[H]
    \centering
    \includegraphics[width=1\linewidth, clip, trim=0 5 0 5]{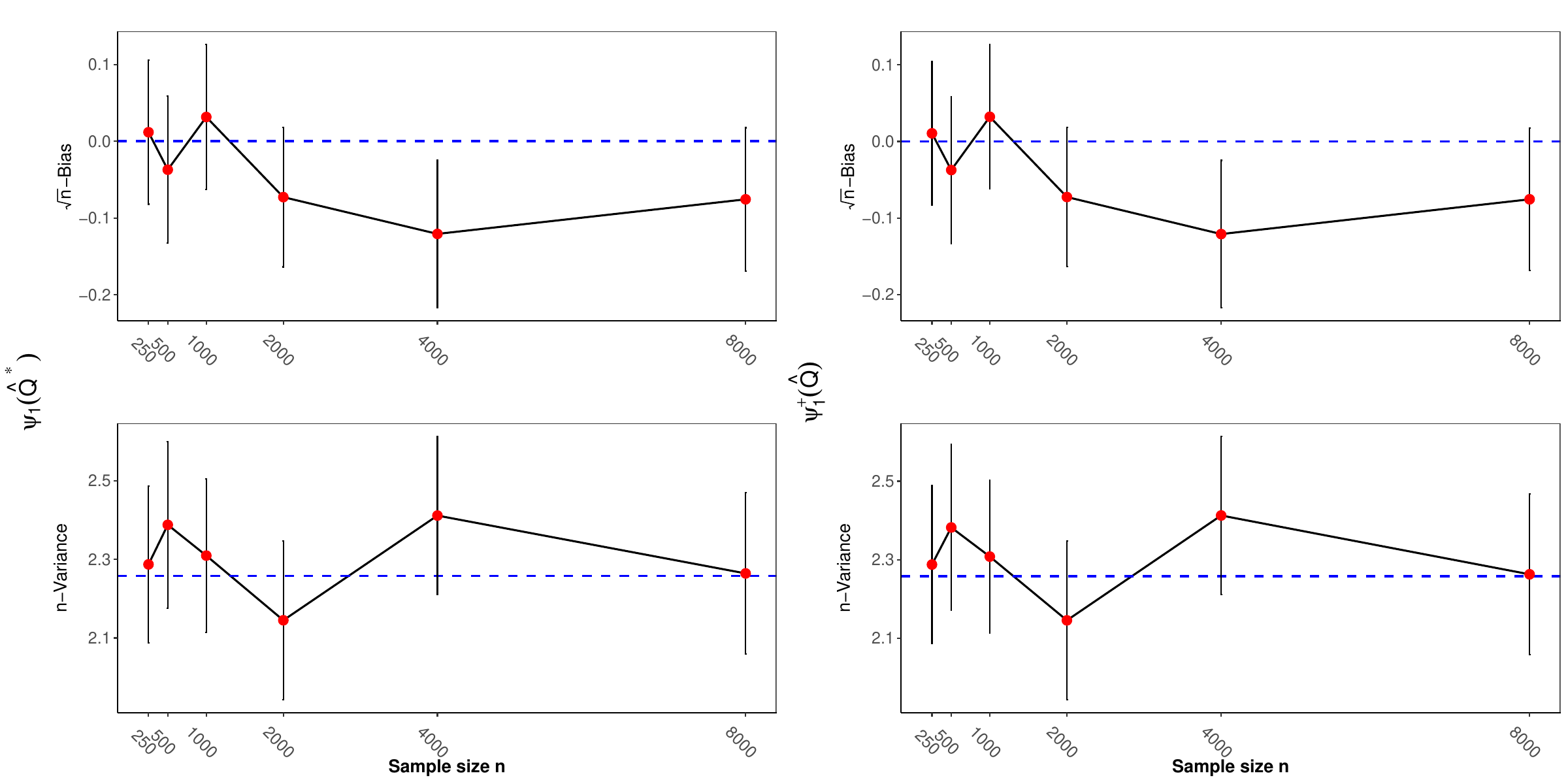}
        \caption{Simulation results validating the $\sqrt{n}$-consistency behaviors of the ATT estimators, under \textbf{univariate binary mediator}: (left) TMLEs; (right) one-step estimators. 
        }
    \label{fig:att_binary}
\end{figure}

\begin{figure}[H]
    \centering
    \includegraphics[width=1\linewidth, clip, trim=0 5 0 5]{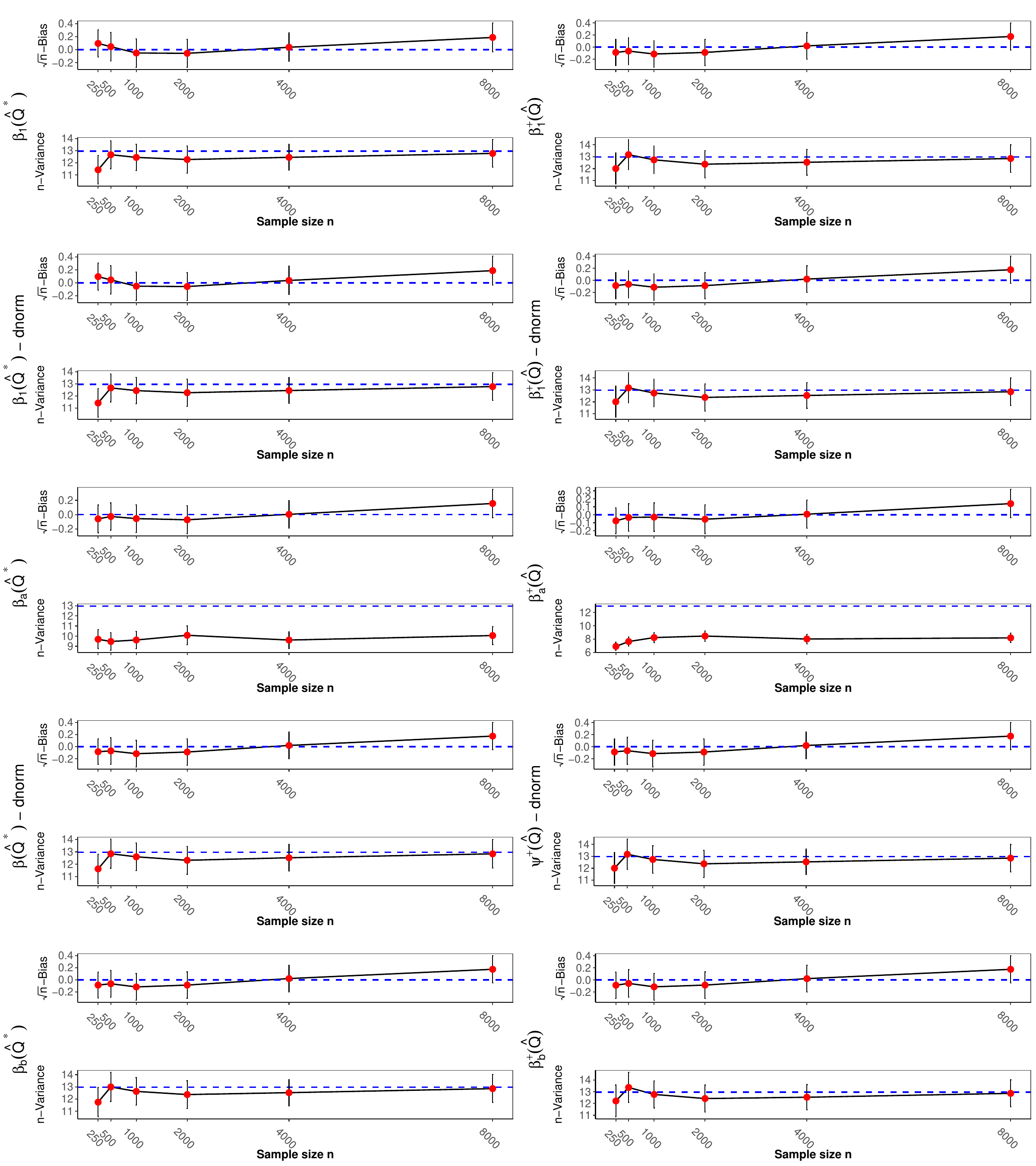}
        \caption{Simulation results validating the $\sqrt{n}$-consistency behaviors of the ATT estimators, under \textbf{univariate continuous mediator}: (left) TMLEs; (right) one-step estimators.  
        }
    \label{fig:att_continuous}
\end{figure}

\begin{figure}[H]
    \centering
    \includegraphics[width=1\linewidth, clip, trim=0 5 0 5]{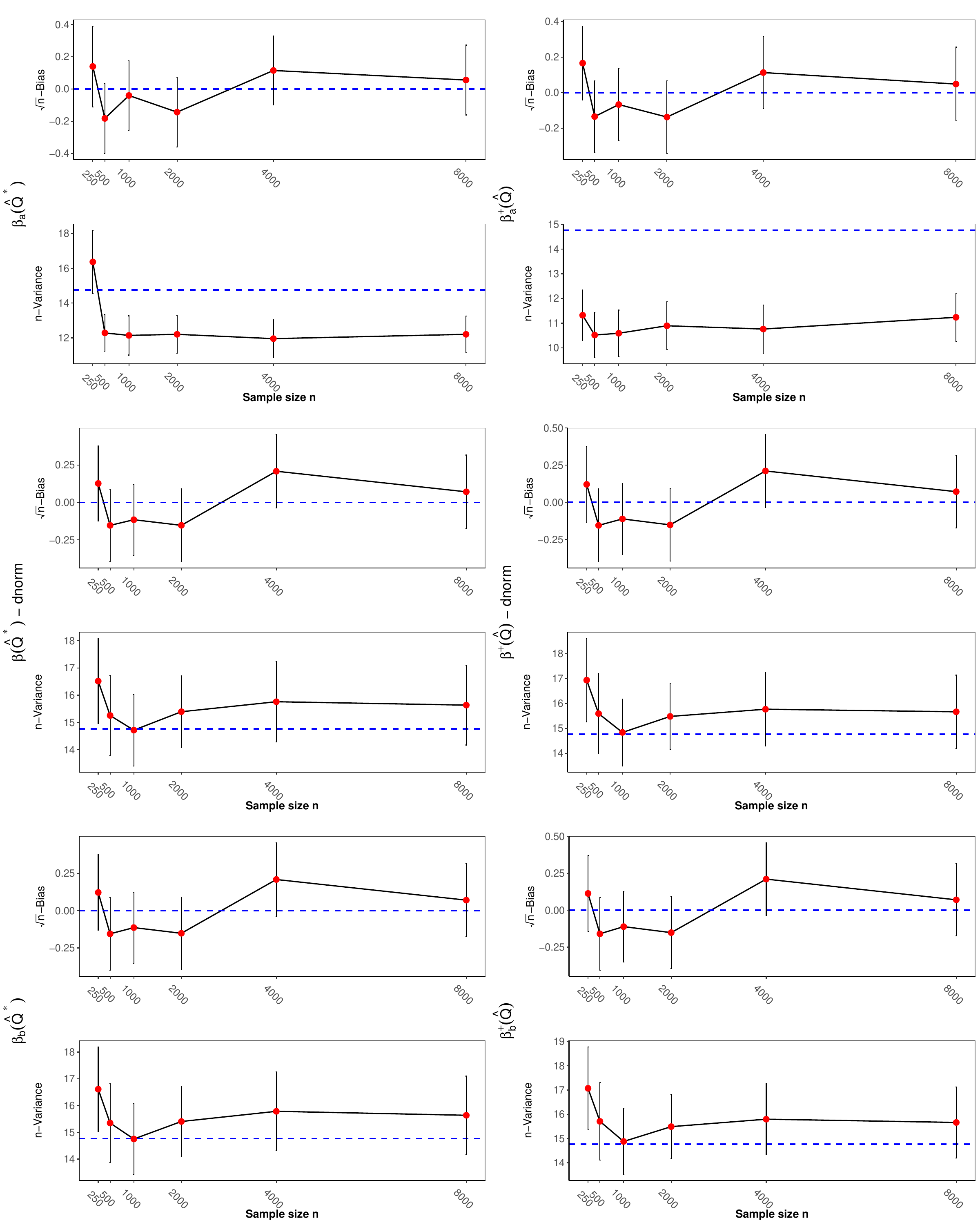}
        \caption{Simulation results validating the $\sqrt{n}$-consistency behaviors of the ATT estimators, under \textbf{bivariate continuous mediators}: : (left) TMLEs; (right) one-step estimators. 
        }
    \label{fig:att_d2}
\end{figure}

\begin{figure}[H]
    \centering
    \includegraphics[width=1\linewidth, clip, trim=0 5 0 5]{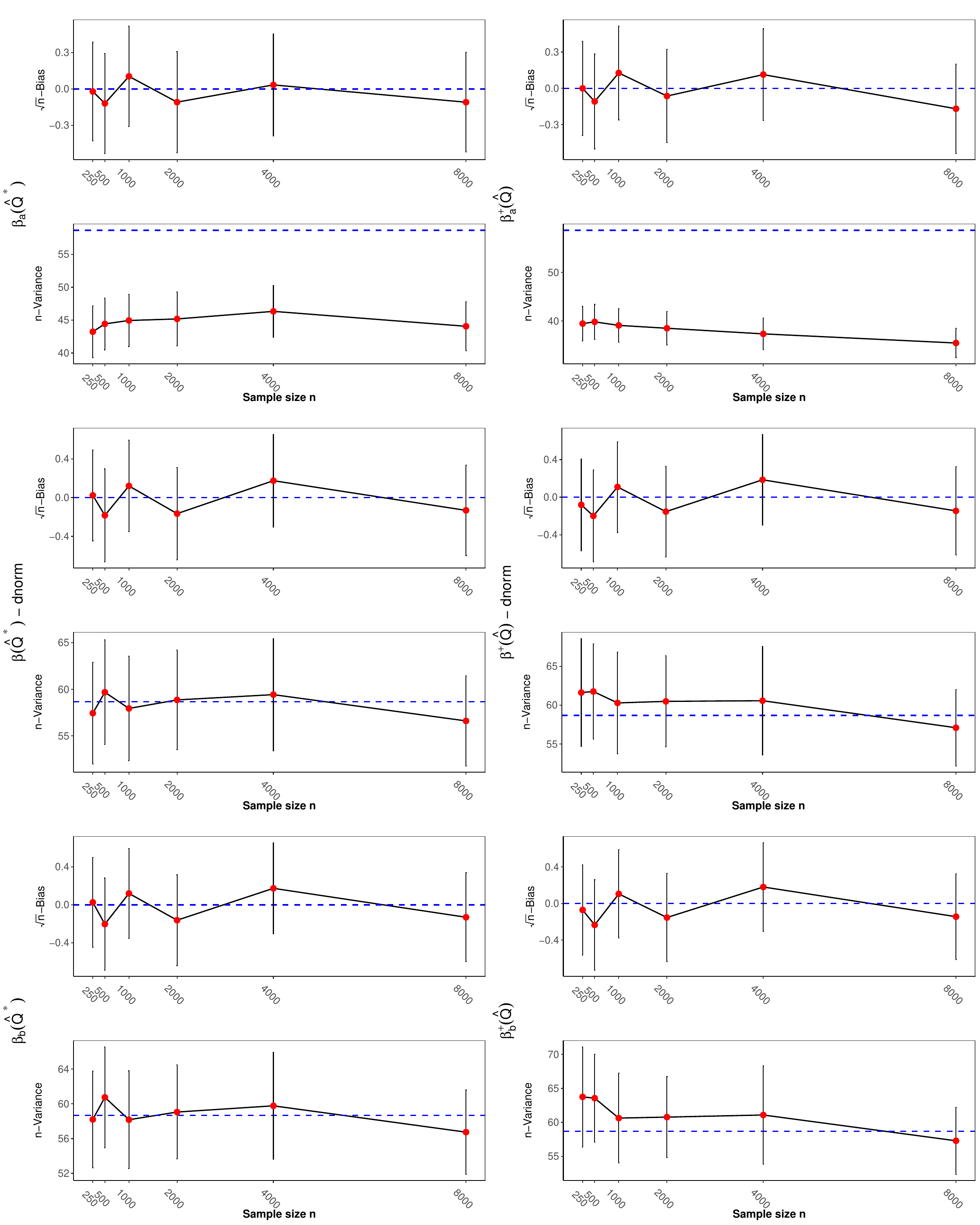}
        \caption{Simulation results validating the $\sqrt{n}$-consistency behaviors of the ATT  estimators, under \textbf{quadrivariate continuous mediators}: : (left) TMLEs; (right) one-step estimators. 
        }
    \label{fig:att_d4}
\end{figure}

\textbf{Summary.} We evaluated the asymptotic properties of our ATE and ATT estimators established in Section~\ref{sec:asymptotic}. Specifically, we verified that (i) the $\sqrt{n}$-bias decayed at the expected rate, and (ii) the $n$-scaled variance converged to the efficient variance $\P[\Phi(\Q)^2]$. Simulations included univariate binary, univariate continuous, bivariate continuous, and four-dimensional continuous mediators. Nuisance parameters were estimated using either fully parametric models or hybrid approaches combining parametric and kernel-based methods. Each scenario was replicated 1000 times at sample sizes ranging from 250 to 8000. Further details are provided below. Results (ATE: Figs~\eqref{fig:binary}--\eqref{fig:d4}; ATT: Figs~\eqref{fig:att_binary}--\eqref{fig:att_d4}) confirmed that the estimators exhibited the expected large-sample behaviors. We also compared linear versus nonlinear (expit-based) submodels for the regression of the continuous outcome (see Appendix~\ref{app:tmle_binaryY}) within the TMLE framework. Performance was evaluated in terms of bias, standard deviation (SD), mean squared error (MSE), 95\% confidence interval (CI) coverage, and CI width. Comparisons were conducted for univariate binary, univariate continuous, and bivariate continuous mediators, across sample sizes of 500, 1,000, and 2,000. Additional details are provided below. Results (ATE: Table~\ref{table:TMLEs_ATE}; ATT: Table~\ref{table:TMLEs_ATT}) confirmed that both linear and nonlinear TMLEs yielded valid inference under correct model specification.

Detailed description of the DGPs used in Simulation 1 are provided below. 

\vspace{-1.25cm}
\begin{align}
    X &\sim \operatorname{Uniform}(0,1) \sp , \notag \\
    A &\sim \operatorname{Binomial}\big(0.3+0.2X\big) \sp , \notag \\ 
    U &\sim \operatorname{Normal}(1+A+X,1) \sp ,  \notag \\
    \text{\small (univariate binary) } M &\sim \operatorname{Binomial}\big(\operatorname{expit}(-1+A+X)\big) \sp , \notag \\ 
    \text{\small (univariate continuous) } M &\sim \operatorname{Normal}(1+A+X, 1) \sp , \notag \\
    \text{\small (bivariate continuous) } M 
    &\sim 
    \operatorname{Normal}\left(
        \begin{bmatrix}
        1+A+X\\
        -1-0.5A+2X 
        \end{bmatrix},
        \begin{bmatrix}
        2 & 1\\
        1 & 3
        \end{bmatrix}
        \right) \sp ,  \label{dgp:sim1} \\
    \text{\small (quadrivariate continuous) } M 
    &\sim
    \operatorname{Normal}\left(
        \begin{bmatrix}
        1+A+X\\
        -1-0.5A+2X \\
        -1+2A+X\\
         1+0.5A-X
        \end{bmatrix},
        \begin{bmatrix}
        5 & -1 & 0 & 2\\
        -1 & 6 & 1 & 0\\
        0 & 1 & 4 & 3 \\
        2 & 0 & 3 & 7
        \end{bmatrix}
        \right) \sp , \notag \\ 
    Y &\sim \operatorname{Normal}(U+M+X, 1) \sp . \notag 
\end{align}

With \textit{univariate binary} mediator, estimating the mediator density $f_M$ through regressions is relatively straightforward. Consequently, $\psi_1(\hat{\Q}^\star)$, $\psi_1^{+}(\hat{\Q})$, $\beta_1(\hat{\Q}^\star)$, and $\beta_1^{+}(\hat{\Q})$ are identified as the most suitable estimators. 
With a \textit{univariate continuous} mediator, we evaluate a total of ten estimators for the ATE and ten for the ATT. For the estimators $\psi_1(\hat{\Q}^\star)$, $\psi_1^{+}(\hat{\Q})$, $\beta_1(\hat{\Q}^\star)$, and $\beta_1^{+}(\hat{\Q})$, we use the \texttt{np} package in \textsf{R} to directly estimate the mediator density. For $\psi_{2a}(\hat{\Q}^\star)$, $\psi_{2a}^{+}(\hat{\Q})$, $\beta_a(\hat{\Q}^\star)$, and $\beta_a^{+}(\hat{\Q})$, we use the \texttt{densratio} package for density ratio estimation. For $\psi_{2b}(\hat{\Q}^\star)$, $\psi_{2b}^{+}(\hat{\Q})$, $\beta_b(\hat{\Q}^\star)$, and $\beta_b^{+}(\hat{\Q})$, we estimate the density ratio using Bayes’ rule. In addition, we consider four variants for both the ATE and ATT that assume the mediator follows a conditional Normal distribution. In these variants, the mediator density is estimated parametrically under this assumption. Specifically, $\psi_1(\hat{\Q}^\star)$-dnorm and $\beta_1(\hat{\Q}^\star)$-dnorm are variants of $\psi_1(\hat{\Q}^\star)$ and $\beta_1(\hat{\Q}^\star)$, respectively, where the mediator density is estimated under a Normal assumption. Similarly, $\psi_{2}(\hat{\Q}^\star)$-dnorm and $\beta(\hat{\Q}^\star)$-dnorm are variants of $\psi_{2a}(\hat{\Q}^\star)$ and $\beta_a(\hat{\Q}^\star)$ (equivalently, variants of $\psi_{2b}(\hat{\Q}^\star)$ and $\beta_b(\hat{\Q}^\star)$) with the same modification. The corresponding changes are also applied to the one-step estimator variants.
With \textit{multivariate mediators}, direct estimation of mediator densities can be challenging and computationally demanding. In applications, estimators that circumvent density estimation are preferred. Therefore, we only consider $\psi_{2a}(\hat{\Q}^\star)$, $\psi_{2a}^+(\hat{\Q})$, $\beta_a(\hat{\Q}^\star)$, $\beta_a^{+}(\hat{\Q})$ $\psi_{2b}(\hat{\Q}^\star)$, $\psi_{2b}^+(\hat{\Q})$, $\beta_b(\hat{\Q}^\star)$, $\beta_b^{+}(\hat{\Q})$, along with the variations  where \texttt{dnorm} is used for mediator density ratio estimation, yielding a total of six estimators for both ATE and ATT estimation. 

Figs~\eqref{fig:binary}--\eqref{fig:d4} present the results establishing the $\sqrt{n}$-consistency of the proposed estimators for ATE, and Figs~\eqref{fig:att_binary}--\eqref{fig:att_d4} are the corresponding results for ATT. In order, figures correspond to the settings with univariate binary, univariate continuous, bivariate continuous, and quadrivariate continuous mediators. In these figures, the left panel presents the $\sqrt{n}$-scaled bias and $n-$scaled variance as a function of sample size for the TMLE estimators, while the right panel presents results from the corresponding one-step estimators. The true variance in the variance plots is empirically calculated under the true DGP with a sample size of $n=10^5$. Additionally, $95\%$ confidence interval for each point estimate is derived and depicted as vertical bars in both the bias and variance plots. Sample standard deviation over $1000$ multiple simulations is adopted for computing the confidence interval for each point estimate. 

According to these figures, TMLE and one-step estimators are highly comparable under correct model specifications. We observe that estimators relying on nonparametric kernel density estimation or mediator density ratio estimation, as implemented via the \texttt{densratio} method, may face challenges in converging to the expected values. This issue is evident in both univariate and multivariate continuous mediator settings, even as the sample size grows. Overall, estimators based on the Bayes' rule to estimate the density ratios are recommended due to their consistent performance in achieving the expected convergence results for most of the simulations.

We further compared TMLEs for the ATE and ATT using linear versus nonlinear submodels in the setting with a univariate continuous outcome. Linear submodels took the form $\tilde{\mu} = \hat{\mu} + \varepsilon_Y$, while nonlinear submodels followed the expit form detailed in \eqref{app:eq:submodels_binaryY} (Appendix~\ref{app:tmle_binaryY}). Results for ATE and ATT are reported in Tables~\ref{table:TMLEs_ATE} and \ref{table:TMLEs_ATT}, respectively. Across submodel types, bias decreased with increasing sample size, and all estimators achieved nominal 95\% CI coverage under correct nuisance specification—confirming the validity of both linear and nonlinear TMLEs.

\subsection{Simulation 2: Weak overlap} \label{app:sims:overlap}

In this simulation, we compared the finite-sample characteristics of our proposed estimators for ATE and ATT in a setting with weak overlap. We generated the treatment variable according to $\operatorname{Binomial}\left(0.001 + 0.998X\right)$, while the rest of the DGPs, as specified in \eqref{dgp:sim1}, remained unchanged. 

Nuisance parameters were estimated as follows. Linear regressions and logistic regressions were employed to estimate $\mu(M, A, X)$ and $\pi(A\mid X)$, respectively. Logistic regression was utilized for estimating $f_M(M\mid A, X)$ under univariate binary mediator. For estimators $\psi_1(\hat{\Q}^\star)$, $\psi_1^+(\hat{\Q})$, $\beta_1(\hat{\Q}^\star)$, and $\beta_1^+(\hat{\Q})$ in the case of a univariate continuous mediator, nonparametric kernel density estimation was applied to estimate $f_M(M\mid A, X)$ using the \texttt{np} package in \textsf{R}. For estimators $\psi_{2a}(\hat{\Q}^\star)$, $\psi_{2a}^+(\hat{\Q})$, $\beta_{a}(\hat{\Q}^\star)$, and $\beta_{a}^+(\hat{\Q})$ mediator density ratio was estimated via the \texttt{densratio} package in \textsf{R}. For estimators $\psi_{2b}(\hat{\Q}^\star)$, $\psi_{2b}^+(\hat{\Q})$, $\beta_{b}(\hat{\Q}^\star)$, and $\beta_{b}^+(\hat{\Q})$, the mediator density ratio was estimated using the reformulation presented in \eqref{eq:bayes}, where $\lambda(A\mid X,M)$ was estimated through logistic regressions.

Similar to Simulation 1, we evaluated the estimators based on bias, standard deviation (SD), mean squared error (MSE), 95\% confidence interval (CI) coverage, and average 95\% CI width. ATE estimation results are shown in Table~\ref{table:weakoverlap_ATE} in the main manuscript. The ATT estimation results are provided in Table~\ref{table:weakoverlap_ATT}. Across all settings, TMLE and one-step estimators exhibited similar bias; however, TMLE typically achieved substantially lower SD, resulting in smaller overall MSE. This increased stability was also reflected in the CI width, which was generally narrower for TMLE, while maintaining comparable or more conservative coverage. These patterns held across both the smallest sample size ($n = 500$) and the largest ($n = 2000$).

\subsection{Simulation 3: Model misspecification} \label{app:sims:misspecification}

Our third simulation explored the behavior of TMLEs and one-step estimators for both ATE and ATT in response to model misspecification, with a focus on univariate binary and univariate continuous mediators. We generated data as follows:

\vspace{-1.cm}
\begin{equation}\label{dgp:sim1-misspecification}
\begin{array}{ll}
    X \sim \operatorname{Uniform}(0,1), & \text{\small (binary) } M \sim \operatorname{Binomial}\big(\operatorname{expit}(-1+A+X-AX)\big),\vspace{0.1cm}\\
    A \sim \operatorname{Binomial}\big(\operatorname{expit}(-1+X)\big), & \text{\small (continuous) } M \sim \operatorname{Normal}(1+A+X-AX, 2), \vspace{0.1cm}\\
    U \sim \operatorname{Normal}(1+A+X-AX,2), & Y \sim \operatorname{Normal}(U+M+X-MX, 2).\vspace{0.1cm}
\end{array}
\end{equation}

This simulation focused on quantifying the impact of nuisance parameter estimation on the final estimation of ATE and ATT. Comparisons between one-step and TMLE estimators were not the primary aim. Instead, we evaluated how a given estimator performs under inconsistent versus flexible estimation of $\Q$. For the misspecified setting, we used main-effects linear regression models that excluded interaction terms present in the data-generating process. For flexible estimation, we employed the Super Learner algorithm \citep{van2007super}, an ensemble method that uses cross-validation to combine multiple candidate learners. These included intercept-only regression, generalized linear models (GLMs), Bayesian GLMs, multivariate adaptive regression splines, generalized additive models (GAMs), random forests, support vector machine (SVM), Bayesian Additive Regression Trees (BART), and extreme gradient boosting (XGBoost). Unlike the parametric models, these candidates are capable of capturing the interactions in the data. However, because many of them involve complex algorithms, they may violate the Donsker condition required by our theorems. To address this, we also implemented cross-fitted versions of each estimator.

We found that when misspecified working models were used for nuisance estimation, causal effect estimates were biased and CI coverage was poor across all sample sizes (see Table~\ref{table:misspecification_ATE} in the main manuscript for ATE and Table~\ref{table:misspecification_ATT} for ATT). In contrast, super learner-based estimators exhibited minimal bias across settings. CI coverage for these estimators generally improved with sample size, though some undercoverage was observed for the $\psi_1$ formulation of both the one-step and TMLE. These results suggest that for complex data-generating processes, flexible nuisance estimation—such as super learner—is recommended to mitigate bias from model misspecification. In this simulation, combining super learner with cross-fitting did not yield substantial gains in estimation performance.

\subsection{Simulation 4: Cross-fitting} \label{app:sims:cross-fit}

We examined the role of cross-fitting by focusing on random forests, which are known to perform poorly without sample splitting in high-dimensional settings \citep{double17chernozhukov, biau2012analysis}. We generated ten uniformly distributed confounders and introduced complex interactions and nonlinear terms between treatment, mediator, and covariates, as follows. Simulations used binary and continuous univariate mediators, with 1,000 replicates and sample sizes of 500, 1,000, and 2,000. 

\vspace{-1.25cm}
\begin{align}
    &X_k \sim \operatorname{Uniform}(0,1), \sp k\in\{1,\ldots,10\} \sp , 
    \notag \\
    &A \sim \mathrm{Binomial}(\mathrm{expit}(V_A \sp [1 \sp X \sp X^2]^T)) \sp , 
    \notag \\
    \text{\textcolor{white}{bin} Binary } &\text{mediator: } \notag \\
     &U \sim \operatorname{Normal}\left(V_U \sp [1 \sp A \sp X \sp AX_{1-5}]^T, 2\right) \sp ,
     \notag \\
     &M \sim\mathrm{Binomial}\left(\mathrm{expit}\left(V_M \sp [1 \sp A \sp X \sp AX_{1-5} \sp X_{6-10}^2]^T \right)\right) \sp , \label{dgp:cross_fititng}
     \\
     &Y \sim\operatorname{Normal}\left(V_Y \sp [U \sp M \sp X \sp MX_{1-5} \sp M^2 \sp X_{6-10}^2]^T, 2 \right) \sp , 
     \notag \\
    \text{Continuous } &\text{mediator: } \notag \\
     &U \sim \operatorname{Normal}\left(V_U \sp [1 \sp A \sp X \sp AX_{1-5}]^T, 1\right) \sp ,
     \notag \\
     &M \sim\operatorname{Normal}\left((V_M \sp [1 \sp A \sp X \sp AX_{1-5} \sp X_{6-10}^2]^T \right),1) \sp , 
     \notag \\
    &Y \sim\operatorname{Normal}\left(V_Y \sp [U \sp M \sp X \sp MX_{1-5} \sp M^2 \sp X_{6-10}^2]^T, 1 \right) \sp ,  \notag 
\end{align}

\vspace{-0.5cm}
where 

\vspace{-1.5cm}
{\small 
\begin{align*}
    &V_A=0.1\times[0.48, 0.07, 1, -1, -0.34, -0.12, 0.3, -0.35, 1, -0.1, 0.46,0.33, 0, 
    \\
    &\hspace{4cm} 0.45, 0.1, -0.32, -0.08, -0.2, 0.5, 0.5, -0.03] \sp , \\
    &V_U=[-2, -1, -1, 2, 3, 0.5, 3, 2, -1, 1, -3, 1.5, -3, -2,  1,  3,  1.5] \sp , \\
    &V_M=0.025\times[3, 1.5, -1.5, -1.5, -1, -2, -3, -3, -1.5, 2, 1.5, 3,  1.5, 2, 0.5, 0.5, 3, \\
    &\hspace{4cm} -0.2, -0.33, 0.5, 0.3, -0.5] \sp , \\
    &V_Y=[1, -2, -3, -1.5, 1, 0.5, -2, 1.5, -2, -3, -3, -1.5, -1, 0.5, 3, 1.5, 0.5,3, 1, 1.5, -2, 3, -1]\\
    &X=[X_1,X_2,X_3,X_4,X_5,X_6,X_7,X_8,X_9,X_10] \sp , \\
    &X_{1-5}=[X_1,X_2,X_3,X_4,X_5] \sp , \\
    &X_{6-10}=[X_6,X_7,X_8,X_9,X_{10}] \sp . 
\end{align*}
}

We implemented random forests using a standard set of tuning parameters: $500$ trees were grown to a minimum node size of five observations for a continuous outcome and one observation for a binary variable.  Cross-fitted ATE results are provided in Table~\ref{table:crossfitting_ATE}. As shown in Table~\ref{table:crossfitting_ATE}, cross-fitted ATE estimators consistently outperformed their non-cross-fitted counterparts, exhibiting lower bias and SD and substantially better CI coverage. Without cross-fitting, performance degraded as sample size increased. These results underscore the importance of cross-fitting in high-dimensional or complex modeling settings. Cross-fitted ATT results are provided in Table~\ref{table:crossfitting_ATT}. 


We also repeated the simulation using a second set of tuning parameters. Specifically, we adopted a sparser random forest with $200$ trees and a minimum node size of $1$. Cross-fitted ATE and ATT results, under the sparser tuning parameter set, are provided in Table~\ref{apptable:crossfitting_ate} and Table~\ref{apptable:crossfitting_att}, respectively. 

Tables~\ref{apptable:crossfitting_ate} and \ref{apptable:crossfitting_att} reveal a comparative analysis using a more sensitive random forest algorithm by increasing the variability of predictions. According to these results, the estimation performance of random forest is inferior, as evidenced by smaller CI coverage when compared with results produced by denser random forests (with $500$ trees). In contrast, results yielded by performing sample splitting in conjunction with the sparser random forest remains highly comparable to those shown in Tables~\ref{table:crossfitting_ATE} and \ref{table:crossfitting_ATT}. These findings imply that in high-dimensional settings or scenarios where high estimation variance is anticipated from nuisance estimates, cross-fitting proves beneficial in reducing estimation bias and enhancing the stability of results.

\subsection{Simulation 5: Model evaluation} \label{app:sims:tests}

Our fifth simulation evaluated the performance of proposed tests in scenarios that they are designed for. Performance was evaluated using type I error and power, which were calculated as the proportion of rejecting the null hypothesis during $200$ simulation replicates for each test scenario. In each replicate, data were generated from a specific DGP, and the tests were applied. The rejection proportion corresponds to the type I error or power, depending on whether the DGP satisfies the front-door assumptions or not.

To evaluate type I error, we generated data from causal models that satisfy the front-door assumptions. We considered two model settings that differs in how $Z$ relates to the other variables. In one setting, labeled as ``DAG1'', $Z$ has direct effects on both $A$ and $M$. In another setting, labeled ``DAG2'', $Z$ has a direct effect on $A$ and shares unmeasured confounding with $M$.

To evaluate power, we generated data from causal models that violate the front-door model assumptions. We considered two distinct settings, each representing a different type of violation. In both cases, $Z$ had direct effects on both $A$ and $M$. In a setting, labeled ``DAG3'', violations arose from unmeasured confounding between $A$ and $M$, as well as between $M$ and $Y$. In another setting, labeled ``DAG4'', the violation occurred through a direct effect of $A$ on $Y$ that was not mediated by $M$. The relationships among all variables in causal models DAG1-DAG4 are depicted in Fig.~\ref{fig:test_graphs}. 

Under these four causal models, we designed three experimental settings to achieve different objectives by varying the types of variables (binary or continuous) for $(X, Z, A, M, Y)$ and considering different forms of DGPs, including linear, quadratic, and interaction terms. The first setting aimed to verify that the proposed tests have approximately $0.05$ type I error and show increasing power with larger sample sizes in the settings they were designed for. With the second setting, we aimed to demonstrate the advantage of the DR-CCM test in providing valid inference even when some nuisance models are misspecified, by comparing its performance to other tests. The third setting aimed to showcase the flexibility of the proposed primal and dual tests in handling continuous mediators and incorporating machine learning models for nuisance estimation by examining scenarios with complex DGP forms, where both the anchor variable and the mediator are continuous.

In the first experimental setting, we evaluated the CCM, dual, and primal tests across three distinct variable-type configurations. In configuration 1, all variables, $X$, $Z$, $A$, $M$, and $Y$, were binary. In configuration 2, the outcome $Y$ was continuous, while all other variables remained binary. In configuration 3, both $X$ and $Y$ were continuous, while $Z$, $A$, and $M$ remained binary. Each test was evaluated at sample sizes of $500$, $1000$, $2000$, $4000$, and $10000$.  

With the first experimental setting, we found that all tests maintained type I error rates near the nominal level of 0.05 and exhibited increasing power with larger sample sizes. These results are summarized in Appendix Table~\ref{table:test_three}.  

\vspace{0.5cm}
The DGPs for the \textbf{first experimental setting} with \textbf{variable-type configuration 1} (where all variables are binary) are displayed in \eqref{appeq:setting1_allbinary_dag1}.

\vspace{-0.5cm}
\begin{spacing}{1.5}
\begin{align}
    &\text{({DAG1})} \notag \\ 
    &U_1\sim\binomial(0.7),
  \notag  \\
    &X\sim\binomial(0.3),\quad Z\sim \binomial(\expit(-0.5 + 0.5X))
  \notag  \\
    &A\sim\binomial(\expit(-0.5 -1.1Z + 1.3U_1 + 0.5 X +1.75 U_1 Z - 1.2 U_1 X - 1.5ZX -1.8 U_1 Z X)),
   \notag \\
    &M\sim\binomial(\expit(-0.5 -A + 1.1 Z -0.5 X - 1.25 AZ + 1.5 AX - 1.5 ZX -1.7 AZX)),
   \notag \\
    &Y\sim \binomial(\expit(-0.5 -0.5 M + U_1 + 0.5 X - 1.2 MU_1 +1.5  MX - 1.5 U_1X -1.7 MU_1X)) \sp.
\notag \\ \hline 
    &\text{({DAG2})} \notag \\ 
    &U_i\sim\binomial(0.5), \ i\in\{1,2\},
    \notag \\
    &X\sim\binomial(0.3),\quad Z\sim \binomial(\expit(-0.5 +  X + 1.5 U_2 +1.5 X U_2)),
    \notag \\
    &A\sim\binomial(\expit(-1 + Z + 1.5 X + U_1 +1.5 ZX - 1.5 U_1 Z +1.5 U_1 X -1.7 U_1 Z X)),
    \notag \\
    &M\sim\binomial(\expit(-1 + A + 1.5 X +  U_2 +1.5 AX - 1.5 A U_2 + 1.5 U_2 X -1.7 A U_2 X)),
    \notag \\
    &Y\sim \binomial(\expit(-1 + 0.2 M + 1.2 X +  U1 + 1.5 X U_1 - 1.5 MX +1.5 M U_1 -1.7 M U_1 X))\sp .
\notag \\ \hline
    &\text{({DAG3})} \notag \\ 
    &U_i\sim\binomial(0.5), \ i\in\{1,2\},
    \notag \\
    &X\sim\binomial(0.5),\quad Z\sim \binomial(\expit(-0.5 +  0.5X)),
    \notag \\
    &A\sim\binomial(\expit(-0.5 +  Z + 1.5 X +  U_1 +  U_2 -1.5 ZX + 1.5  ZU_1 -1.5  ZU_2 -1.5  XU_1  
    \notag \\
    &\hspace{1cm}+ 1.5  XU_2 -1.5  U_1U_2 -1.7  ZXU_1 + 1.2  ZXU_2 -1.7  ZU_1U2_ -1.7  XU_1U_2 + 
                                       1.4  ZXU_1U_2)),
    \notag \\
    &M\sim\binomial(\expit(-1 +   A + 1.5  Z +   X +   U_1 -1.5  AZ + 1.5  AX -1.5  AU_1 -1.5  ZX 
    \notag \\
    &\hspace{1cm}+ 1.5  ZU_1 -1.5  XU_1 -1.7  AZX + 1.2  AZU_1 -1.7  AXU_1 -1.7  ZXU_1 +
  1.4  AZXU_1)),
    \notag \\
    &Y\sim \binomial(\expit(-0.5 + 0.5  M + 0.2  X + 1.2  U_1 -1.5  U_2 -  MX -1.5  MU_1 +   MU_2 + 1.2  XU_1 \notag \\
    &\hspace{1cm}+0.5  XU_2 +  U_1U_2 + 
  1.1  MXU_1 -0.75  MXU_2 -  MU_1U_2 -0.2  XU_1U_2 +0.5  MXU_1U_2)) \sp. 
\notag  \\ \hline
&\text{({DAG4})} \notag  \\ 
    &\P(U_1, X, Z, A, M) \text{ aligns with DAG1} \sp , \notag  \\ 
    &Y\sim \binomial(\expit(-1 -0.2  M + 1.5  A + 0.5  X + 0.2  U_1 -1.2  MA + 0.5  MX + 0.3  MU_1 -  AX  
    \notag \\ 
    &\hspace{1cm}+ 0.5  AU_1-0.5  XU_1 + 0.5  MAX -0.5  MAU_1 + 0.2  MXU_1 -0.5  AXU_1 +  MAXU_1)) \sp. \label{appeq:setting1_allbinary_dag1}
\end{align}

\end{spacing} 


\vspace{0.5cm}
The DGPs for the \textbf{first experimental setting} with \textbf{variable-type configuration 2} (where $Y$ is continuous, while all other variables are binary) are displayed in \eqref{appeq:setting1_yconti_dag1}.  

\vspace{-1.cm}
\begin{spacing}{1.5}
\begin{align}
    &\text{({DAG1})}\notag  \\ 
    &U_1\sim\binomial(0.7),
  \notag  \\
    &X\sim\binomial(0.3),\quad Z\sim \binomial(\expit(-0.5 + 0.5X)),
  \notag  \\
    &A\sim\binomial(\expit(-0.5-1.1Z+1.3 U_1 + 0.5X)),
  \notag  \\
    &M\sim\binomial(\expit(-0.5-A+1.1 Z -0.5X)),
   \notag \\
    &Y\sim\operatorname{Normal}(-0.5-0.5M+U_1+0.5X-1.2M U_1,0.5) \sp.
\notag \\  \hline 
    &\text{({DAG2})} \notag \\ 
    &\P(U_1,U_2, X, Z, A, M) \text{ aligns with DAG2 in \eqref{appeq:setting1_allbinary_dag1}}  \sp , \notag \\
    &Y\sim\operatorname{Normal}(-1 + 0.2 M + 1.2 X +  U_1 + 1.5 X U_1 - 1.5 MX +1.5 M U_1 -1.7 M U_1 X,1) \sp.
\notag \\ \hline 
    &\text{({DAG3})} \notag \\
    &\P(U_1,U_2, X, Z, A, M) \text{ aligns with DAG3 in \eqref{appeq:setting1_allbinary_dag1}} \sp ,   \notag \\ 
    &Y\sim\operatorname{Normal}(-0.5 + 0.5  M + 0.2  X + 1.2  U_1 -1.5  U_2 -  MX -1.5  MU_1 +   MU_2 + 1.2  XU_1 \notag \\
    &\hspace{1cm}+0.5  XU_2 +  U_1U_2 + 
  1.1  MXU_1 -0.75  MXU_2 -  MU_1U_2 -0.2  XU_1U_2 +0.5  MXU_1U_2,1) \sp.
\notag \\ \hline 
    &\text{({DAG4})}  \notag \\
    &\P(U_1,U_2, X, Z) \text{ aligns with DAG2 in \eqref{appeq:setting1_allbinary_dag1}} \sp , \notag\\  
    &\P(A, M \mid  U, X, Z) \text{ aligns with DAG1 in \eqref{appeq:setting1_allbinary_dag1}}  \sp , \notag\\  
    &Y\sim\operatorname{Normal}(-1 -0.2  M + 1.5  A + 0.5  X + 0.2  U_1 -1.2  MA + 0.5  MX + 0.3  MU_1 -  AX 
    \notag \\
    &\hspace{1cm} + 0.5  AU_1 -0.5  XU_1  + 0.5  MAX -0.5  MAU_1 + 0.2  MXU_1 -0.5  AXU_1 +  MAXU_1,1) \sp. \label{appeq:setting1_yconti_dag1}
\end{align}
\end{spacing} 

\vspace{0.5cm}
The DGPs for the \textbf{first experimental setting} with \textbf{variable-type configuration 3} (where $(X,Y)$ are continuous, while all other variables are binary) are displayed in \eqref{appeq:setting1_xyconti_dag1}.  

\vspace{-1.cm}
\begin{spacing}{1.5}
\begin{align}
    &\text{({DAG1})} \notag \\ 
    &\P(U_1,Z,M) \text{ aligns with DAG1 in \eqref{appeq:setting1_yconti_dag1}} \sp , \notag \\  
    &X\sim\uniform(0,1),\quad A\sim\binomial((1-0.5 Z+1.3 U_1 + 0.5X)/4),
   \notag  \\
    &Y\sim\operatorname{Normal}(-0.5-0.5M+0.5X,1) \sp.
\notag \\ \hline 
    &\text{({DAG2})} \notag \\ 
    &U_i\sim\binomial(0.5), \ i\in\{1,2\},
\notag     \\
    &X\sim\uniform(0,1),\quad Z\sim \binomial((1+X+1.5U_2)/4),
\notag     \\
    &A\sim\binomial((1-0.5 Z+U_1 + 1.5X)/4),\quad M\sim \binomial(\expit(-1+A+1.5X+U_2)),
 \notag    \\
    &Y\sim\operatorname{Normal}(-1 + 0.2 M + 1.2 X,1)\sp.
\notag \\ \hline 
    &\text{({DAG3})} \notag  \\ 
    &\P(U_1,U_2,Z) \text{ aligns with DAG3 in \eqref{appeq:setting1_allbinary_dag1}} \sp , \notag  \\ 
    &X\sim\uniform(0,1),\quad A\sim\binomial(\expit(-0.5+ Z+U_1 + 1.5X)),
   \notag  \\
    &M\sim\binomial(\expit(-1+A+1.5 Z+X+U_1-U_2)),
  \notag   \\
    &Y\sim\operatorname{Normal}(-0.5+0.5M+0.2X+1.2 U_1 -1.5 U_2,0.5)\sp .
\notag  \\ \hline 
    &\text{({DAG4})} \notag  \\ 
    &U_1\sim\binomial(0.7),
    \notag  \\
    &X\sim\N(1,1),\quad A\sim \binomial(\expit(-0.5-1.1 Z+1.3 U_1+0.5 X)),
   \notag  \\
    &\P(Z \mid X) \text{ aligns with DAG1 in \eqref{appeq:setting1_allbinary_dag1}} \sp ,   \notag  \\ 
    &M\sim \binomial(\expit(-0.5-A+1.1 Z-0.5X)) \sp , \notag \\ 
    &Y\sim\operatorname{Normal}(-1 -0.2  M + 1.5  A + 0.5  X + 0.2  U_1 ,1)\sp . \label{appeq:setting1_xyconti_dag1}
\end{align}
\end{spacing}

\vspace{0.5cm}
In the second experimental setting, to demonstrate the advantage of the DR-CCM test, we compared its performance with that of the dual, and primal tests in a setting where the outcome regression could not be correctly specified using simple linear models. We considered a variable-type configuration in which both $X$ and $Y$ were continuous, while all other variables remained binary.  Quadratic term $X^2$ and interaction term $MX$ were added to the data-generating distribution of $Y$ such that the outcome regression can no longer be correctly specified by simple linear models, creating condition of model misspecification to showcase the double robustness property of the DR-CCM test. As in previous evaluations, performance was assessed under the four causal models (DAG1–DAG4) across sample sizes of $500$, $1000$, $2000$, $4000$, and $10000$.

In the second experimental setting, we observed that DR-CCM test was the only test among the four that consistently achieved type I error rates close to 0.05 while demonstrating increased power with larger sample sizes. In contrast, the other tests yielded increased type I error with larger sample sizes. These findings are presented in Table~\ref{table:test_DR}. 

\vspace{0.5cm}
The DGPs for the \textbf{second experimental setting} are displayed in \eqref{appeq:setting2_dag1}. 

\vspace{-1.cm}
\begin{spacing}{1.5}
\begin{align}
    &\text{({DAG1})} \notag \\ 
    &\P(U_1,X,Z, A, M) \text{ aligns with DAG4 in \eqref{appeq:setting1_xyconti_dag1}}  \sp , \notag \\  
    &Y\sim\operatorname{Normal}(-0.5-0.5M+U_1+0.5X+1.2 X^2-1.5MX)\sp.
\notag \\ \hline 
    &\text{({DAG2})} \notag \\ 
    &U_i\sim\binomial(0.5), \ i\in\{1,2\},
   \notag  \\
    &X\sim\N(1,0.5),\quad Z\sim \binomial(\expit(-0.5+X+1.5U_2)),
  \notag   \\
    &A\sim \binomial(\expit(-1+A+1.5X+U_1)),\quad M\sim \binomial(\expit(-1+A+1.5X+U_2)),
  \notag   \\
    &Y\sim\operatorname{Normal}(-1+0.2M+1.2X+1.2 X^2+1.5MX+U_1,0.5)\sp.
\notag \\ \hline 
    &\text{({DAG3})} \notag \\ 
    &\P(U_1,U_2,Z) \text{ aligns with DAG3 in \eqref{appeq:setting1_allbinary_dag1}} \sp , \notag \\  
    &X\sim\N(1,1),\quad A\sim\binomial(\expit(-0.5 + Z + 1.5 X + U_1)),
  \notag   \\
    &M\sim\binomial(\expit(-1 + A + 1.5 Z + X+0.5 X^2 +  U_1 -U_2)),
   \notag  \\
    &Y\sim\operatorname{Normal}(-0.5+0.5M+0.2X+1.2 X^2+1.2 U_1 -1.5 U_2,1)\sp. 
\notag \\ \hline 
    &\text{({DAG4})} \notag \\ 
    &U_1\sim\binomial(0.5),\quad X\sim\N(1,1),
   \notag  \\
    &Z\sim\binomial(\expit(-0.5+0.5X)),\quad A\sim \binomial(\expit(-0.5-1.1 Z+1.3 U_1+0.5 X)),
    \notag  \\
    &M\sim \binomial(\expit(-0.5-A+1.1 Z-0.5X+0.5 X^2)),
    \notag \\
    &Y\sim\operatorname{Normal}(-1 -0.2  M + 1.5  A + 0.5  X + 0.2  U_1 + 0.5 MX+1.2 X^2,1)\sp. \label{appeq:setting2_dag1}
\end{align}
\end{spacing}


\vspace{0.5cm}
In the third experimental setting, we further evaluated the performance of the dual and primal tests, with and without the use of flexible machine learning methods for model fitting, in a configuration where all variables except $A$ were univariate continuous, and the outcome regression could not be correctly specified using simple linear models. When employing flexible methods, we used the Super Learner algorithm with two learners: the generalized linear model (\texttt{SL.glm}) and random forests (\texttt{SL.ranger}). This evaluation was conducted under three sample sizes: $500$, $1000$, and $2000$. 

With the third experimental setting, we found that incorporating Super Learners for nuisance model estimation helped keep type I error around 0.05 for both the dual and primal tests. In comparison, the tests without Super Learners had inflated type I errors, often exceeding 0.1 and increasing with sample sizes. While Super Learners helped keep type I error at the desired level, it came at the cost of reduced power relative to their non–Super Learner counterparts. These results are summarized in Table~\ref{table:test_SL}. 

\vspace{0.5cm}
The DGPs for the \textbf{third experimental setting} are displayed in \eqref{appeq:setting3_dag1}. 

\vspace{-1.cm}
\begin{spacing}{1.5}
\begin{align}
    &\text{({DAG1})} \notag  \\ 
    &U_1\sim \binomial(0.7),\quad X\sim\operatorname{Normal}(1,1),
   \notag  \\
    &Z\sim\operatorname{Normal}(-0.5+0.5X,0.5), \quad A\sim\binomial(\expit(-0.5 -1.1 Z + 0.5 X)),
   \notag  \\
    &M\sim\operatorname{Normal}(-0.5-A+1.1Z-0.5X,0.5),\quad Y\sim\operatorname{Normal}(-0.5-0.5M+U_1+0.5X,2)\sp.
\notag \\ \hline 
        &\text{({DAG2})} \notag \\ 
        &U_1\sim\binomial(0.5),\quad U_2\sim\N(1,1),
       \notag  \\
        &X\sim\operatorname{Normal}(1,1),\quad Z\sim\operatorname{Normal}(-0.5+1.5U_2,0.5),
       \notag  \\
        &A\sim\binomial(\expit(-1 + Z + 1.5 X)),\quad M\sim\operatorname{Normal}(-1 + A + 1.5 X +  U_2,0.5),
      \notag   \\
        &Y\sim\operatorname{Normal}(-1 + 0.2 M + 1.2 X + U_1,1)\sp.
\notag \\ \hline 
        &\text{({DAG3})} \notag \\ 
        &U_i\sim\binomial(0.5),\ i\in\{1,2\},
       \notag  \\
        &X\sim\N(1,1),\quad Z\sim\operatorname{Normal}(-0.5+0.5X,0.5),
      \notag  \\
        &A\sim\binomial(\expit(-0.5 + Z + 1.5 X + U_1 + U_2)),\quad M\sim\operatorname{Normal}(-1 + A + 1.5 Z +  X + U_1,0.5),
      \notag   \\
        &Y\sim\N(-0.5 + 0.5  M + 0.2  X + 1.2  U_1 -1.5  U_2,1)\sp.
\notag \\ \hline 
        &\text{({DAG4})}\notag  \\ 
        &U_1\sim\binomial(0.7),\quad X\sim \uniform(0.5,1),
       \notag  \\
        &Z\sim \uniform(0,X),\quad A\sim \binomial((1 -0.5 Z + 1.3 U_1 + 0.5  X )/4),
      \notag   \\
        &M\sim\operatorname{Normal}(-0.5 - A + 1.1  Z -0.5 X,0.2),
      \notag   \\
        &Y\sim\operatorname{Normal}(-1 -0.2 M + 10 A +3AM -0.5 X +0.2 U_1,2)\sp. \label{appeq:setting3_dag1}
    \end{align}
\end{spacing}

\subsection{Simulation 6: Efficiency gain}\label{app:sims:eff}
This simulation investigated the efficiency gains from leveraging the Verma constraint, considering scenarios where $Z$ is either univariate binary or continuous. The DGP for binary $Z$ is given in \eqref{appeq:eff_binaryZ_interAZ} and the DGP for continuous $Z$ is given in \eqref{appeq:eff_contiZ}.

\vspace{-0.5cm}
\begin{equation}\label{appeq:eff_binaryZ_interAZ}
\begin{aligned}
&U \sim\operatorname{Normal}( 0.6 A - 0.3 Z + 0.2AZ, 1), \\
&Z \sim \binomial(0.5),\quad A \sim \binomial( 0.3 + 0.3 Z ), \\
&M \sim \binomial( \expit( -1.1 + 1.3A + 1.4Z + 1.6AZ ) ), \\
&Y \sim\operatorname{Normal}( 2.3U + 0.3M, 1 )\sp. \\
\end{aligned}
\end{equation}

\begin{equation}\label{appeq:eff_contiZ}
\begin{aligned}
&U \sim\operatorname{Normal}( 1 + A -\expit( 0.3 + 0.2 Z ), 1 ), \\
&Z \sim\operatorname{Normal}(1, 1),\quad A \sim \binomial( \expit( 0.3 + 0.2 Z ) ), \\
&M \sim \binomial( \expit( -1 + A + Z + 2AZ) ),\\
&Y \sim\operatorname{Normal}( U + M, 1 )\sp.
\end{aligned}
\end{equation}

For binary $Z$, results are shown in Fig.~\ref{fig:sim6_binary_interAZ}. The estimator $\psi_{\mathrm{opt}}^{+}(\hat{\Q})$ exhibited lower asymptotic variance than both $\psi_{z^*=1}^+(\hat{\Q})$ and $\psi_{z^*=0}^+(\hat{\Q})$. It also reduced the variance by 1.7-fold compared with the one-step ``non-Verma" estimator, i.e., the one-step estimator $\psi_{1}^+$ from Section~\ref{sec:est_ATE} where $Z$ is considered as part of baseline covariates. 

For continuous $Z$, results are shown in Fig.~\ref{fig:sim6_continuous_interAZ}. We considered three choices for $\tilde{p}(Z)$: the true density $p(Z)$, $\operatorname{Uniform}(-1,1)$, and a truncated Normal distribution $\operatorname{TN}(0.6,0.7)$ with truncation at the $0.1\%$ and $99.9\%$ quantiles of $\operatorname{Normal}(1,1)$. The truncated Normal choice yielded the lowest variance of $3.07$, corresponding to a 23-fold reduction relative to the estimator using $\p(Z)$ and a 7-fold reduction relative to the one-step non-Verma estimator, i.e.,  the one-step estimator $\psi_{1}^+$ from Section~\ref{sec:est_ATE} where $Z$ is considered as part of baseline covariates. . The estimator using the Uniform distribution achieved the second lowest variance of $5.53$.

\pagebreak
\section{Details on real data application}
\label{app:real} 

\subsection{Effect of mobile stroke unit care on functional outcomes} 
\label{app:real_data_msu_dispatch}

In \citep{piccininni2023effect}, only single mediator variable $M_2$ was adopted, and $M_2$ was categorized into three categories for easier estimation. Categorization was achieved using $M_2$'s first quantile and median value as (1) $\leq 48$ minutes (1st quantile), (2) $48-75$ minutes (between 1st quantile and median), and (3) $>75$ minutes (median) or no thrombolysis received. To compare with their result, we conducted the analysis under various scenarios by handling the outcome and mediator variables in various ways. First, we treated both $M_2$ and $Y$ as continuous. Then, we binarized them using different cut-off points. For the outcome $Y$, we applied three cut-off points: 2 (slight disability), 3 (moderate disability), and 4 (moderately severe disability). The binarization was as follows: (1) slight disability or less ($Y=0$) vs. worse than slight disability ($Y=1$), (2) moderate disability or less ($Y=0$) vs. worse than moderate disability, and (3) moderately severe disability or less ($Y=0$) vs. worse than moderately severe disability ($Y=1$). For the mediator $M_2$, we used two cut-offs: (1) $\leq 48$ minutes ($M=0$) vs. $>48$ minutes ($M=1$), and (2) $\leq 75$ minutes ($M=0$) vs. $>75$ minutes or no thrombolysis received ($M=1$). This resulted in six binary outcome-mediator scenarios.

We evaluated both ATE and ATT across all scenarios. For the ATE, we used the one-step estimator $\psi^+_{1}(\hat{\Q})$ and TMLE $\psi_{1}(\hat{\Q}^\star)$ when $M_2$ was binarized, and $\psi^+_{2b}(\hat{\Q})$ and  $\psi_{2b}(\hat{\Q}^\star)$ when $M_2$ was treated as continuous. For the ATT, we used the corresponding one-step and TMLE estimators, $\beta^+_{1}(\hat{\Q})$ and $\beta_{1}(\hat{\Q}^\star)$ for binary $M_2$, and $\beta^+_{b}(\hat{\Q})$ and $\beta_{b}(\hat{\Q}^\star)$ for continuous $M_2$. Super learner with five-fold cross-fitting was adopted for nuisance estimation to capture potential complex relationships among variables, such as interactions and nonlinear relationships. The super learner’s candidate algorithms included intercept-only regression, generalized linear models, multivariate adaptive regression splines, and random forests. Missing data was handled with 10-fold multiple imputations. 

Our analysis yields ATE estimates indicating that adopting MSU care improves patients’ 3-month functional outcomes, consistent with the findings of \citep{piccininni2023effect}. The ATT estimates are uniformly negative, suggesting that the subpopulation receiving MSU care benefits more from the treatment than the overall population. These conclusions hold across different approaches to handling $M$ and $Y$, although statistical significance varies. TMLE and one-step estimators produce consistent and comparable results across all analyses. In addition, larger effect sizes are observed at lower cutoff values of $Y$ for both ATE and ATT, indicating that the benefits of MSU care are more pronounced in reducing mild disabilities, with the treated subpopulation experiencing greater gains than the overall population. Detailed estimates for both the ATE and ATT under each approach are reported in Appendix Table~\ref{apptable:bproud}.

\vspace{0.25cm}
\noindent\textbf{ATT estimates under the main modeling approach:}
In the main analysis, variables were modeled in their original forms, with $M_1$ treated as binary, $M_2$ as continuous, and $Y$ as ordinal. To facilitate comparison with estimates in Appendix Table~\ref{apptable:bproud} and the ATE estimates presented in the main text, we estimated the ATT under the same specification. Using the one-step estimator $\beta_{b}^{+}(\hat{\Q})$, we obtained an estimate of $-0.199$ (95\% CI: $(-0.392,-0.006)$), and the TMLE counterpart $\beta_{b}(\hat{\Q}^\star)$ yielded $-0.169$ (95\% CI: $(-0.406,0.069)$). These results are consistent with the ATT estimates reported in Appendix Table~\ref{apptable:bproud}. The magnitude of the effect varies across modeling approaches, with the main specification and the approach that treats $M_2$ and $Y$ as continuous variables yielding larger effects than approaches that binarize these variables.


\subsection{Effect of academic performance on future annual income} 
\label{app:real_data_academic}

Utilizing our front-door estimation framework, we investigated how early academic achievements influence future annual income. The data for this analysis was sourced from the Life Course Study, which spans from 1971 to 2002 and are publicly available through the Finnish Social Science Data Archive \citep{fsd}. These data originate from a longitudinal study of $634$ individuals born between 1964 and 1968 in Jyväskylä, Finland. The study aimed to understand how abilities, social background, and educational achievements shape an individual's life path. The data collection occurred in four phases. The first phase in the 1970s gathered initial information such as age, gender, family socioeconomic status, and results from the Illinois Test of Psycholinguistic Abilities (ITPA), assessing verbal intelligence in Finnish children aged 3-9. The second phase in the 1980s focused on academic achievements and performance. In 1991, the third phase collected data on occupational progress and higher education choices of the participants. Finally, the 2002 phase, as the subjects neared middle age, involved collecting information on their income, educational levels, and occupational status.

We were interested in estimating the causal effect of early academic performance ($A$) on an individual's annual income ($Y$). We used a binary measure of academic performance based on whether an individual's sixth-grade all-subject grade averages were above or below the median for the population. Our hypothesis is that early academic performance influences annual income by shaping educational and career paths, quantifiable through eight mediators ($M_1-M_8$), detailed in Table~\ref{tab:realdata}. We also controlled for family socio-economic status, intelligence (measured by ITPA score), age, and gender ($X_1-X_4$). 

We estimated both the ATE and the ATTE to assess the effect in the overall population as well as among individuals with above-median early academic performance. 

Given the dimension of the mediators and due to the fact that the mediators include binary, categorical, and continuous-valued variables, we elected to use our proposed estimators that avoid mediator density estimation. Due to the potential for interactions and non-linear relationships, we wished to estimate nuisance parameters flexibly, and thus adopted a super learner approach combined with 5 folds cross-fitting. The candidate estimators included in the super learner include intercept-only regression, generalized linear models, multivariate adaptive regression splines, random forests, and XGBoost. For simplicity, we managed missing data in the variables mentioned by employing  single imputation.


Our analysis underscores the role of strong academic foundations in shaping future income, likely mediated through higher educational attainment and more advantageous career paths. However, the interpretation of these estimates depends on the validity of the no direct effect assumption—namely, that the effect of academic performance on income operates entirely through the eight measured mediators ($M_1$–$M_8$). Due to the lack of a valid anchor variable in this application, we cannot empirically test the front-door assumptions required for identifying the ATE. As such, we present two interpretations based on whether the no direct effect assumption is believed to hold.

If the assumption holds, our TMLE estimator for ATE, $\psi_{2b}(\hat{\Q}^\star)$, indicates that individuals with above-median academic performance in early stages earn, on average, \texteuro{}3239.18 more in future annual income (95\% CI: \texteuro{}725.35, \texteuro{}5753.00) than their below-median counterparts. The one-step estimator $\psi_{2b}^+(\hat{\Q}^\star)$ provides a similar estimate of \texteuro{}3378.29 (95\% CI: \texteuro{}857.74, \texteuro{}5898.84). 

Focusing on the subpopulation with above-median early academic performance (i.e., the ATT), the TMLE estimator $\beta_b(\hat{\Q}^\star)$ suggests that these individuals earn on average \texteuro{}2,671.88 more in future annual income (95\% CI: \texteuro{}627.30, \texteuro{}4716.46) than they would have earned had they exhibited below-median performance. The one-step estimator, $\beta_b(\hat{\Q})$, provides a comparable estimate of \texteuro{}2,572.98 (95\% CI: \texteuro{}898.80, \texteuro{}4,247.16).

If the full mediation assumption (i.e., no direct effect of $A$ on $Y$) is violated—such as when academic performance influences income through unmeasured pathways—then neither the ATE nor the ATT is identifiable and the reported effects may be biased. In such settings, one may instead consider the PIIE and PIIE-T estimands. The PIIE captures the effect of shifting the observed mediators under an intervention in the overall population (see Section~\ref{sec:prelim}), whereas the PIIE-T captures the effect of a control intervention within the treated subpopulation (see Appendix~\ref{app:alternative_interpretations}), which in our case consists of individuals with above-median academic performance. 

The PIIE is numerically identical to the ATE,  but it has a distinct interpretation. The TMLE estimate suggests that shifting everyone’s educational and career paths to the values they would have taken under above-median academic performance would increase the average income by \texteuro{}1380.37 (95\% CI: \texteuro{}-360.72, \texteuro{}3121.45), relative to the observed average income. Conversely, shifting everyone to the mediator values corresponding to below-median academic performance would decrease average income by \texteuro{}1858.81 (95\% CI: \texteuro{}596.41, \texteuro{}3121.22), compared to the observed average income. The one-step estimator yields similar results, with an estimated increase of \texteuro{}1529.63 (95\% CI: \texteuro{}-257.07, \texteuro{}3316.33) under above-median academic performance and a decrease of \texteuro{}1848.66 (95\% CI: \texteuro{}638.75, \texteuro{}3058.57) under below-median performance. 

The PIIE-T is numerically identical to the ATT reported above, but it has a distinct interpretation. The TMLE estimate indicates that if the educational and career trajectories of individuals with above-median academic performance were shifted to the values they would have taken under below-median performance, their average income would decrease by \texteuro{}2,671.88 relative to the observed level. The one-step estimator admits the same interpretation.

\clearpage

\pagebreak  

\begin{figure}[t] 
	\begin{center}
    \scalebox{0.45}{
    \begin{tikzpicture}[>=stealth, node distance=1.7cm]
        \tikzstyle{format} = [thick, circle, minimum size=1.0mm, inner sep=2pt]
        \tikzstyle{square} = [draw, thick, minimum size=4.5mm, inner sep=2pt]
    
    \begin{scope}[xshift=0cm, yshift=0cm]
		\path[->, thick]
		
		node[] (a) {$A$}
            node[left of=a] (z) {$Z$}
            node[above left of=z] (x) {$X$}
		node[right of=a] (m) {$M$}
		node[above of=m, yshift=-0.2cm] (u) {\red{$U_1$}}
		node[right of=m] (y) {$Y$}

            (x) edge[blue] (z)
            (x) edge[blue, bend left=0] (a)
            (x) edge[blue, bend left=10] (m)
            (x) edge[blue, bend left=20] (y)
            (z) edge[blue] (a)
            (z) edge[blue,bend left=25] (m)
		(a) edge[blue] (m) 
		(m) edge[blue] (y) 
		(u) edge[blue] (y) 
		(u) edge[blue] (a)

        node[below of=a, xshift=0cm, yshift=0.85cm] (t3) {(DAG1)} ; 
		
	\end{scope}
   
	\begin{scope}[xshift=8cm, yshift=0cm]
		\path[->, thick]
		
		node[] (a) {$A$}
            node[left of=a] (z) {$Z$}
            node[above left of=z] (x) {$X$}
		node[right of=a] (m) {$M$}
		node[above of=m, yshift=-0.2cm] (u1) {\red{$U_1$}}
		node[right of=m] (y) {$Y$}
            node[above of=a, yshift=-0.2cm] (u2) {\red{$U_2$}}

            (x) edge[blue] (z)
            (x) edge[blue, bend left=0] (a)
            (x) edge[blue, bend left=10] (m)
            (x) edge[blue, bend left=20] (y)
            (z) edge[blue] (a)
		(a) edge[blue] (m) 
            (u2) edge[blue] (m)
            (u2) edge[blue] (z)
		(m) edge[blue] (y) 
		(u1) edge[blue] (y) 
		(u1) edge[blue] (a)

        node[below of=a, xshift=0cm, yshift=0.85cm] (t3) {(DAG2)} ; 
		
	\end{scope}
				
	\begin{scope}[xshift=16cm, yshift=0cm]
		\path[->, thick]
		
		node[] (a) {$A$}
            node[left of=a] (z) {$Z$}
            node[above left of=z] (x) {$X$}
		node[right of=a] (m) {$M$}
		node[above of=m, yshift=-0.2cm] (u1) {\red{$U_1$}}
		node[right of=m] (y) {$Y$}
            node[above of=a, yshift=-0.2cm] (u2) {\red{$U_2$}}

            (x) edge[blue] (z)
            (x) edge[blue, bend left=0] (a)
            (x) edge[blue, bend left=10] (m)
            (x) edge[blue, bend left=20] (y)
            (z) edge[blue] (a)
		(a) edge[blue] (m) 
            (u2) edge[blue] (a)
            (u2) edge[blue] (y)
		(m) edge[blue] (y) 
		(u1) edge[blue] (y)
            (u1) edge[blue] (m)
		(u1) edge[blue] (a)

        node[below of=a, xshift=0cm, yshift=0.85cm] (t3) {(DAG3)} ;
		
	\end{scope}

    \begin{scope}[xshift=24cm, yshift=0cm]
		\path[->, thick]
		
		node[] (a) {$A$}
            node[left of=a] (z) {$Z$}
            node[above left of=z] (x) {$X$}
		node[right of=a] (m) {$M$}
		node[above of=m, yshift=-0.2cm] (u) {\red{$U_1$}}
		node[right of=m] (y) {$Y$}

            (x) edge[blue] (z)
            (x) edge[blue, bend left=0] (a)
            (x) edge[blue, bend left=10] (m)
            (x) edge[blue, bend left=20] (y)
            (z) edge[blue] (a)
            (z) edge[blue,bend left=25] (m)
		(a) edge[blue] (m) 
		(m) edge[blue] (y) 
		(u) edge[blue] (y) 
		(u) edge[blue] (a)
            (a) edge[blue,bend right=25] (y)

        node[below of=a, xshift=0cm, yshift=0.85cm] (t3) {(DAG4)} ; 
		
	\end{scope}
	\end{tikzpicture}
	}
	\caption{DAGs used in simulations on model evaluations: DAG1 and DAG2 correspond to scenarios where the front-door assumptions hold, while DAG3 and DAG4 depict scenarios where the assumptions are violated.} 
	\label{fig:test_graphs}
	\end{center}
\end{figure}

\begin{figure}[t]
    \centering
    \includegraphics[width=0.9\linewidth, clip, trim=0 5 0 5]{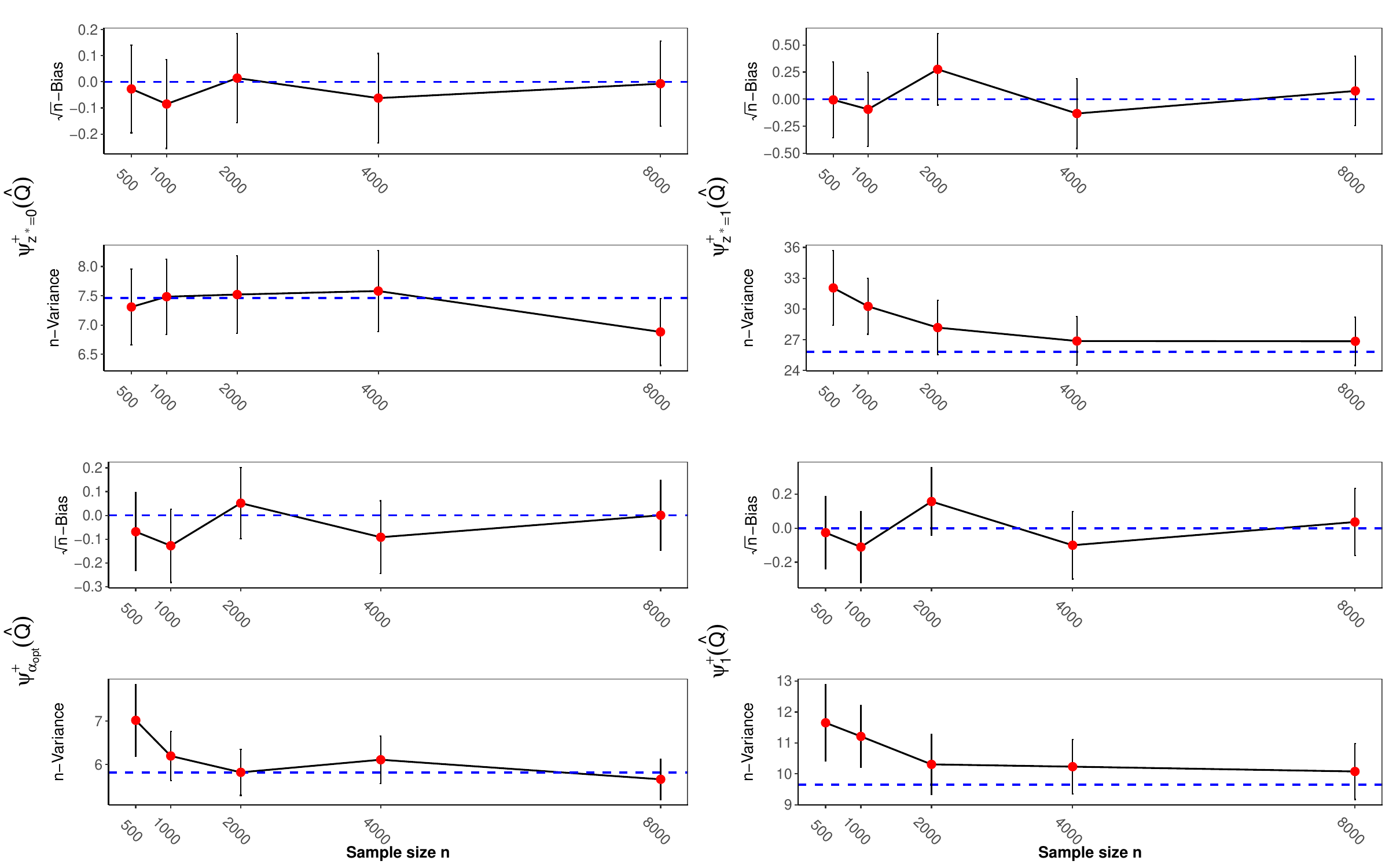}
    \caption{Simulation results demonstrating efficiency gains in ATE estimation under binary $Z$ when utilizing the Verma constraint.}
    \label{fig:sim6_binary_interAZ}
\end{figure}

\begin{figure}[t]
    \centering
    \includegraphics[width=0.9\linewidth, clip, trim=0 5 0 5]{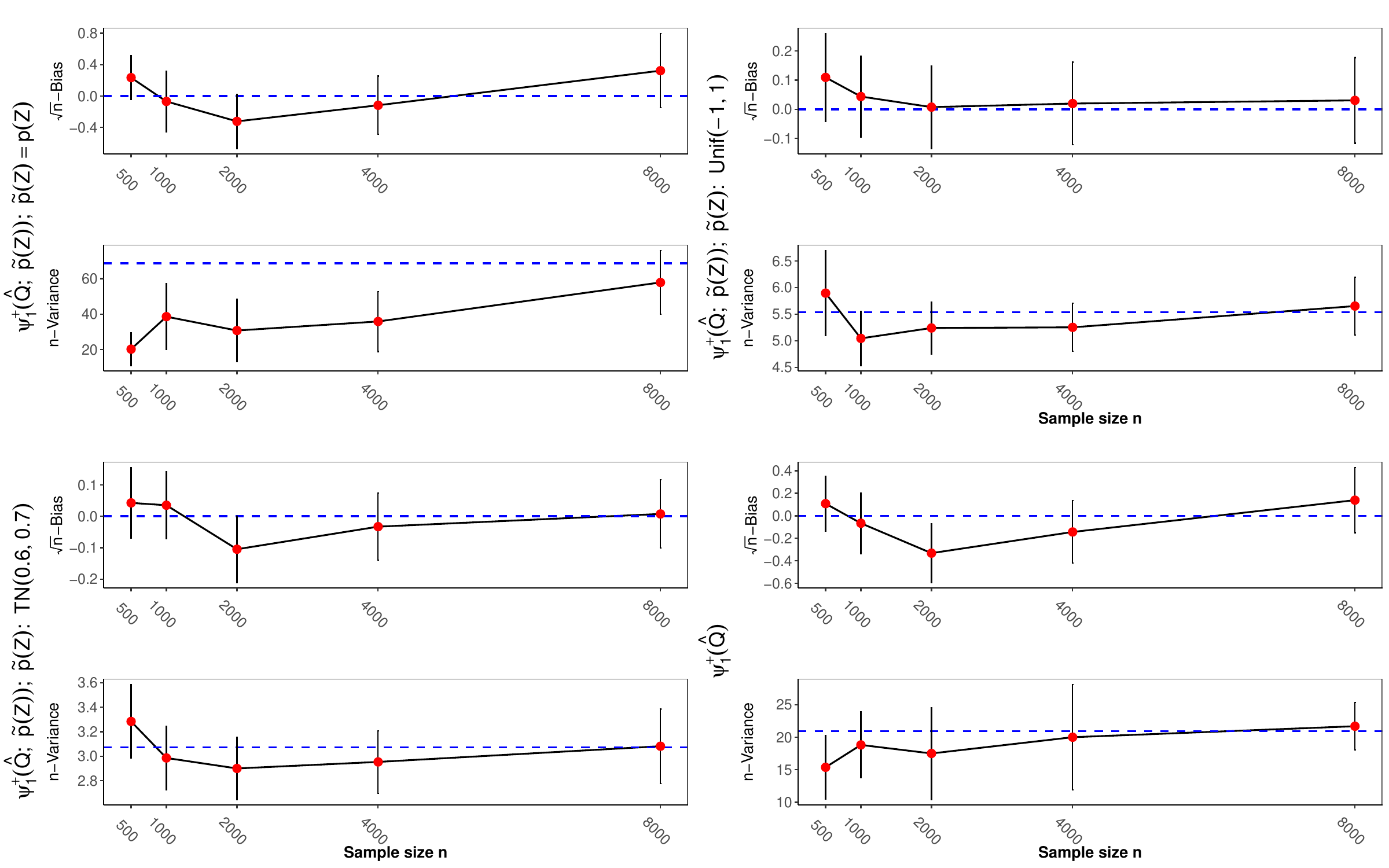}
        \caption{Simulation results demonstrating efficiency gains in ATE estimation under continuous $Z$ when utilizing the Verma constraint. Here, ``TN'' denotes a truncated Normal distribution with mean $0.6$ and standard deviation $0.7$, truncated at the $0.1\%$ and $99.9\%$ quantiles of $\operatorname{Normal}(1,1)$.}
    \label{fig:sim6_continuous_interAZ}
\end{figure}

\clearpage

  \providecommand{\huxb}[2]{\arrayrulecolor[RGB]{#1}\global\arrayrulewidth=#2pt}
  \providecommand{\huxvb}[2]{\color[RGB]{#1}\vrule width #2pt}
  \providecommand{\huxtpad}[1]{\rule{0pt}{#1}}
  \providecommand{\huxbpad}[1]{\rule[-#1]{0pt}{#1}}

{
\renewcommand{\huxtpad}[1]{\rule{0pt}{0.3ex}}
\renewcommand{\huxbpad}[1]{\rule[-0.05ex]{0pt}{0.05ex}}
\renewcommand{\arraystretch}{0.6}
\begin{table}[H]
\begin{center}
\captionsetup{justification=centering,singlelinecheck=off}
\caption{Performance of ATE TMLEs under linear vs. expit outcome submodels across mediator types.}
\setlength{\tabcolsep}{0pt}
\resizebox{\textwidth}{!}{
\begin{tabular}{l l l l l l l l l l l l l l l l}
\noalign{\hrule height 1pt} 


\multicolumn{1}{!{\huxvb{0, 0, 0}{0}}l!{\huxvb{0, 0, 0}{0}}}{\huxtpad{1pt + 1em}\raggedright \hspace{1pt} \textbf{{\fontsize{8pt}{9.6pt}\selectfont }} \hspace{1pt}\huxbpad{1pt}} &
\multicolumn{1}{c!{\huxvb{0, 0, 0}{0}}}{\huxtpad{1pt + 1em}\centering \hspace{1pt} \textbf{{\fontsize{8pt}{9.6pt}\selectfont }} \hspace{1pt}\huxbpad{1pt}} &
\multicolumn{2}{c!{\huxvb{0, 0, 0}{0}}}{\huxtpad{1pt + 1em}\centering \hspace{1pt} \textbf{{\fontsize{8pt}{9.6pt}\selectfont Univariate Binary}} \hspace{1pt}\huxbpad{1pt}} &
\multicolumn{6}{c!{\huxvb{0, 0, 0}{0}}}{\huxtpad{1pt + 1em}\centering \hspace{1pt} \textbf{{\fontsize{8pt}{9.6pt}\selectfont Univariate Continuous}} \hspace{1pt}\huxbpad{1pt}} &
\multicolumn{6}{c!{\huxvb{0, 0, 0}{0}}}{\huxtpad{1pt + 1em}\centering \hspace{1pt} \textbf{{\fontsize{8pt}{9.6pt}\selectfont Bivariate Continuous}} \hspace{1pt}\huxbpad{1pt}} \tabularnewline
\end{tabular}}\label{table:TMLEs_ATE}
\par\end{center}

\end{table}
}

  \providecommand{\huxb}[2]{\arrayrulecolor[RGB]{#1}\global\arrayrulewidth=#2pt}
  \providecommand{\huxvb}[2]{\color[RGB]{#1}\vrule width #2pt}
  \providecommand{\huxtpad}[1]{\rule{0pt}{#1}}
  \providecommand{\huxbpad}[1]{\rule[-#1]{0pt}{#1}}

{
\renewcommand{\huxtpad}[1]{\rule{0pt}{0.3ex}}
\renewcommand{\huxbpad}[1]{\rule[-0.05ex]{0pt}{0.05ex}}
\renewcommand{\arraystretch}{0.6}
\begin{table}[H]
\begin{center}
\captionsetup{justification=centering,singlelinecheck=off}
\caption{Performance of ATT TMLEs under linear vs. expit outcome submodels across mediator types.}
 \setlength{\tabcolsep}{0pt}
\resizebox{\textwidth}{!}{
\begin{tabular}{l l l l l l l l l l l l l l l l}

\noalign{\hrule height 1pt}

\multicolumn{1}{!{\huxvb{0, 0, 0}{0}}l!{\huxvb{0, 0, 0}{0}}}{\huxtpad{1pt + 1em}\raggedright \hspace{1pt} \textbf{{\fontsize{8pt}{9.6pt}\selectfont }} \hspace{1pt}\huxbpad{1pt}} &
\multicolumn{1}{c!{\huxvb{0, 0, 0}{0}}}{\huxtpad{1pt + 1em}\centering \hspace{1pt} \textbf{{\fontsize{8pt}{9.6pt}\selectfont }} \hspace{1pt}\huxbpad{1pt}} &
\multicolumn{2}{c!{\huxvb{0, 0, 0}{0}}}{\huxtpad{1pt + 1em}\centering \hspace{1pt} \textbf{{\fontsize{8pt}{9.6pt}\selectfont Univariate Binary}} \hspace{1pt}\huxbpad{1pt}} &
\multicolumn{6}{c!{\huxvb{0, 0, 0}{0}}}{\huxtpad{1pt + 1em}\centering \hspace{1pt} \textbf{{\fontsize{8pt}{9.6pt}\selectfont Univariate Continuous}} \hspace{1pt}\huxbpad{1pt}} &
\multicolumn{6}{c!{\huxvb{0, 0, 0}{0}}}{\huxtpad{1pt + 1em}\centering \hspace{1pt} \textbf{{\fontsize{8pt}{9.6pt}\selectfont Bivariate Continuous}} \hspace{1pt}\huxbpad{1pt}} \tabularnewline
\end{tabular}}\label{table:TMLEs_ATT}
\par\end{center}

\end{table}
}

  \providecommand{\huxb}[2]{\arrayrulecolor[RGB]{#1}\global\arrayrulewidth=#2pt}
  \providecommand{\huxvb}[2]{\color[RGB]{#1}\vrule width #2pt}
  \providecommand{\huxtpad}[1]{\rule{0pt}{#1}}
  \providecommand{\huxbpad}[1]{\rule[-#1]{0pt}{#1}}

{\renewcommand{\huxtpad}[1]{\rule{0pt}{0.3ex}}
\renewcommand{\huxbpad}[1]{\rule[-0.05ex]{0pt}{0.05ex}}
\renewcommand{\arraystretch}{0.9}
\begin{table}[t]
\begin{center}

\captionsetup{justification=centering,singlelinecheck=off}
\caption{Comparison of ATT TMLE and one-step estimators under weak overlap across mediator types.}
 \setlength{\tabcolsep}{0pt}
\resizebox{\textwidth}{!}{
\begin{tabular}{l l l l l l l l l l l l l l}

\noalign{\hrule height 1pt} 

\multicolumn{1}{!{\huxvb{0, 0, 0}{0}}l!{\huxvb{0, 0, 0}{0}}}{\huxtpad{1pt + 1em}\raggedright \hspace{1pt} \textbf{{\fontsize{6pt}{7.2pt}\selectfont }} \hspace{1pt}\huxbpad{1pt}} &
\multicolumn{1}{c!{\huxvb{0, 0, 0}{0}}}{\huxtpad{1pt + 1em}\centering \hspace{1pt} {\fontsize{6pt}{7.2pt}\selectfont } \hspace{1pt}\huxbpad{1pt}} &
\multicolumn{2}{c!{\huxvb{0, 0, 0}{0}}}{\huxtpad{1pt + 1em}\centering \hspace{1pt} {\fontsize{6pt}{7.2pt}\selectfont Univariate Binary} \hspace{1pt}\huxbpad{1pt}} &
\multicolumn{6}{c!{\huxvb{0, 0, 0}{0}}}{\huxtpad{1pt + 1em}\centering \hspace{1pt} {\fontsize{6pt}{7.2pt}\selectfont Univariate Continuous} \hspace{1pt}\huxbpad{1pt}} &
\multicolumn{4}{c!{\huxvb{0, 0, 0}{0}}}{\huxtpad{1pt + 1em}\centering \hspace{1pt} {\fontsize{6pt}{7.2pt}\selectfont Bivariate Continuous} \hspace{1pt}\huxbpad{1pt}} \tabularnewline
\end{tabular}}\label{table:weakoverlap_ATT}
\par\end{center}

\end{table}
}

\providecommand{\huxb}[2]{\arrayrulecolor[RGB]{#1}\global\arrayrulewidth=#2pt}
  \providecommand{\huxvb}[2]{\color[RGB]{#1}\vrule width #2pt}
  \providecommand{\huxtpad}[1]{\rule{0pt}{#1}}
  \providecommand{\huxbpad}[1]{\rule[-#1]{0pt}{#1}}

\begin{table}[t]
\begin{center}

\captionsetup{justification=centering,singlelinecheck=off}
\caption{Performance of ATT estimators under model misspecifications across mediator types.}
 \setlength{\tabcolsep}{0pt}
\resizebox{\textwidth}{!}{
\begin{tabular}{l l l l l l l l l l l l l l l l l l l l}

\thickline

\multicolumn{1}{!{\huxvb{0, 0, 0}{0}}l!{\huxvb{0, 0, 0}{0}}}{\huxtpad{-1pt + 1em}\raggedright \hspace{-1pt} \textbf{{\fontsize{7pt}{8.4pt}\selectfont }} \hspace{-1pt}\huxbpad{-1pt}} &
\multicolumn{1}{c!{\huxvb{0, 0, 0}{0}}}{\huxtpad{-1pt + 1em}\centering \hspace{-1pt} \textbf{{\fontsize{7pt}{8.4pt}\selectfont }} \hspace{-1pt}\huxbpad{-1pt}} &
\multicolumn{9}{c!{\huxvb{0, 0, 0}{0}}}{\huxtpad{-1pt + 1em}\centering \hspace{-1pt} \textbf{{\fontsize{7pt}{8.4pt}\selectfont TMLEs}} \hspace{-1pt}\huxbpad{-1pt}} &
\multicolumn{9}{c!{\huxvb{0, 0, 0}{0}}}{\huxtpad{-1pt + 1em}\centering \hspace{-1pt} \textbf{{\fontsize{7pt}{8.4pt}\selectfont One-step estimators}} \hspace{-1pt}\huxbpad{-1pt}} \tabularnewline
\end{tabular}}\label{table:misspecification_ATT}
\par\end{center}

\end{table}

  \providecommand{\huxb}[2]{\arrayrulecolor[RGB]{#1}\global\arrayrulewidth=#2pt}
  \providecommand{\huxvb}[2]{\color[RGB]{#1}\vrule width #2pt}
  \providecommand{\huxtpad}[1]{\rule{0pt}{#1}}
  \providecommand{\huxbpad}[1]{\rule[-#1]{0pt}{#1}}

\begin{table}[t]
\begin{center}
\captionsetup{justification=centering,singlelinecheck=off}
\caption{Impact of cross-fitting on ATE TMLE and one-step estimators using random forests (RF: 500 trees; min node size = 5 for continuous, 1 for binary; CF: 5-fold cross-fitting).}
 \setlength{\tabcolsep}{0pt}
\resizebox{\textwidth}{!}{
\begin{tabular}{l l l l l l l l l l l l l l}

\thickline

\multicolumn{1}{!{\huxvb{0, 0, 0}{0}}l!{\huxvb{0, 0, 0}{0}}}{\huxtpad{-1pt + 1em}\raggedright \hspace{-1pt} \textbf{{\fontsize{6.5pt}{7.8pt}\selectfont }} \hspace{-1pt}\huxbpad{-1pt}} &
\multicolumn{1}{c!{\huxvb{0, 0, 0}{0}}}{\huxtpad{-1pt + 1em}\centering \hspace{-1pt} \textbf{{\fontsize{6.5pt}{7.8pt}\selectfont }} \hspace{-1pt}\huxbpad{-1pt}} &
\multicolumn{6}{c!{\huxvb{0, 0, 0}{0}}}{\huxtpad{-1pt + 1em}\centering \hspace{-1pt} \textbf{{\fontsize{6.5pt}{7.8pt}\selectfont TMLEs}} \hspace{-1pt}\huxbpad{-1pt}} &
\multicolumn{6}{c!{\huxvb{0, 0, 0}{0}}}{\huxtpad{-1pt + 1em}\centering \hspace{-1pt} \textbf{{\fontsize{6.5pt}{7.8pt}\selectfont One-step estimators}} \hspace{-1pt}\huxbpad{-1pt}} \tabularnewline
\end{tabular}}\label{table:crossfitting_ATE}
\par\end{center}

\end{table}

\providecommand{\huxb}[2]{\arrayrulecolor[RGB]{#1}\global\arrayrulewidth=#2pt}
  \providecommand{\huxvb}[2]{\color[RGB]{#1}\vrule width #2pt}
  \providecommand{\huxtpad}[1]{\rule{0pt}{#1}}
  \providecommand{\huxbpad}[1]{\rule[-#1]{0pt}{#1}}

\begin{table}[t]
\begin{center}

\captionsetup{justification=centering,singlelinecheck=off}
\caption{Impact of cross-fitting on ATT TMLE and one-step estimators using random forests (RF: 500 trees; min node size = 5 for continuous, 1 for binary; CF: 5-fold cross-fitting).}
 \setlength{\tabcolsep}{0pt}
\resizebox{\textwidth}{!}{
\begin{tabular}{l l l l l l l l l l l l l l}

\thickline

\multicolumn{1}{!{\huxvb{0, 0, 0}{0}}l!{\huxvb{0, 0, 0}{0}}}{\huxtpad{0pt + 1em}\raggedright \hspace{0pt} \textbf{{\fontsize{6.5pt}{7.8pt}\selectfont }} \hspace{0pt}\huxbpad{0pt}} &
\multicolumn{1}{c!{\huxvb{0, 0, 0}{0}}}{\huxtpad{0pt + 1em}\centering \hspace{0pt} \textbf{{\fontsize{6.5pt}{7.8pt}\selectfont }} \hspace{0pt}\huxbpad{0pt}} &
\multicolumn{6}{c!{\huxvb{0, 0, 0}{0}}}{\huxtpad{0pt + 1em}\centering \hspace{0pt} \textbf{{\fontsize{6.5pt}{7.8pt}\selectfont TMLEs}} \hspace{0pt}\huxbpad{0pt}} &
\multicolumn{6}{c!{\huxvb{0, 0, 0}{0}}}{\huxtpad{0pt + 1em}\centering \hspace{0pt} \textbf{{\fontsize{6.5pt}{7.8pt}\selectfont One-step estimators}} \hspace{0pt}\huxbpad{0pt}} \tabularnewline[-0.5pt]

\cline{3-14}

\multicolumn{1}{!{\huxvb{0, 0, 0}{0}}l!{\huxvb{0, 0, 0}{0}}}{\huxtpad{0pt + 1em}\raggedright \hspace{0pt} \textit{\textbf{{\fontsize{6.5pt}{7.8pt}\selectfont }}} \hspace{0pt}\huxbpad{0pt}} &
\multicolumn{1}{c!{\huxvb{0, 0, 0}{0}}}{\huxtpad{0pt + 1em}\centering \hspace{0pt} \textit{{\fontsize{6.5pt}{7.8pt}\selectfont }} \hspace{0pt}\huxbpad{0pt}} &
\multicolumn{2}{c!{\huxvb{0, 0, 0}{0}}}{\huxtpad{0pt + 1em}\centering \hspace{0pt} \textit{{\fontsize{6.5pt}{7.8pt}\selectfont Univariate Binary}} \hspace{0pt}\huxbpad{0pt}} &
\multicolumn{4}{c!{\huxvb{0, 0, 0}{0}}}{\huxtpad{0pt + 1em}\centering \hspace{0pt} \textit{{\fontsize{6.5pt}{7.8pt}\selectfont Univariate Continuous}} \hspace{0pt}\huxbpad{0pt}} &
\multicolumn{2}{c!{\huxvb{0, 0, 0}{0}}}{\huxtpad{0pt + 1em}\centering \hspace{0pt} \textit{{\fontsize{6.5pt}{7.8pt}\selectfont Univariate Binary}} \hspace{0pt}\huxbpad{0pt}} &
\multicolumn{4}{c!{\huxvb{0, 0, 0}{0}}}{\huxtpad{0pt + 1em}\centering \hspace{0pt} \textit{{\fontsize{6.5pt}{7.8pt}\selectfont Univariate Continuous}} \hspace{0pt}\huxbpad{0pt}} \tabularnewline
\end{tabular}}\label{table:crossfitting_ATT}
\par\end{center}

\end{table}

  \providecommand{\huxb}[2]{\arrayrulecolor[RGB]{#1}\global\arrayrulewidth=#2pt}
  \providecommand{\huxvb}[2]{\color[RGB]{#1}\vrule width #2pt}
  \providecommand{\huxtpad}[1]{\rule{0pt}{#1}}
  \providecommand{\huxbpad}[1]{\rule[-#1]{0pt}{#1}}

\begin{table}[t]
\begin{center}

\captionsetup{justification=centering,singlelinecheck=off}
\caption{Impact of cross-fitting on ATE TMLE and one-step estimators using random forests (RF: 200 trees; min node size = 1; CF: 5-fold cross-fitting).}
 \setlength{\tabcolsep}{0pt}
\resizebox{\textwidth}{!}{
\begin{tabular}{l l l l l l l l l l l l l l}

\thickline

\multicolumn{1}{!{\huxvb{0, 0, 0}{0}}l!{\huxvb{0, 0, 0}{0}}}{\huxtpad{0pt + 1em}\raggedright \hspace{0pt} \textbf{{\fontsize{6.5pt}{7.8pt}\selectfont }} \hspace{0pt}\huxbpad{0pt}} &
\multicolumn{1}{c!{\huxvb{0, 0, 0}{0}}}{\huxtpad{0pt + 1em}\centering \hspace{0pt} \textbf{{\fontsize{6.5pt}{7.8pt}\selectfont }} \hspace{0pt}\huxbpad{0pt}} &
\multicolumn{6}{c!{\huxvb{0, 0, 0}{0}}}{\huxtpad{0pt + 1em}\centering \hspace{0pt} \textbf{{\fontsize{6.5pt}{7.8pt}\selectfont TMLEs}} \hspace{0pt}\huxbpad{0pt}} &
\multicolumn{6}{c!{\huxvb{0, 0, 0}{0}}}{\huxtpad{0pt + 1em}\centering \hspace{0pt} \textbf{{\fontsize{6.5pt}{7.8pt}\selectfont One-step estimators}} \hspace{0pt}\huxbpad{0pt}} \tabularnewline[-0.5pt]

\cline{3-14}

\multicolumn{1}{!{\huxvb{0, 0, 0}{0}}l!{\huxvb{0, 0, 0}{0}}}{\huxtpad{0pt + 1em}\raggedright \hspace{0pt} \textit{\textbf{{\fontsize{6.5pt}{7.8pt}\selectfont }}} \hspace{0pt}\huxbpad{0pt}} &
\multicolumn{1}{c!{\huxvb{0, 0, 0}{0}}}{\huxtpad{0pt + 1em}\centering \hspace{0pt} \textit{{\fontsize{6.5pt}{7.8pt}\selectfont }} \hspace{0pt}\huxbpad{0pt}} &
\multicolumn{2}{c!{\huxvb{0, 0, 0}{0}}}{\huxtpad{0pt + 1em}\centering \hspace{0pt} \textit{{\fontsize{6.5pt}{7.8pt}\selectfont Univariate Binary}} \hspace{0pt}\huxbpad{0pt}} &
\multicolumn{4}{c!{\huxvb{0, 0, 0}{0}}}{\huxtpad{0pt + 1em}\centering \hspace{0pt} \textit{{\fontsize{6.5pt}{7.8pt}\selectfont Univariate Continuous}} \hspace{0pt}\huxbpad{0pt}} &
\multicolumn{2}{c!{\huxvb{0, 0, 0}{0}}}{\huxtpad{0pt + 1em}\centering \hspace{0pt} \textit{{\fontsize{6.5pt}{7.8pt}\selectfont Univariate Binary}} \hspace{0pt}\huxbpad{0pt}} &
\multicolumn{4}{c!{\huxvb{0, 0, 0}{0}}}{\huxtpad{0pt + 1em}\centering \hspace{0pt} \textit{{\fontsize{6.5pt}{7.8pt}\selectfont Univariate Continuous}} \hspace{0pt}\huxbpad{0pt}} \tabularnewline
\end{tabular}}\label{apptable:crossfitting_ate}
\par\end{center}

\end{table}

  \providecommand{\huxb}[2]{\arrayrulecolor[RGB]{#1}\global\arrayrulewidth=#2pt}
  \providecommand{\huxvb}[2]{\color[RGB]{#1}\vrule width #2pt}
  \providecommand{\huxtpad}[1]{\rule{0pt}{#1}}
  \providecommand{\huxbpad}[1]{\rule[-#1]{0pt}{#1}}

\begin{table}[t]
\begin{center}

\captionsetup{justification=centering,singlelinecheck=off}
\caption{Impact of cross-fitting on ATT TMLE and one-step estimators using random forests (RF: 200 trees; min node size = 1; CF: 5-fold cross-fitting).}
 \setlength{\tabcolsep}{0pt}
\resizebox{\textwidth}{!}{
\begin{tabular}{l l l l l l l l l l l l l l}

\thickline

\multicolumn{1}{!{\huxvb{0, 0, 0}{0}}l!{\huxvb{0, 0, 0}{0}}}{\huxtpad{0pt + 1em}\raggedright \hspace{0pt} \textbf{{\fontsize{6.5pt}{7.8pt}\selectfont }} \hspace{0pt}\huxbpad{0pt}} &
\multicolumn{1}{c!{\huxvb{0, 0, 0}{0}}}{\huxtpad{0pt + 1em}\centering \hspace{0pt} \textbf{{\fontsize{6.5pt}{7.8pt}\selectfont }} \hspace{0pt}\huxbpad{0pt}} &
\multicolumn{6}{c!{\huxvb{0, 0, 0}{0}}}{\huxtpad{0pt + 1em}\centering \hspace{0pt} \textbf{{\fontsize{6.5pt}{7.8pt}\selectfont TMLEs}} \hspace{0pt}\huxbpad{0pt}} &
\multicolumn{6}{c!{\huxvb{0, 0, 0}{0}}}{\huxtpad{0pt + 1em}\centering \hspace{0pt} \textbf{{\fontsize{6.5pt}{7.8pt}\selectfont One-step estimators}} \hspace{0pt}\huxbpad{0pt}} \tabularnewline[-0.5pt]

\cline{3-14}

\multicolumn{1}{!{\huxvb{0, 0, 0}{0}}l!{\huxvb{0, 0, 0}{0}}}{\huxtpad{0pt + 1em}\raggedright \hspace{0pt} \textit{\textbf{{\fontsize{6.5pt}{7.8pt}\selectfont }}} \hspace{0pt}\huxbpad{0pt}} &
\multicolumn{1}{c!{\huxvb{0, 0, 0}{0}}}{\huxtpad{0pt + 1em}\centering \hspace{0pt} \textit{{\fontsize{6.5pt}{7.8pt}\selectfont }} \hspace{0pt}\huxbpad{0pt}} &
\multicolumn{2}{c!{\huxvb{0, 0, 0}{0}}}{\huxtpad{0pt + 1em}\centering \hspace{0pt} \textit{{\fontsize{6.5pt}{7.8pt}\selectfont Univariate Binary}} \hspace{0pt}\huxbpad{0pt}} &
\multicolumn{4}{c!{\huxvb{0, 0, 0}{0}}}{\huxtpad{0pt + 1em}\centering \hspace{0pt} \textit{{\fontsize{6.5pt}{7.8pt}\selectfont Univariate Continuous}} \hspace{0pt}\huxbpad{0pt}} &
\multicolumn{2}{c!{\huxvb{0, 0, 0}{0}}}{\huxtpad{0pt + 1em}\centering \hspace{0pt} \textit{{\fontsize{6.5pt}{7.8pt}\selectfont Univariate Binary}} \hspace{0pt}\huxbpad{0pt}} &
\multicolumn{4}{c!{\huxvb{0, 0, 0}{0}}}{\huxtpad{0pt + 1em}\centering \hspace{0pt} \textit{{\fontsize{6.5pt}{7.8pt}\selectfont Univariate Continuous}} \hspace{0pt}\huxbpad{0pt}} \tabularnewline
\end{tabular}}\label{apptable:crossfitting_att}
\par\end{center}

\end{table}

\providecommand{\huxb}[2]{\arrayrulecolor[RGB]{#1}\global\arrayrulewidth=#2pt}
  \providecommand{\huxvb}[2]{\color[RGB]{#1}\vrule width #2pt}
  \providecommand{\huxtpad}[1]{\rule{0pt}{#1}}
  \providecommand{\huxbpad}[1]{\rule[-#1]{0pt}{#1}}

{\renewcommand{\huxtpad}[1]{\rule{0pt}{0.3ex}}
\renewcommand{\huxbpad}[1]{\rule[-0.05ex]{0pt}{0.05ex}}
\renewcommand{\arraystretch}{0.6}
\begin{table}[H]
\begin{center}

\captionsetup{justification=centering,singlelinecheck=off}
\caption{Comparisons of the CCM, dual, and primal tests on type I error and power under model misspecification.}
 \setlength{\tabcolsep}{0pt}
\resizebox{0.7\textwidth}{!}{
\begin{tabular}{l l l l l l l l l l l l l}

\hhline{>{\huxb{0, 0, 0}{1}}->{\huxb{0, 0, 0}{1}}->{\huxb{0, 0, 0}{1}}->{\huxb{0, 0, 0}{1}}->{\huxb{0, 0, 0}{1}}->{\huxb{0, 0, 0}{1}}->{\huxb{0, 0, 0}{1}}->{\huxb{0, 0, 0}{1}}->{\huxb{0, 0, 0}{1}}->{\huxb{0, 0, 0}{1}}->{\huxb{0, 0, 0}{1}}->{\huxb{0, 0, 0}{1}}->{\huxb{0, 0, 0}{1}}-}

\multicolumn{1}{!{\huxvb{0, 0, 0}{0}}l!{\huxvb{0, 0, 0}{0}}}{\huxtpad{-1pt + 1em}\raggedright \hspace{-1pt} \textbf{{\fontsize{8pt}{9.6pt}\selectfont }} \hspace{-1pt}\huxbpad{-1pt}} &
\multicolumn{4}{c!{\huxvb{0, 0, 0}{0}}}{\huxtpad{-1pt + 1em}\centering \hspace{-1pt} \textbf{{\fontsize{8pt}{9.6pt}\selectfont All binary}} \hspace{-1pt}\huxbpad{-1pt}} &
\multicolumn{4}{c!{\huxvb{0, 0, 0}{0}}}{\huxtpad{-1pt + 1em}\centering \hspace{-1pt} \textbf{{\fontsize{8pt}{9.6pt}\selectfont Y continuous}} \hspace{-1pt}\huxbpad{-1pt}} &
\multicolumn{4}{c!{\huxvb{0, 0, 0}{0}}}{\huxtpad{-1pt + 1em}\centering \hspace{-1pt} \textbf{{\fontsize{8pt}{9.6pt}\selectfont Y,X continuous}} \hspace{-1pt}\huxbpad{-1pt}} \tabularnewline
\end{tabular}}\label{table:test_three}
\par\end{center}

\end{table}
}
\providecommand{\huxb}[2]{\arrayrulecolor[RGB]{#1}\global\arrayrulewidth=#2pt}
  \providecommand{\huxvb}[2]{\color[RGB]{#1}\vrule width #2pt}
  \providecommand{\huxtpad}[1]{\rule{0pt}{#1}}
  \providecommand{\huxbpad}[1]{\rule[-#1]{0pt}{#1}}
 
{\renewcommand{\huxtpad}[1]{\rule{0pt}{0.3ex}}
\renewcommand{\huxbpad}[1]{\rule[-0.05ex]{0pt}{0.05ex}}
\renewcommand{\arraystretch}{0.6}
\begin{table}[H]
\begin{center}

\captionsetup{justification=centering,singlelinecheck=off}
\caption{Comparative analysis of dual and primal tests using linear vs Super Learners for complex DGPs.}
 \setlength{\tabcolsep}{0pt}
\resizebox{0.6\textwidth}{!}{\begin{tabular}{l l l l l l l l l}

\hhline{>{\huxb{0, 0, 0}{1}}->{\huxb{0, 0, 0}{1}}->{\huxb{0, 0, 0}{1}}->{\huxb{0, 0, 0}{1}}->{\huxb{0, 0, 0}{1}}->{\huxb{0, 0, 0}{1}}->{\huxb{0, 0, 0}{1}}->{\huxb{0, 0, 0}{1}}->{\huxb{0, 0, 0}{1}}-}
\arrayrulecolor{black}

\multicolumn{1}{!{\huxvb{0, 0, 0}{0}}l!{\huxvb{0, 0, 0}{0}}}{\huxtpad{-1pt + 1em}\raggedright \hspace{-1pt} \textbf{{\fontsize{8pt}{9.6pt}\selectfont }} \hspace{-1pt}\huxbpad{-1pt}} &
\multicolumn{8}{c!{\huxvb{0, 0, 0}{0}}}{\huxtpad{-1pt + 1em}\centering \hspace{-1pt} \textbf{{\fontsize{8pt}{9.6pt}\selectfont Y,M,Z,X continuous}} \hspace{-1pt}\huxbpad{-1pt}} \tabularnewline[-0.5pt]

\hhline{>{\huxb{255, 255, 255}{0.4}}->{\huxb{0, 0, 0}{0.4}}->{\huxb{0, 0, 0}{0.4}}->{\huxb{0, 0, 0}{0.4}}->{\huxb{0, 0, 0}{0.4}}->{\huxb{0, 0, 0}{0.4}}->{\huxb{0, 0, 0}{0.4}}->{\huxb{0, 0, 0}{0.4}}->{\huxb{0, 0, 0}{0.4}}-}
\arrayrulecolor{black}

\multicolumn{1}{!{\huxvb{0, 0, 0}{0}}l!{\huxvb{0, 0, 0}{0}}}{\huxtpad{-1pt + 1em}\raggedright \hspace{-1pt} \textit{{\fontsize{8pt}{9.6pt}\selectfont N}} \hspace{-1pt}\huxbpad{-1pt}} &
\multicolumn{4}{c!{\huxvb{0, 0, 0}{0}}}{\huxtpad{-1pt + 1em}\centering \hspace{-1pt} \textit{{\fontsize{8pt}{9.6pt}\selectfont Type I error}} \hspace{-1pt}\huxbpad{-1pt}} &
\multicolumn{4}{c!{\huxvb{0, 0, 0}{0}}}{\huxtpad{-1pt + 1em}\centering \hspace{-1pt} \textit{{\fontsize{8pt}{9.6pt}\selectfont Power}} \hspace{-1pt}\huxbpad{-1pt}} \tabularnewline[-0.5pt]

\hhline{>{\huxb{255, 255, 255}{0.4}}->{\huxb{0, 0, 0}{0.4}}->{\huxb{0, 0, 0}{0.4}}->{\huxb{0, 0, 0}{0.4}}->{\huxb{0, 0, 0}{0.4}}->{\huxb{0, 0, 0}{0.4}}->{\huxb{0, 0, 0}{0.4}}->{\huxb{0, 0, 0}{0.4}}->{\huxb{0, 0, 0}{0.4}}-}
\arrayrulecolor{black}

\multicolumn{1}{!{\huxvb{0, 0, 0}{0}}l!{\huxvb{0, 0, 0}{0}}}{\huxtpad{-1pt + 1em}\raggedright \hspace{-1pt} {\fontsize{8pt}{9.6pt}\selectfont } \hspace{-1pt}\huxbpad{-1pt}} &
\multicolumn{2}{c!{\huxvb{0, 0, 0}{0}}}{\huxtpad{-1pt + 1em}\centering \hspace{-1pt} {\fontsize{8pt}{9.6pt}\selectfont DAG1} \hspace{-1pt}\huxbpad{-1pt}} &
\multicolumn{2}{c!{\huxvb{0, 0, 0}{0}}}{\huxtpad{-1pt + 1em}\centering \hspace{-1pt} {\fontsize{8pt}{9.6pt}\selectfont DAG2} \hspace{-1pt}\huxbpad{-1pt}} &
\multicolumn{2}{c!{\huxvb{0, 0, 0}{0}}}{\huxtpad{-1pt + 1em}\centering \hspace{-1pt} {\fontsize{8pt}{9.6pt}\selectfont DAG3} \hspace{-1pt}\huxbpad{-1pt}} &
\multicolumn{2}{c!{\huxvb{0, 0, 0}{0}}}{\huxtpad{-1pt + 1em}\centering \hspace{-1pt} {\fontsize{8pt}{9.6pt}\selectfont DAG4} \hspace{-1pt}\huxbpad{-1pt}} \tabularnewline[-0.5pt]

\hhline{>{\huxb{255, 255, 255}{0.4}}->{\huxb{0, 0, 0}{0.4}}->{\huxb{0, 0, 0}{0.4}}->{\huxb{0, 0, 0}{0.4}}->{\huxb{0, 0, 0}{0.4}}->{\huxb{0, 0, 0}{0.4}}->{\huxb{0, 0, 0}{0.4}}->{\huxb{0, 0, 0}{0.4}}->{\huxb{0, 0, 0}{0.4}}-}
\arrayrulecolor{black}

\multicolumn{1}{!{\huxvb{0, 0, 0}{0}}l!{\huxvb{0, 0, 0}{0}}}{\huxtpad{-1pt + 1em}\raggedright \hspace{-1pt} \textit{{\fontsize{8pt}{9.6pt}\selectfont Nuisance models}} \hspace{-1pt}\huxbpad{-1pt}} &
\multicolumn{1}{c!{\huxvb{0, 0, 0}{0}}}{\huxtpad{-1pt + 1em}\centering \hspace{-1pt} {\fontsize{8pt}{9.6pt}\selectfont Linear} \hspace{-1pt}\huxbpad{-1pt}} &
\multicolumn{1}{c!{\huxvb{0, 0, 0}{0}}}{\huxtpad{-1pt + 1em}\centering \hspace{-1pt} {\fontsize{8pt}{9.6pt}\selectfont SL} \hspace{-1pt}\huxbpad{-1pt}} &
\multicolumn{1}{c!{\huxvb{0, 0, 0}{0}}}{\huxtpad{-1pt + 1em}\centering \hspace{-1pt} {\fontsize{8pt}{9.6pt}\selectfont Linear} \hspace{-1pt}\huxbpad{-1pt}} &
\multicolumn{1}{c!{\huxvb{0, 0, 0}{0}}}{\huxtpad{-1pt + 1em}\centering \hspace{-1pt} {\fontsize{8pt}{9.6pt}\selectfont SL} \hspace{-1pt}\huxbpad{-1pt}} &
\multicolumn{1}{c!{\huxvb{0, 0, 0}{0}}}{\huxtpad{-1pt + 1em}\centering \hspace{-1pt} {\fontsize{8pt}{9.6pt}\selectfont Linear} \hspace{-1pt}\huxbpad{-1pt}} &
\multicolumn{1}{c!{\huxvb{0, 0, 0}{0}}}{\huxtpad{-1pt + 1em}\centering \hspace{-1pt} {\fontsize{8pt}{9.6pt}\selectfont SL} \hspace{-1pt}\huxbpad{-1pt}} &
\multicolumn{1}{c!{\huxvb{0, 0, 0}{0}}}{\huxtpad{-1pt + 1em}\centering \hspace{-1pt} {\fontsize{8pt}{9.6pt}\selectfont Linear} \hspace{-1pt}\huxbpad{-1pt}} &
\multicolumn{1}{c!{\huxvb{0, 0, 0}{0}}}{\huxtpad{-1pt + 1em}\centering \hspace{-1pt} {\fontsize{8pt}{9.6pt}\selectfont SL} \hspace{-1pt}\huxbpad{-1pt}} \tabularnewline[-0.5pt]

\hhline{>{\huxb{255, 255, 255}{0.4}}->{\huxb{0, 0, 0}{0.4}}->{\huxb{0, 0, 0}{0.4}}->{\huxb{0, 0, 0}{0.4}}->{\huxb{0, 0, 0}{0.4}}->{\huxb{0, 0, 0}{0.4}}->{\huxb{0, 0, 0}{0.4}}->{\huxb{0, 0, 0}{0.4}}->{\huxb{0, 0, 0}{0.4}}-}
\arrayrulecolor{black}

\multicolumn{3}{!{\huxvb{0, 0, 0}{0}}l!{\huxvb{0, 0, 0}{0}}}{\huxtpad{-1pt + 1em}\raggedright \hspace{-1pt} {\fontsize{8pt}{9.6pt}\selectfont \textbf{Dual test}} \hspace{-1pt}\huxbpad{-1pt}} &
\multicolumn{1}{c!{\huxvb{0, 0, 0}{0}}}{\huxtpad{-1pt + 1em}\centering \hspace{-1pt} {\fontsize{8pt}{9.6pt}\selectfont } \hspace{-1pt}\huxbpad{-1pt}} &
\multicolumn{1}{c!{\huxvb{0, 0, 0}{0.4}}!{\huxvb{0, 0, 0}{0.4}}}{\huxtpad{-1pt + 1em}\centering \hspace{-1pt} {\fontsize{8pt}{9.6pt}\selectfont } \hspace{-1pt}\huxbpad{-1pt}} &
\multicolumn{1}{c!{\huxvb{0, 0, 0}{0}}}{\huxtpad{-1pt + 1em}\centering \hspace{-1pt} {\fontsize{8pt}{9.6pt}\selectfont } \hspace{-1pt}\huxbpad{-1pt}} &
\multicolumn{1}{c!{\huxvb{0, 0, 0}{0}}}{\huxtpad{-1pt + 1em}\centering \hspace{-1pt} {\fontsize{8pt}{9.6pt}\selectfont } \hspace{-1pt}\huxbpad{-1pt}} &
\multicolumn{1}{c!{\huxvb{0, 0, 0}{0}}}{\huxtpad{-1pt + 1em}\centering \hspace{-1pt} {\fontsize{8pt}{9.6pt}\selectfont } \hspace{-1pt}\huxbpad{-1pt}} &
\multicolumn{1}{c!{\huxvb{0, 0, 0}{0}}}{\huxtpad{-1pt + 1em}\centering \hspace{-1pt} {\fontsize{8pt}{9.6pt}\selectfont } \hspace{-1pt}\huxbpad{-1pt}} \tabularnewline[-0.5pt]

\hhline{>{\huxb{0, 0, 0}{0.4}}||}
\arrayrulecolor{black}

\multicolumn{1}{!{\huxvb{0, 0, 0}{0}}l!{\huxvb{0, 0, 0}{0}}}{\huxtpad{-1pt + 1em}\raggedright \hspace{-1pt} {\fontsize{8pt}{9.6pt}\selectfont 500} \hspace{-1pt}\huxbpad{-1pt}} &
\multicolumn{1}{c!{\huxvb{0, 0, 0}{0}}}{\huxtpad{-1pt + 1em}\centering \hspace{-1pt} {\fontsize{8pt}{9.6pt}\selectfont 0.065} \hspace{-1pt}\huxbpad{-1pt}} &
\multicolumn{1}{c!{\huxvb{0, 0, 0}{0}}}{\huxtpad{-1pt + 1em}\centering \hspace{-1pt} {\fontsize{8pt}{9.6pt}\selectfont 0.045} \hspace{-1pt}\huxbpad{-1pt}} &
\multicolumn{1}{c!{\huxvb{0, 0, 0}{0}}}{\huxtpad{-1pt + 1em}\centering \hspace{-1pt} {\fontsize{8pt}{9.6pt}\selectfont 0.11} \hspace{-1pt}\huxbpad{-1pt}} &
\multicolumn{1}{c!{\huxvb{0, 0, 0}{0.4}}!{\huxvb{0, 0, 0}{0.4}}}{\huxtpad{-1pt + 1em}\centering \hspace{-1pt} {\fontsize{8pt}{9.6pt}\selectfont 0.055} \hspace{-1pt}\huxbpad{-1pt}} &
\multicolumn{1}{c!{\huxvb{0, 0, 0}{0}}}{\huxtpad{-1pt + 1em}\centering \hspace{-1pt} {\fontsize{8pt}{9.6pt}\selectfont 0.465} \hspace{-1pt}\huxbpad{-1pt}} &
\multicolumn{1}{c!{\huxvb{0, 0, 0}{0}}}{\huxtpad{-1pt + 1em}\centering \hspace{-1pt} {\fontsize{8pt}{9.6pt}\selectfont 0.05} \hspace{-1pt}\huxbpad{-1pt}} &
\multicolumn{1}{c!{\huxvb{0, 0, 0}{0}}}{\huxtpad{-1pt + 1em}\centering \hspace{-1pt} {\fontsize{8pt}{9.6pt}\selectfont 0.35} \hspace{-1pt}\huxbpad{-1pt}} &
\multicolumn{1}{c!{\huxvb{0, 0, 0}{0}}}{\huxtpad{-1pt + 1em}\centering \hspace{-1pt} {\fontsize{8pt}{9.6pt}\selectfont 0.07} \hspace{-1pt}\huxbpad{-1pt}} \tabularnewline[-0.5pt]

\hhline{>{\huxb{0, 0, 0}{0.4}}||}
\arrayrulecolor{black}

\multicolumn{1}{!{\huxvb{0, 0, 0}{0}}l!{\huxvb{0, 0, 0}{0}}}{\huxtpad{-1pt + 1em}\raggedright \hspace{-1pt} {\fontsize{8pt}{9.6pt}\selectfont 1000} \hspace{-1pt}\huxbpad{-1pt}} &
\multicolumn{1}{c!{\huxvb{0, 0, 0}{0}}}{\huxtpad{-1pt + 1em}\centering \hspace{-1pt} {\fontsize{8pt}{9.6pt}\selectfont 0.055} \hspace{-1pt}\huxbpad{-1pt}} &
\multicolumn{1}{c!{\huxvb{0, 0, 0}{0}}}{\huxtpad{-1pt + 1em}\centering \hspace{-1pt} {\fontsize{8pt}{9.6pt}\selectfont 0.06} \hspace{-1pt}\huxbpad{-1pt}} &
\multicolumn{1}{c!{\huxvb{0, 0, 0}{0}}}{\huxtpad{-1pt + 1em}\centering \hspace{-1pt} {\fontsize{8pt}{9.6pt}\selectfont 0.155} \hspace{-1pt}\huxbpad{-1pt}} &
\multicolumn{1}{c!{\huxvb{0, 0, 0}{0.4}}!{\huxvb{0, 0, 0}{0.4}}}{\huxtpad{-1pt + 1em}\centering \hspace{-1pt} {\fontsize{8pt}{9.6pt}\selectfont 0.05} \hspace{-1pt}\huxbpad{-1pt}} &
\multicolumn{1}{c!{\huxvb{0, 0, 0}{0}}}{\huxtpad{-1pt + 1em}\centering \hspace{-1pt} {\fontsize{8pt}{9.6pt}\selectfont 0.685} \hspace{-1pt}\huxbpad{-1pt}} &
\multicolumn{1}{c!{\huxvb{0, 0, 0}{0}}}{\huxtpad{-1pt + 1em}\centering \hspace{-1pt} {\fontsize{8pt}{9.6pt}\selectfont 0.08} \hspace{-1pt}\huxbpad{-1pt}} &
\multicolumn{1}{c!{\huxvb{0, 0, 0}{0}}}{\huxtpad{-1pt + 1em}\centering \hspace{-1pt} {\fontsize{8pt}{9.6pt}\selectfont 0.595} \hspace{-1pt}\huxbpad{-1pt}} &
\multicolumn{1}{c!{\huxvb{0, 0, 0}{0}}}{\huxtpad{-1pt + 1em}\centering \hspace{-1pt} {\fontsize{8pt}{9.6pt}\selectfont 0.105} \hspace{-1pt}\huxbpad{-1pt}} \tabularnewline[-0.5pt]

\hhline{>{\huxb{0, 0, 0}{0.4}}||}
\arrayrulecolor{black}

\multicolumn{1}{!{\huxvb{0, 0, 0}{0}}l!{\huxvb{0, 0, 0}{0}}}{\huxtpad{-1pt + 1em}\raggedright \hspace{-1pt} {\fontsize{8pt}{9.6pt}\selectfont 2000} \hspace{-1pt}\huxbpad{-1pt}} &
\multicolumn{1}{c!{\huxvb{0, 0, 0}{0}}}{\huxtpad{-1pt + 1em}\centering \hspace{-1pt} {\fontsize{8pt}{9.6pt}\selectfont 0.135} \hspace{-1pt}\huxbpad{-1pt}} &
\multicolumn{1}{c!{\huxvb{0, 0, 0}{0}}}{\huxtpad{-1pt + 1em}\centering \hspace{-1pt} {\fontsize{8pt}{9.6pt}\selectfont 0.05} \hspace{-1pt}\huxbpad{-1pt}} &
\multicolumn{1}{c!{\huxvb{0, 0, 0}{0}}}{\huxtpad{-1pt + 1em}\centering \hspace{-1pt} {\fontsize{8pt}{9.6pt}\selectfont 0.175} \hspace{-1pt}\huxbpad{-1pt}} &
\multicolumn{1}{c!{\huxvb{0, 0, 0}{0.4}}!{\huxvb{0, 0, 0}{0.4}}}{\huxtpad{-1pt + 1em}\centering \hspace{-1pt} {\fontsize{8pt}{9.6pt}\selectfont 0.055} \hspace{-1pt}\huxbpad{-1pt}} &
\multicolumn{1}{c!{\huxvb{0, 0, 0}{0}}}{\huxtpad{-1pt + 1em}\centering \hspace{-1pt} {\fontsize{8pt}{9.6pt}\selectfont 0.84} \hspace{-1pt}\huxbpad{-1pt}} &
\multicolumn{1}{c!{\huxvb{0, 0, 0}{0}}}{\huxtpad{-1pt + 1em}\centering \hspace{-1pt} {\fontsize{8pt}{9.6pt}\selectfont 0.11} \hspace{-1pt}\huxbpad{-1pt}} &
\multicolumn{1}{c!{\huxvb{0, 0, 0}{0}}}{\huxtpad{-1pt + 1em}\centering \hspace{-1pt} {\fontsize{8pt}{9.6pt}\selectfont 0.725} \hspace{-1pt}\huxbpad{-1pt}} &
\multicolumn{1}{c!{\huxvb{0, 0, 0}{0}}}{\huxtpad{-1pt + 1em}\centering \hspace{-1pt} {\fontsize{8pt}{9.6pt}\selectfont 0.145} \hspace{-1pt}\huxbpad{-1pt}} \tabularnewline[-0.5pt]

\hhline{>{\huxb{0, 0, 0}{0.4}}||}
\arrayrulecolor{black}

\multicolumn{3}{!{\huxvb{0, 0, 0}{0}}l!{\huxvb{0, 0, 0}{0}}}{\huxtpad{-1pt + 1em}\raggedright \hspace{-1pt} {\fontsize{8pt}{9.6pt}\selectfont \textbf{Primal test}} \hspace{-1pt}\huxbpad{-1pt}} &
\multicolumn{1}{c!{\huxvb{0, 0, 0}{0}}}{\huxtpad{-1pt + 1em}\centering \hspace{-1pt} {\fontsize{8pt}{9.6pt}\selectfont } \hspace{-1pt}\huxbpad{-1pt}} &
\multicolumn{1}{c!{\huxvb{0, 0, 0}{0.4}}!{\huxvb{0, 0, 0}{0.4}}}{\huxtpad{-1pt + 1em}\centering \hspace{-1pt} {\fontsize{8pt}{9.6pt}\selectfont } \hspace{-1pt}\huxbpad{-1pt}} &
\multicolumn{1}{c!{\huxvb{0, 0, 0}{0}}}{\huxtpad{-1pt + 1em}\centering \hspace{-1pt} {\fontsize{8pt}{9.6pt}\selectfont } \hspace{-1pt}\huxbpad{-1pt}} &
\multicolumn{1}{c!{\huxvb{0, 0, 0}{0}}}{\huxtpad{-1pt + 1em}\centering \hspace{-1pt} {\fontsize{8pt}{9.6pt}\selectfont } \hspace{-1pt}\huxbpad{-1pt}} &
\multicolumn{1}{c!{\huxvb{0, 0, 0}{0}}}{\huxtpad{-1pt + 1em}\centering \hspace{-1pt} {\fontsize{8pt}{9.6pt}\selectfont } \hspace{-1pt}\huxbpad{-1pt}} &
\multicolumn{1}{c!{\huxvb{0, 0, 0}{0}}}{\huxtpad{-1pt + 1em}\centering \hspace{-1pt} {\fontsize{8pt}{9.6pt}\selectfont } \hspace{-1pt}\huxbpad{-1pt}} \tabularnewline[-0.5pt]

\hhline{>{\huxb{0, 0, 0}{0.4}}||}
\arrayrulecolor{black}

\multicolumn{1}{!{\huxvb{0, 0, 0}{0}}l!{\huxvb{0, 0, 0}{0}}}{\huxtpad{-1pt + 1em}\raggedright \hspace{-1pt} {\fontsize{8pt}{9.6pt}\selectfont 500} \hspace{-1pt}\huxbpad{-1pt}} &
\multicolumn{1}{c!{\huxvb{0, 0, 0}{0}}}{\huxtpad{-1pt + 1em}\centering \hspace{-1pt} {\fontsize{8pt}{9.6pt}\selectfont 0.165} \hspace{-1pt}\huxbpad{-1pt}} &
\multicolumn{1}{c!{\huxvb{0, 0, 0}{0}}}{\huxtpad{-1pt + 1em}\centering \hspace{-1pt} {\fontsize{8pt}{9.6pt}\selectfont 0.05} \hspace{-1pt}\huxbpad{-1pt}} &
\multicolumn{1}{c!{\huxvb{0, 0, 0}{0}}}{\huxtpad{-1pt + 1em}\centering \hspace{-1pt} {\fontsize{8pt}{9.6pt}\selectfont 0.13} \hspace{-1pt}\huxbpad{-1pt}} &
\multicolumn{1}{c!{\huxvb{0, 0, 0}{0.4}}!{\huxvb{0, 0, 0}{0.4}}}{\huxtpad{-1pt + 1em}\centering \hspace{-1pt} {\fontsize{8pt}{9.6pt}\selectfont 0.055} \hspace{-1pt}\huxbpad{-1pt}} &
\multicolumn{1}{c!{\huxvb{0, 0, 0}{0}}}{\huxtpad{-1pt + 1em}\centering \hspace{-1pt} {\fontsize{8pt}{9.6pt}\selectfont 0.915} \hspace{-1pt}\huxbpad{-1pt}} &
\multicolumn{1}{c!{\huxvb{0, 0, 0}{0}}}{\huxtpad{-1pt + 1em}\centering \hspace{-1pt} {\fontsize{8pt}{9.6pt}\selectfont 0.115} \hspace{-1pt}\huxbpad{-1pt}} &
\multicolumn{1}{c!{\huxvb{0, 0, 0}{0}}}{\huxtpad{-1pt + 1em}\centering \hspace{-1pt} {\fontsize{8pt}{9.6pt}\selectfont 0.555} \hspace{-1pt}\huxbpad{-1pt}} &
\multicolumn{1}{c!{\huxvb{0, 0, 0}{0}}}{\huxtpad{-1pt + 1em}\centering \hspace{-1pt} {\fontsize{8pt}{9.6pt}\selectfont 0.445} \hspace{-1pt}\huxbpad{-1pt}} \tabularnewline[-0.5pt]

\hhline{>{\huxb{0, 0, 0}{0.4}}||}
\arrayrulecolor{black}

\multicolumn{1}{!{\huxvb{0, 0, 0}{0}}l!{\huxvb{0, 0, 0}{0}}}{\huxtpad{-1pt + 1em}\raggedright \hspace{-1pt} {\fontsize{8pt}{9.6pt}\selectfont 1000} \hspace{-1pt}\huxbpad{-1pt}} &
\multicolumn{1}{c!{\huxvb{0, 0, 0}{0}}}{\huxtpad{-1pt + 1em}\centering \hspace{-1pt} {\fontsize{8pt}{9.6pt}\selectfont 0.18} \hspace{-1pt}\huxbpad{-1pt}} &
\multicolumn{1}{c!{\huxvb{0, 0, 0}{0}}}{\huxtpad{-1pt + 1em}\centering \hspace{-1pt} {\fontsize{8pt}{9.6pt}\selectfont 0.06} \hspace{-1pt}\huxbpad{-1pt}} &
\multicolumn{1}{c!{\huxvb{0, 0, 0}{0}}}{\huxtpad{-1pt + 1em}\centering \hspace{-1pt} {\fontsize{8pt}{9.6pt}\selectfont 0.145} \hspace{-1pt}\huxbpad{-1pt}} &
\multicolumn{1}{c!{\huxvb{0, 0, 0}{0.4}}!{\huxvb{0, 0, 0}{0.4}}}{\huxtpad{-1pt + 1em}\centering \hspace{-1pt} {\fontsize{8pt}{9.6pt}\selectfont 0.08} \hspace{-1pt}\huxbpad{-1pt}} &
\multicolumn{1}{c!{\huxvb{0, 0, 0}{0}}}{\huxtpad{-1pt + 1em}\centering \hspace{-1pt} {\fontsize{8pt}{9.6pt}\selectfont 0.935} \hspace{-1pt}\huxbpad{-1pt}} &
\multicolumn{1}{c!{\huxvb{0, 0, 0}{0}}}{\huxtpad{-1pt + 1em}\centering \hspace{-1pt} {\fontsize{8pt}{9.6pt}\selectfont 0.165} \hspace{-1pt}\huxbpad{-1pt}} &
\multicolumn{1}{c!{\huxvb{0, 0, 0}{0}}}{\huxtpad{-1pt + 1em}\centering \hspace{-1pt} {\fontsize{8pt}{9.6pt}\selectfont 0.5} \hspace{-1pt}\huxbpad{-1pt}} &
\multicolumn{1}{c!{\huxvb{0, 0, 0}{0}}}{\huxtpad{-1pt + 1em}\centering \hspace{-1pt} {\fontsize{8pt}{9.6pt}\selectfont 0.485} \hspace{-1pt}\huxbpad{-1pt}} \tabularnewline[-0.5pt]

\hhline{>{\huxb{0, 0, 0}{0.4}}||}
\arrayrulecolor{black}

\multicolumn{1}{!{\huxvb{0, 0, 0}{0}}l!{\huxvb{0, 0, 0}{0}}}{\huxtpad{-1pt + 1em}\raggedright \hspace{-1pt} {\fontsize{8pt}{9.6pt}\selectfont 2000} \hspace{-1pt}\huxbpad{-1pt}} &
\multicolumn{1}{c!{\huxvb{0, 0, 0}{0}}}{\huxtpad{-1pt + 1em}\centering \hspace{-1pt} {\fontsize{8pt}{9.6pt}\selectfont 0.185} \hspace{-1pt}\huxbpad{-1pt}} &
\multicolumn{1}{c!{\huxvb{0, 0, 0}{0}}}{\huxtpad{-1pt + 1em}\centering \hspace{-1pt} {\fontsize{8pt}{9.6pt}\selectfont 0.035} \hspace{-1pt}\huxbpad{-1pt}} &
\multicolumn{1}{c!{\huxvb{0, 0, 0}{0}}}{\huxtpad{-1pt + 1em}\centering \hspace{-1pt} {\fontsize{8pt}{9.6pt}\selectfont 0.155} \hspace{-1pt}\huxbpad{-1pt}} &
\multicolumn{1}{c!{\huxvb{0, 0, 0}{0.4}}!{\huxvb{0, 0, 0}{0.4}}}{\huxtpad{-1pt + 1em}\centering \hspace{-1pt} {\fontsize{8pt}{9.6pt}\selectfont 0.06} \hspace{-1pt}\huxbpad{-1pt}} &
\multicolumn{1}{c!{\huxvb{0, 0, 0}{0}}}{\huxtpad{-1pt + 1em}\centering \hspace{-1pt} {\fontsize{8pt}{9.6pt}\selectfont 0.975} \hspace{-1pt}\huxbpad{-1pt}} &
\multicolumn{1}{c!{\huxvb{0, 0, 0}{0}}}{\huxtpad{-1pt + 1em}\centering \hspace{-1pt} {\fontsize{8pt}{9.6pt}\selectfont 0.205} \hspace{-1pt}\huxbpad{-1pt}} &
\multicolumn{1}{c!{\huxvb{0, 0, 0}{0}}}{\huxtpad{-1pt + 1em}\centering \hspace{-1pt} {\fontsize{8pt}{9.6pt}\selectfont 0.39} \hspace{-1pt}\huxbpad{-1pt}} &
\multicolumn{1}{c!{\huxvb{0, 0, 0}{0}}}{\huxtpad{-1pt + 1em}\centering \hspace{-1pt} {\fontsize{8pt}{9.6pt}\selectfont 0.42} \hspace{-1pt}\huxbpad{-1pt}} \tabularnewline[-0.5pt]

\hhline{>{\huxb{0, 0, 0}{1}}->{\huxb{0, 0, 0}{1}}->{\huxb{0, 0, 0}{1}}->{\huxb{0, 0, 0}{1}}->{\huxb{0, 0, 0}{1}}->{\huxb{0, 0, 0}{1}}->{\huxb{0, 0, 0}{1}}->{\huxb{0, 0, 0}{1}}->{\huxb{0, 0, 0}{1}}-}
\arrayrulecolor{black}
\end{tabular}}\label{table:test_SL}
\par\end{center}

\end{table}
}


 \providecommand{\huxb}[2]{\arrayrulecolor[RGB]{#1}\global\arrayrulewidth=#2pt}
  \providecommand{\huxvb}[2]{\color[RGB]{#1}\vrule width #2pt}
  \providecommand{\huxtpad}[1]{\rule{0pt}{#1}}
  \providecommand{\huxbpad}[1]{\rule[-#1]{0pt}{#1}}

\begin{center}
\begin{table}[t]
\captionsetup{justification=centering,singlelinecheck=off}
\caption{One-step and TMLE estimates of the average treatment effect and average treatment effect on the treated of additional mobile stroke unit (MSU) care on modified Rankin scale (mRS) score}
\setlength{\tabcolsep}{0pt}
\renewcommand{\arraystretch}{1}
\resizebox{1\textwidth}{!}{\begin{tabular}{l l l l l l l}

\hhline{>{\huxb{0, 0, 0}{1}}->{\huxb{0, 0, 0}{1}}->{\huxb{0, 0, 0}{1}}->{\huxb{0, 0, 0}{1}}->{\huxb{0, 0, 0}{1}}->{\huxb{0, 0, 0}{1}}->{\huxb{0, 0, 0}{1}}-}

\multicolumn{1}{!{\huxvb{0, 0, 0}{0}}m{0.153061224489796\textwidth}!{\huxvb{0, 0, 0}{0}}}{\hspace{1pt}\parbox[c]{0.153061224489796\textwidth-1pt-1pt}{\huxtpad{1pt + 1em}\raggedright \textbf{{\fontsize{7pt}{8.4pt}\selectfont $M$, $Y$ type}}\huxbpad{1pt}}} &
\multicolumn{2}{m{0.112244897959184\textwidth+2\tabcolsep}!{\huxvb{0, 0, 0}{0}}}{\hspace{1pt}\parbox[c]{0.112244897959184\textwidth+2\tabcolsep-1pt-1pt}{\huxtpad{1pt + 1em}\raggedright \textbf{{\fontsize{7pt}{8.4pt}\selectfont Cutoffs}}\huxbpad{1pt}}} &
\multicolumn{2}{m{0.36734693877551\textwidth+2\tabcolsep}!{\huxvb{0, 0, 0}{0}}}{\hspace{1pt}\parbox[c]{0.36734693877551\textwidth+2\tabcolsep-1pt-1pt}{\huxtpad{1pt + 1em}\centering \textbf{{\fontsize{7pt}{8.4pt}\selectfont One-step estimator}}\huxbpad{1pt}}} &
\multicolumn{2}{m{0.36734693877551\textwidth+2\tabcolsep}!{\huxvb{0, 0, 0}{0}}}{\hspace{1pt}\parbox[c]{0.36734693877551\textwidth+2\tabcolsep-1pt-1pt}{\huxtpad{1pt + 1em}\centering \textbf{{\fontsize{7pt}{8.4pt}\selectfont TMLE}}\huxbpad{1pt}}} \tabularnewline[-0.5pt]

\hhline{>{\huxb{0, 0, 0}{0.4}}->{\huxb{0, 0, 0}{0.4}}->{\huxb{0, 0, 0}{0.4}}->{\huxb{0, 0, 0}{0.4}}->{\huxb{0, 0, 0}{0.4}}->{\huxb{0, 0, 0}{0.4}}->{\huxb{0, 0, 0}{0.4}}-}

\multicolumn{1}{!{\huxvb{0, 0, 0}{0}}m{0.153061224489796\textwidth}!{\huxvb{0, 0, 0}{0}}}{\hspace{1pt}\parbox[c]{0.153061224489796\textwidth-1pt-1pt}{\huxtpad{1pt + 1em}\raggedright {\fontsize{7pt}{8.4pt}\selectfont }\huxbpad{1pt}}} &
\multicolumn{1}{m{0.0561224489795918\textwidth}!{\huxvb{0, 0, 0}{0}}}{\hspace{1pt}\parbox[c]{0.0561224489795918\textwidth-1pt-1pt}{\huxtpad{1pt + 1em}\raggedright {\fontsize{7pt}{8.4pt}\selectfont $M$}\huxbpad{1pt}}} &
\multicolumn{1}{m{0.0561224489795918\textwidth}!{\huxvb{0, 0, 0}{0}}}{\hspace{1pt}\parbox[c]{0.0561224489795918\textwidth-1pt-1pt}{\huxtpad{1pt + 1em}\raggedright {\fontsize{7pt}{8.4pt}\selectfont $Y$}\huxbpad{1pt}}} &
\multicolumn{1}{m{0.183673469387755\textwidth}!{\huxvb{0, 0, 0}{0}}}{\hspace{1pt}\parbox[c]{0.183673469387755\textwidth-1pt-1pt}{\huxtpad{1pt + 1em}\centering {\fontsize{7pt}{8.4pt}\selectfont ATE}\huxbpad{1pt}}} &
\multicolumn{1}{m{0.183673469387755\textwidth}!{\huxvb{0, 0, 0}{0}}}{\hspace{1pt}\parbox[c]{0.183673469387755\textwidth-1pt-1pt}{\huxtpad{1pt + 1em}\centering {\fontsize{7pt}{8.4pt}\selectfont ATT}\huxbpad{1pt}}} &
\multicolumn{1}{m{0.183673469387755\textwidth}!{\huxvb{0, 0, 0}{0}}}{\hspace{1pt}\parbox[c]{0.183673469387755\textwidth-1pt-1pt}{\huxtpad{1pt + 1em}\centering {\fontsize{7pt}{8.4pt}\selectfont ATE}\huxbpad{1pt}}} &
\multicolumn{1}{m{0.183673469387755\textwidth}!{\huxvb{0, 0, 0}{0}}}{\hspace{1pt}\parbox[c]{0.183673469387755\textwidth-1pt-1pt}{\huxtpad{1pt + 1em}\centering {\fontsize{7pt}{8.4pt}\selectfont ATT}\huxbpad{1pt}}} \tabularnewline[-0.5pt]

\hhline{>{\huxb{0, 0, 0}{0.4}}->{\huxb{0, 0, 0}{0.4}}->{\huxb{0, 0, 0}{0.4}}->{\huxb{0, 0, 0}{0.4}}->{\huxb{0, 0, 0}{0.4}}->{\huxb{0, 0, 0}{0.4}}->{\huxb{0, 0, 0}{0.4}}-}

\multicolumn{1}{!{\huxvb{0, 0, 0}{0}}m{0.153061224489796\textwidth}!{\huxvb{0, 0, 0}{0}}}{\hspace{1pt}\parbox[c]{0.153061224489796\textwidth-1pt-1pt}{\huxtpad{1pt + 1em}\raggedright {\fontsize{7pt}{8.4pt}\selectfont Continuous}\huxbpad{1pt}}} &
\multicolumn{1}{m{0.0561224489795918\textwidth}!{\huxvb{0, 0, 0}{0}}}{\hspace{1pt}\parbox[c]{0.0561224489795918\textwidth-1pt-1pt}{\huxtpad{1pt + 1em}\raggedright {\fontsize{7pt}{8.4pt}\selectfont -}\huxbpad{1pt}}} &
\multicolumn{1}{m{0.0561224489795918\textwidth}!{\huxvb{0, 0, 0}{0}}}{\hspace{1pt}\parbox[c]{0.0561224489795918\textwidth-1pt-1pt}{\huxtpad{1pt + 1em}\raggedright {\fontsize{7pt}{8.4pt}\selectfont -}\huxbpad{1pt}}} &
\multicolumn{1}{m{0.183673469387755\textwidth}!{\huxvb{0, 0, 0}{0}}}{\hspace{1pt}\parbox[c]{0.183673469387755\textwidth-1pt-1pt}{\huxtpad{1pt + 1em}\centering {\fontsize{7pt}{8.4pt}\selectfont -0.031 (-0.4, 0.339)}\huxbpad{1pt}}} &
\multicolumn{1}{m{0.183673469387755\textwidth}!{\huxvb{0, 0, 0}{0}}}{\hspace{1pt}\parbox[c]{0.183673469387755\textwidth-1pt-1pt}{\huxtpad{1pt + 1em}\centering {\fontsize{7pt}{8.4pt}\selectfont -0.236 (-0.516, 0.044)}\huxbpad{1pt}}} &
\multicolumn{1}{m{0.183673469387755\textwidth}!{\huxvb{0, 0, 0}{0}}}{\hspace{1pt}\parbox[c]{0.183673469387755\textwidth-1pt-1pt}{\huxtpad{1pt + 1em}\centering {\fontsize{7pt}{8.4pt}\selectfont -0.048 (-0.465, 0.368)}\huxbpad{1pt}}} &
\multicolumn{1}{m{0.183673469387755\textwidth}!{\huxvb{0, 0, 0}{0}}}{\hspace{1pt}\parbox[c]{0.183673469387755\textwidth-1pt-1pt}{\huxtpad{1pt + 1em}\centering {\fontsize{7pt}{8.4pt}\selectfont -0.175 (-0.375, 0.025)}\huxbpad{1pt}}} \tabularnewline[-0.5pt]

\hhline{}

\multicolumn{1}{!{\huxvb{0, 0, 0}{0}}m{0.153061224489796\textwidth}!{\huxvb{0, 0, 0}{0}}}{\hspace{1pt}\parbox[c]{0.153061224489796\textwidth-1pt-1pt}{\huxtpad{1pt + 1em}\raggedright {\fontsize{7pt}{8.4pt}\selectfont Binary}\huxbpad{1pt}}} &
\multicolumn{1}{m{0.0561224489795918\textwidth}!{\huxvb{0, 0, 0}{0}}}{\hspace{1pt}\parbox[c]{0.0561224489795918\textwidth-1pt-1pt}{\huxtpad{1pt + 1em}\raggedright {\fontsize{7pt}{8.4pt}\selectfont 48}\huxbpad{1pt}}} &
\multicolumn{1}{m{0.0561224489795918\textwidth}!{\huxvb{0, 0, 0}{0}}}{\hspace{1pt}\parbox[c]{0.0561224489795918\textwidth-1pt-1pt}{\huxtpad{1pt + 1em}\raggedright {\fontsize{7pt}{8.4pt}\selectfont 2}\huxbpad{1pt}}} &
\multicolumn{1}{m{0.183673469387755\textwidth}!{\huxvb{0, 0, 0}{0}}}{\hspace{1pt}\parbox[c]{0.183673469387755\textwidth-1pt-1pt}{\huxtpad{1pt + 1em}\centering {\fontsize{7pt}{8.4pt}\selectfont -0.046 (-0.084, -0.009)}\huxbpad{1pt}}} &
\multicolumn{1}{m{0.183673469387755\textwidth}!{\huxvb{0, 0, 0}{0}}}{\hspace{1pt}\parbox[c]{0.183673469387755\textwidth-1pt-1pt}{\huxtpad{1pt + 1em}\centering {\fontsize{7pt}{8.4pt}\selectfont -0.056 (-0.083, -0.028)}\huxbpad{1pt}}} &
\multicolumn{1}{m{0.183673469387755\textwidth}!{\huxvb{0, 0, 0}{0}}}{\hspace{1pt}\parbox[c]{0.183673469387755\textwidth-1pt-1pt}{\huxtpad{1pt + 1em}\centering {\fontsize{7pt}{8.4pt}\selectfont -0.048 (-0.084, -0.012)}\huxbpad{1pt}}} &
\multicolumn{1}{m{0.183673469387755\textwidth}!{\huxvb{0, 0, 0}{0}}}{\hspace{1pt}\parbox[c]{0.183673469387755\textwidth-1pt-1pt}{\huxtpad{1pt + 1em}\centering {\fontsize{7pt}{8.4pt}\selectfont -0.056 (-0.084, -0.029)}\huxbpad{1pt}}} \tabularnewline[-0.5pt]

\hhline{}

\multicolumn{1}{!{\huxvb{0, 0, 0}{0}}m{0.153061224489796\textwidth}!{\huxvb{0, 0, 0}{0}}}{\hspace{1pt}\parbox[c]{0.153061224489796\textwidth-1pt-1pt}{\huxtpad{1pt + 1em}\raggedright {\fontsize{7pt}{8.4pt}\selectfont Binary}\huxbpad{1pt}}} &
\multicolumn{1}{m{0.0561224489795918\textwidth}!{\huxvb{0, 0, 0}{0}}}{\hspace{1pt}\parbox[c]{0.0561224489795918\textwidth-1pt-1pt}{\huxtpad{1pt + 1em}\raggedright {\fontsize{7pt}{8.4pt}\selectfont 48}\huxbpad{1pt}}} &
\multicolumn{1}{m{0.0561224489795918\textwidth}!{\huxvb{0, 0, 0}{0}}}{\hspace{1pt}\parbox[c]{0.0561224489795918\textwidth-1pt-1pt}{\huxtpad{1pt + 1em}\raggedright {\fontsize{7pt}{8.4pt}\selectfont 3}\huxbpad{1pt}}} &
\multicolumn{1}{m{0.183673469387755\textwidth}!{\huxvb{0, 0, 0}{0}}}{\hspace{1pt}\parbox[c]{0.183673469387755\textwidth-1pt-1pt}{\huxtpad{1pt + 1em}\centering {\fontsize{7pt}{8.4pt}\selectfont -0.024 (-0.062, 0.014)}\huxbpad{1pt}}} &
\multicolumn{1}{m{0.183673469387755\textwidth}!{\huxvb{0, 0, 0}{0}}}{\hspace{1pt}\parbox[c]{0.183673469387755\textwidth-1pt-1pt}{\huxtpad{1pt + 1em}\centering {\fontsize{7pt}{8.4pt}\selectfont -0.042 (-0.068, -0.015)}\huxbpad{1pt}}} &
\multicolumn{1}{m{0.183673469387755\textwidth}!{\huxvb{0, 0, 0}{0}}}{\hspace{1pt}\parbox[c]{0.183673469387755\textwidth-1pt-1pt}{\huxtpad{1pt + 1em}\centering {\fontsize{7pt}{8.4pt}\selectfont -0.028 (-0.063, 0.007)}\huxbpad{1pt}}} &
\multicolumn{1}{m{0.183673469387755\textwidth}!{\huxvb{0, 0, 0}{0}}}{\hspace{1pt}\parbox[c]{0.183673469387755\textwidth-1pt-1pt}{\huxtpad{1pt + 1em}\centering {\fontsize{7pt}{8.4pt}\selectfont -0.041 (-0.067, -0.016)}\huxbpad{1pt}}} \tabularnewline[-0.5pt]

\hhline{}

\multicolumn{1}{!{\huxvb{0, 0, 0}{0}}m{0.153061224489796\textwidth}!{\huxvb{0, 0, 0}{0}}}{\hspace{1pt}\parbox[c]{0.153061224489796\textwidth-1pt-1pt}{\huxtpad{1pt + 1em}\raggedright {\fontsize{7pt}{8.4pt}\selectfont Binary}\huxbpad{1pt}}} &
\multicolumn{1}{m{0.0561224489795918\textwidth}!{\huxvb{0, 0, 0}{0}}}{\hspace{1pt}\parbox[c]{0.0561224489795918\textwidth-1pt-1pt}{\huxtpad{1pt + 1em}\raggedright {\fontsize{7pt}{8.4pt}\selectfont 48}\huxbpad{1pt}}} &
\multicolumn{1}{m{0.0561224489795918\textwidth}!{\huxvb{0, 0, 0}{0}}}{\hspace{1pt}\parbox[c]{0.0561224489795918\textwidth-1pt-1pt}{\huxtpad{1pt + 1em}\raggedright {\fontsize{7pt}{8.4pt}\selectfont 4}\huxbpad{1pt}}} &
\multicolumn{1}{m{0.183673469387755\textwidth}!{\huxvb{0, 0, 0}{0}}}{\hspace{1pt}\parbox[c]{0.183673469387755\textwidth-1pt-1pt}{\huxtpad{1pt + 1em}\centering {\fontsize{7pt}{8.4pt}\selectfont 0 (-0.035, 0.036)}\huxbpad{1pt}}} &
\multicolumn{1}{m{0.183673469387755\textwidth}!{\huxvb{0, 0, 0}{0}}}{\hspace{1pt}\parbox[c]{0.183673469387755\textwidth-1pt-1pt}{\huxtpad{1pt + 1em}\centering {\fontsize{7pt}{8.4pt}\selectfont -0.019 (-0.04, 0.002)}\huxbpad{1pt}}} &
\multicolumn{1}{m{0.183673469387755\textwidth}!{\huxvb{0, 0, 0}{0}}}{\hspace{1pt}\parbox[c]{0.183673469387755\textwidth-1pt-1pt}{\huxtpad{1pt + 1em}\centering {\fontsize{7pt}{8.4pt}\selectfont -0.004 (-0.036, 0.027)}\huxbpad{1pt}}} &
\multicolumn{1}{m{0.183673469387755\textwidth}!{\huxvb{0, 0, 0}{0}}}{\hspace{1pt}\parbox[c]{0.183673469387755\textwidth-1pt-1pt}{\huxtpad{1pt + 1em}\centering {\fontsize{7pt}{8.4pt}\selectfont -0.015 (-0.035, 0.005)}\huxbpad{1pt}}} \tabularnewline[-0.5pt]

\hhline{}

\multicolumn{1}{!{\huxvb{0, 0, 0}{0}}m{0.153061224489796\textwidth}!{\huxvb{0, 0, 0}{0}}}{\hspace{1pt}\parbox[c]{0.153061224489796\textwidth-1pt-1pt}{\huxtpad{1pt + 1em}\raggedright {\fontsize{7pt}{8.4pt}\selectfont Binary}\huxbpad{1pt}}} &
\multicolumn{1}{m{0.0561224489795918\textwidth}!{\huxvb{0, 0, 0}{0}}}{\hspace{1pt}\parbox[c]{0.0561224489795918\textwidth-1pt-1pt}{\huxtpad{1pt + 1em}\raggedright {\fontsize{7pt}{8.4pt}\selectfont 75}\huxbpad{1pt}}} &
\multicolumn{1}{m{0.0561224489795918\textwidth}!{\huxvb{0, 0, 0}{0}}}{\hspace{1pt}\parbox[c]{0.0561224489795918\textwidth-1pt-1pt}{\huxtpad{1pt + 1em}\raggedright {\fontsize{7pt}{8.4pt}\selectfont 2}\huxbpad{1pt}}} &
\multicolumn{1}{m{0.183673469387755\textwidth}!{\huxvb{0, 0, 0}{0}}}{\hspace{1pt}\parbox[c]{0.183673469387755\textwidth-1pt-1pt}{\huxtpad{1pt + 1em}\centering {\fontsize{7pt}{8.4pt}\selectfont -0.031 (-0.053, -0.008)}\huxbpad{1pt}}} &
\multicolumn{1}{m{0.183673469387755\textwidth}!{\huxvb{0, 0, 0}{0}}}{\hspace{1pt}\parbox[c]{0.183673469387755\textwidth-1pt-1pt}{\huxtpad{1pt + 1em}\centering {\fontsize{7pt}{8.4pt}\selectfont -0.031 (-0.057, -0.005)}\huxbpad{1pt}}} &
\multicolumn{1}{m{0.183673469387755\textwidth}!{\huxvb{0, 0, 0}{0}}}{\hspace{1pt}\parbox[c]{0.183673469387755\textwidth-1pt-1pt}{\huxtpad{1pt + 1em}\centering {\fontsize{7pt}{8.4pt}\selectfont -0.035 (-0.058, -0.012)}\huxbpad{1pt}}} &
\multicolumn{1}{m{0.183673469387755\textwidth}!{\huxvb{0, 0, 0}{0}}}{\hspace{1pt}\parbox[c]{0.183673469387755\textwidth-1pt-1pt}{\huxtpad{1pt + 1em}\centering {\fontsize{7pt}{8.4pt}\selectfont -0.03 (-0.056, -0.005)}\huxbpad{1pt}}} \tabularnewline[-0.5pt]

\hhline{}

\multicolumn{1}{!{\huxvb{0, 0, 0}{0}}m{0.153061224489796\textwidth}!{\huxvb{0, 0, 0}{0}}}{\hspace{1pt}\parbox[c]{0.153061224489796\textwidth-1pt-1pt}{\huxtpad{1pt + 1em}\raggedright {\fontsize{7pt}{8.4pt}\selectfont Binary}\huxbpad{1pt}}} &
\multicolumn{1}{m{0.0561224489795918\textwidth}!{\huxvb{0, 0, 0}{0}}}{\hspace{1pt}\parbox[c]{0.0561224489795918\textwidth-1pt-1pt}{\huxtpad{1pt + 1em}\raggedright {\fontsize{7pt}{8.4pt}\selectfont 75}\huxbpad{1pt}}} &
\multicolumn{1}{m{0.0561224489795918\textwidth}!{\huxvb{0, 0, 0}{0}}}{\hspace{1pt}\parbox[c]{0.0561224489795918\textwidth-1pt-1pt}{\huxtpad{1pt + 1em}\raggedright {\fontsize{7pt}{8.4pt}\selectfont 3}\huxbpad{1pt}}} &
\multicolumn{1}{m{0.183673469387755\textwidth}!{\huxvb{0, 0, 0}{0}}}{\hspace{1pt}\parbox[c]{0.183673469387755\textwidth-1pt-1pt}{\huxtpad{1pt + 1em}\centering {\fontsize{7pt}{8.4pt}\selectfont -0.033 (-0.058, -0.008)}\huxbpad{1pt}}} &
\multicolumn{1}{m{0.183673469387755\textwidth}!{\huxvb{0, 0, 0}{0}}}{\hspace{1pt}\parbox[c]{0.183673469387755\textwidth-1pt-1pt}{\huxtpad{1pt + 1em}\centering {\fontsize{7pt}{8.4pt}\selectfont -0.021 (-0.053, 0.011)}\huxbpad{1pt}}} &
\multicolumn{1}{m{0.183673469387755\textwidth}!{\huxvb{0, 0, 0}{0}}}{\hspace{1pt}\parbox[c]{0.183673469387755\textwidth-1pt-1pt}{\huxtpad{1pt + 1em}\centering {\fontsize{7pt}{8.4pt}\selectfont -0.036 (-0.061, -0.01)}\huxbpad{1pt}}} &
\multicolumn{1}{m{0.183673469387755\textwidth}!{\huxvb{0, 0, 0}{0}}}{\hspace{1pt}\parbox[c]{0.183673469387755\textwidth-1pt-1pt}{\huxtpad{1pt + 1em}\centering {\fontsize{7pt}{8.4pt}\selectfont -0.03 (-0.058, -0.003)}\huxbpad{1pt}}} \tabularnewline[-0.5pt]

\hhline{}

\multicolumn{1}{!{\huxvb{0, 0, 0}{0}}m{0.153061224489796\textwidth}!{\huxvb{0, 0, 0}{0}}}{\hspace{1pt}\parbox[c]{0.153061224489796\textwidth-1pt-1pt}{\huxtpad{1pt + 1em}\raggedright {\fontsize{7pt}{8.4pt}\selectfont Binary}\huxbpad{1pt}}} &
\multicolumn{1}{m{0.0561224489795918\textwidth}!{\huxvb{0, 0, 0}{0}}}{\hspace{1pt}\parbox[c]{0.0561224489795918\textwidth-1pt-1pt}{\huxtpad{1pt + 1em}\raggedright {\fontsize{7pt}{8.4pt}\selectfont 75}\huxbpad{1pt}}} &
\multicolumn{1}{m{0.0561224489795918\textwidth}!{\huxvb{0, 0, 0}{0}}}{\hspace{1pt}\parbox[c]{0.0561224489795918\textwidth-1pt-1pt}{\huxtpad{1pt + 1em}\raggedright {\fontsize{7pt}{8.4pt}\selectfont 4}\huxbpad{1pt}}} &
\multicolumn{1}{m{0.183673469387755\textwidth}!{\huxvb{0, 0, 0}{0}}}{\hspace{1pt}\parbox[c]{0.183673469387755\textwidth-1pt-1pt}{\huxtpad{1pt + 1em}\centering {\fontsize{7pt}{8.4pt}\selectfont -0.015 (-0.036, 0.005)}\huxbpad{1pt}}} &
\multicolumn{1}{m{0.183673469387755\textwidth}!{\huxvb{0, 0, 0}{0}}}{\hspace{1pt}\parbox[c]{0.183673469387755\textwidth-1pt-1pt}{\huxtpad{1pt + 1em}\centering {\fontsize{7pt}{8.4pt}\selectfont -0.008 (-0.033, 0.018)}\huxbpad{1pt}}} &
\multicolumn{1}{m{0.183673469387755\textwidth}!{\huxvb{0, 0, 0}{0}}}{\hspace{1pt}\parbox[c]{0.183673469387755\textwidth-1pt-1pt}{\huxtpad{1pt + 1em}\centering {\fontsize{7pt}{8.4pt}\selectfont -0.017 (-0.037, 0.004)}\huxbpad{1pt}}} &
\multicolumn{1}{m{0.183673469387755\textwidth}!{\huxvb{0, 0, 0}{0}}}{\hspace{1pt}\parbox[c]{0.183673469387755\textwidth-1pt-1pt}{\huxtpad{1pt + 1em}\centering {\fontsize{7pt}{8.4pt}\selectfont -0.016 (-0.042, 0.009)}\huxbpad{1pt}}} \tabularnewline[-0.5pt]

\hhline{>{\huxb{0, 0, 0}{1}}->{\huxb{0, 0, 0}{1}}->{\huxb{0, 0, 0}{1}}->{\huxb{0, 0, 0}{1}}->{\huxb{0, 0, 0}{1}}->{\huxb{0, 0, 0}{1}}->{\huxb{0, 0, 0}{1}}-}

\multicolumn{7}{!{\huxvb{0, 0, 0}{0}}m{1\textwidth+12\tabcolsep}!{\huxvb{0, 0, 0}{0}}}{\hspace{1pt}\parbox[c]{1\textwidth+12\tabcolsep-1pt-1pt}{\huxtpad{1pt + 1em}\raggedright {\fontsize{7pt}{8.4pt}\selectfont * ATE is estimated using the one-step estimator $\psi^+_1(\hat{Q})$ and the TMLE $\psi_1(\hat{Q}^\star)$ when $M$ is binary, 
  and $\psi^+_{2b}(\hat{Q})$ and $\psi_{2b}(\hat{Q}^\star)$ when $M$ is continuous. 
  ATT is estimated using the corresponding one-step and TMLE estimators: $\beta^+_1(\hat{Q})$ and $\beta_1(\hat{Q}^\star)$ for binary $M$, 
  and $\beta^+_{b}(\hat{Q})$ and $\beta_{b}(\hat{Q}^\star)$ for continuous $M$. Estimates are reported as point estimates with corresponding 95\% confidence intervals.}\huxbpad{1pt}}} \tabularnewline[-0.5pt]

\hhline{}

\end{tabular}}\label{apptable:bproud}
\par

\end{table}
\end{center}

\begin{table}[!t]
\caption{Variable descriptions used in real data analysis (from the Finnish Social Science Data Archive.) Summary statistics contain information about mean and standard deviation for continuous variables and category frequency for categorical variables. \vspace{-0.2cm} }
\label{tab:realdata}
\begin{center}
{\small
\begin{tabular}{p{0.1\textwidth}p{0.75\textwidth}p{0.1\textwidth}}
    \hline 
    \textbf{Variable}  & \textbf{Definition; Summary statistic}  &  \textbf{Year} 
    \\ \hline 
    $X_1$ & Socio-economic status as the total family taxable income in years 1983-84; {$21619.54  \ (9806.7)$} & 1983-84 
    \\ 
    $X_2$ & ITPA score; {$35.87 \ (5.97)$} & 1971-72 
    \\ 
    $X_3$ & Gender; male ($49.68\%$), female ($50.32\%$)  & 1971-91 
    \\ 
    $X_4$ & Age; {$25.17 \ (1.2)$} & 1991
    \\ \hline 
    $A$ & 6th-grade all-subject grade averages compared to median; above ($44.95\%$), below ($55.05\%$) & 1984
    \\ \hline 
    $M_1$ & Undergraduate degree; yes ($24.13\%$), no ($75.87\%$) & 1991
    \\
    $M_2$ & Highest educational field (categorised in accordance with Statistics Finland’s Classification of Education 1988); {science ($90.06\%$), art ($9.94\%$)}  & 1991
    \\ 
    $M_3$ & Age at the start of the highest attained educational qualification; {$19.33 \  (2.53)$}  & 1991 
    \\ 
    $M_4$ & Length of formal education in months after comprehensive/upper secondary school (including education in progress; {$28.55 \ (14.62)$} & 1991
    \\ 
    $M_5$ & Number of different fields of education (including education in progress); {$1.14 \ (0.5)$} & 1991
    \\ 
    $M_6$ & Educational qualification required for current job; no ($22.56\%$), somewhat ($19.87\%$), yes ($57.57\%$) & 1991 
    \\ 
    $M_7$ & Total length of the spells of unemployment greater than one year; no ($84.07\%$), yes ($15.93\%$) & 1991
    \\ 
    $M_8$ & Age when started working; {$21.34 (2.4)$}  & 1991
    \\ \hline 
    $Y$ & Respondent's earned income in euros in year 2000; {20541.93 (14462.12)} & 2002
    \\ \hline 
\end{tabular}
}
\end{center}
\end{table}

\end{document}